\documentclass[preprint,showpacs,amsmath,amssymb,aps,pr,superscriptaddress,floatfix]{revtex4}

\usepackage{amsmath}
\usepackage{amssymb}

\usepackage{graphicx,color}
\usepackage{dcolumn}
\usepackage{bm}

\def\iomn{i\omega_n}
\def\inun{i\nu_n}
\def\vq{{\bf q}}
\def\vk{{\bf k}}

\newcommand{\jmbra}[1]{\ensuremath{\langle #1|}}
\newcommand{\jmket}[1]{\ensuremath{|#1\rangle}}

\newcommand{\be}{\begin{equation}}
\newcommand{\ee}{\end{equation}}
\newcommand{\ba}{\begin{eqnarray*}}
\newcommand{\ea}{\end{eqnarray*}}
\newcommand{\bea}{\begin{eqnarray}}
\newcommand{\eea}{\end{eqnarray}}

\newcommand{\vc}[1]{{\bf #1}}

\renewcommand{\Re}{\mathrm{Re}}

\begin{document}


\title{Dynamical screening in correlated electron systems \\ -- from lattice models to realistic materials}
\author{Philipp Werner}
\affiliation{Department of Physics, University of Fribourg, Chemin du Mus\'ee 3, 1700 Fribourg, Switzerland}
\author{Michele Casula}
\affiliation{CNRS and Institut de Min\'eralogie, de Physique des
  Mat\'eriaux et de Cosmochimie,
Universit\'e Pierre et Marie Curie, 4 place Jussieu, 75252 Paris, France}

\date{January 30, 2016}

\begin{abstract}
Recent progress in treating the dynamically screened nature of the Coulomb interaction in strongly correlated lattice models and materials is reviewed with a focus on computational schemes based on the dynamical mean field approximation. We discuss approximate and exact methods for the solution of impurity models with retarded interactions, and explain how these models appear as auxiliary problems in various extensions of the dynamical mean field formalism. The current state of the field is illustrated with results from recent applications of these schemes to $U$-$V$ Hubbard models and correlated materials. 
\end{abstract}

\maketitle 

\newpage

\tableofcontents



\newpage

\section{Introduction}
\label{introduction}

Screening describes the way the electrons rearrange themselves to reduce the inter-particle interaction and total energy of the system, while correlation describes how the electrons move collectively in the system. Both concepts are closely related and connected. In Hartree-Fock theory, the electrons are not correlated except for 
like-spin particles, which are constrained by the Pauli exclusion principle. 
A consistent framework to improve upon this theory was proposed by Hedin in his seminal 1965 paper \cite{Hedin1965}.  
He expanded the self-energy up to the first-order in the dynamically screened 
interaction $W$, and thereby laid the foundation of the GW theory \cite{ferdi_gw,Hedin1999,Onida2002}. 
The screened interaction $W$ is the bare interaction reduced by the
dielectric function, determined by the charge-charge response function, which
in turn is sensitive to electronic correlations. 
Screening is thus a dynamical process, as it depends on how the
electrons 
respond to a given perturbation. It is obviously sensitive to the nature of the underlying electronic states.
One of the fingerprints of screening are the plasmon satellites, collective excitations
associated with long-range charge fluctuations, which are seen in
electron spectroscopy (such as, for instance, photoelectron spectroscopy (PES) \cite{PhysRevB.43.11971} and electron
energy loss spectroscopy (EELS) \cite{marton1962}). 

In correlated electron systems, where local atomic-like
interactions are dominant, screening is also crucial to set the actual
value of the Coulomb repulsion, and thus to determine the level of
correlation in the system.
Describing within the same framework both
the long-range nature of screening and its role in tuning the local
repulsion $U$ has been a long-standing theoretical challenge. 
The GW approximation, while neglecting vertex corrections,
takes into account reasonably well dynamical 
long-range screening effects \cite{PhysRevLett.77.2268,Guzzo2011}, but it usually fails for large and local
Coulomb repulsions, which are better described within a Hubbard model.
This pushed Hedin to write in his review \cite{Hedin1999}: ``Clearly, the GW
approximation describing long-range charge fluctuations, and a Hubbard
model focusing on local on-site correlations, are two extremes.''

These two extremes can now be merged into a coherent and unified
picture, thanks to recent progress in Green's function
embedding schemes, which 
have been developed as extended dynamical mean field theories.
In dynamical mean field theory (DMFT) the 
correspondence between the local lattice Green's
function $G_\textrm{loc}$ and the 
solution of an auxiliary Anderson impurity
problem 
is realized through 
the dynamical Weiss field ${\cal G}_0$,
which mimics the effect of the lattice environment on the impurity site. 
In extended dynamical mean field theory (EDMFT),
the embedding is extended to the local screened interaction
$W_\textrm{loc}$ which matches the impurity screened interaction 
and determines the dynamical bosonic field ${\cal U}$, which
represents the effect of non-local interactions on  
the auxiliary Anderson problem. 
Thus, the EDMFT embedding procedure maps non-local
interaction effects onto a local dynamical screening, 
and allows to treat these 
local interactions in a non-perturbative way. Indeed, since the
advent of continuous time quantum Monte Carlo (CTQMC) algorithms \cite{Gull2011}, solving
the Anderson impurity problem with retarded - aka frequency dependent -
interactions has become feasible. This has been a major advance in the field.
However, in the EDMFT framework the resulting 
self-energy is still local.

A further step forward is represented by the GW+DMFT theory, 
proposed in 2003 by Biermann and coworkers \cite{Biermann2003}, 
where 
the embedding is performed at the Green's function level between 
the local self-energy coming from the EDMFT solution of the impurity
problem  
and the
non-local self-energy, taken from GW. 
Therefore, non-locality is
included at the GW level, and the method can be applied to study
first-principles Hamiltonians. 
Different schemes along
these lines have become popular also in other contexts,
such as in quantum chemistry \cite{Zgid2011,Kananenka2015,Lan2015}, where the quantum system
is divided into two parts, one of which is treated at the perturbative level, e.~g. by the self-consistent second-order Green's function method (GF2), while the
other is solved at a higher level by the configuration 
interaction (CI) method \cite{Zgid2012,Go2015}. However, what distinguishes GW+DMFT from
other approaches is the \emph{double} embedding in both the Green's function $G$
and the dynamically screened interaction $W$. This has many advantages, as
we will see in this review; one of the most important is that it promotes
the Hubbard $U$ parameter to the Weiss field ${\cal U}$, which is 
determined self-consistently. Therefore, the underlying Hubbard
Hamiltonian with retarded interactions is no longer a model, but is an
auxiliary system, making the GW+DMFT scheme a truly \emph{ab initio}
approach. At the same time, non-locality is kept in both the self-energy $\Sigma$
and the electronic polarization $P$, which are related to $G$ and $W$, respectively.
The GW+DMFT self-consistency in $G_\textrm{loc}$ and $W_\textrm{loc}$
closes the gap between the two extremes quoted by Hedin.

The general framework sketched above is conceptually
appealing but still far from being a black-box machinery capable of 
solving generic correlated electron systems and first-principles
Hamiltonians. However, in the last few years, 
significant 
progress has been made toward the final goal of a self-consistent description of
dynamical screening effects in strongly correlated materials.
In this review, we are going to introduce the theoretical
and numerical tools which made this progress possible, such as, for
instance, 
the CTQMC algorithm for frequency
dependent interactions. Moreover, we are going to present the latest
applications, both at the model level and 
from first principles.
Model applications are useful to understand the limitations of the approximations
underlying the EDMFT or GW+DMFT approaches, while the
latest applications to real materials are extremely interesting,
as they suggest new promising directions to approximate GW+DMFT
schemes, 
which are computationally more affordable, and can be applied to a wider 
class of \emph{ab initio} systems.

In our review we will not only explain the general
theory, but also focus on practical aspects, with a twofold
purpose: First, making tighter
theoretical connections between various methods which deal with
dynamical screening in \emph{ab initio} correlated electron systems.
Second, presenting methodological details, which help the reader
implement the methods in a computer program. 
Throughout the review, we will use eV as energy units in the \emph{ab
  initio} applications, while the energy units will be set to
the bandwidth, or some related hopping scale, in model applications. 

The general organization of this review is as follows.
In Sec.~\ref{downfolding} we explain how
the screening is modeled from first principles 
in strongly correlated
materials.  We show that the retarded interactions emerge naturally from
a multi-scale approach, where the coupling parameters are dynamically screened by
higher-energy degrees of freedom, which are traced out in the derivation of the low-energy
Hamiltonian. 

In Sec.~\ref{solving_models}, we introduce the main theoretical
framework used throughout this review 
(Sec.~\ref{dmft}),
namely the DMFT, employed
to solve Hubbard-like low-energy Hamiltonians 
with retarded local interactions $U$. We describe
several solvers capable of tackling the Anderson impurity problem with
frequency dependent $U$,
with a particular emphasis on the CTQMC 
method (Sec.~\ref{ctqmc}), which is formally exact and efficient in the case of
density-density interactions. Different approximations are proposed
for solving the impurity problem, such as the dynamic atomic-limit approximation
(DALA) in Sec.~\ref{dala} and the Lang-Firsov approach 
in Sec.~\ref{lang_firsov}. 
These methods are explained first for the
solution of the Holstein-Hubbard model (Sec.~\ref{holstein_hubbard}),
where the dynamic nature of the local interaction $U$ 
derives from 
bosons (plasmons), locally coupled to the correlated
electrons. This formalism 
can be generalized 
to deal with arbitrary \emph{ab
initio} $U(\omega)$ (Sec.~\ref{sec:generalU}) and with
multi-band/multi-orbital systems typical of realistic
materials (Sec.~\ref{subsec:multiorbital_systems}), thanks to the
mapping to a continuum of bosonic modes. The physics behind the
generalized Holstein-Hubbard model can be made more transparent by the
derivation of an effective static model, with renormalized hoppings and
interactions due to screening effects, detailed in Sec.~\ref{sec:static}.

With the tools of Sec.~\ref{solving_models}, one would
like to go beyond the local interaction picture, which could be too
rough for realistic materials. Dealing with non-local interactions 
is a prerequisite for treating 
long-range screening effects, which lead to
collective phenomena in solids. 
In Sec.~\ref{selfconsistent_screening}, we present different
strategies for treating long-range interactions in strongly correlated
systems, 
both for extended $U$-$V$ Hubbard models (Sec.~\ref{sub:UVhubbard}) and realistic
materials (Sec.~\ref{subsec:realistic_materials}). 
Thse approaches are 
based on the Green's function embedding
theory, where both the local Green's function and the local screened 
interaction are self-consistently determined by the many-body solution
of an auxiliary single-impurity problem, which provides both the local electronic
self-energy and local polarization of the physical system. In this
way, non-local interactions give rise to additional local screening for
the impurity problem with dynamic ${\cal U}$. The Green's function
embedding can be done at the extended-DMFT level
(Sec.~\ref{sec:edmft}), or at the GW+DMFT level
(Sec.~\ref{sec:gw+dmft}). In the latter scheme, the non-local self-energies and non-local
polarizations, computed within the GW approximation, are
self-consistently added to the corresponding local quantities,
evaluated in a non-perturbative way at the DMFT level. Additional
approximations are necessary to apply the GW+DMFT framework to
realistic materials (Sec.~\ref{sec:realistic_materials_gw+dmft}), due
to the large number of degrees of freedom of \emph{ab initio}
Hamiltonians. These approximate schemes are still under active
development. In this review, we report the frozen cRPA
polarization (Sec.~\ref{sec:URPA}), the SEX+DMFT
(Sec.~\ref{sec:sex+dmft}), and DMFT schemes bases on effective
Hamiltonians renormalized by the non-local GW self-energy, such as the
DMFT@nonlocal-GW (Sec.~\ref{sec:dmft@nonlocal-GW}) and the
quasi-particle self-consistent GW (QSGW) + DMFT
(Sec.~\ref{sec:qsgw+dmft}). On the model side, we briefly introduce
the recently developed dual boson approach in
Sec.~\ref{subsec:dual_boson}. 

As the CTQMC impurity solver works in imaginary time, an 
analytic continuation to real frequencies is necessary to obtain spectral
functions which can be compared to experiment. One needs to take particular
care in the case of retarded interactions, in order to resolve features
such as satellites arising from screening plasmons. A detailed
procedure to 
incorporate these high-energy structures 
is explained 
in Sec.~\ref{sec:fermi_spectral_function}
for the fermionic spectral function. 
The properly normalized kernel for the bosonic spectral functions is derived in Sec.~\ref{sec:bose_spectral_function}.

Applications to both model systems and realistic materials are
presented in Secs.~\ref{sec:applications_models} and 
\ref{sec:realistic_applications}, respectively.
On the model side, the $U$-$V$ Hubbard Hamiltonian
(Secs.~\ref{sec:applications_edmft_UVhubbard} and
\ref{sec:UVhubbard_gw+dmft}), photo-doped Mott
insulators (Sec.~\ref{sec:applications_edmft_photo_doped_Mott}), and
adatom systems on semi-conductor surfaces
(Sec.~\ref{sec:applications_gw+dmft_adatoms}) are discussed. 
On the \emph{ab initio} side, we present results for the SrVO$_3$
transition metal oxide (Secs.~\ref{srvo3_lda+dmft} and \ref{sub:gw+dmft_srvo3}),
for the BaFe$_2$As$_2$ (Sec.~\ref{bafe2as2}) and BaCo$_2$As$_2$
(Sec.~\ref{sec:sex+dmft_baco2as2}) pnictides, and for the 
La$_2$CuO$_4$ cuprate (Sec.~\ref{sec:mott_gap_cuprates}).

Conclusions and perspectives are drawn in Sec.~\ref{conclusions}.

\section{Downfolding}
\label{downfolding}

Computational methods play an important role in the study of strongly correlated systems. 
Many interesting materials, such as the high-T$_c$ cuprates, have been studied in great detail using a broad range of experimental 
probes, which results in a detailed quantitative knowledge of the correlated electronic structure. 
A good theoretical description should be consistent with these experimental findings, and provide insights into the underlying physical mechanism. Ideally, the accuracy of the computational approach will enable quantitative predictions of material properties. 
However, even the simplest model used to investigate the physics of the cuprates, the 
two-dimensional (2D) single-band Hubbard model, does not admit a closed-form solution in the most
interesting parameter regimes, which are relevant for the experimental situations. For
instance, the solution of this model near the Mott transition can be obtained in an
approximate yet accurate way only if the most advanced numerical techniques are employed \cite{Maier2005,Sorella2008,Werner2009,Gull2009,Chang2010,Zhang2013,Gukelberger2015,Chan2015,simons_foundation_2D_Hubbard}.

Moreover, the complexity and richness of strongly correlated materials puts limitations on the predictive power of simple models. 
The link between the model parameters and the experimental conditions, such as doping, pressure, chemical substitution, and temperature, may be difficult to establish.  Yet, the validity of a theory is based on its ability to explain unambiguously and quantitatively a variety of situations. Therefore, a systematic procedure for the derivation of the model parameters is essential.
Numerical approaches play a crucial role in the calculation of model parameters, which are seamlessly connected to the physics of 
actual compounds. 

The idea is to use a multi-scale \emph{ab initio}
scheme \cite{Imada_Miyake_2010,Motoaki2013}, which is able to treat and predict material properties
at a level of accuracy depending on the target energy. The common
scheme is to first compute the \emph{full} band structure
by density functional theory (DFT) \cite{Hohenberg64,Kohn65,Jones15}, which takes
into account electrons ranging from the deepest core levels to the
virtual empty states, and covers an energy range of several tens or hundreds of eV. As it is well
know, DFT is not the most accurate theory for strongly correlated
materials, but it is able to reproduce general trends starting from the
first-principle Hamiltonian, including several degrees of freedom.
Its failure 
is usually related to the description of  
low-energy states, i.e. the ones near the Fermi level. In the case of strong correlations,
they are poorly described by most of the functionals. At the same time, 
these states are primarily responsible for a material's macroscopic
properties. Thus, we would like to improve upon the DFT 
description, by applying more accurate, but computationally heavier, many-body
methods to this low-energy sector. In order to do this, we need to derive a 
low-energy model by integrating out the high-energy degrees of freedom
from the DFT solution. This intermediate, crucial step, is called
\emph{downfolding}. The last step is the solution of the resulting 
model using advanced and hopefully accurate many-body techniques.

In order to \emph{downfold} the DFT Hamiltonian, one needs to find a basis set
which spans the low-energy space. Usually, one takes the maximally
localized Wannier functions, $\phi_{m\textbf{R}}$ for the $m$-th
orbital centered in the unit cell $\textbf{R}$,  which give a reliable
representation of the correlated orbitals and their low-energy
bands \cite{PhysRevB.56.12847,PhysRevB.77.085122}. Then, one projects the \emph{ab
  initio} Hamiltonian onto the 
basis set elements. The Coulomb matrix elements, which define the bare $U$, are therefore 
\begin{equation}
U^\textrm{bare}_{mn}(\vc{R}) = \langle  \phi_{m\textbf{0}} \phi_{n\textbf{R}} | v |
\phi_{m\textbf{0}}\phi_{n\textbf{R}} \rangle = \int d\vc{r}~
d\vc{r}^\prime \frac{|\phi_{m\textbf{0}}(\vc{r})|^2 |\phi_{n\textbf{R}}(\vc{r}^\prime)|^2}{|\vc{r}-\vc{r}^\prime|} , 
\label{U_bare_ab_initio}
\end{equation}
where $v$ is the Coulomb potential. Without loss of generality, 
we restrict our discussion here to interactions of density-density form. 

\begin{figure}[t]
\begin{center} 
\includegraphics[width=0.8\textwidth]{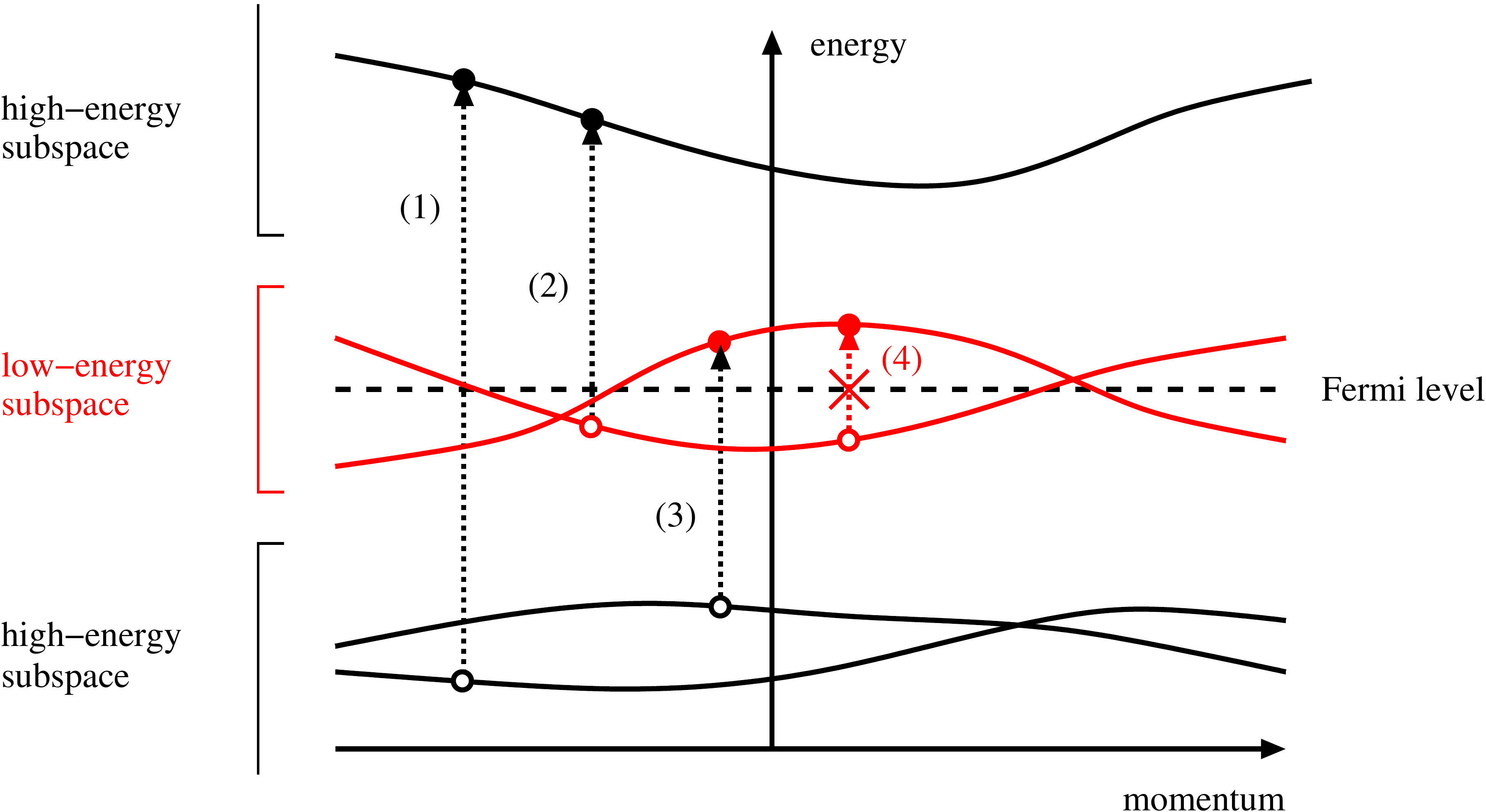}
\end{center}
\caption{ 
Schematic illustration of the screening processes in the constrained RPA framework.
The black bands are the so-called
``screening bands'', while the red bands are assumed to be strongly correlated. Only the transitions within the low-energy subspace
(excitations of type (4)) contribute to the low-energy polarization $P^L$, while
the ``complement'' $P^H$ includes excitations of type (1), (2), and (3), i.e., excitations involving the high-energy bands.
}
\label{cRPA_screening_channels}
\end{figure}

The above expression neglects the fundamental effect of
screening from high-energy electrons on the low-energy manifold. In
the random phase approximation (RPA) framework, the fully screened 
interaction is given by $W=v/(1 - v P)$, where $P$ is the polarization
function, 
defined as 
$P(\vc{r},\vc{r}^\prime,t) = - 2 i G_0(\vc{r},\vc{r}^\prime,t)
G_0(\vc{r}^\prime,\vc{r},-t)$.
In the latter definition, the factor of 2 comes from the spin
summation in the spin-degenerate case, and $G_0$ is the zero
temperature DFT Green's function, which reads:
\begin{equation}
G_0(\vc{r},\vc{r}^\prime,t) = \begin{cases} i \sum_l^\textrm{occ}
  \Psi_l(\vc{r})\Psi_l^*(\vc{r}^\prime) e^{-i \epsilon_l t}, &
  \mbox{if } t<0 \\ 
-i \sum_l^\textrm{unocc} \Psi_l(\vc{r})\Psi_l^*(\vc{r}^\prime) e^{-i
  \epsilon_l t}, & \mbox{if } t>0
\end{cases} 
\end{equation}
where the $\{\psi_l, \epsilon_l\}$ are the
one-particle Bloch eigenfunctions 
and eigenvalues corresponding to the system's band structure. 
In the frequency domain, and for temperature $T=0$, 
the above expression for the polarization becomes 
\begin{equation}
P(\vc{r},\vc{r}^\prime,\omega) = 2 \sum_i^\textrm{occ}
\sum_j^\textrm{unocc} \psi_i(\vc{r}) \psi_i^*(\vc{r}^\prime)
\psi_j^*(\vc{r}) \psi_j(\vc{r}^\prime) \left \{ \frac{1}{\omega -
    \epsilon_j + \epsilon_i + i0^+} -  \frac{1}{\omega +
    \epsilon_j - \epsilon_i - i0^+} \right \}.
\label{polarization_function}
\end{equation} 
One can then separate $P$ into the two contributions $P = P^H + P^L$, with $P^L$ defined by
polarization channels fully contained in the low-energy (L) 
window around
the Fermi level, and $P^H$ its complement. Note that $P^H$
includes not only particle-hole 
excitations within the high-energy (H)
sector, but also those connecting the
high- and the low-energy sector. 
A schematic illustration of the division between $P^H$ and $P^L$ is
reported in Fig.~\ref{cRPA_screening_channels}.

In the constrained RPA (cRPA)
theory \cite{PhysRevB.70.195104,PhysRevB.74.125106}, the
\emph{partially screened} 
interaction is given by $W^L= v/(1-v P^H)$. The physical
interpretation is transparent: $W^L$ is 
the bare interaction 
screened by scattering
processes leaking from the low-energy sector.
If $W^L$ is further screened by the polarization $P^L$ of the low-energy sector, the fully screened interaction is recovered: $W=W^L/(1-W^L P^L)$.
 
$W^L$ yields the matrix elements of the partially screened $U$:
\begin{equation}
U_{mn} (\vc{R}, \omega) =  \Big\langle  \phi_{m\textbf{0}}
\phi_{n\textbf{R}} \Big| \frac{v}{1-v P^H(\omega)} \Big|
\phi_{m\textbf{0}}\phi_{n\textbf{R}} \Big\rangle. 
\label{screened_U_ab_initio}
\end{equation}
The above $U$ defines the electron-electron interaction felt by the
low-energy electrons in the
\emph{downfolded} model, i.e., after the high-energy degrees of freedom have been
integrated out. 
The important thing to note here is that the partially screened $U$ becomes \emph{frequency dependent},
a direct consequence of 
the frequency dependence of $P^H$ \cite{Springer1998}. 
In the time-domain, this implies that the effective interaction becomes retarded. 
This \emph{dynamical screening} is the main focus of our review. We will show that it
can be explicitly taken into account in the solution of the low-energy
models, and that it will affect the results in nontrivial ways. 

\begin{figure}[t]
\begin{center}
\includegraphics[angle=0,width=0.49\textwidth]{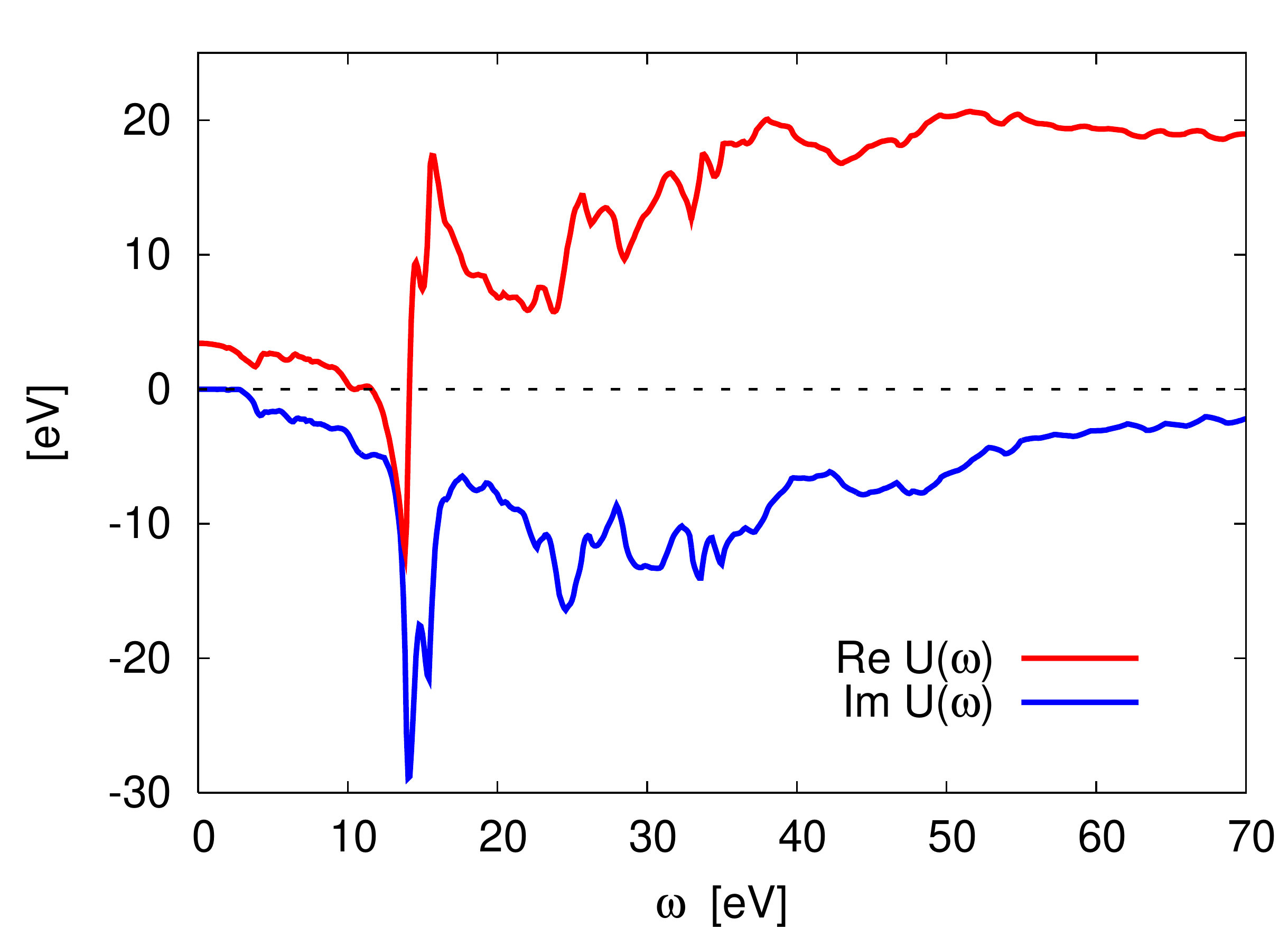}
\includegraphics[angle=0,width=0.49\textwidth]{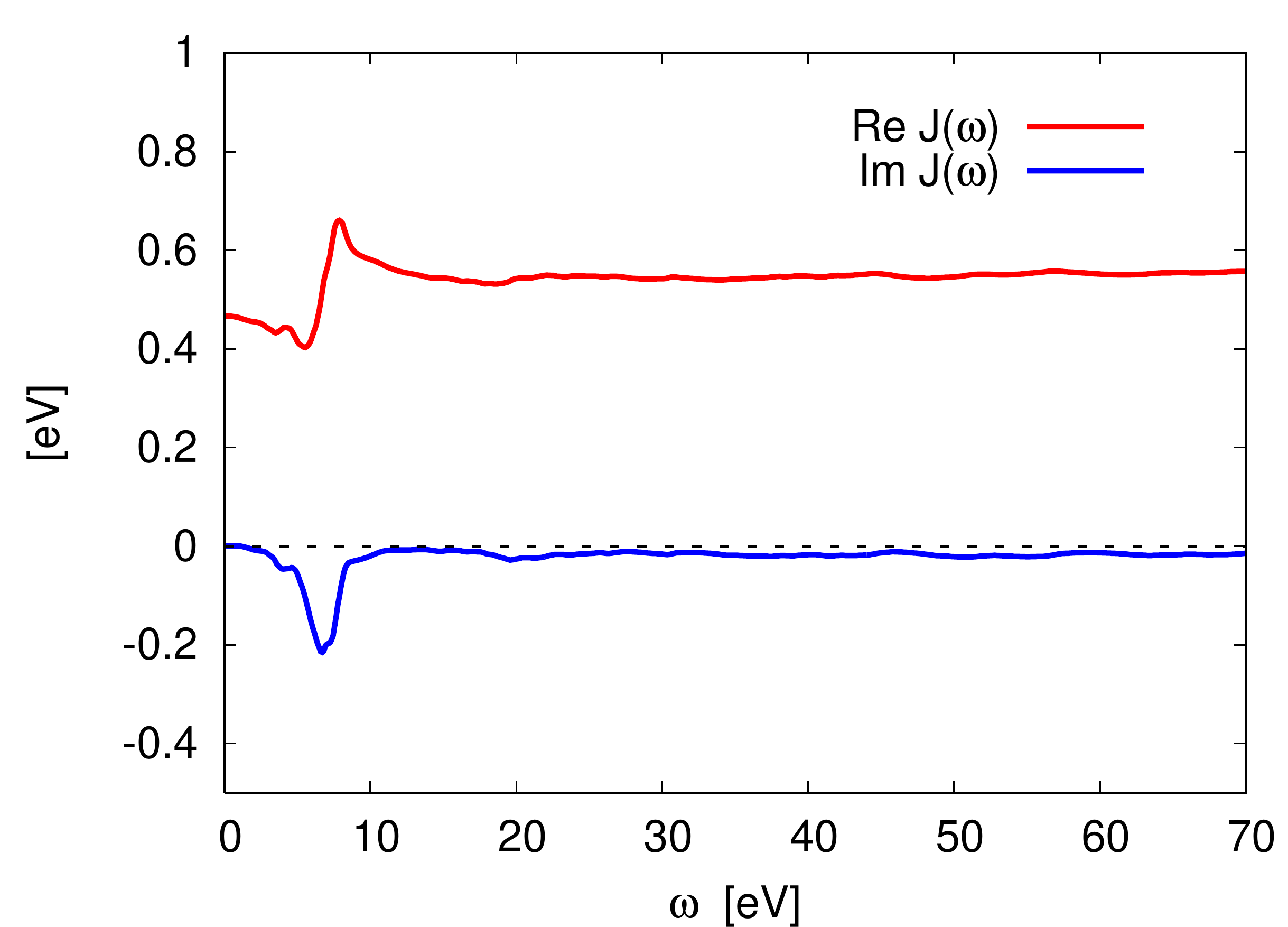}
\end{center}
\caption{
cRPA $U(\omega)\equiv U_{nn}(\omega)$ and $J(\omega)\equiv J_{nm}(\omega)$ ($n\ne m$) for the 3-band model of SrVO$_3$. (From~Ref.~\cite{Sakuma_private}.)}
\label{Uw_srvo3_crpa}
\end{figure}

It is interesting to remark here that the frequency dependence is
particularly strong for monopole-monopole interactions (direct terms
of the RPA expansion), while the
screening is less effective for multipolar charge distributions
(``exchange'' terms in RPA), like the ones related to the estimate of the 
couplings $J$:
\begin{equation}
J_{mn} (\vc{R}, \omega) =  \Big\langle  \phi_{m\textbf{0}}
\phi_{n\textbf{R}} \Big| \frac{v}{1-v P^H(\omega)} \Big|
\phi_{n\textbf{0}}\phi_{m\textbf{R}} \Big\rangle. 
\label{screened_J_ab_initio}
\end{equation}
It turns out that $U_{mn}$ is reduced by an order of magnitude with
respect to $U^\textrm{bare}_{mn}$, while $J_{mn}$ is almost unaffected by screening (change of typically less than 20\% \cite{Sasioglu2011}). As an illustration, we plot the cRPA results for the 3-band model of SrVO$_3$ in Fig.~\ref{Uw_srvo3_crpa}. 

We would like to stress that the cRPA values of $U$ and $J$ are not
adjusted by fitting schemes or empirical arguments. They are
evaluated from
first-principles \cite{Miyake2008,nakamura-2008,Tomczak2009,Miyake2010,Tomczak2010,Nomura2012,Vaugier2012,Sakuma2013_U},
according to a rigorous procedure, which 
has been recently extended also to the more involved case of entangled
bands \cite{PhysRevB.65.035109,miyake:155134}, where the separation 
between the low- and high-energy sectors is not sharp, due to the hybridization between correlated orbitals and 
more extended ones. 

Although the cRPA scheme seems plausible, it has
some limitations, which originate from two sources: First, it is
built on the RPA method which is not exact; second, the underlying
DFT band structure is not exact either. While the first limitation can
lead to underscreening or overscreening depending on the particular case,
the second one is most severe in materials which are close to a 
Mott transition (which leads to strong modifications in the starting DFT band
structure, and rearrangements of the screening bands), 
or in the
case where the ligands are not included in the model (as usually the $p-d$
hybridization is not well reproduced by DFT). A way to overcome the
latter limitation is to use a better \emph{ab initio} scheme to solve
the full Hamiltonian, e. g. by replacing DFT by GW, while
the former one can be reduced by improving upon the RPA theory. A possible approach is
the recently developed constrained functional renormalization group
(cfRG) method \cite{Kinza2015}. Testing the reliability of the cRPA method is a
subject of current research \cite{Shinaoka2015}.

Besides the cRPA method, other methods
have been proposed to evaluate the local electron-electron couplings
from \emph{ab initio} calculations, such as the linear response
approach \cite{Cococcioni2005,Hsu2009} and the constrained
DFT \cite{constrainedLDA,Anisimov2009}. They however neglect the explicit 
frequency dependence of the Hubbard parameters. A slight variation
of the cRPA method, which retains its full frequency dependence but is
based on orbitals rather than bands, can be found in Ref.~\cite{Kutepov2010}.
Along the same lines, i.e. working in the maximally localized orbitals
space, Nomura \emph{et al.} proposed to evaluate the partially screened $U$ by
undressing the fully screened coupling via the \emph{local}
polarization (i.e. by also including non-local processes in the
low-energy manifold as active screening channels) \cite{Nomura2012_U}. This should give a
$U$ particularly suited for embedding 
theories such as DMFT, where only local correlations are taken into
account. However, in 
our review, we will 
consider an alternative strategy, which is 
to include non-local processes in an explicit way, using 
extended-DMFT frameworks.

While the two-body part of the \emph{downfolded} Hamiltonian is
provided by the matrix elements of the partially screened interaction, 
Eqs.~(\ref{screened_U_ab_initio}) and (\ref{screened_J_ab_initio}), the one-body part is given by 
\begin{equation}
t_{mn}(\vc{R}) = \langle \phi_{m\textbf{0}} | H^\textrm{DFT} |
\phi_{n\textbf{R}} \rangle,
\label{t_dft}
\end{equation}
where $ H^\textrm{DFT}$ is the DFT Hamiltonian determined in the first
step. This $t_{mn}(\vc{R})$ already includes
some correlation effects, as described by
$H^\textrm{DFT}$. Therefore, one needs to correct the
one-body part, i.e., subtract the low-energy correlation effects from
the density functional, to avoid their \emph{double counting} (DC). The
so-called ``double-counting correction'' is common to all methods
which augment the DFT band structure by explicit electron-electron interaction terms, such as DFT+$U$ \cite{Anisimov1991,Anisimov1997} or DFT+DMFT \cite{Kotliar2006}.
This is the weakest point of the downfolding procedure, as an exact expression for the 
DC term is hard to derive. 
This is because the exchange correlation potential $V_\textrm{xc}$ in
$H^\textrm{DFT}$ is a nonlinear functional of the total density, which
makes it impossible to separate it into low- and high-energy
contributions.

Several forms of approximate DC corrections have been proposed in the  
literature.  
One of the most successful
for strongly correlated systems
has proven to be the fully localized limit (FLL) form \cite{Aichhorn2011}, which reads 
\begin{equation}
\Sigma^\textrm{FLL DC}_{mm^\prime\sigma} = \delta_{mm^\prime} \left ( U
  \left (n_d - \frac{1}{2} \right ) - J \left( n^\sigma_d
    - \frac{1}{2} \right ) \right ),
\label{double_counting}
\end{equation}
where $n^\sigma_d$ is the spin-resolved occupancy
of the correlated orbitals, and $n_d = n^\uparrow_d +
n^\downarrow_d$. In the above equation, $U=\frac{1}{N^2}
\sum_{mn} U_{mn}(\vc{0},0)$ is the average local static Coulomb
interaction ($N$ is the number of correlated bands), 
while the Hund's coupling $J$ is related to the couplings
in Eq.~(\ref{screened_J_ab_initio}) through the relation $J = U -
\frac{1}{N(N-1)} \sum_{m \ne n} \left( U_{mn}(\vc{0},0) -
  J_{mn}(\vc{0},0) \right)$.
The form in Eq.~(\ref{double_counting}) has been
used in the \emph{ab initio} applications presented in this review,
unless otherwise stated.

Very recently, Haule proposed an  
improved DC
scheme, in which a $V^\text{imp}_\textrm{xc}$ potential for the
impurity system is constructed in terms of the local Green's functions 
and a local energy functional built on 
screened interactions \cite{Haule2015}. 
He found that this procedure
leads to double counting shifts in good agreement with an approximate DC
correction obtained from Eq.~(\ref{double_counting}), by replacing $n$
with its closest integer value 
$n^0=[n_d]$ 
(nominal DFT occupancy \cite{Haule2010}), i.e.
\begin{equation}
\Sigma^\textrm{nominal DC}_{mm^\prime\sigma} = \delta_{mm^\prime} \left ( U
  \left (n^0 - \frac{1}{2} \right ) - \frac{J}{2} \left( n^0 - 1 \right ) \right ).
\label{double_counting_haule}
\end{equation}

Another interesting solution of the double counting problem has been
proposed recently in Ref.~\cite{Motoaki2013}. It is based on the replacement of
the exchange-correlation functional $V_\textrm{xc}$ by the perturbative expansion of the potential to
first order in the fully screened interaction $W$, 
as it is done in the $GW$ approach \cite{Hedin1965}. 
In contrast to $V_\textrm{xc}$, in the
$GW$ framework one can easily separate the low-energy contributions
($\Sigma^L=G^LW$) from the high-energy
ones ($\Sigma^H=G^HW$), where $G^H$ ($G^L$) is the unperturbed Green's function
living in the high- (low-) energy subspace. The off-diagonal matrix
elements connecting the two subspaces are disregarded in the
quasiparticle approximation. This leaves us with the operator
$H^\textrm{DFT}-V_\textrm{xc}+\Sigma^H$, which is well defined in the
high-energy space, although being frequency dependent. To get rid of
the frequency dependence, the dynamic part included in the $\Sigma^H$
self-energy is treated as a first-order variation
around the Fermi energy, and condensed into a bandwidth
renormalization factor $Z^\textrm{H}=(1 - \frac{\partial \textrm{Re}
  \Sigma^H}{\partial \omega})^{-1}$. The corresponding DC-free 
hoppings read 
\begin{equation}
t^\textrm{eff}_{mn}(\vc{R}) = \langle \phi_{m\textbf{0}} | H^\textrm{eff} |
\phi_{n\textbf{R}} \rangle,
\label{t_eff}
\end{equation}
with $H^\textrm{eff} = Z^H \left ( H^\textrm{DFT} - V_\textrm{xc} +
  \textrm{Re}\Sigma^\textrm{H}(0) \right)$. It turns out that the
effective band structure in Eq.~(\ref{t_eff}) has a larger bandwidth
than the one in Eq.~(\ref{t_dft}). Therefore, a proper treatment of the
double counting correction leads to a much more complex rearrangement
of the low-energy bands than the simple rigid (orbital independent)
shift of the correlated manifold implied by
Eq.~(\ref{double_counting}). In most of the applications presented in this
review, we are however going to use the latter correction, even though it is less accurate.  
It is simpler to compute, because it does not require a
single-shot GW calculation of the full problem.

Usually, in the downfolded model one considers only \emph{local}
contributions in the 
two-body part, i.e. for $\vc{R} = \vc{0}$ in
Eqs.~(\ref{screened_U_ab_initio}) and (\ref{screened_J_ab_initio}), as the screening reduces
the range of the interaction, making $U_{mn} (\vc{R}, \omega)$ a
localized function in space. Of course the localization of these
matrix elements depends on the type of system and on the correlation
strength. 
In fact, the elimination of the low-energy metallic screening (by
unscreening $W$ with $P^L$) makes the partially screened interactions longer-ranged than the fully screened ones. Therefore, Hubbard-type approximations with only on-site interactions could be problematic.
In fact, the restriction to Hubbard-like terms in most existing calculations is mainly due to the
difficulty of dealing with long-ranged interactions in the solution of the low-energy model. 
Downfolded extended $U$-$V$ models
have however been taken into account in recent DMFT studies~\cite{Hansmann2013}, and one of the purposes of this review is 
to explain the techniques which enable these simulations. 

There 
has also been a vast amount of work focusing on the explicit solution of extended lattice models. These calculations keep the full spatial dependence of the Hamiltonian, both in the one- and two-body parts, but neglect dynamical effects. Successful methods of this kind are for instance lattice and diagrammatic quantum Monte Carlo (although affected by the
sign problem and/or by finite-size effects) \cite{Imada1989,Zhang2003,Sorella2005,Tahara2008,Sorella2012,Zhang2013}, exact diagonalization (which is limited to very small sizes), and methods based on 
matrix product states and their variants (most appropriate for low-dimensional
systems) \cite{Jordan2008,Corboz2014,Corboz2015}.
For models with purely local interactions, the exact diagonalization approach is also useful within
cluster embedding theories \cite{Maier2005}, such as the variational cluster approximation
(VCA) \cite{Potthoff2003}, cellular dynamical mean field theory
(CDMFT) \cite{Kotliar2001}, or the dynamical 
cluster approximation (DCA) \cite{Hettler1998}).  
Here, we focus our attention on Green's function methods,  
and Green's function embedding approaches, such as the DMFT, 
where
significant progress has been made in the consistent treatment of 
dynamically screened interaction parameters and the corresponding retardation effects.

From the point of view of including dynamically screened non-local
interactions, a very appealing framework is 
the GW+DMFT theory, which treats the $\vc{k}$-dependent electron correlations
at the GW level, while the local diagrams are summed up to infinite
order by the DMFT solver. This is justified by the
interaction-range reduction from screening, which makes
the $U$ matrix 
more 
localized in $\vc{R}$.
Therefore, a perturbative expansion of the self-energy may be a good
approximation for the non-local $U$ terms. However, we recall that a
$\vc{k}$-dependent $\Sigma$ can also arise from purely local
interactions, particularly in low dimensional systems (1D and 2D), where the
strength of the resulting $\vc{k}$-modulation can seriously challenge the
validity of the perturbative nature of GW. 
As it will be explained in
Sec.~\ref{sec:gw+dmft}, in the GW+DMFT formalism the dynamically
screened interaction $U(\omega)$ is replaced by an auxiliary local interaction
$\mathcal{U}(\omega)$, which becomes frequency dependent due to screening
provided not only by high-energy states (related to
\emph{downfolding}) but also by non-local
contributions which enter the local polarization function (related to
two-particle \emph{embedding}).
This double nature of screening is captured by the GW+DMFT 
self-consistency construction, where the effective hybridization
$\mathcal{G}_0$ and effective local interaction $\mathcal{U}$ are 
fixed by a double constraint imposed on the local Green's function
and the local polarization, respectively.

\section{Solving low-energy models with dynamically screened $U(\omega)$}
\label{solving_models}

Once a low energy lattice model with a few correlated orbitals and a
frequency-dependent (retarded) monopole interaction $U$ has been
derived, we need a suitable method to solve it. Here, we focus on
dynamical mean field based methods, which are adequate for the
description of systems with a large coordination number, or systems in
the strongly correlated (Mott insulating) regime, where the local
physics is dominant. We will start by introducing the dynamical mean
field approximation 
(Sec.~\ref{dmft}), 
and
then discuss various strategies for dealing 
with the retarded interaction originating from the coupling to a
single bosonic mode (Sec.~\ref{holstein_hubbard}). These techniques will then be generalized to
arbitrary $U(\omega)$, in Sec.~\ref{sec:generalU}. We also discuss the proper definition of
effective low-energy models with static interactions
(Sec.~\ref{sec:static}).  

\begin{figure}[t]
\begin{center}
\includegraphics[angle=0, width=0.9\textwidth]{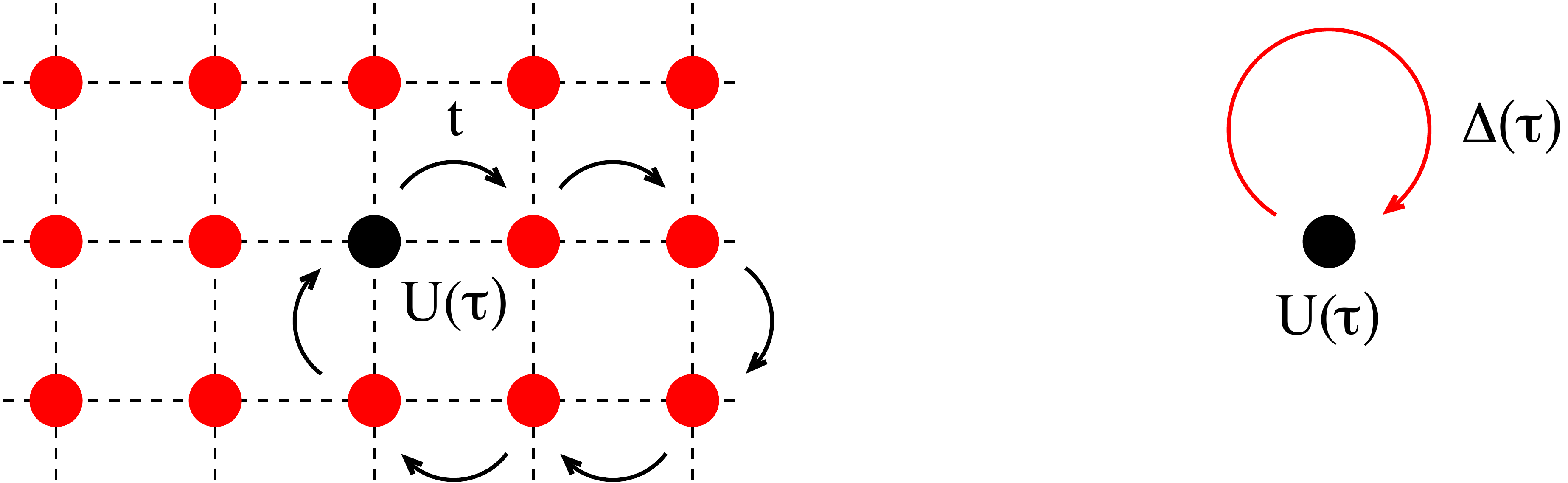}
\caption{
Mapping of the Hubbard model to an effective single-site model (quantum impurity model).
}
\label{meanfield}
\end{center}
\end{figure}

\subsection{Dynamical mean field theory}
\label{dmft}

For simplicity, we discuss the dynamical mean field approximation \cite{Georges1996} for a simple, but fundamentally important model for correlated electron materials, the single-orbital Hubbard model with a retarded on-site interaction $U(\tau)$. Written in terms of fermionic creation and annihilation operators $d^\dagger$ and $d$, the action of the lattice model is given by 
\begin{align}
S_\text{Hubbard}=& - t\sum_{\langle ij\rangle\sigma}\int_0^\beta d\tau \big[ d^\dag_{i\sigma}(\tau) d^{}_{j\sigma}(\tau) + d^\dag_{j\sigma}(\tau) d^{}_{i\sigma}(\tau) \big]-\mu\sum_{i\sigma}\int_0^\beta d\tau n_{i\sigma}(\tau)\nonumber\\
&+\frac12 \sum_i\int_0^\beta d\tau d\tau' U(\tau-\tau')n_i(\tau) n_i(\tau'),
\end{align}
with $n_\sigma=d^\dagger_\sigma d_\sigma$,
$n=n_\uparrow+n_\downarrow$, $t$ the hopping amplitude and $\beta=1/T$
the inverse temperature. The partition function of the lattice model
is $Z=\text{Tr}[\mathcal{T}e^{-S_\text{Hubbard}}]$, with $\mathcal{T}$
the time-ordering 
operator. 
Note that the retarded interaction also couples electrons of the same spin. 
In the spirit of mean-field approximations, we now focus on one particular site of the lattice (black dot in the left panel of Fig.~\ref{meanfield}) and replace the remaining degrees of freedom of the model by a hybridization term, which describes the hopping of electrons in and out of the impurity. The effective single-site problem then becomes an Anderson impurity model \cite{Georges1992} with retarded interaction, 
\begin{align}
S_\text{imp}=&\sum_\sigma \int_0^\beta d\tau d\tau' d^\dag_\sigma(\tau)\Delta_\sigma(\tau-\tau')d_\sigma(\tau')-\mu\sum_{\sigma}\int_0^\beta d\tau n_{\sigma}(\tau)\nonumber\\
&+\frac12 \int_0^\beta d\tau d\tau' U(\tau-\tau')n(\tau) n(\tau'). 
\label{Simp}
\end{align}
The hybridization function $\Delta(\tau)$ plays the role of the {\it dynamical mean field}, which is computed self-consistently in such a way that the 
Anderson impurity model mimics the lattice environment as closely as possible.  
More precisely, the self-consistent solution is constructed such that the impurity Green's function $G_\text{imp}(i\omega_n)$ reproduces the {\it local} lattice Green's function $G_\text{loc}(i\omega_n)\equiv G_{i,i}(i\omega_n)$. 
If $G(\vc{k},i\omega_n)$ is the momentum-dependent lattice Green's function of the Hubbard model, we thus seek a hybridization function such that  
\begin{equation}
\int (d\vc{k}) G(\vc{k},i\omega_n) \equiv G_\text{imp}(i\omega_n),\label{selfconsistency}
\end{equation}
where $\int (d\vc{k})$ denotes a normalized integral over the Brillouin zone. 

It is also useful to introduce the Green's function of the non-interacting impurity (``Weiss Green's function") $\mathcal G_0$, 
which is related to the hybridization function by
\begin{equation}
[\mathcal G_{0\sigma}]^{-1}(i\omega_n)=i\omega_n+\mu-\Delta_\sigma(i\omega_n). 
\label{G0def}
\end{equation} 
Depending on the method used, it is more natural to work with the Weiss Green's function $\mathcal G_0$ rather than the hybridization function $\Delta$.

\subsubsection{DMFT approximation}
\label{subsec:dmft_approx}

We obtain the solution of Eq.~(\ref{selfconsistency}) iteratively. 
However,
it is not immediately clear how we can use this self-consistency condition to update the dynamical mean field. To 
define a practical procedure, we have to relate the left-hand-side of Eq.~(\ref{selfconsistency}) to impurity model quantities. 
This step involves, 
as the essential approximation of the DMFT method, a simplification of the momentum-dependence of the lattice self-energy. 

The self-energy describes the effect of interactions on the propagation of electrons. In the non-interacting model,  the lattice Green's function is 
$G_0(\vc{k},i\omega_n)=[i\omega_n+\mu-\epsilon_\vc{k}]^{-1}$, 
with $\epsilon_\vc{k}$
the Fourier transform of the hopping matrix. The Green's function of the interacting model is $G(\vc{k},i\omega_n)=[i\omega_n+\mu-\epsilon_\vc{k}-\Sigma(\vc{k},i\omega_n)]^{-1}$ with $\Sigma(\vc{k},i\omega_n)$ the lattice self-energy.
Therefore 
\begin{equation}
\Sigma(\vc{k},i\omega_n)=G^{-1}_0(\vc{k},i\omega_n)-G^{-1}(\vc{k},i\omega_n). 
\end{equation} 
Similarly, we obtain the impurity self-energy as 
\begin{equation}
\Sigma_\text{imp}(i\omega_n)=\mathcal G^{-1}_{0}(i\omega_n)-G^{-1}_\text{imp}(i\omega_n), 
\end{equation}
with $\mathcal G^{-1}_{0}$ defined in Eq.~(\ref{G0def}).
The DMFT approximation is the identification of the lattice self-energy with the momentum-independent impurity self-energy,
\begin{equation}
\Sigma(\vc{k},i\omega_n)\approx \Sigma_\text{imp}(i\omega_n).
\end{equation}
This approximation enables us to rewrite the self-consistency equation (\ref{selfconsistency}) as
\begin{equation}
\int (d\vc{k}) [i\omega_n+\mu-\epsilon_\vc{k}-\Sigma_\text{imp}(i\omega_n)]^{-1} \equiv G_\text{imp}(i\omega_n).
\label{selfconsistency2}
\end{equation}
Since both $G_\text{imp}(i\omega_n)$ and $\Sigma_\text{imp}(i\omega_n)$ are determined by the  
hybridization function $\Delta(\tau)$ (or $\mathcal{G}_0(\tau)$), Eq.~\eqref{selfconsistency2} defines a self-consistency condition for these functions.

\subsubsection{DMFT self-consistency loop} 
\label{subsec:dmft_loop}
 
We now formulate the self-consistency loop for the Weiss Green's function  $\mathcal G_0(i\omega_n)$. 
Starting from an arbitrary initial 
$\mathcal G_0(i\omega_n)$ (for example, 
the local Green's function of the non-interacting lattice model), 
we iterate the following steps until convergence: 
 
\begin{enumerate}
\item Solve the impurity problem, that is, compute the impurity Green's function $G_\text{imp}(i\omega_n)$ for the given 
$\mathcal G_0(i\omega_n)$,
\item Extract the self-energy of the impurity model: $\Sigma_\text{imp}(i\omega_n)=\mathcal G^{-1}_0(i\omega_n)-G^{-1}_\text{imp}(i\omega_n)$,
\item Identify the lattice self-energy with the impurity self-energy, $\Sigma(\vc{k},i\omega_n)=\Sigma_\text{imp}(i\omega_n)$ (DMFT approximation), and compute the local lattice Green's function $G_\text{loc}(i\omega_n)=\int (d\vc{k})[i\omega_n+\mu-\epsilon_\vc{k}-\Sigma_\text{imp}(i\omega_n)]^{-1}$,
\item Apply the DMFT self-consistency condition, $G_\text{loc}(i\omega_n)=G_\text{imp}(i\omega_n)$, and use it to define a new Weiss Green's function $\mathcal G^{-1}_{0}(i\omega_n)=G^{-1}_\text{loc}(i\omega_n)+\Sigma_\text{imp}(i\omega_n)$.
\end{enumerate}

The computationally expensive step is the solution of the impurity problem (Step (i)). When the loop converges, the Weiss Green's function contains information about the topology of the lattice (through the density of states), and about the phase (metal, Mott insulator, antiferromagnetic insulator, \ldots). The impurity thus behaves, at least to some extent, as if it were a site of the lattice. 

Obviously, a single-site impurity model does not capture all the physics. In particular, the DMFT approximation neglects all spatial fluctuations. These fluctuations are important, for example, in low-dimensional systems. 
The DMFT formalism is believed to provide a qualitatively correct description of three-dimensional unfrustrated lattice models. 
It becomes exact in the limit of infinite dimension \cite{Metzner1989,MuellerHartmann1989} or infinite coordination number (where spatial fluctuations are negligible), in the non-interacting limit ($U=0$ implies $\Sigma=0$), and in the atomic limit ($t=0$ implies $\Delta=0$).

\subsection{Holstein-Hubbard model}
\label{holstein_hubbard}

In a Hamiltonian formulation, a lattice model with dynamically screened $U(\omega)$ can be represented by coupling the electron density on a given site to a continuum of bosonic modes with frequency $\omega$, with appropriately chosen coupling strengths $g_\omega$. If there is only a single bosonic mode with frequency $\omega_0$ and coupling $g$, this corresponds to the Holstein-Hubbard model 
\begin{align}
H_\text{HH} =& - t\sum_{\langle ij\rangle\sigma}( d^\dag_{i\sigma} d^{}_{j\sigma} + d^\dag_{j\sigma} d^{}_{i\sigma})+U\sum_i n_{i\uparrow} n_{i\downarrow}-\mu\sum_{i\sigma}n_{i\sigma}\nonumber\\
&+ g\sum_i (b_i^\dagger+b_i)(n_{i\uparrow}+n_{i\downarrow}-1)+\omega_0\sum_i b^\dagger_i b_i.\label{HH}
\end{align}
Here, $b^\dagger$ denotes the boson creation operator. In the DMFT approximation, this lattice model is mapped onto a single-site impurity model with Hamiltonian $H_\text{imp}=H_\text{loc}+H_\text{hyb}+H_\text{bath}$, where
\begin{align}
H_\text{loc}&=U n_\uparrow n_\downarrow
-\mu(n_\uparrow+n_\downarrow)+g(n_\uparrow+n_\downarrow-1)(b^\dagger+b)+\omega_0b^\dagger
b,
\label{Hloc}
\\
H_\text{hyb}&= \sum_{p\sigma} \Big[V_{p\sigma}d^\dagger_\sigma
a_{p\sigma}+V^*_{p\sigma}a^\dagger_{p\sigma}d_\sigma\Big],
\label{Hhyb}
\\
H_\text{bath}&=
\sum_{p\sigma}\varepsilon_{p}a^\dagger_{p\sigma}a_{p\sigma}.
\label{Hbath}
\end{align} 
In terms of the harmonic oscillator position and momentum operators 
\begin{equation}
X=(b^\dagger+b)/\sqrt{2},\quad P=i(b^\dagger-b)/\sqrt{2}, \label{bosonX}
\end{equation}
the local Hamiltonian can (up to a constant) be written as 
\begin{align}
H_\text{loc}&=U n_\uparrow n_\downarrow -\mu(n_\uparrow+n_\downarrow)+\sqrt{2}g(n_\uparrow+n_\downarrow-1)X+\frac{\omega_0}{2}(X^2+P^2),\label{Hloc_X}
\end{align}
so the physics of this model is that the charge couples to bosons
describing either local lattice distortions (phonons) or local density
fluctuations (plasmons). 

\begin{figure}[t]
\begin{center}
\includegraphics[angle=0, width=0.49\textwidth]{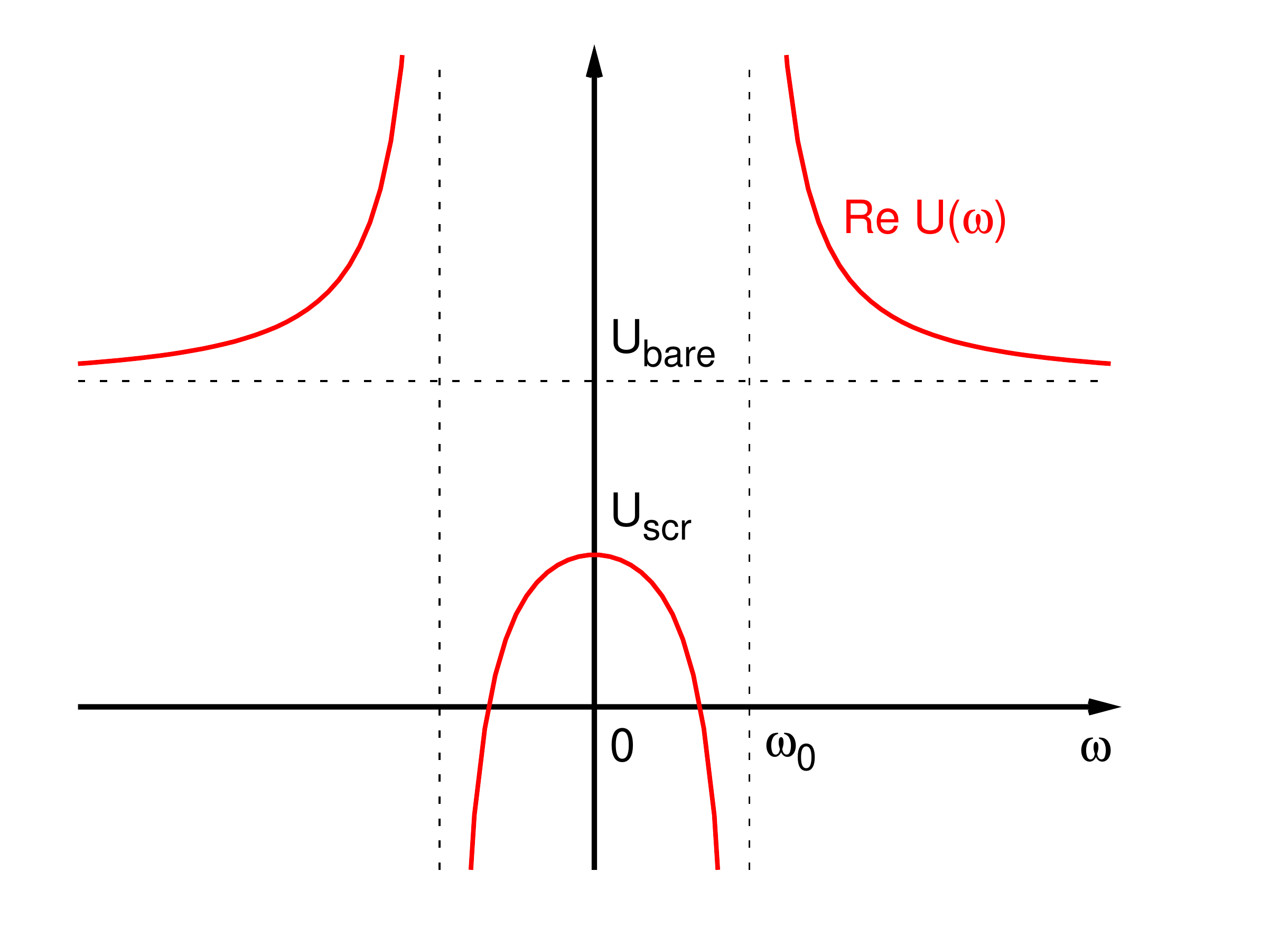}\hfill
\includegraphics[angle=0, width=0.49\textwidth]{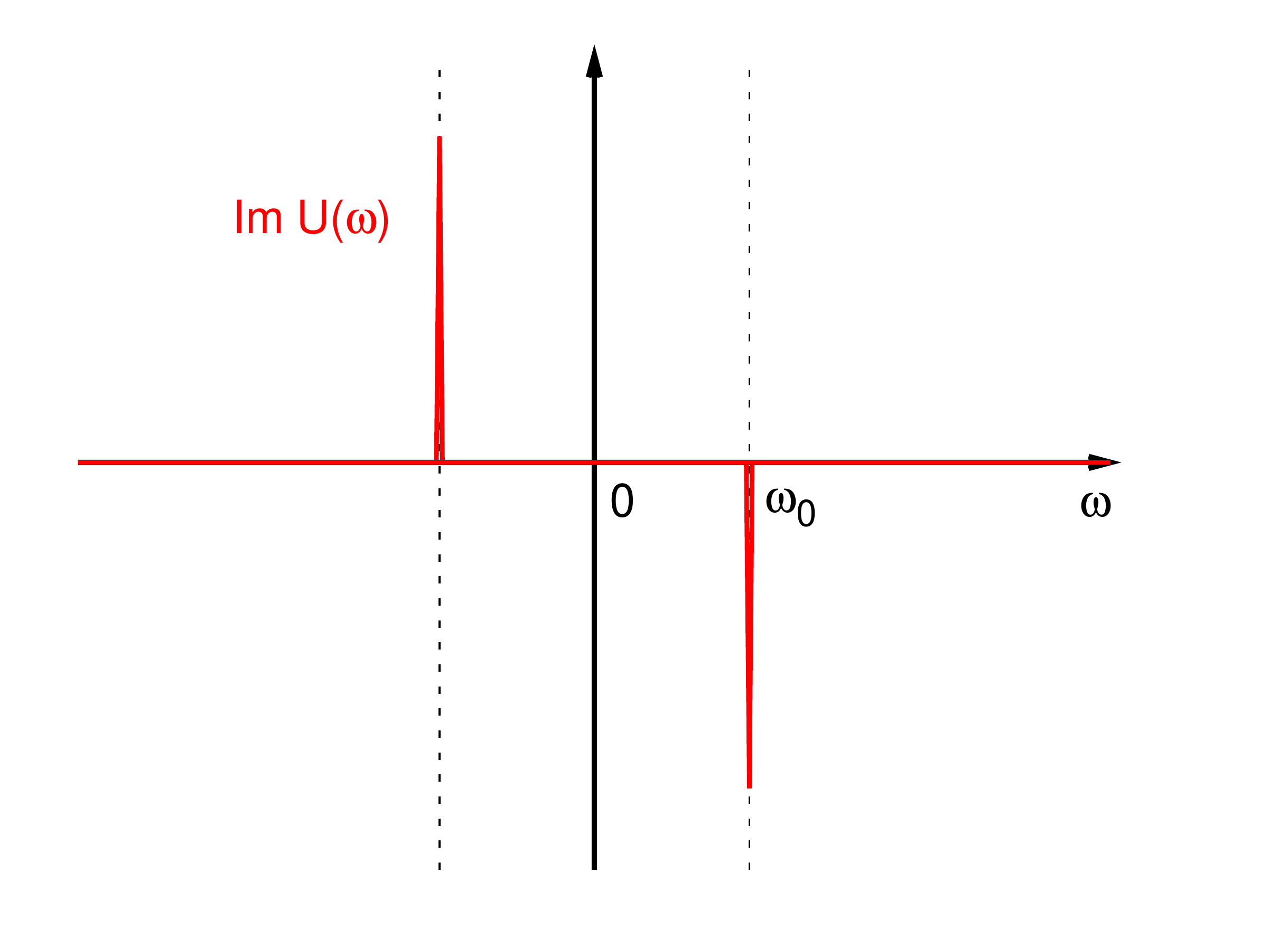}
\caption{
Retarded interaction corresponding to the Holstein-Hubbard model with on-site interaction $U=U_\text{bare}$, bosonic frequency $\omega_0$ and electron-boson coupling $g$. The difference between bare and screened interaction is $\lambda=2g^2/\omega_0.$
}
\label{fig_Ueff_HH}
\end{center}
\end{figure}

The hybridization term $H_\text{hyb}$ and the fermionic bath $H_\text{bath}$ are defined such that the parameters $V_{p\sigma}$ and $\varepsilon_p$ encode the hybridization function
\begin{equation}
\Delta_\sigma(i\omega_n)=\sum_{p} \frac{|V_{p\sigma}|^2}{i\omega_n-\varepsilon_p}.
\end{equation}
On the other hand, integrating out the bosons yields the frequency-dependent interaction
\begin{equation}
U(i\nu_n)=U+\frac{2g^2\omega_0}{(i\nu_n)^2-\omega_0^2},\label{u_hh_matsubara}
\end{equation}
or, after analytical continuation to the real-frequency axis, 
\begin{align}
\text{Re} U(\omega)&=U+\frac{2g^2\omega_0}{\omega^2-\omega_0^2}, \\
\text{Im} U(\omega)&=-g^2\pi(\delta(\omega-\omega_0)-\delta(\omega+\omega_0)).\label{ImU}
\end{align}
The real part of the frequency-dependent interaction, illustrated in the left panel of Fig.~\ref{fig_Ueff_HH} therefore has poles at the boson energy $\pm \omega_0$ and ranges from the unscreened interaction $U_\text{bare}=U$ at high frequencies to a static (screened) interaction 
\begin{equation}
U_\text{src}=U-2g^2/\omega_0 
\label{Uscr}
\end{equation}
at $\omega=0$. 

While the DMFT approximation simplifies the problem considerably, by mapping the Holstein-Hubbard lattice model onto an auxiliary single-site impurity model, this effective model is still a complicated interacting many-body system. The electron-boson coupling introduces additional energy scales, besides the bandwidth and Kondo scale of the Anderson impurity model, namely the boson frequency $\omega_0$ and the effective coupling strength $\lambda=2g^2/\omega_0$. (In the high-frequency limit, the Holstein-Hubbard model simplifies to the Hubbard model with interaction $U_\text{scr}=U-\lambda$.) Even in the DMFT approximation, and in the absence of long-range order, the Holstein-Hubbard model features a rich phase diagram with metallic, Mott insulating and bipolaronic insulating phases (Sec.~\ref{sec:phasediagram}) \cite{Koller2004,Koller2004_prb,GunSang2004,Koller2005,Barone2006}. Antiferromagnetic, charge-ordered, superconducting and supersolid phases can also be found \cite{Freericks1993,Murakami2013,Murakami2014} if symmetry breaking is allowed.      
In the following, we will discuss efficient, yet accurate numerical approaches for solving the Holstein-Hubbard impurity problem, and also show how these techniques can be generalized to models with a coupling to a continuum of bosonic modes (or arbitrary retarded interactions). In fact, in the context of DMFT based \emph{ab initio} simulations of correlated materials, the numerical challenge of treating dynamically screened interactions has been a major bottleneck which has hampered the implementation of advanced LDA+DMFT or GW+DMFT schemes for many years. The techniques introduced in the following sections eliminate this bottleneck.   

\subsubsection{DALA approach}
\label{dala}

An approximate, but often very useful scheme is the so-called dynamic atomic-limit approximation (DALA) \cite{Casula2012maxent}.  
The main idea behind the DALA is to assume the following \emph{ansatz} for the
Green's function in imaginary time:
\begin{eqnarray}
G(\tau) & = &  G^\textrm{static}(\tau, \mu, U_\textrm{scr}) \exp\left(
\frac{1}{\beta} \sum_{n \ne 0} \frac{U(i\nu_n)-U_\textrm{scr}}{\nu_n^2} \left( e^{i \nu_n \tau}-1 \right) \right),
\label{atomic_limit_factorization}
\end{eqnarray}
where $G^\textrm{static}(\tau,\mu,U_\textrm{scr})$ is the ``standard'' Green's
function of the model with static interaction $U_\textrm{scr}$. The factor which multiplies this Green's function takes into account the
dynamic nature of the interaction. 
For the Holstein-Hubbard model, $U(i\nu_n)$ is defined in Eq.~(\ref{u_hh_matsubara})  
and $U_\textrm{scr}$ is the static
screened interaction, defined in Eq.~(\ref{Uscr}) as the zero-frequency
limit of $U(\omega)$.
This \emph{ansatz} can be extended to
arbitrary retarded interactions, fillings and number of orbitals. The latter dependence
enters only through $G^\textrm{static}$. The above
\emph{ansatz} is exact in the atomic limit, that is, when the
hybridization function $\Delta$ is set to zero and the
impurity is isolated from the bath. In that limit the Green's function
exactly factorizes into the Bose factor 
\begin{equation}
F_\textrm{DALA}(\tau)=\exp\left(\frac{1}{\beta}
\sum_{n \ne 0} \frac{U(i\nu_n)-U_\textrm{scr}}{\nu_n^2} \left( e^{i \nu_n
    \tau}-1 \right) \right),
\label{atomic_limit_F}
\end{equation}
depending only on dynamic $U$ 
quantities, and a fermionic part $G^\textrm{static}$, which is the 
Green's
function of the atomic limit with static screened $U_\textrm{scr}$. 
Other trivial exact limits are the static and
non-interacting regimes, where $F_\textrm{DALA}=1$.

Away from the atomic limit, one can still define the Bose factor $F$ by
\begin{equation}
F(\tau)= \frac{G(\tau)}{G^\textrm{static}(\tau)},
\end{equation}
where $G(\tau)$ is the exact Green's function of the dynamic $U(\omega)$
problem and $G^\textrm{static}(\tau)$ is the exact Green's function of the Hubbard
model with the same hybridization $\Delta$ as $G$ but with the static
Hubbard repulsion $U_\textrm{scr}$. 
One may then approximate the exact $F$ by its 
atomic limit form $F_\textrm{DALA}$ in Eq.~(\ref{atomic_limit_F}). On the other hand, 
$G^\textrm{static}(\tau)$ can be obtained quite accurately by the solution
of the Anderson model with static $U_\textrm{scr}$ via ``standard''
methods. Therefore, in the DALA framework, the Green's function $G$ of
the full problem is approximated by the 
product $G^\textrm{static} F_\textrm{DALA}$.

Several remarks are in order here: (i) the DALA approach is non-perturbative 
in both the electron-electron screened interaction
(through the $G^\textrm{static}$ solution) \emph{and} the electron-boson
interaction, as one may argue from the exponential form of
$F_\textrm{DALA}$. In fact, $F_\textrm{DALA}$ can also be seen as a
cumulant expansion of the Green's function in the retarded
interaction around $\Delta=0$.  A similar formalism has been used to
compute the multiple plasmon satellites in silicon within the GW framework \cite{Guzzo2011}.
(ii) It has been shown that the DALA
works quite well even away from the atomic limit, as long as
$U_\textrm{scr}$ and $\omega_0$ are large compared to $\Delta$. The
approximation is 
problematic 
when the hybridization becomes essential 
for the low-energy low-temperature properties of the system. In
particular, the DALA 
does not fulfill the Friedel sum rule, which sets the zero frequency
limit of the exact Green's function, $\text{Im}G(i 0^+) = -4/D$, in a half-filled system with a  
semicircular density of states of bandwidth $D$. On the other hand,
the DALA gives the right high-energy properties of the Green's function.
(iii) Once the DALA solution of the Anderson impurity model with
retarded interaction (Eq.~(\ref{Simp})) is plugged into the DMFT cycle,
the high-energy tails of the impurity Green's function, correctly
modulated by the Bose factor $F_\textrm{DALA}$, can have a strong
impact on the self-consistent solution. In general, in the strongly
correlated regime, the Mott transition line modified by retardation
effects is quite accurately described by the DALA-DMFT solution of the
lattice problem, as shown by a direct comparison with more advanced
and accurate methods such as CTQMC (see Sec.~\ref{ctqmc} and
Fig.~\ref{phasediagram_g}). 
(iv) The very simple structure of the factorization
\emph{ansatz} used in the DALA framework is suggestive of the
physics of the system. Indeed, the analytic form of $F_\textrm{DALA}$
is the basis for a very effective analytic continuation method, described
in Sec.~\ref{sec:maxent}, which can produce the correct strength and position of the
plasmon/phonon satellites in the spectral function of $G$. These high energy features are fingerprints of the 
dynamically screened electron-electron interaction and the DALA
inspired analytic continuation can successfully resolve them.

\subsubsection{Lang-Firsov approach}
\label{lang_firsov}

To introduce the Lang-Firsov (LF) approach, let us start with
$H_\textrm{imp}$, the single-band single-boson impurity Hamiltonian, introduced in
Sec.~\ref{holstein_hubbard}. The Lang-Firsov transformation is a
unitary transformation $O$ of the fermion and boson local (impurity) operators, which eliminates 
the explicit electron-boson interaction and introduces dressed
fermion quasiparticles \cite{Lang1962}. It is defined as follows:
\begin{eqnarray}
O          & =  & \exp \left ( - \frac{g}{\omega_0} (b^\dag + b) \left( n_\uparrow + n_\downarrow - 1 \right) \right ),\\
\tilde{b} & = &  O^{-1} b O = b - \frac{g}{\omega_0} \left ( n_\uparrow + n_\downarrow - 1 \right ), \\ 
\tilde{d} & = &  O^{-1} d O = \exp \left ( - \frac{g}{\omega_0} ( b^\dag - b) \right ) d,
\end{eqnarray}
where $\tilde{b}$ and $\tilde{d}$ are the transformed boson and
fermion local operators, respectively. Therefore, the transformed impurity
Hamiltonian $\tilde{H}_\textrm{imp}= O^{-1} H_\textrm{imp} O$ reads
\begin{eqnarray}
\tilde{H}_\textrm{imp}  & = & \sum_\sigma (-\mu+g^2/\omega_0) d^\dag_\sigma d_\sigma 
+ (U - 2 g^2 / \omega_0)  n_\uparrow n_\downarrow  + \omega_0 b^\dag b
\nonumber  \\
& +  & \sum_{p\sigma} \left [ V_{p\sigma} \exp \left (  \frac{g}{\omega_0} (b^\dag - b) \right)  d^\dag_\sigma  a_{p\sigma} 
                                             +  V^*_{p\sigma} \exp
                                             \left
                                               (-\frac{g}{\omega_0} (b^\dag - b) \right)
                                             a^\dag_{p\sigma}
                                             d_\sigma \right]
                                           \nonumber \\ 
& + &  \sum_{p\sigma} \varepsilon_p  a^\dag_{p\sigma} a_{p\sigma}.
\label{LFH}
\end{eqnarray}
A nontrivial effect of the LF transformation is that it changes 
the bare local Hubbard repulsion to the screened value 
$U_\textrm{scr}$ (Eq.~(\ref{Uscr})). The interaction between dressed quasiparticles
$\tilde{d}^\dagger|0\rangle$ is \emph{reduced} by the presence of bosons. 
Analogously, we can write the impurity Green's function expressed in
the transformed coordinates as
\begin{equation}
G(\tau)  =  \textit{Z}^{-1} \langle T \tilde{d}^\dag(0) \tilde{d}(\tau) \exp(-\beta \tilde{H}) \rangle.
\label{green_mother} 
\end{equation}
While the LF transformation reveals the role of bosons as mediators
of the effective screened electron-electron 
interaction, which is the core of the screening theory presented in
this review, the transformed $H_\text{imp}$ in Eq.~(\ref{LFH}) is not immediately useful, as
it shifts all the complexity of the electron-boson interaction into 
``electronic polarons'', the dressed quasiparticles, i.e.
electrons coupled to their surrounding polarization cloud.
 We need to make some
approximation, which exploits the form of Eq.~(\ref{LFH}). A widely used
one is to project the full Fock space onto the zero boson mode, the
so-called LF approximation. The dynamic nature of the bosons is thereby
reduced to their lowest harmonic level. It is clear that this
approximation is good when $\omega_0 \gg E^*$, with $E^*$ the
relevant energy scale (bandwidth, hybridization, or Hubbard $U$
repulsion) of the purely fermionic part. In this regime, the fermionic
degrees of freedom are well separated from the bosonic high-energy ones.
The LF approximation is exact in the
antiadiabatic $\omega_0 \rightarrow \infty$ limit.
It turns out that many realistic cases fall into the antiadiabatic
regime, as the characteristic screening frequency is usually the
plasma frequency $\omega_0$ of the homogeneous electron gas evaluated at the same
average density as the real system. As shown in
Ref.~\cite{Casula2012maxent}, $\omega_0 \ge 15$ eV  in many correlated
materials, which is normally larger than the energy scale of the
correlated bands.

Once $\tilde{H}_\textrm{imp}$ in Eq.~(\ref{LFH}) is projected onto the boson vacuum,  the
resulting LF approximated Anderson model becomes
a standard (non-retarded) one with
renormalized hybridization, shifted chemical potential, and screened on-site interaction:
\begin{eqnarray}
H_0  &= & \sum_\sigma (-\mu+g^2/\omega_0) d^\dag_\sigma d_\sigma 
+ (U - 2 g^2 / \omega_0)  n_\uparrow n_\downarrow  + \omega_0 b^\dag b
\label{LFH_approx}  \nonumber \\
& +  & \sum_{p\sigma} \left ( V_{p\sigma} \exp \left ( - \frac{g^2}{2 \omega^2_0} \right) d^\dag_\sigma  a_{p\sigma} 
                                             +  V^*_{p\sigma} \exp \left (
                                               -\frac{g^2}{2
                                                 \omega^2_0} \right)
                                             a^\dag_{p\sigma}
                                             d_\sigma \right)
                                           \nonumber \\ 
& + &  \sum_{p\sigma} \varepsilon_p  a^\dag_{p\sigma} a_{p\sigma}.  
\end{eqnarray}
Analogously, the LF approximated form of the Green's function in
Eq.~\eqref{green_mother} is
\begin{eqnarray}
G_{LF}(\tau) &   =  &    \exp \left ( - g^2/\omega^2_0 \right)
\textit{Z}^{-1} \langle T d^\dag(0) d(\tau) \exp(-\beta H_0) \rangle \nonumber \\
& \equiv & \exp \left ( - g^2/\omega^2_0 \right)   G_{H_0}(\tau) .
\label{GLF}
\end{eqnarray}
After integration over the fermionic bath, one obtains the following
Lang-Firsov Green's function for the Anderson-Holstein model in
Eqs.~(\ref{Hloc})-(\ref{Hbath}): 
\begin{equation}
G_{LF}(i \omega_n)  =   \frac{ \exp ( - g^2/\omega^2_0 ) }{ i \omega_n + \mu -  g^2/\omega^2_0 -    
\exp  ( - g^2/\omega^2_0 ) \Delta (i \omega_n) - \Sigma[U_\textrm{scr}] (i \omega_n)},
\label{LFsolution}
\end{equation}
where $\Delta (i \omega_n) = \sum_{p\sigma} |V_{p\sigma}|^2 / ( i \omega_n
- \varepsilon_p ) $ is the 
hybridization function and
$\Sigma[U_\textrm{scr}] (i \omega_n)$ is the self-energy, which depends
on the screened value of $U$.

We note that the same $ \exp \left ( - g^2/\omega^2_0 \right) $
factor renormalizes both the hybridization function and the Green's
function. Therefore, the physical effect of retarding bosons,
at least in the antiadiabatic regime, is threefold: (i) They reduce
the Hubbard repulsion $U$ to its screened value, (ii)
they reduce the hopping
elements between the impurity and the bath, which in the DMFT framework 
means reducing the bandwidth of the related lattice model, and (iii) 
they reduce the weight of the low-energy Green's function, which
mimics the spectral weight transfer 
from the correlated manifold to higher-energy satellites, due to boson
shake-off processes.
All these physical insights provided by the LF approximation will be
used in Sec.~\ref{sec:static} to 
derive an effective static model which properly describes the low-energy
effects of screening in realistic strongly correlated systems.

In contrast to the DALA approximation, the LF
Green's function of Eq.~\eqref{LFsolution} fulfills the Friedel sum
rule. Therefore the LF approximation can be used to correct the
low-energy behavior of the DALA, as suggested in
Ref.~\cite{Casula2012maxent}. However, the combination of the two theories is
quite cumbersome. 
If a numerically exact result is desirable, it is better to use 
the continuous-time
Monte Carlo approach extended to dynamic interactions, which we describe
in the next subsection.

\subsubsection{Continuous-time Monte Carlo approach}
\label{ctqmc}

Continuous-time Monte Carlo simulations rely on an expansion of the partition function into a series of diagrams and the stochastic sampling of (collections) of these diagrams. We represent the partition function as a sum (or, more precisely, integral) of configurations $c$ with weight $w_c$,
\begin{equation}
Z=\sum_c w_c,
\label{Z_c}
\end{equation}
and implement a random walk $c_1\rightarrow c_2\rightarrow c_3\rightarrow \ldots$ in configuration space in such a way that 
each configuration can be reached from any other in a finite number of steps 
({\it ergodicity}) and that {\it detailed balance} is satisfied,
\begin{equation}
|w_{c_1}| p(c_1\rightarrow c_2) = |w_{c_2}| p(c_2\rightarrow c_1).
\label{detailed_balance}
\end{equation}
This assures that each configuration $c$ is visited with a probability proportional to $|w_c|$. One can thus obtain an estimate for the Green's function from a finite number $N$ of measurements:
\begin{equation}
g =\frac{\sum_c w_c g_c}{\sum_c w_c}=\frac{\sum_c |w_c| \text{sign}_c g_c}{\sum_c |w_c|\text{sign}_c}\approx \frac{\sum_{i=1}^N \text{sign}_{c_i} g_{c_i}}{\sum_{i=1}^N \text{sign}_{c_i}}=\frac{\langle \text{sign} \cdot g\rangle_{MC}}{\langle \text{sign}\rangle_{MC}}.
\end{equation}
The error on this estimate decreases like $1/\sqrt{N}$. 

The first step in the derivation of the continuous-time impurity solver \cite{Gull2011} is to rewrite the partition function $Z=\text{Tr}[e^{-\beta H_\text{imp}}]$ as a time ordered exponential using some interaction representation. We split the impurity Hamiltonian into two parts, $H=H_1+H_2$ and define the time-dependent operators in the interaction picture as $O(\tau)=e^{\tau H_1}O e^{-\tau H_1}$. We furthermore introduce the operator $A(\beta)=e^{\beta H_1}e^{-\beta H}$ and write the partition function as $Z=\text{Tr}[e^{-\beta H_1}A(\beta)]$. The operator $A(\beta)$ satisfies $dA/d\beta=e^{\beta H_1}(H_1-H)e^{-\beta H}=-H_2(\beta)A(\beta)$ and can be expressed as $A(\beta)=\mathcal{T}\exp[-\int_0^\beta d\tau H_2(\tau)]$.

In a second step, the time-ordered exponential is expanded into a power series,
\begin{eqnarray}
Z&=&\text{Tr} \Big[e^{-\beta H_1} \mathcal{T}e^{-\int_0^\beta d\tau H_2(\tau)} \Big]\nonumber\\
&=&\sum_{n=0}^\infty \int_0^\beta \!\!\!d\tau_1\ldots \int_{\tau_{n-1}}^\beta \!\!\!d\tau_n \text{Tr}\Big[ e^{-(\beta-\tau_n)H_1}(-H_2) \ldots e^{-(\tau_2-\tau_1)H_1}(-H_2)e^{-\tau_1H_1}\Big],\nonumber
\label{Z_interaction_picture}
\end{eqnarray}
which is a representation of the partition function of the form (\ref{Z_c}), namely the sum of all configurations $c=\{\tau_1, \ldots, \tau_n\}$, $n=0, 1, \ldots$, $\tau_i\in[0,\beta)$ with weight
\begin{equation}
w_c=\text{Tr}\Big[ e^{-(\beta-\tau_n)H_1}(-H_2)\ldots e^{-(\tau_2-\tau_1)H_1}(-H_2)e^{-\tau_1H_1}\Big]d
\tau^n. 
\label{weight}
\end{equation}
In the following we will discuss the so-called ``hybridization-expansion" approach \cite{Werner2006,Werner2006matrix}, which is based on an expansion of $Z$ in powers of the impurity-bath hybridization term $H_\text{hyb}$, and an interaction representation in which the time evolution is determined by the {\it local} part $H_\text{loc}+H_\text{bath}$.

In the case of the Holstein-Hubbard model \cite{Werner2007} the trace in Eq.~(\ref{weight}) is over the Fock states of the impurity, the fermionic bath and the bosonic bath ($\text{Tr}=\text{Tr}_d \text{Tr}_c \text{Tr}_d$), or an equivalent basis. After the expansion in the hybridization operators, the time-evolution from one hybridization event to the next does no longer couple the impurity and the fermionic bath. Since the fermionic bath is noninteracting, the trace over the $c$-states can be evaluated analytically, resulting in two determinants (one for each spin) of matrices $M_\sigma^{-1}$, whose elements are the hybridization functions $\Delta$ evaluated at the time-intervals determined by the hybridization operator positions.   
The 
weight of a Monte Carlo configuration corresponding to a perturbation order $\sum_\sigma 2n_\sigma$ ($n_\sigma$ creation operators $d^\dagger_\sigma(\tau_\sigma)$ and $n_\sigma$ annihilation operators $d_\sigma(\tau'_\sigma)$) can thus be expressed as
\begin{eqnarray}
&&w(\{O_i(\tau_i)\}) = Z_{0,c}\text{Tr}_d \text{Tr}_b\Big[  \mathcal{T} e^{-\beta H_\text{loc}} \prod_\sigma d_\sigma(\tau_{\sigma, n_\sigma})d^\dagger_\sigma(\tau'_{\sigma, n_\sigma}) \ldots\nonumber\\
&&\hspace{45mm}\ldots 
d_\sigma(\tau_{\sigma, 1})d^\dagger_\sigma(\tau'_{\sigma, 1}) \Big] \prod_\sigma (\det M_\sigma^{-1}) (d\tau)^{2n_\sigma},
\label{weight_HH}
\end{eqnarray}
where the the matrix elements are $M_\sigma^{-1}(i,j)=\Delta_\sigma(\tau'_{\sigma,i}-\tau_{\sigma,j})$. We denote the time-ordered sequence of impurity creation and annihilation operators by 
$\{ O_i(\tau_i) \}$ ($1\le i\le 2n$, $n=\sum_\sigma n_\sigma$). 
At this stage, the time-evolution from one operator to the next still includes the coupling to the bosons. In order to evaluate the trace over the boson states, 
we perform the Lang-Firsov \cite{Lang1962} transformation introduced in the previous subsection. Here, we express it in terms of the boson position and momentum operators $X$ and $P$ defined in Eq.~(\ref{bosonX}).  
The unitary transformation specified by $e^{iPX_0}$ shifts $X$ by $X_0$, and if we choose $X_0=(\sqrt{2}g/\omega_0)(n_\uparrow+n_\downarrow-1)$, the transformed local Hamiltonian (\ref{Hloc_X}), 
\begin{eqnarray}
\tilde H_\text{loc} &=& e^{-iPX_0}H_\text{loc} e^{iPX_0}\nonumber\\
&=& \tilde U \tilde n_\uparrow \tilde n_\downarrow-\tilde \mu(\tilde n_\uparrow+\tilde n_\downarrow)+\frac{\omega_0}{2}(X^2+P^2), 
\label{H_loc_transf}
\end{eqnarray}
has no explicit electron-phonon coupling anymore. ${\tilde H}_\text{loc}$ is of the Hubbard form but
with modified chemical potential and interaction strength, 
\begin{eqnarray}
\tilde \mu&=&\mu-g^2/\omega_0,\label{mu_tilde}\\
\tilde U&=&U-2g^2/\omega_0.\label{U_tilde}
\end{eqnarray}
Note that $\tilde U=U_\text{scr}$ (Eq.~(\ref{Uscr})). 
The transformation also affects the impurity creation and annihilation operators, which acquire a boson factor:
\begin{eqnarray}
\tilde d^\dagger_\sigma &=& e^{iPX_0}d^\dagger_\sigma e^{-iPX_0} = e^{\frac{g}{\omega_0}(b^\dagger-b)}d^\dagger_\sigma,\label{dtilde_dag}\\
\tilde d_\sigma &=& e^{iPX_0}d_\sigma e^{-iPX_0} = e^{-\frac{g}{\omega_0}(b^\dagger-b)}d_\sigma.\label{dtilde}
\end{eqnarray}

\begin{figure}[t]
\begin{center}
\includegraphics[angle=0, width=0.6\textwidth]{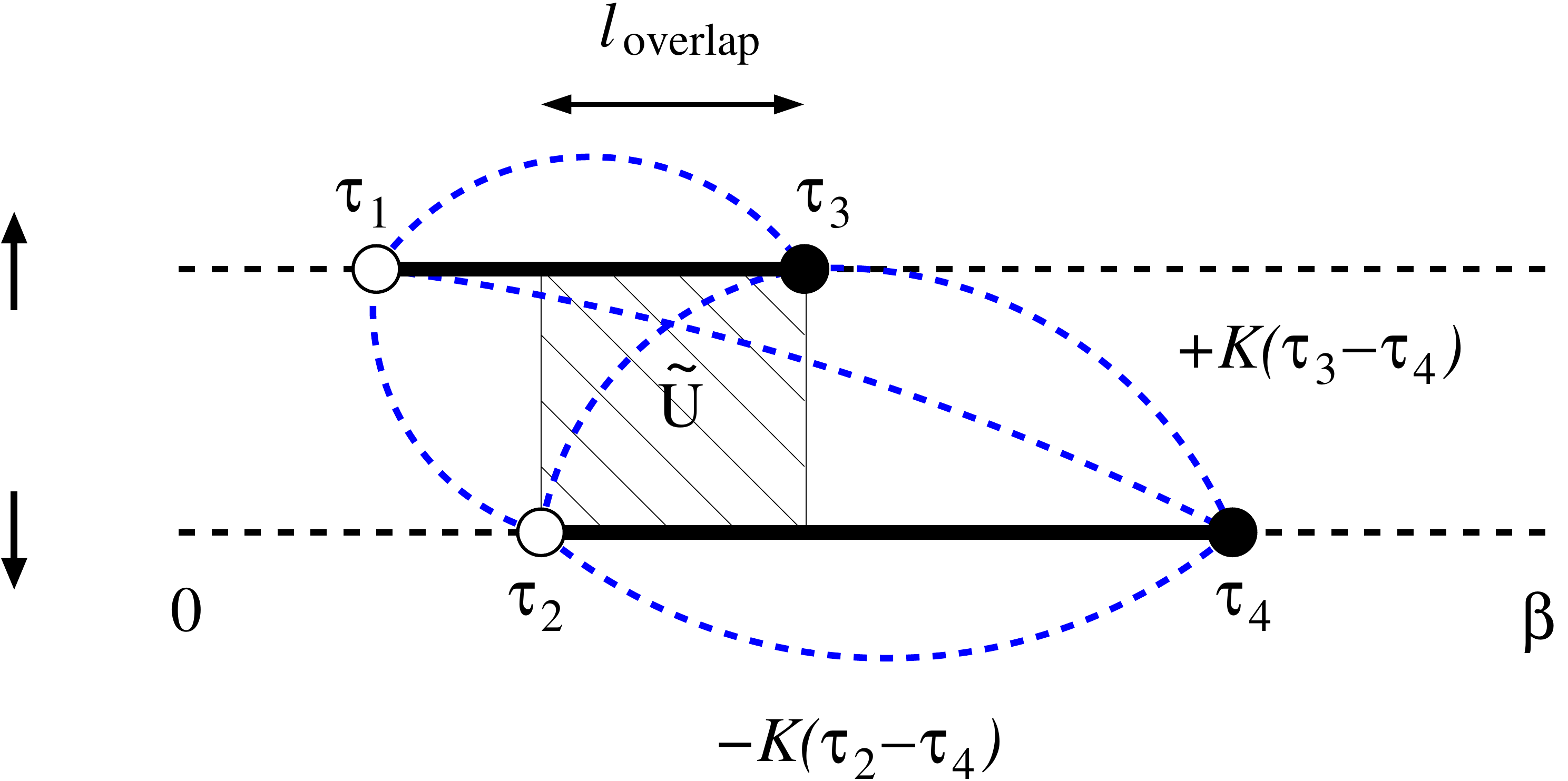}
\caption{Illustration of a segment configuration with one segment for spin up and spin down. The interaction and chemical potential contribution to the configuration weight can be obtained from the lengths and overlaps of the segments. The boson contribution is indicated by the blue dashed lines, which connect all pairs of operators. 
}
\label{segment}
\end{center}
\end{figure}

After this transformation, the electron and boson sectors are decoupled and the 
trace in Eq.~(\ref{weight_HH}) becomes the product of a term
involving electron operators which is identical to the weight appearing in a Hubbard
model simulation (with shifted $\tilde U$ and $\tilde \mu$), and a phonon term which is the expectation value of a product
of exponentials of boson operators, to be evaluated with the noninteracting boson Hamiltonian $\omega_0 b^\dagger b$. 
The phonon contribution $w_b$ to the weight (\ref{weight_HH}) can be written as
\begin{eqnarray}
w_b(\{O_i(\tau_i)\})&=&Z_{0,b}\left<e^{s_{2n}A(\tau_{2n})}e^{s_{2n-1}A(\tau_{2n-1})}\cdots e^{s_1A(\tau_1)}\right>_b, 
\end{eqnarray}
with $Z_{0,b}$ the noninteracting boson partition function, $\langle \ldots \rangle_b=\text{Tr}_b[\ldots]/Z_{0,b}$, $0\le \tau_1<\tau_2<\ldots <\tau_{2n}<\beta$, $s_i=1$ $(-1)$ if the $i^{th}$ operator is a creation (annihilation) operator and $A(\tau)$ given by
\begin{equation}
A(\tau)=\frac{g}{\omega_0}(e^{\omega_0\tau}b^\dagger-e^{-\omega_0\tau}b).
\end{equation}
The expectation value is to be taken in the thermal state of free bosons. 
Using 
$e^{s A(\tau)}=e^{-g^2/(2\omega_0^2)}e^{s(g/\omega_0)e^{\omega_0\tau}b^\dagger}e^{-s(g/\omega_0)e^{-\omega_0\tau}b}$,
one finds the disentangled expression
\begin{eqnarray}
w_b(\{O_i(\tau_i)\})&=&Z_{0,b}e^{-n (g^2/\omega_0^2)}e^{-\sum_{2n\geq i>j\geq 1} s_is_j(g^2/\omega_0^2)e^{-\omega_0(\tau_i-\tau_j)}}\nonumber\\
&&\times \left<e^{\sum_js_j (g/\omega_0)e^{\omega_0 \tau_j}b^\dagger}e^{-\sum_js_j(g/\omega_0)e^{-\omega_0 \tau_j}b}\right>_b.
\end{eqnarray}
For thermal expectation values, we have the formula \cite{mahan} $\langle e^{u b^\dagger}e^{v b}\rangle_b=e^{uv/(e^{\beta \omega_0}-1)}$,  
which gives the final expression for the bosonic weight 
\begin{align}
w_b(\{O_i(\tau_i)\})=&Z_{0,b}\exp\bigg[\sum_{2n\geq i>j\geq 1} s_is_jK(\tau_i-\tau_j)\bigg],\\
K(\tau)=&-\frac{g^2}{\omega_0^2}\frac{ \cosh(\omega_0(\beta/2-\tau))-\cosh(\beta \omega_0/2)}{\sinh(\beta \omega_0/2)}.\label{EqK}
\end{align}
The total weight (\ref{weight_HH}) can be expressed as
\begin{equation}
w(\{O_i(\tau_i)\})=w_b(\{O_i(\tau_i)\})\tilde w_\text{Hubbard}(\{O_i(\tau_i)\}), 
\end{equation}
where $\tilde w_\text{Hubbard}$ denotes the weight of a corresponding configuration in the 
pure Hubbard impurity model (with parameters modified according to Eqs.~(\ref{mu_tilde}) 
and (\ref{U_tilde})). This weight can be efficiently computed using the segment picture \cite{Werner2006} (see Fig.~\ref{segment}). For each spin, we mark the imaginary-time intervals corresponding to an occupied impurity state by a segment. Then, $w_\text{Hubbard}(\{O_i(\tau_i)\})$ can be obtained from the total length $l_\sigma$ of these segments and the total length $l_\text{overlap}$ of the overlaps between up-spin and down-spin segments:
\begin{equation}
w_\text{Hubbard}(\{O_i(\tau_i)\})=s Z_{0,c} e^{\tilde\mu (l_\uparrow+l_\downarrow)-\tilde U l_\text{overlap}}\prod_\sigma (\det M_\sigma^{-1}) (d\tau)^{2n},
\end{equation}
with $s$ a permutation sign related to the time-ordering of the impurity creation and annihilation operators. 

In a segment insertion move, a pair of $\sigma$-operators is added to the sequence $\{O_i(\tau_i)\}$. For the Metropolis test, one has to 
compute the ratio $w_b^\text{new}/w_b^\text{old}$ and a determinant ratio 
$\det M_{\sigma,\text{new}}^{-1}/\det M_{\sigma,\text{old}}^{-1}$, both at a cost 
$O(n)$. The ratio of the traces over the impurity states (evaluated from the segment lengths and overlaps) is also at most $O(n)$. If 
the move is accepted, the matrix $M_{\sigma}$ needs to be updated. This  is the computationally most expensive step requiring an effort $O(n^2)$, but this is identical to the effort
in the Hubbard model without electron-phonon coupling \cite{Gull2011}. Hence, the treatment of the bosons does not affect the scaling of the algorithm.

Observables such as the Green's function or double occupation can be measured as in the case with static interaction~\cite{Werner2006}, i.e. from the elements of the inverse hybridization matrix and the segment overlaps. Specifically, the formula for the Green's function reads
\begin{equation}
G_\sigma(\tau)=\Bigg\langle -\sum_{ij}\frac{1}{\beta}\delta(\tau,\tau_i-\tau'_j)(M_\sigma)_{ij}\Bigg\rangle
\end{equation}
with $\delta(\tau,\tau')=\delta(\tau-\tau')$ for $\tau'>0$, $\delta(\tau,\tau')=-\delta(\tau-\tau'-\beta)$ for $\tau'<0$, and $\tau_i$ ($\tau'_j$) denoting the time of the $i$th annihilation ($j$th creation) operator. Efficient measurement schemes for the self-energy and vertex function \cite{Hafermann2012} on the other hand need to be specifically adapted to the case of retarded interactions. A detailed discussion of these techniques can be found in Ref.~\cite{Hafermann2014}.  

As a final remark, we note that the DALA Bose factor reported in Eq.~(\ref{atomic_limit_F}) of
Sec.~\ref{dala} can be written in terms of the $K$ function (\ref{EqK}) as 
$F_\textrm{DALA}(\tau)=\exp(-K(\tau))$.
In the atomic limit, i.e. zero
hybridization, only the $0$-th order term in the hybridization
expansion survives, leading to a total Green's function $G_\sigma(\tau)$ consisting of the
product of the global factor $F_\textrm{DALA}(\tau)$ and the static Green's function, as
explained in Sec.~\ref{dala}. Therefore, the exact CTQMC hybridization
expansion algorithm with retarded interactions naturally reduces to the
DALA result in the atomic limit.

\subsubsection{Phase diagram}
\label{sec:phasediagram}

\begin{figure}[t]
\begin{center} 
\includegraphics[angle=0, height=0.38\columnwidth]{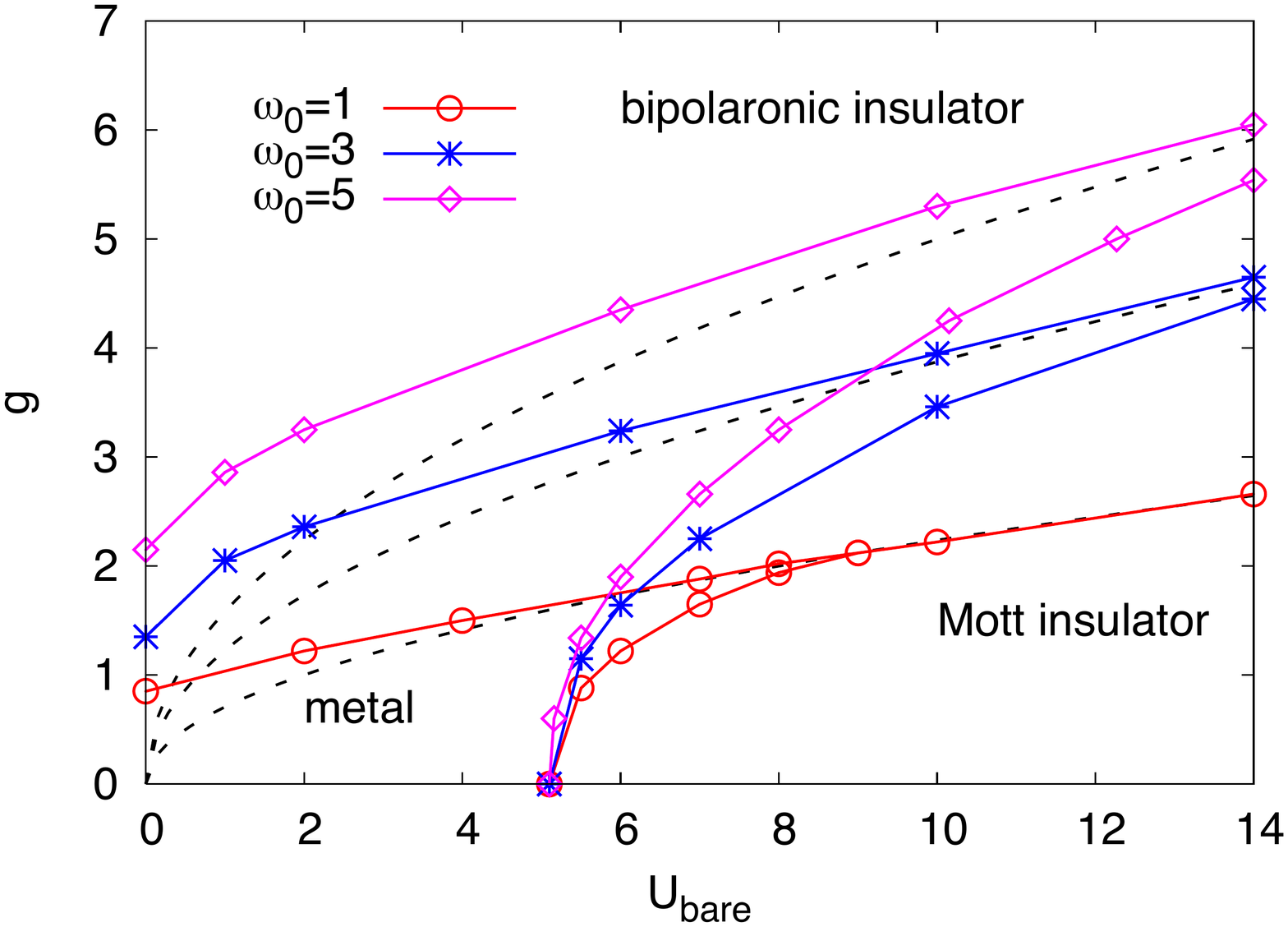}
\includegraphics[angle=0, height=0.353\columnwidth]{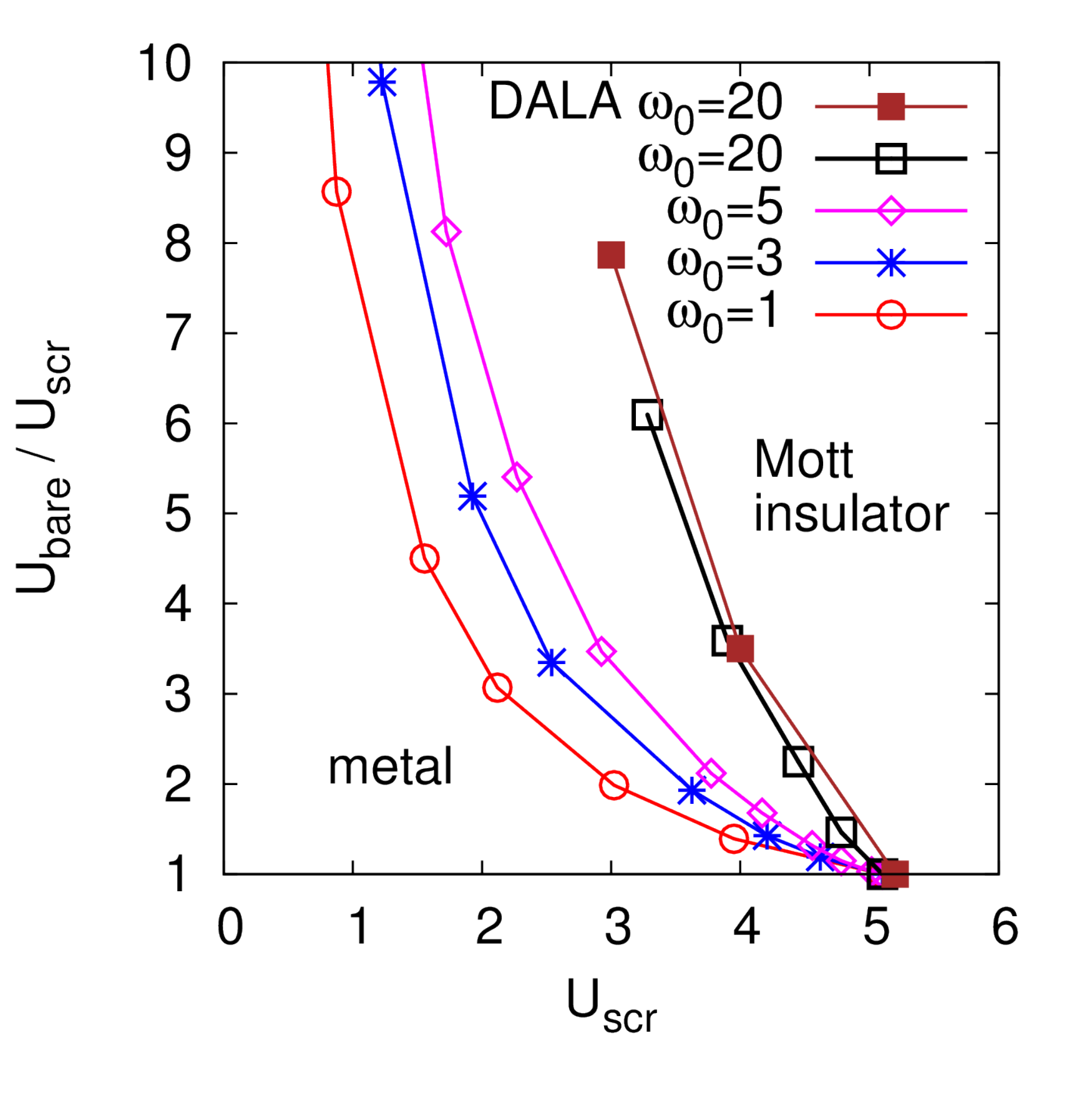}
\caption{Left panel: half-filled metal-insulator phase diagram for
  $\beta=50$ in the space of bare interaction $U_\text{bare}$ and
  boson coupling strength $g$ for different values of the screening
  frequency $\omega_0$ (DMFT solution for a semi-circular density of
  states with bandwidth 4). The dashed lines indicate the boson coupling 
  at which the screened interaction $U_\text{scr}$ changes
  sign. Right panel: phase diagram in the space of screened and bare
  interaction. The approximated DALA results (Sec.~\ref{dala}), corrected by the
  low-energy Lang-Firsov behavior, are also reported for
  $\omega_0=20$ and different coupling strengths. (Adapted from Ref.~\cite{Werner2010}.) 
  }   
\label{phasediagram_g}
\end{center}
\end{figure}
We illustrate the DMFT phase diagram of the half-filled Holstein-Hubbard model in the left panel of Fig.~\ref{phasediagram_g}. The results are for a semi-circular density of states with bandwidth $4$ and inverse temperature $\beta=50$. A paramagnetic solution is enforced.  At weak boson coupling $g$, there is a metallic phase at small $U_\text{bare}$ and a Mott insulating phase for sufficiently large $U_\text{bare}$. As $g$ is increased, the effective on-site interaction is reduced and eventually, the system makes a transition to a bipolaronic insulating phase. The dashed line indicates the value of $g$ at which the screened interaction $U_\text{scr}=U_\text{bare}-2g^2/\omega_0$ vanishes. For small boson frequency, the sign change in $U_\text{scr}$ essentially coincides with the transition from Mott insulator to bipolaronic insulator, while for large boson frequency, there exists a metallic solution near $U_\text{scr}=0$, which separates the two insulators.   

The right panel shows the metal-to-Mott insulator phase boundary in the space of $U_\text{bare}/U_\text{scr}$ and $U_\text{scr}$, for different values of the boson frequency. A larger $U_\text{bare}/U_\text{scr}$ stabilizes the Mott insulator, and this effect is most pronounced for a low screening frequency.

\subsection{General $U(\omega)$}
\label{sec:generalU}

It follows from Eq.~(\ref{ImU}) that the Holstein-Hubbard model corresponds to a frequency-dependent interaction $U(\omega)$, whose imaginary part is a $\delta$-function at $\omega=\pm \omega_0$, with a weight given by $\mp g^2\pi$. An arbitrary $U(\omega)$ can thus be thought of as arising from a Holstein-type coupling to a continuum of bosonic modes with energy $\omega$ and coupling strength $g_\omega$ given by $g_\omega^2=-\text{Im}U(\omega)/\pi$. According to the derivation in Sec.~\ref{ctqmc} each boson contributes an effective ``interaction" $K(\tau_i-\tau_j)=-\frac{g_\omega^2}{\omega^2}\frac{ \cosh(\omega(\beta/2-(\tau_i-\tau_j))-\cosh(\beta \omega/2)}{\sinh(\beta \omega/2)}$ between impurity creation or annihilation operators at imaginary times $\tau_i$ and $\tau_j$ (Eq.~(\ref{EqK})). Hence, the hybridization expansion Monte Carlo simulation for a model with general $U(\omega)$ proceeds exactly as in the case of the Holstein-Hubbard model, but with the $K$-function replaced by \cite{Werner2010}
\begin{align}
K(\tau)&=\int_0^\infty d\omega\frac{\text{Im}U(\omega)}{\pi\omega^2}\frac{ \cosh(\omega(\beta/2-\tau))-\cosh(\beta \omega/2)}{\sinh(\beta \omega/2)}
\label{K}
\end{align}
and the shifted interaction and chemical potential (Eqs.~(\ref{mu_tilde}) and (\ref{U_tilde})) given by 
\begin{eqnarray}
\tilde \mu&=&\mu+\int_0^\infty d\omega \frac{\text{Im}U(\omega)}{\pi\omega},\label{mu_tilde_general}\\
\tilde U&=&U+2\int_0^\infty d\omega \frac{\text{Im}U(\omega)}{\pi\omega}=U_\text{scr}.\label{U_tilde_general}
\end{eqnarray}
The last identity follows from the Kramers-Kronig relation and the anti-symmetry of $\text{Im}U(\omega)$.

These formulas can also be derived from the impurity action with retarded $U(\tau)$, without the detour through the Hamiltonian representation \cite{Ayral2013}. In fact, it is easy to evaluate the interaction contribution $\frac12 \int_0^\beta d\tau d\tau' U(\tau-\tau')n(\tau) n(\tau')$ in the action (\ref{Simp}) for a given segment configuration. In a first step, we split the interaction into its instantaneous contribution and a retarded part:
\begin{align}
U(\tau)&=U_\text{bare}\delta(\tau)+U_\text{ret}(\tau-\tau'),\label{Uret0}\\
U(\omega)&=U_\text{bare}+U_\text{ret}(\omega).
\end{align}
$U_\text{ret}(\omega)$ has the same imaginary part as $U(\omega)$, but its real part approaches $0$ in the high frequency limit. Therefore, we have the spectral representation ($0\le \tau\le \beta$)
\begin{align}
U_\text{ret}(\tau)&=-\frac{1}{\pi}\int_{-\infty}^\infty d\omega \text{Im}U(\omega)\frac{e^{-\omega\tau}}{e^{-\omega\beta}-1}=\frac{1}{\pi}\int_0^\infty d\omega \text{Im}U(\omega)\frac{\cosh(\omega(\beta/2-\tau))}{\sinh(\omega\beta/2)}.\label{Uret}
\end{align}
Plugging Eq.~(\ref{Uret}) into the expression for the interaction contribution and using $n=n_\uparrow+n_\downarrow$, as well as $n_\sigma^2=n_\sigma$, we find
\begin{align}
&\frac12 \int_0^\beta d\tau d\tau' U(\tau-\tau')n(\tau) n(\tau')= U_\text{bare}\int_0^\beta d\tau n_\uparrow(\tau)n_\downarrow(\tau)+\frac{U_\text{bare}}{2}\int_0^\beta d\tau (n_\uparrow(\tau)+n_\downarrow(\tau))\nonumber\\
&\hspace{5mm}+\int_0^\beta d\tau d\tau' n_\uparrow(\tau)U_\text{ret}(\tau-\tau')n_\downarrow(\tau')+\sum_\sigma \frac{1}{2}\int_0^\beta d\tau d\tau' n_\sigma(\tau)U_\text{ret}(\tau-\tau')n_\sigma(\tau').
\end{align}
The first term yields $U_\text{bare}l_\text{overlap}$ and the second term a shift of the chemical potential. The contribution from the retarded interaction, which acts both between same and opposite spin electrons, is given by the last two terms. Since $n_\sigma$ is 1 only on a segment of spin $\sigma$, and zero otherwise, we can express the retarded contribution as
\begin{align}
\sum_{k_1\ne k_2}\int_{k_1}d\tau_1\int_{k_2}d\tau_2 U_\text{ret}(\tau_1-\tau_2)+\frac{1}{2}\sum_{k}\int_{k}d\tau_1\int_{k}d\tau_2 U_\text{ret}(\tau_1-\tau_2),
\end{align}
Where $k_i$ denotes the segment $i$. The first term is the inter-segment contribution and the second term the intra-segment contribution of the retarded interaction. Now, let as assume that we know a $\beta$-periodic function $h(\tau)$, which is even, and satisfies $\frac{d^2h}{d\tau^2}=U_\text{ret}(\tau)$ in the interval $0<\tau<\beta$. With the help of this function, we can express the integral over a pair of segments as $\int_{k_1}d\tau_1\int_{k_2}d\tau_2 U_\text{ret}(\tau_1-\tau_2)=-h(\tau_1^e-\tau_2^e)+h(\tau_1^e-\tau_2^s)+h(\tau_1^s-\tau_2^e)-h(\tau_1^s-\tau_2^s)$, where $\tau_i^s$ ($\tau_i^e$) denotes the start and end points of segment $k_i$. Similarly, the double integral over segment $k$ evaluates to $h(\tau_e-\tau_s)-h(0)$. Hence, the retarded interaction energy for a segment configuration with $n$ creation and $n$ annihilation operators can be written as 
\begin{align}
-\sum_{i>j} s_i s_j (h(\tau_i-\tau_j)-h(0)),
\end{align}
where we assumed that the operators are time-ordered ($\tau_i>\tau_j$ for $i>j$) and the sign $s_i$ is $+1$ if the $i$th operator is a creation operator, and $-1$ if it is an annihilation operator.

From the expression (\ref{Uret}) we find by double-integration that $h(\tau)-h(0)=K(\tau)$, the function defined in Eq.~(\ref{K}). Hence, the nonlocal interaction contribution to the weight becomes $\exp\!\big[\sum_{i>j}s_is_j K(\tau_i-\tau_j)\big]$, as derived in section~\ref{ctqmc}.

There is one subtlety which we need to consider: due to the slope-discontinuity in the $\beta$-periodic function $h(\tau)$ at $\tau=0$, the second derivative yields a delta-function contribution with weight 
\begin{align}
-\frac{2}{\pi}\int_0^\infty d\omega \frac{\text{Im}U(\omega)}{\omega}=-\text{Re}U_\text{ret}(\omega=0)=-(U_\text{scr}-U_\text{bare}).
\end{align}
So, if we use this $h(\tau)$ in the formula for the retarded interaction contribution, we have to subtract $-(U_\text{scr}-U_\text{bare})\delta(\tau)$ from the instantaneous interaction. The total instantaneous interaction thus becomes $(U_\text{bare}+(U_\text{scr}-U_\text{bare}))\delta(\tau)=U_\text{scr}\delta(\tau)$, which means that the overlap contribution to the weight of a segment configuration has to be evaluated with the screened interaction. A similar shift is introduced to the chemical potential, so that the end result is identical to Eqs.~(\ref{mu_tilde_general}) and (\ref{U_tilde_general}).

\subsection{Effective static model}
\label{sec:static}

Even though there exist efficient numerical methods for treating dynamically screened monopole interactions, it is conceptually interesting to ask how to define a low-energy model with static interactions, which properly captures the low-energy physics. To derive this model, we again first consider the simple case of the Holstein-Hubbard model (\ref{HH}). Applying the Lang-Firsov transformation \cite{Lang1962} $H_\text{HH}\rightarrow H_\text{LF}=e^S H_\text{HH} e^{-S}$ with $S=\frac{g}{\omega_0}\sum_i (n_{i\uparrow}+n_{i\downarrow}-1)(b_{i}^\dagger+b_i)$ yields a decoupled Hamiltonian in terms of the polaron operators (\ref{dtilde_dag}) and (\ref{dtilde}):
\begin{equation}
H_\text{LF}= - t\sum_{\langle ij\rangle\sigma}( \tilde d^\dag_{i\sigma} \tilde d^{}_{j\sigma} + \tilde d^\dag_{j\sigma} \tilde d^{}_{i\sigma})+U_\text{scr}\sum_i n_{i\uparrow} n_{i\downarrow}-\mu\sum_{i\sigma}n_{i\sigma}+\sum_i \omega_0 b_i^\dagger b_i,
\end{equation} 
with $U_\text{scr}=U-2g^2/\omega_0$. As shown in Ref.~\cite{Casula2012static}, the low-energy effective model can now be defined by projecting this Hamiltonian onto the zero-boson subspace: 
\begin{equation}
H_\text{eff}=\langle 0| H_\text{LF} | 0\rangle=- Z_Bt\sum_{\langle ij\rangle\sigma}( d^\dag_{i\sigma} d^{}_{j\sigma} + d^\dag_{j\sigma} d^{}_{i\sigma})+U_\text{scr}\sum_i n_{i\uparrow} n_{i\downarrow}-\mu\sum_{i\sigma}n_{i\sigma}.
\label{H_eff}
\end{equation}
$H_\text{eff}$ is of the Hubbard form and involves the original fermionic operators $d^\dagger$ and $d$, the screened interaction $U_\text{scr}$, and a hopping term reduced by a factor 
\begin{equation}
Z_B=\exp[-g^2/\omega_0^2].\label{Z_B}
\end{equation} 
Hence, a crucial effect of the dynamical screening is an effective bandwidth reduction. 

The projection onto the zero boson subspace is a good approximation if the screening frequency $\omega_0$ is large. As an illustration of this, we list in Tab.~\ref{tab_eff_static} the critical interaction strengths for the Mott transition in a DMFT simulation of the Holstein-Hubbard model with semi-circular density of states of bandwidth 4, at inverse temperature $\beta=100$. The result from the exact DMFT treatment based on the method explained in Sec.~\ref{ctqmc} is compared to the result of a DMFT treatment of the effective static model (\ref{H_eff}), with the bandwidth renormalization (\ref{Z_B}).

\begin{table}[t]
\caption{Critical interaction strength $U_{0,\textrm{crit}}^\textrm{exact}$ for the Mott transition obtained from the single-site DMFT approximation to model (\ref{HH}) at inverse temperature $\beta=100$, compared to the estimate $U_{0,\textrm{crit}}^\textrm{eff}$ from the solution of the effective static model (\ref{H_eff}) for different values of the screening frequency $\omega_0$ and coupling strength $g$. Also shown is the Lang-Firsov renormalization factor (\ref{Z_B}). (From Ref.~\cite{Casula2012static}.)}
\label{tab_eff_static}
\centering
\vspace{3mm}
\begin{tabular}{ l | l | l | l | l l}
{$\omega_0$} \hspace{8mm} & {$g$} \hspace{12mm} & {$Z_B$} \hspace{8mm} & {$U_{\textrm{crit}}^\textrm{exact}$ } \hspace{4mm} & {$U_{\text{crit}}^\textrm{eff}$}\hspace{4mm}  \\
\hline
1.5 & 0.820 & 0.74 &  2.103 &  1.891 \\
1.5 & 2.010  & 0.17 & 0.613  & 0.423 \\
2.5 & 1.330 & 0.75 &  2.085 & 1.921 \\
2.5 & 2.770 & 0.29 & 0.861 & 0.747 \\
10.0 &  3.725 & 0.87&  2.225 & 2.220 \\
10.0 &  6.465 & 0.66&  1.640 & 1.679 \\
\end{tabular}
\end{table}

In a model with a general $U(\omega)$ we can again view the frequency dependence as arising from a coupling to a continuum of bosonic modes with frequency $\omega$ and coupling strength $g_\omega^2=-\text{Im}U(\omega)/\pi$. The screened interaction is then given by Eq.~(\ref{U_tilde_general}) (or simply $\text{Re} U(\omega=0)$), and the formula for the Lang-Firsov renormalization factor becomes
\begin{equation}
Z_B=\exp\Big[\frac{1}{\pi}\int_0^\infty d\omega \frac{\text{Im}U(\omega)}{\omega^2}\Big].\label{Z_B_general}
\end{equation}

In a model with strongly correlated ``$d$" states and itinerant ``$p$" states, and associated hopping parameters ${\cal T}_{pp}$, ${\cal T}_{pd}$ and ${\cal T}_{dd}$, the Lang-Firsov transformation and subsequent projection onto the zero-boson sector leads to a renormalization of each $d$ operator by a factor $\sqrt{Z_B}=\langle 0 | \exp(\frac{g}{\omega_0} (b_i - b_i^{\dagger})) | 0 \rangle$. Hence, the hopping part of the one-particle Hamiltonian is renormalized as
\begin{equation}
\left( p^\dagger d^\dagger \right) \left( \begin{array}{cc}
{\cal T}_{pp}  & \sqrt{Z_B} {\cal T}_{pd} \\
\sqrt{Z_B} {\cal T}_{pd}^\dagger & Z_B {\cal T}_{dd} \end{array} \right) 
\left( \begin{array}{c} 
p \\
d \end{array} \right).
\label{matrix_transformation}
\end{equation}
This expression 
shows that the bandwidth reduction implied by the effective model cannot simply be translated into an effective increase of the on-site interaction in the multi-band situation  
typically considered in first-principles calculations.

\subsection{Multiorbital systems} 
\label{subsec:multiorbital_systems}

Up to now, we have discussed dynamically screened on-site interactions in a single-orbital model. The extension of these techniques to multi-orbital systems is straight-forward if the total charge on a given site is screened, that is, if the electron-boson coupling in a Hamiltonian formulation is of the type
\begin{equation}
H_\text{el-b}=\sum_{i,\lambda} g_{i,\lambda} n^\text{tot}_i(b^\dagger_{i,\lambda}+b_{i,\lambda}),
\label{H_el_b}
\end{equation}
with $i$ the site index, $b_{i,\lambda}$ the annihilation operator for the oscillator of frequency $\omega_{i,\lambda}$ at site $i$ and $g_{i,\lambda}$ the corresponding coupling strength. The total charge on site $i$ is $n^\text{tot}_i=\sum_{\alpha=1}^{N_\text{orbitals}}\sum_\sigma n_{i,\alpha,\sigma}$. Within DMFT, one then has to solve a multi-orbital impurity system with an analogous electron-boson coupling. In the hybridization expansion formalism introduced in Sec.~\ref{ctqmc}, each hybridization operator changes the total charge on the multi-orbital impurity by $\pm 1$ and hence excites the bosons. After integrating out the bosons, one again finds an ``interaction" of the form (\ref{K}) between each operator pair, with a sign that depends on the operator types (creation or annihilation operators). Now, $U(\omega)$ denotes the dynamically screened monopole interaction. In the trace calculation, the intra- and inter-orbital interaction parameters are replaced by the static values, in analogy to Eq.~(\ref{U_tilde_general}). Note that this procedure works both in the case of density-density interactions (segment formalism \cite{Werner2006}), and for models with rotationally invariant interactions (matrix formalism \cite{Werner2006matrix}), because spin-flips and pair-hoppings do not change the total charge on the impurity and hence do not couple to the bosons. 

A more complicated situation arises if the Hund coupling has a significant frequency dependence, such as in alkali-doped fullerides  \cite{Capone2009,Nomura2012}, where Jahn-Teller screening may even produce an overscreened (negative) static $J$. In this case, the changes in the orbital occupation associated with spin-flips and pair-hoppings will couple to 
bosonic modes 
and the rotationally invariant system becomes much more difficult to simulate. After integrating out the phonons, the interaction part of the action reads \cite{Steiner2015}
\begin{align}
S_\text{int}&=\frac{1}{2} \sum_{\begin{subarray}{l} \alpha, \sigma \\ \beta, \sigma' \end{subarray}} \int_0^{\beta} d \tau \int_0^{\beta} d \tau'  
	n_{\alpha, \sigma} ( \tau ) U_{\alpha, \sigma}^{\beta, \sigma'} ( \tau - \tau' ) n_{\beta, \sigma'} ( \tau' )\nonumber\\
	&+\frac{1}{2} \sum_{\begin{subarray}{l} \alpha, \beta, \sigma,\sigma' \\ \alpha<\beta\end{subarray}}  \int_0^{\beta} d \tau \int_0^{\beta} d \tau'  J(\tau-\tau') \Big[X^{\alpha\beta}_\sigma(\tau)X^{\beta\alpha}_{\sigma'}(\tau')\nonumber\\
	&+X^{\beta\alpha}_\sigma(\tau)X^{\alpha\beta}_{\sigma'}(\tau')+X^{\alpha\beta}_\sigma(\tau)X^{\alpha\beta}_{\sigma'}(\tau')+X^{\beta\alpha}_\sigma(\tau)X^{\beta\alpha}_{\sigma'}(\tau')\Big],
\label{action_jret}
\end{align}
with $X^{\alpha\beta}_\sigma =  c^\dagger_{\alpha, \sigma} c_{\beta, \sigma}$ and retarded density-density interaction and Hund coupling parameters. 

A possible algorithm based on a double-expansion in the hybridization and Hund coupling has recently been presented in Ref.~\cite{Steiner2015}. The full problem with retarded spin-flips and pair-hoppings however suffers from a sign problem, so that in practice it seems necessary to treat these operators as instantaneous and keep the retardation only in the density-density component. Note that even the simulation of such a simplified model requires a double-expansion approach. To illustrate this, let us consider an on-site electron-boson coupling term of the form $H_\text{el-b}=\sum_{\alpha,\lambda} g_{\alpha,\lambda} n_{\alpha}(b^\dagger_{\lambda}+b_{\lambda})$ and the multi-orbital and multi-boson version of the Lang-Firsov transformation (\ref{dtilde_dag}), (\ref{dtilde}): $\tilde d^\dagger_{\alpha,\sigma} = e^{\sum_\lambda\frac{g_{\alpha,\lambda}}{\omega_\lambda}(b^\dagger_{\lambda}-b_{\lambda})}d^\dagger_{\alpha,\sigma}$, $\tilde d_{\alpha,\sigma} = e^{-\sum_\lambda\frac{g_{\alpha,\lambda}}{\omega_\lambda}(b^\dagger_{\lambda}-b_{\lambda})}d_{\alpha,\sigma}$. This transformation leaves density operators unchanged, $\tilde n_\alpha=n_\alpha$, while a pair hopping operator $d^\dagger_{\alpha,\uparrow}d^\dagger_{\alpha,\downarrow}d_{\beta\uparrow}d_{\beta,\downarrow}$ acquires a factor $e^{\sum_\lambda \frac{2(g_{\alpha,\lambda}-g_{\beta,\lambda})}{\omega_\lambda}(b^\dagger_\lambda-b_\lambda)}$, which for example in the case of a Jahn-Teller coupling ($g_\alpha=-g_\beta$) is nonzero. Hence, we cannot keep such non-density-density terms in the time-evolution operators, but need to expand in them.

\section{Towards a self-consistent description of screening}
\label{selfconsistent_screening}

In the previous sections, we have discussed the cRPA technique for
computing low-energy effective models for correlated materials, and
different approximate and exact schemes for treating the resulting
dynamically screened interactions within DMFT. We also mentioned the
fact that the elimination of low-energy screening processes in cRPA
generically leads to rather long-ranged interactions. In this section,
we would like to improve the DMFT description by (i) including the
screening effect from long-ranged Coulomb interactions, and (ii)
considering nonlocal correlations. First, we will focus on the single
orbital extended Hubbard model (Sec.~\ref{sub:UVhubbard}), which
allows us to introduce these 
advanced schemes in the simplest possible set-up, and then discuss how
some of these techniques can be implemented in an \emph{ab initio}
context (Sec.~\ref{subsec:realistic_materials}). The goal is to
provide a self-consistent description of 
screening and correlations within the low-energy window defined by the
cRPA downfolding, and in the future perhaps even within a larger
window containing some of the screening bands.  

\subsection{$U$-$V$ Hubbard model}
\label{sub:UVhubbard}

\subsubsection{Extended DMFT} 
\label{sec:edmft}

The screening from nonlocal Coulomb interactions can be described by the so-called extended DMFT (EDMFT) formalism \cite{Sengupta1995,Si1996,Sun2002,Ayral2013}. This method is still based on an effective single-site impurity model, but involves two self-consistently computed dynamical mean fields: the Weiss Green's function $\mathcal G_0(i\omega_n)$ (or hybridization function $\Delta(i\omega_n)$), which controls the hopping of electrons in and out of the impurity site, and the dynamical on-site interaction $\mathcal U(i\nu_n)$, which incorporates the effect of screening. While $\mathcal G_0$ is $\beta$-antiperiodic, $\mathcal U$ is $\beta$-periodic. We denote the fermionic Matsubara frequencies by $\omega_n$ and the bosonic Matsubara frequencies by $\nu_n$.   

For simplicity, we derive the EDMFT formalism for the $U$-$V$ Hubbard model
\begin{align}
H_\text{UV-Hubbard}&=-t\sum_{\langle ij\rangle\sigma}( d^\dag_{i\sigma} d^{}_{j\sigma}+d^\dag_{j\sigma} d^{}_{i\sigma})-\mu\sum_{i\sigma}n_{i\sigma}+U\sum_i n_{i\uparrow} n_{i\downarrow}
+\frac{V}{2}\sum_{\langle ij\rangle} n_i n_j,
\label{H_UV}
\end{align}
with $n_i=n_{i\uparrow}+n_{i\downarrow}$. Here, we assume that the hoppings and off-site interactions are between nearest-neighbor sites, although the generalization to arbitrary hoppings and longer range interactions is straightforward. We furthermore take the bare interactions $U$ and $V$ as static. If the $U$-$V$ Hubbard model is a low-energy effective theory derived from some downfolding procedure, then these bare parameters can themselves have a frequency-dependence. Again, the generalization of the following formalism to the case of frequency-dependent bare interactions is straightforward, and merely involves the replacement of $U$ by $U(\tau)$ or $U(i\nu_n)$ and similarly for $V$.

We start by writing the action of the lattice model in terms of Grassmann fields $d^*$, $d$ as
\begin{align}
S&=\int_0^\beta d\tau\Big[\sum_{ij\sigma} d^*_{i\sigma}(\tau)((\partial_\tau-\mu)\delta_{ij}+t_{ij})d_{j\sigma}(\tau)+U\sum_i n_{i\uparrow}(\tau)n_{i\downarrow}(\tau)+\frac{1}{2}\sum_{ij}v_{ij}n_i(\tau)n_{j}(\tau)\Big]\\
&=\int_0^\beta d\tau \Big[\sum_{ij\sigma} d^*_{i\sigma}(\tau)((\partial_\tau-\tilde \mu)\delta_{ij}+t_{ij})d_{j\sigma}(\tau)+\frac{1}{2}\sum_{ij}\tilde v_{ij}n_i(\tau)n_{j}(\tau)\Big].
\label{S_lattice_grassmann}
\end{align}
In the second expression, we have written the interaction contributions in terms of $\tilde v_{ij} = U\delta_{ij}+v_{ij}$, and shifted the chemical potential as $\tilde \mu=\mu+\frac{U}{2}$. Note that for the $U$-$V$ Hubbard model, $t_{ij}=-t\delta_{\langle ij\rangle}$ and $v_{ij}=V \delta_{\langle ij\rangle}$, but it is more convenient to use the general notation. 

Since we want to map the lattice model onto a single-site impurity model, we next decouple the interaction term by a Hubbard-Stratonovich transformation, thereby replacing the (on-site and off-site) interaction by a local coupling to a real, $\beta$-periodic field $\phi$ \cite{negele}:
\begin{align}
&\exp\Bigg[-\frac{1}{2}\int_0^\beta d\tau \sum_{ij}n_i(\tau)\tilde v_{ij} n_j(\tau)\Bigg]=((2\pi)^N\det v)^{-1/2}\int \mathcal{D}[\phi_1, \ldots, \phi_N] \nonumber\\
&\hspace{30mm}\times \exp\Bigg[-\int_0^\beta d\tau \Big\{ \frac{1}{2}\sum_{ij}\phi_i(\tau)(\tilde v^{-1})_{ij} \phi_j(\tau)+i\sum_j\phi_j(\tau)n_j(\tau)\Big\}\Bigg].
\label{HS}
\end{align}
After this decoupling, the action of the lattice model can be written as
\begin{align}
S&=\int_0^\beta d\tau \Big[-\sum_{ij\sigma} d^*_{i\sigma}(\tau)(G_0^{-1})_{ij}d_{j\sigma}(\tau)+\frac{1}{2}\sum_{ij}\phi_i(\tau)(\tilde v^{-1})_{ij} \phi_j(\tau)+i\sum_j\phi_j(\tau)n_j(\tau) \Big],
\end{align}
where we have introduced the inverse of the noninteracting lattice Green's function, $(G_0^{-1})_{ij}=((-\partial_\tau+\tilde \mu)\delta_{ij}-t_{ij})$ to simplify the first term. In EDMFT, this lattice model is self-consistently mapped onto an impurity model with action
\begin{align}
S_{\text{EDMFT}}^\phi =&-\int_0^\beta d\tau d\tau' \sum_\sigma d^*_{\sigma}(\tau) \mathcal{G}^{-1}(\tau-\tau') d_{\sigma}(\tau')+\frac{1}{2}\int_0^\beta d\tau d\tau' \phi(\tau)\mathcal{U}^{-1}(\tau-\tau')\phi(\tau')\nonumber\\
&+i\int_0^\beta d\tau \phi(\tau)n(\tau).
\label{S_EDMFT_phi}
\end{align}
A detailed derivation can be found in Refs.~\cite{Sun2002} and
\cite{Ayral2013}. For the present purpose it suffices to note that in
addition to the fermionic Weiss field $\mathcal G$, there also appears
a $\beta$-periodic Weiss field $\mathcal U$, which has to be adjusted
in such a way that the local dynamics of the Hubbard-Stratonovich
field is well reproduced. In the self-consistency loop, one computes
the impurity Green's functions 
\begin{align}
G_\text{imp}&=-\langle \mathcal{T} d(\tau)d^*(0)\rangle_{S_\text{EDMFT}},\label{G_EDMFT}\\
W_\text{imp}&=\langle \mathcal{T} \phi(\tau)\phi(0)\rangle_{S_\text{EDMFT}},\label{W_EDMFT}
\end{align}
and identifies them with the corresponding local lattice Green's functions
$G_{i,i}^\text{latt}=-\langle \mathcal{T} d_i(\tau) d_i^*(0)\rangle_{S_\text{latt}}$
and $W_{i,i}^\text{latt}=\langle \mathcal{T} \phi_i(\tau) \phi_i(0)\rangle_{S_\text{latt}}$, where the latter are computed using a local approximation for the lattice self-energies $\Sigma$ and $P$:
\begin{align}
\Sigma(\vc{k},i\omega_n)\approx \Sigma_\text{imp}(i\omega_n),& \quad \Sigma_\text{imp}(i\omega_n)={\mathcal G}^{-1}(i\omega_n)-G_\text{imp}(i\omega_n),\\
P(\vc{k},i\nu_n)\approx P_\text{imp}(i\nu_n),& \quad P_\text{imp}(i\nu_n)={\mathcal U}^{-1}(i\nu_n)-W_\text{imp}(i\nu_n).
\end{align}
Physically, $P(\vc{k},i\nu_n)$ ($P_\text{imp}(i\nu_n)$) represents the lattice (impurity) polarization. 
For the actual calculations, it is convenient to integrate out the $\phi$-field from the impurity action (\ref{S_EDMFT_phi}), to obtain
\begin{align}
S_{\text{EDMFT}} =&-\int_0^\beta d\tau d\tau' \sum_\sigma d^*_{\sigma}(\tau) \mathcal{G}^{-1}(\tau-\tau') d_{\sigma}(\tau')\nonumber\\
&+\frac{1}{2}\int_0^\beta d\tau d\tau' n(\tau)\mathcal{U}(\tau-\tau')n(\tau')-\frac{1}{2}\text{Tr}\ln \mathcal{U}.
\label{S_EDMFT}
\end{align}
This impurity problem with retarded density-density interaction can be solved for example using the hybridization expansion Monte Carlo method discussed in Sec.~\ref{ctqmc}. While the fermionic Green's function~(\ref{G_EDMFT}) can be measured directly, the evaluation of the bosonic Green's function~(\ref{W_EDMFT}) requires an intermediate step. From Eq.~(\ref{S_EDMFT_phi}) it follows that $\frac{\delta \text{ln}Z}{\delta \mathcal{U}^{-1}}=-2\mathcal U \frac{\delta \text{ln}Z}{\delta \mathcal{U}} \mathcal U$, while Eq.~(\ref{S_EDMFT}) implies $\frac{\delta \text{ln}Z}{\delta \mathcal{U}}=\frac{1}{2}\langle \mathcal{T} n(\tau)n(0)\rangle_{S_\text{EDMFT}}-\frac{1}{2\mathcal U}$. Combining the two expressions yields the measurement formula
\begin{align}
W_\text{imp}&=\mathcal{U}-\mathcal{U}\chi_\text{loc}\mathcal{U}, \quad \chi_\text{loc}=\langle \mathcal{T} n(\tau)n(0)\rangle_{S_\text{EDMFT}}.
\label{W_chi}
\end{align}
The density density correlation function $\chi_\text{loc}$ can be easily evaluated in the hybridization expansion Monte Carlo method discussed in Sec.~\ref{ctqmc}.  

The fermionic self-consistency loop in an EDMFT calculation is identical to usual DMFT:
\begin{enumerate}
\item Compute $G_\text{imp}(i\omega_n)$ for the given $S_\text{EDMFT}$,
\item Extract fermionic self-energy: $\Sigma_\text{imp}(i\omega_n)=\mathcal G^{-1}_0(i\omega_n)-G^{-1}_\text{imp}(i\omega_n)$,
\item Use DMFT approximation $\Sigma(\vc{k},i\omega_n)=\Sigma_\text{imp}(i\omega_n)$ to compute the local lattice Green's function $G_\text{loc}(i\omega_n)=\int (d\vc{k})[(G_0)^{-1}-\Sigma_\text{imp}(i\omega_n)]^{-1}$,
\item Use DMFT self-consistency condition $G_\text{loc}(i\omega_n)=G_\text{imp}(i\omega_n)$ to define a new Weiss Green's function $\mathcal G^{-1}_{0}(i\omega_n)=G^{-1}_\text{loc}(i\omega_n)+\Sigma_\text{imp}(i\omega_n)$,
\end{enumerate}
while the bosonic self-consistency loop is analogous, 
\begin{enumerate}
\item Compute $W_\text{imp}(i\nu_n)$ for the given $S_\text{EDMFT}$ (Eq.~(\ref{W_chi})),
\item Extract bosonic self-energy: $P_\text{imp}(i\nu_n)=\mathcal U^{-1}(i\nu_n)-W^{-1}_\text{imp}(i\nu_n)$,
\item Use DMFT approximation $P(\vc{k},i\nu_n)=P_\text{imp}(i\nu_n)$ to compute the local lattice Green's function $W_\text{loc}(i\nu_n)=\int (d\vc{k})[\frac{1}{2}(\tilde v)^{-1}-P_\text{imp}(i\nu_n)]^{-1}$,
\item Use DMFT self-consistency condition $W_\text{loc}(i\nu_n)=W_\text{imp}(i\nu_n)$ to define a new retarded interaction $\mathcal U^{-1}(i\nu_n)=W^{-1}_\text{loc}(i\nu_n)+P_\text{imp}(i\nu_n)$.
\end{enumerate}
In an EDMFT calculation, these two loops are typically solved in parallel, i.e., both the Weiss Green's function $\mathcal G$ and the  retarded interaction $\mathcal{U}$ is updated before the next impurity calculation is started. 
We finally remark that in the case of frequency dependent $U$ and $V$,
the bosonic Dyson equation which allows to update $W_\text{loc}$ from
$P_\text{imp}$ will have frequency dependent $(\tilde
v)^{-1}$. Therefore, retarded $U$ and $V$ are readily included in this
framework, as already mentioned before.

\subsubsection{GW+DMFT}
\label{sec:gw+dmft}

We next discuss the implementation of the GW+DMFT method for the $U$-$V$ Hubbard model (\ref{H_UV}), for which fully self-consistent calculations have recently been implemented \cite{Ayral2012,Ayral2013,Huang2014}. GW+DMFT is based on the EDMFT framework, but involves momentum dependent fermionic and bosonic self-energies, which are obtained by combining the (local) EDMFT self-energies with the nonlocal components of the GW self-energies:
\begin{align}
\label{UV_total_sigma}
\Sigma^\text{GW+DMFT}_{jk}(i\omega_n)&=\Sigma^\text{EDMFT}_{jj}(i\omega_n)\delta_{jk}+\Sigma^{GW}_{jk}(i\omega_n)(1-\delta_{jk}),\\
\label{UV_total_polarization}
P^\text{GW+DMFT}_{jk}(i\nu_n)&=P^\text{EDMFT}_{jj}(i\nu_n)\delta_{jk}+P^{GW}_{jk}(i\nu_n)(1-\delta_{jk}).
\end{align}  
We note that this is not the only combination which avoids a double
counting of self-energy diagrams. In fact, the subtraction of all the
local GW diagrams also removes contributions (e.~g. with nonlocal
polarization bubbles) which are not accounted for in the EDMFT
self-energy. An alternative strategy would be to remove the subset
of GW diagrams which contains only local propagators:
$\Sigma^\text{GW+DMFT}_{jk}(i\omega_n)=\Sigma^\text{EDMFT}_{jj}(i\omega_n)\delta_{jk}+\Sigma^{GW}_{jk}(i\omega_n)-\Sigma^{GW}_{jj}[G_{ii}](i\omega_n)\delta_{jk}$,
and similarly for $P$. In the case of the two-dimensional Hubbard model in the
weak-coupling regime, 
both double counting corrections were found to produce similar results 
\cite{Gukelberger2015}, but
in more general situations, 
the effect of different double counting schemes  
has not yet been studied
systematically. 

The computational steps in the self-consistent GW+DMFT calculation are the following:
\begin{enumerate}
\item Start, e. g., from the converged EDMFT solution
  ($\Sigma_\text{imp}(i\omega_n)$, $P_\text{imp}(i\nu_n)$), 
and define approximate lattice self-energies:
\begin{equation}
\Sigma(\vc{k},i\omega_n) = \Sigma_\text{imp}(i\omega_n), \quad P(\vc{k},i\nu_n) = P_\text{imp}(i\nu_n),
\end{equation}

\item Update the lattice Green's functions:
\begin{align}
G(\vc{k},i\omega_n)&=[G_0^{-1}(\vc{k},i\omega_n)-\Sigma(\vc{k},i\omega_n)]^{-1},\\
W(\vc{k},i\nu_n)&=[\tfrac12 \tilde v(\vc{k})^{-1}-P(\vc{k},i\nu_n)]^{-1},
\end{align}

\item Compute the local lattice Green's functions and the new Weiss fields $\mathcal{G}_0$ and $\mathcal U$:
\begin{align}
G_\text{loc}(i\omega_n) = \int (d\vc{k}) G(\vc{k},i\omega_n) \quad &\rightarrow \quad \mathcal{G}_0^{-1}(i\omega_n)=G_\text{loc}(i\omega_n)^{-1}+\Sigma_\text{imp}(i\omega_n),\\
W_\text{loc}(i\nu_n) = \int (d\vc{k}) W(\vc{k},i\nu_n) \quad &\rightarrow \quad \mathcal{U}^{-1}(i\nu_n)=W_\text{loc}(i\nu_n)^{-1}+P_\text{imp}(i\nu_n),
\end{align}
\item Solve the impurity problem, i.e. compute $G_\text{imp}$ and $\chi_\text{loc}\rightarrow W_\text{imp}=\mathcal U-\mathcal U \chi_\text{loc}\mathcal{U}$. Compute the fermionic and bosonic self-energies
\begin{align}
\Sigma_\text{imp}(i\omega_n)&=\mathcal{G}_0^{-1}(i\omega_n)-G_\text{imp}^{-1}(i\omega_n),\\
P_\text{imp}(i\nu_n)&=\mathcal{U}^{-1}(i\nu_n)-W_\text{imp}^{-1}(i\nu_n),
\end{align}
\item Calculate the GW+DMFT self-energies:
\begin{itemize}
\item Calculate the GW self-energies,
\begin{align}
\Sigma^{GW}(\vc{k},i\omega_n)&=-\frac{T}{N_\vc{k}}\sum_\vc{q}\sum_{\nu_m} G(\vc{q},i\omega_n-i\nu_m)W(\vc{k}-\vc{q},i\nu_m),\label{GW-selfenergy}\\
P^{GW}(\vc{k},i\nu_n)&=2\frac{T}{N_\vc{k}}\sum_\vc{q} \sum_{\omega_m} G(\vc{q},i\omega_m)G(\vc{q}-\vc{k},i\omega_m-i\nu_n),\label{P_GW}
\end{align}
\item Extract the nonlocal parts,
\begin{align}
\Sigma^{GW}_\text{nonlocal}(\vc{k},i\omega_n)&=\Sigma^{GW}(\vc{k},i\omega_n)-\int (d\vc{k}) \Sigma^{GW}(\vc{k},i\omega_n),\label{GW-nonlocal}\\
P^{GW}_\text{nonlocal}(\vc{k},i\nu_n)&=P^{GW}(\vc{k},i\nu_n)-\int (d\vc{k}) P^{GW}(\vc{k},i\nu_n),
\end{align}
\item Combine GW and EDMFT self-energies 
\begin{align} 
\Sigma(\vc{k},i\omega_n)&=\Sigma_\text{imp}(i\omega_n)+\Sigma^{GW}_\text{nonlocal}(\vc{k},i\omega_n),\\
P(\vc{k},i\nu_n)&=P_\text{imp}(i\nu_n)+P^{GW}_\text{nonlocal}(\vc{k},i\nu_n),
\end{align}
\end{itemize} 
\item Go back to (ii) until convergence is reached.
\end{enumerate}

Note that the GW self-energy defined in Eq.~(\ref{GW-selfenergy}) contains a Hartree contribution involving the bare local interaction. This contribution is however removed in the definition of $\Sigma^{GW}_\text{nonlocal}$ (Eq.~(\ref{GW-nonlocal})).

\begin{figure}[t]
\begin{center} 
\includegraphics[angle=0, width=0.47\columnwidth]{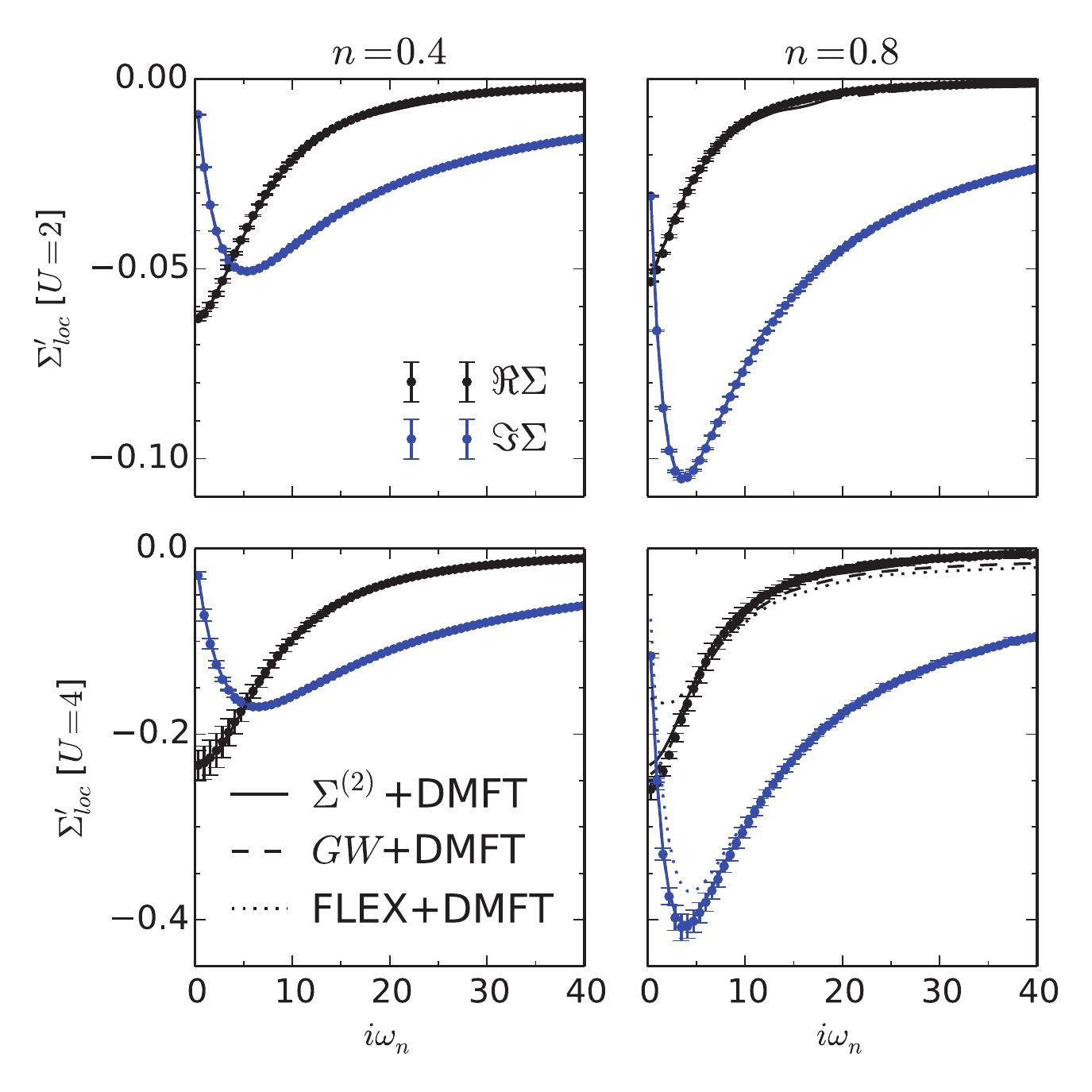}\hfill
\includegraphics[angle=0, width=0.47\columnwidth]{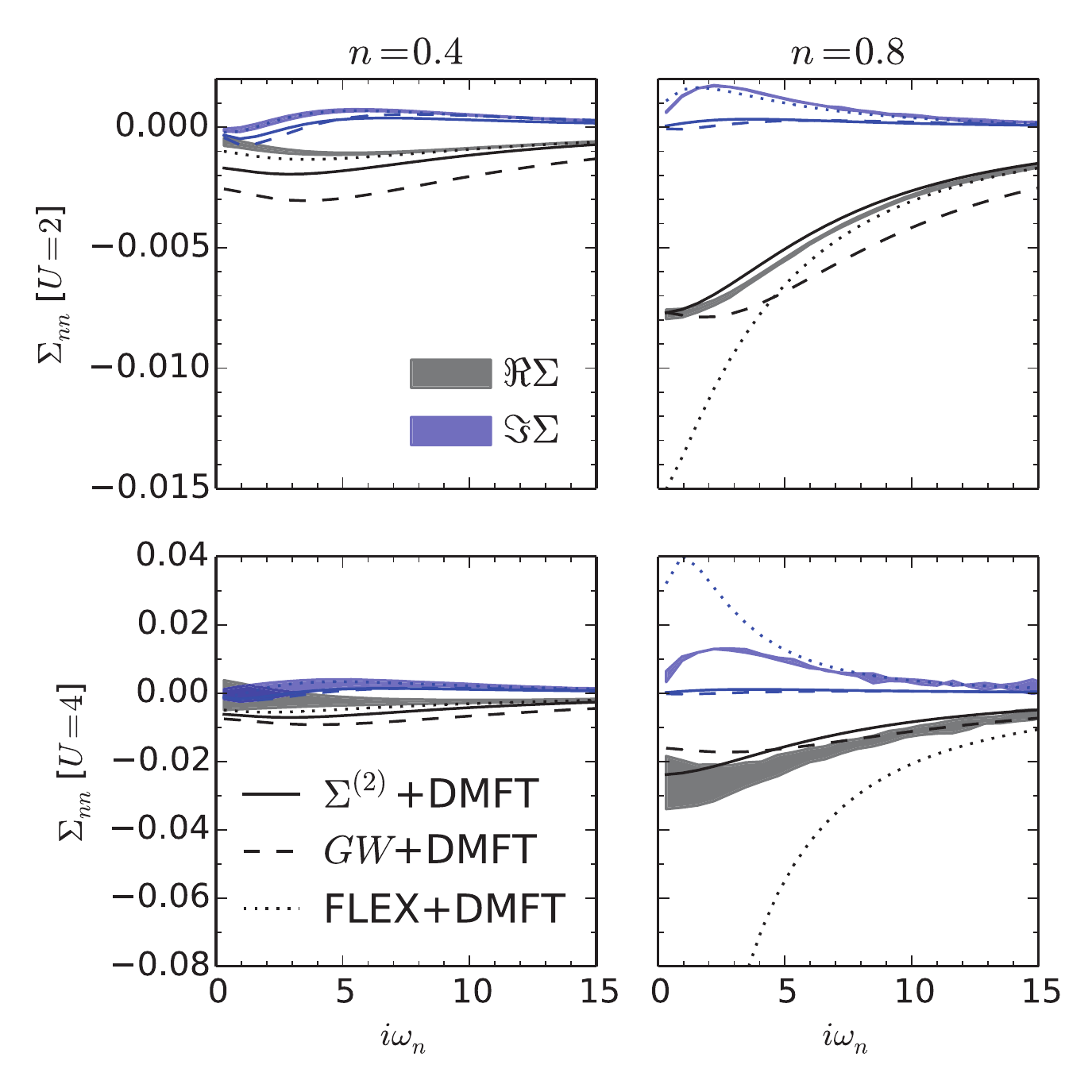}
\caption{Comparison of the local and nonlocal self-energies from different many-body perturbation theory + DMFT approximations to the numerically exact diagrammatic Monte Carlo results for the square lattice Hubbard model with bandwidth $8$. The left panels show the imaginary part of the local self-energy for indicated values of the interaction $U$ and filling $n$. The error bars are estimated from different cutoff orders in the diagrammatic sampling. The right panels show the imaginary part of the non-local self-energy for nearest neighbor sites, with the gray and blue shaded bands corresponding to the diagrammatic Monte Carlo result. Note the much smaller y-axis scale compared to the left panels. (From Ref.~\cite{Gukelberger2015}.) }   
\label{diagmc}
\end{center}
\end{figure}

An interesting question is how accurately the GW+DMFT method captures the momentum and energy dependence of the nonlocal self-energies. While a complete picture is lacking, some systematic tests against numerically exact diagrammatic Monte Carlo results \cite{Prokofev2007} have recently been performed for the two-dimensional square-lattice Hubbard model in the weak-coupling regime \cite{Gukelberger2015}. It was shown that the DMFT approximation provides a very good description of $\Sigma_\text{loc}(i\omega_n)$, and that the GW+DMFT result is of comparable accuracy (Fig.~\ref{diagmc}). While the nonlocal components are of the correct order of magnitude in the weak-coupling regime, their relative errors are large. Apparently, the GW approximation does not capture the correct momentum dependence at weak $U$ and away from half-filling, and the result can only be expected to get worse in the intermediate coupling regime and closer to half-filling. In particular, GW+DMFT does not reproduce the strong differentiation between node and antinode which is found in cluster DMFT simulations \cite{Werner2009,Gull2009} in the intermediate coupling regime. Figure~\ref{diagmc} also shows the comparison to alternative many-body perturbation theory + DMFT schemes, namely the bare second order perturbation theory ($\Sigma^{(2)}$) + DMFT \cite{Sun2002} and the fluctuation exchange approximation (FLEX) + DMFT methods. They are of similar accuracy, but also fail to correctly capture the nonlocal components. $\Sigma^{(2)}$+DMFT at least ensures the correct high-frequency behavior of the local self-energy. 

While GW+DMFT produces rather poor results for the momentum dependence of the two-dimensional Hubbard model, it should be kept in mind that (i) DMFT based methods are by construction most appropriate for high-dimensional systems, and (ii) that the main advantage of the GW+DMFT lies in the self-consistent description of the screening, and thus in the possibility to self-consistently compute the appropriate ``Hubbard-$U$" parameters in an \emph{ab initio} simulation.    

Finally, let us note that 
the factor of two in the GW polarization (\ref{P_GW}) comes from the sum over spin orientations. For the Hubbard model, with its spin-dependent instantaneous on-site interaction, the RPA polarization diagrams should in fact only include odd numbers of bubbles with alternating spin. To avoid unphysical diagrams, one should implement a spin-dependent GW formalism, which involves $2\times 2$ matrices in spin space. While the polarization $P$ is diagonal, the Hubbard interaction becomes an off-diagonal matrix $U\sigma_x$. Therefore, in a spin-dependent GW calculation for the Hubbard model, the self-energy $\Sigma$ is constructed with the following diagonal element of the screened interaction:
\begin{equation}
W_{\sigma\sigma}(\vc{k},i\nu_n)=\frac{U^2P(\vc{k},i\nu_n)}{1-[UP(\vc{k},i\nu_n)]^2}.
\end{equation} 
Systematic tests of the spin-dependent and spin-independent GW schemes for the two-dimensional Hubbard model \cite{Gukelberger2015} have shown that the spin-dependent formulation indeed cures the most obvious deficiencies of the spin-independent scheme. However, in models with nonlocal interactions and realistic material simulations within GW or GW+DMFT, this issue becomes less relevant.

\subsubsection{Dual boson}
\label{subsec:dual_boson}

\begin{figure}[t]
\begin{center}
\includegraphics[angle=0, width=0.8\columnwidth]{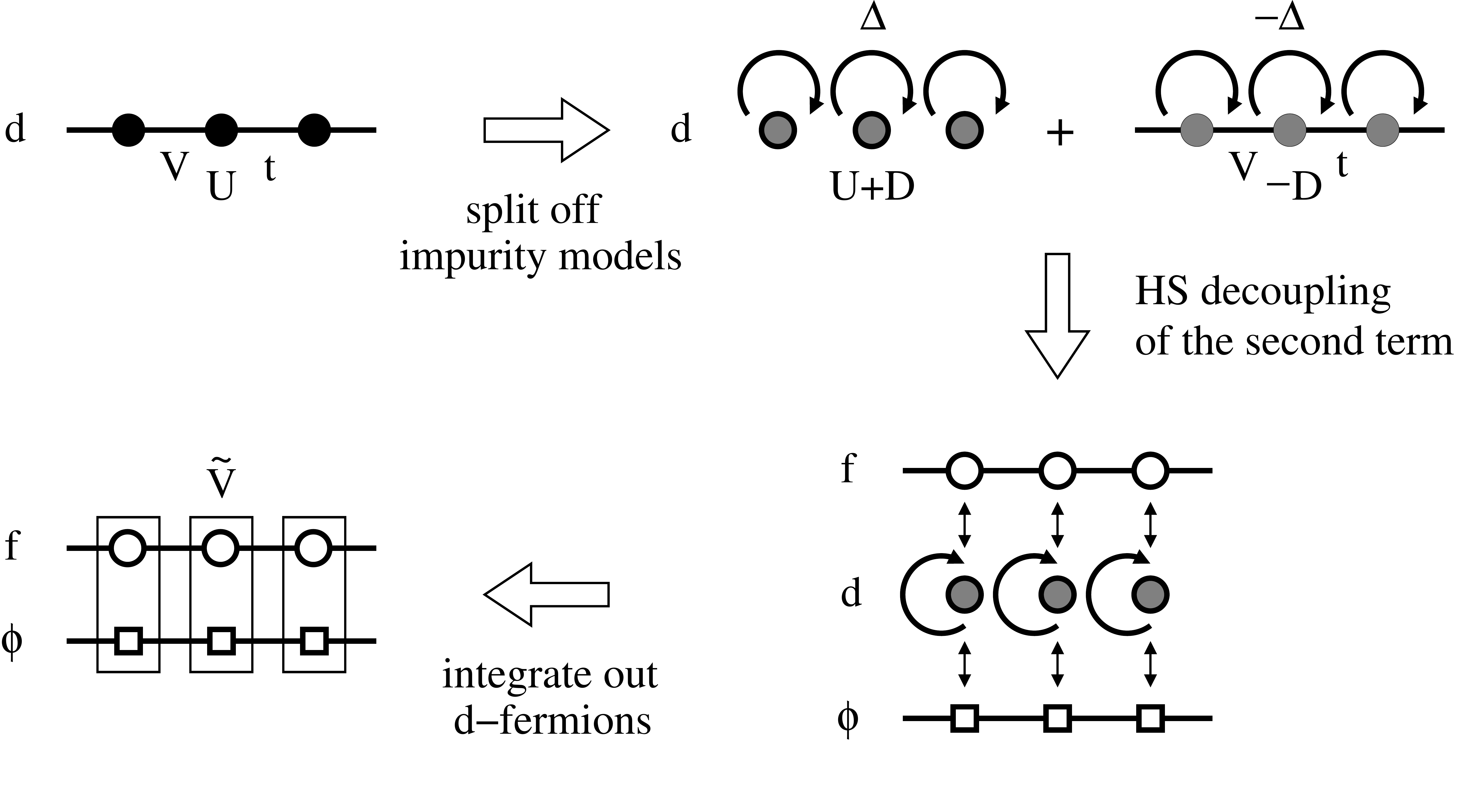}\hfill
\caption{Illustration of the dual boson formalism. In the first step,
  the lattice model is split into a collection of impurity models with
  retarded interaction $D$ and a rest. In the second step, the rest
  term is decoupled by Hubbard-Stratonovich transformations, which
  introduce a local coupling to dual fermions ($f$) and dual bosons
  ($\phi$). In the last step, the original $d$-fermions are integrated
  out, which yields a dual theory with an interaction $\tilde V$, which
  is of arbitrary order in $f$, $f^*$ and $\phi$.
}   
\label{fig_db}
\end{center}
\end{figure}

The dual boson formalism \cite{Rubtsov2012,Stepanov2015} is a systematic extension of EDMFT, which incorporates momentum-dependent correlations and enables a consistent description of collective excitations. This method is still under active development, so we will content ourselves here with a sketch of the main ideas, and references to the original papers, where the mathematical details can be found. As 
in the previous sections, 
we will consider the $U$-$V$ Hubbard model and start the discussion by rewriting the Grassmann path-integral for the lattice action (\ref{S_lattice_grassmann}) in the Matsubara formalism:
\begin{align}
S=&-T\sum_{j\omega_n\sigma} d^*_{j\sigma}(i\omega_n)[i\omega_n+\tilde \mu] d_{j\sigma}(i\omega_n)+T\sum_{\langle jl\rangle \omega_n\sigma} t_{jl} d^*_{j\sigma}(i\omega_n) d_{l\sigma}(i\omega_n)\nonumber\\
&+\frac{UT}{2}\sum_{j\nu_n} n_j(i\nu_n)n_j(i\nu_n)+\frac{VT}{2}\sum_{\langle jl\rangle \nu_n} n_j(-i\nu_n)n_l(i\nu_n),
\label{S_matsubara}
\end{align}
where the angular brackets denote the sum over nearest neighbors, and we have split $\tilde v_{ij}=U\delta_{ij}+V\delta_{\langle ij\rangle}$ into the on-site and nearest-neighbor contributions.  
We next rewrite Eq.~(\ref{S_matsubara}) as a sum of EDMFT-type impurity actions and a rest
\begin{align}
S &= \sum_j S_{\text{imp},j}+S_\text{rest},\\
S_\text{imp}&=-T\sum_{\omega_n\sigma}d^*_\sigma(i\omega_n)[i\omega_n+\tilde \mu-\Delta_\sigma(i\omega_n)]d_\sigma(i\omega_n)+\frac{T}{2}\sum_{\nu_n}n(i\nu_n)[U+D(i\nu_n)]n(i\nu_n),
\end{align}
with at this stage an unspecified hybridization function $\Delta$ and retarded interaction $D$ (see illustration in Fig.~\ref{fig_db}). In the next step, $S_\text{rest}$, which contains hopping and hybridization terms, the nonlocal interactions and a local retarded interaction, is decoupled using Hubbard-Stratonovich transformations. The decoupling of the interactions is analogous to EDMFT (Eq.~(\ref{HS})) and introduces the bosonic fields $\phi_j$, which in the present context are called `dual bosons'. At the same time, the fermionic hopping and hybridization terms in $S_\text{rest}$ are decoupled by an appropriate Hubbard-Stratonovich transformation, which introduces auxiliary fermions $f_j$, called `dual fermions' \cite{Rubtsov2008}, and replaces the nonlocal quadratic term in the $d$-operators by a local coupling between $d$- and $f$-fermions (see bottom right panel of Fig.~\ref{fig_db}). In the final step, the $d$-electrons are integrated out, which generates a dual action for the $f$ and $\phi$ variables, with a complicated interaction $\tilde V(f_j,f^*_j, \phi_j)$, which can be related to (high-order) vertices of the impurity model $S_\text{imp}$. The explicit form of the dual action is

\begin{align}
\tilde S &=-T\sum_{\vc{k}\omega_n\sigma} f^*_{\vc{k}\sigma}(i\omega_n) \mathcal{\tilde G}_{0,\vc{k}\sigma}^{-1} (i\omega_n)f_{\vc{k}\sigma}(i\omega_n)-\frac{T}{2}\sum_{\vc{k}\nu_n} \phi_\vc{k}(i\nu_n)\mathcal{\tilde X}^{-1}_{0,\vc{k}}(i\nu_n)\phi_{\vc{k}}(i\nu_n)+\sum_j \tilde V(f_j, f^*_j,\phi_j),
\label{S_db}
\end{align}
where the bare dual propagators, expressed in terms of the impurity Green's function $g(i\omega_n)=-\langle d^*(i\omega_n)d(i\omega_n)\rangle_{S_\text{imp}}$ and impurity charge susceptibility $\chi(i\nu_n)=-\langle n(i\nu_n)n(-i\nu_n)\rangle_{S_\text{imp}}$ are \cite{Stepanov2015}
\begin{align}
\mathcal{\tilde G}_{0,\vc{k}\sigma} (i\omega_n) &= [g^{-1}(i\omega_n)+\Delta(i\omega_n)-\epsilon_\vc{k}]^{-1}-g(i\omega_n), \\
\mathcal{\tilde X}_{0,\vc{k}}(i\nu_n) &= [\chi(i\nu_n)^{-1}+D(i\nu_n)-V_\vc{k}]^{-1}-\chi(i\nu_n),
\end{align}
with $\epsilon_\vc{k}$ and $V_\vc{k}$ the Fourier transforms of the hopping and nearest neighbor interaction. 

The dual action (\ref{S_db}) is treated in perturbation theory, where in practice one only retains low-order or ladder-type diagrams in $\tilde V$. The main idea is to exploit the freedom of choosing $\Delta(i\omega_n)$ and $D(i\nu_n)$, and to define these quantities in such a way that the strong correlation effects are captured at the level of the impurity model $S_\text{imp}$, which can be solved exactly using the technique described in Sec.~\ref{ctqmc}, so that only weaker correlations must be treated by the dual perturbation theory. For example, the EDMFT solution can be incorporated as the zeroth order of this dual perturbation theory, in which case the dual boson formalism becomes a systematic expansion around EDMFT. Better choices for $\Delta$ and $D$,
which take into account a feedback from the lattice solution onto the impurity problem 
 may exist, as has been recently discussed in Ref.~\cite{Stepanov2015}.   

From the dual perturbation theory, one obtains a dual self-energy $\tilde \Sigma_\vc{k}(i\omega_n)$ and a dual polarization $\tilde P_{\vc{k}}(i\nu_n)$. These can then be used to obtain the lattice Green's function $G_\vc{k}(i\omega_n)$ and lattice susceptibility $\mathcal{X}_\vc{k}(i\nu_n)$ of the original $d$-fermions:
\begin{align}
G_\vc{k}^{-1}(i\omega_n) &= [g(i\omega_n)+g(i\omega_n)\tilde \Sigma_\vc{k}(i\omega_n)g(i\omega_n)]^{-1}+\Delta(i\omega_n)-\epsilon_\vc{k},\\
\mathcal{X}^{-1}_\vc{k}(i\nu_n) &= [\chi(i\nu_n)+\chi(i\nu_n)\tilde P_{\vc{k}}(i\nu_n)\chi(i\nu_n) ]^{-1}+D(i\nu_n)-V_\vc{k}.
\end{align}
While an appropriately formulated dual boson theory is self-consistent both on the single-particle and two-particle level \cite{Stepanov2015}, in contrast to GW+DMFT, this appealing feature comes at the cost of having to calculate and manipulate vertex functions. This makes it challenging to apply this formalism to realistic multiband systems.

\subsection{Realistic materials}
\label{subsec:realistic_materials}

\subsubsection{GW+DMFT}
\label{sec:realistic_materials_gw+dmft}

The full implementation of the
GW+DMFT scheme as proposed in Ref.~\cite{Biermann2003} is doable in practice
only for simple Hamiltonians, such as the $U$-$V$ extended
Hubbard model discussed above. This represents already a significant step
forward with respect to the situation just a few years ago, when
only \emph{static} and \emph{non-self-consistent} GW+DMFT applications had
been performed. The methods which allow to treat the
dynamically screened nature of $\mathcal U$, inherent in the GW+DMFT
formalism, and to determine it self-consistently in some particular cases,
have been discussed in Secs.~\ref{holstein_hubbard} and \ref{sec:generalU}.
However, in an \emph{ab initio} framework, the local dynamical impurity
problem is too large to be solved in a reasonable computer
time. Indeed, the local basis set of an \emph{ab initio} Hamiltonian
can be very large, and in the genuine GW+DMFT formulation \emph{all}
local orbitals should be taken into account in the embedded site, as they all
contribute to the screening of the effective local interaction $\mathcal U$. 

A practical GW+DMFT implementation recently introduced is the so-called
``orbital-separated'' scheme, where only the correlated orbitals are
kept in the impurity problem, whose size becomes then manageable by
state-of-the-art CTQMC solvers,
provided the low energy model is a single-site multiorbital system, and not a cluster. 
On one hand, it is reasonable to
include only the most correlated orbitals in the impurity problem. On the other
hand, separating the self-energy contribution of the local, correlated manifold from the non-local \emph{or} non-correlated one
has no unique solution, 
as there is no rigorous free energy functional which
generates this separation unambiguously, as discussed in Ref.~\cite{Tomczak2014}. 
Therefore, in defining the local self energy, 
one needs to make an \emph{ad hoc} choice. The one which seems the most general is 
\begin{equation}
\Sigma^\text{xc}(\vk,\iomn)_{LL'} = \Sigma_{GW}^\text{xc}(\vk,\iomn)_{LL'}
- \sum_\vk [\Sigma_{GW}^{\text{xc},d}(\vk,\iomn)]_{LL'}
+ [\Sigma^{\text{xc},d}_\text{imp}(\iomn)]_{LL'},
\label{ab_initio_total_sigma}
\end{equation}
where $L$ is the full-orbital index, and $d$ denotes the projection onto
the low-energy correlated space. The corresponding equation for the
total polarization is
\begin{equation}
P(\vk,\inun)_{\alpha\beta} = P_{GW}(\vk,\inun)_{\alpha\beta}
- \sum_\vq [P_{GW}^d(\vk,\inun)]_{\alpha\beta}+ [P_\text{imp}^{d}(\inun)]_{\alpha\beta},
\label{ab_initio_total_polarization}
\end{equation}
where $\alpha$ and $\beta$ are indices of a two-particle basis, constructed
from the full one-body basis set $L$. Equations~(\ref{ab_initio_total_sigma})
and (\ref{ab_initio_total_polarization}) are analogous to
Eqs.~(\ref{UV_total_sigma}) and (\ref{UV_total_polarization}), written
for the GW+DMFT calculation of the $U$-$V$ Hubbard model. 
$\Sigma^{\text{xc},d}_\text{imp}$ and $P_\text{imp}^{d}$ are the (fermionic) self-energy and
polarization (bosonic self-energy) computed as solutions of the impurity
problem in the correlated local basis (see the extended DMFT
description in Sec.~\ref{sec:edmft}). From
$\Sigma^\text{xc}(\vk,\iomn)_{LL'}$ and $P(\vk,\inun)_{\alpha\beta}$ one
obtains the dressed Green's function $G(\vk,\iomn)_{LL'}$ and the
fully screened interaction $W(\vk,\inun)_{\alpha\beta}$ by
standard procedures. As in step (ii) of the GW+DMFT loop in
Sec.~\ref{sec:gw+dmft}, one then computes the ``Weiss'' fields
$\mathcal G_0$ and $\mathcal U$, after projection of the local $G^\textrm{loc}$ and
$W^\textrm{loc}$ onto $d$. Thus, the orbital-separated framework
follows the usual GW+DMFT self-consistency loop, where the
non-perturbative many-body
solution is provided only in a correlated subspace, and the
convergence is reached when the local lattice Green's function
and polarization projected to the correlated subspace become identical to the
impurity Green's function and impurity polarization, respectively.

Significant effort has also been devoted to the development of simplified and
numerically more efficient approaches, based on some
approximations. On the one hand, the double self-consistency in the Green's
function and polarization has been replaced by only one based on the Green's
function.  
In these calculations, the polarization is frozen to the cRPA value, but the corresponding retarded interaction is kept in the impurity problem. 
This approximate scheme is detailed in Sec.~\ref{sec:URPA}.
On the other hand, a series of approximations has been proposed, 
which simplify the frequency \emph{and} spatial dependence of
the GW self-energy, which is one of the heaviest ingredients to compute.  
We will give a short survey of these simplified 
methods, which in order of increasing complexity are 
(i) SEX+DMFT (Sec.~\ref{sec:sex+dmft}), (ii) DMFT@nonlocal-GW (Sec.~\ref{sec:dmft@nonlocal-GW}), and (iii) quasi-particle self-consistent GW (QSGW) + DMFT (Sec.~\ref{sec:qsgw+dmft}).

\subsubsection{Frozen polarization: $\mathcal U$  replaced by the cRPA $U(\omega)$}
\label{sec:URPA}

Instead of
computing explicitly $P_\textrm{imp}$ to update the
bosonic Weiss field $\mathcal U$ at each GW+DMFT iteration, 
one can approximate it by its RPA value,
i.e. $P_\textrm{imp}=2 G^\textrm{loc,d} G^\textrm{loc,d}$, 
with $G^\textrm{loc,d}$ the local starting Green's function, taken from LDA
and projected onto the correlated manifold \cite{Tomczak2012,Tomczak2014}. Therefore, the total
polarization in Eq.~(\ref{ab_initio_total_polarization}) can be
written as $P = 2 G_\textrm{LDA} G_\textrm{LDA}$. Its value is frozen
during the self-consistency cycle, which is performed only on $G$. This also 
implies that the interaction $\mathcal U$ of the impurity model is frozen. Moreover, instead of
evaluating this interaction as
$\mathcal{U}^{-1}(i\nu_n)=W_\text{loc}(i\nu_n)^{-1}+P_\text{imp}(i\nu_n)$
(step (ii) of the GW+DMFT loop),
which involves \emph{local} quantities only, $\mathcal{U}$ is calculated as 
\begin{equation}
\mathcal{U} = \left[ \sum_q [ W^{-1} + P^d ]^{-1} \right]_d,
\label{simplified_GW_DMFT_U}
\end{equation}
where $W$ is the fully screened interaction. Hence, $W$ is undressed by $P^d$,
which is the RPA polarization function containing electron-hole
processes in the $d$ manifold only. Note that the ``locality''
operation ($\sum_q$) is performed after undressing $W$, and
the matrix is projected onto the $d$-manifold only at the end. 
Equation~(\ref{simplified_GW_DMFT_U}) is the partially screened
cRPA value of
$U_{mn}(\vc{0},\omega)$ in Eq.~(\ref{screened_U_ab_initio}), where $n$,
$m$ are indices of the $d$ subspace. Therefore, in this approach,  
$\mathcal U$ is kept fixed at the cRPA value. 
At a 
first glance, this might seem a very rough approximation, with
respect to the double loop on $G$ and $W$. 
However, if compared to available electron energy loss
spectroscopy (EELS) measurements, 
the RPA polarization function computed from the LDA band
structure looks usually reasonable.
This suggests that the cRPA estimate of $\mathcal U$ is
quantitatively correct, particularly at not-so-low frequencies, where the
interplay with the low-energy correlated manifold is supposed to be
weak.
In the impurity calculation, 
the 
cRPA 
frequency dependence of $\mathcal U$ is taken into account
and the self-consistent solution for $G$ is hence affected by
retardation effects contained in the impurity model.

As far as $\Sigma_{GW}^\text{xc}$ is concerned, one needs to 
carry out a one-shot GW calculation on top of the LDA band
structure to compute the initial non-local 
self-energy part. In the simplified implementation based on a constant $P$,
the non-local part does not change, while the local part is changed according to
the solution of the dynamic impurity model. Thus, the resulting lattice
Green's function is $G(\vc{k},i\omega)=[i\omega + \mu - H_0 -
\Sigma(\vc{k},i\omega)]^{-1}$, where
$H_0=H_\textrm{LDA}-V^\textrm{xc}_\textrm{LDA}$ is the LDA Hamiltonian
without the exchange-correlation potential, and
$\Sigma(\vc{k},i\omega)$ is the one defined in Eq.~(\ref{ab_initio_total_sigma}).

We note that if instead of
replacing $V_\textrm{xc}$ by $\Sigma_{GW}^\text{xc}$ one keeps
$V_\textrm{xc}$ and adds a local $\Sigma$ only, 
the above scheme reduces to the DFT+DMFT+$U(\omega)$ approach. From this perspective, the
DFT+DMFT+$U(\omega)$ can be regarded as an embryo of the GW+DMFT
method, which lacks non-locality and a proper treatment of double counting, but where the dynamical nature of both local and non-local
screening effects is taken into account in the effective interaction
of the impurity problem via the cRPA estimate of $U(\omega)$.

In the DFT+DMFT+$U(\omega)$ scheme, one could ask what is the correct double
counting term in the presence of a retarded $U$. 
By assuming that the spectroscopic high-energy features described by the
coupling with plasmons cancel out in the zero-temperature mean-field
solution to recover the potentially exact DFT ground 
state energy, it turns out \cite{Werner2012,Biermann2014} that the appropriate double counting is the
same as the  one introduced in Sec.~\ref{downfolding} for static $U$,
which in this case takes the value of the \emph{screened} 
static limit of $U(\omega)$ ($U=U(\omega \rightarrow 0)$). 

\subsubsection{SEX+DMFT}
\label{sec:sex+dmft}

In the GW approach, the COHSEX approximation\cite{Hedin1965} is a way
to simplify greatly the 
calculation of the self-energy, by separating it into two static
contributions $\Sigma = \Sigma_\textrm{SEX} + \Sigma_\textrm{COH}$, where:
\begin{eqnarray}
\Sigma_\textrm{SEX}(\vc{r},\vc{r}^\prime) & = & - \sum_i^\textrm{occ}
\psi_i(\vc{r}) \psi_i^*(\vc{r}^\prime) W(\vc{r},\vc{r}^\prime,0) \label{sex}\\
\Sigma_\textrm{COH}(\vc{r},\vc{r}^\prime) & = & \frac{1}{2}
\delta(\vc{r}-\vc{r}^\prime) \left( W(\vc{r},\vc{r}^\prime,0) - v(\vc{r}-\vc{r}^\prime) \right),
\label{coh}
\end{eqnarray}
called screened exchange and Coulomb hole, respectively \cite{ferdi_gw,Hedin1999,Onida2002}. $W$ is the
static fully screened interaction. While the first term significantly improves upon the
exchange contribution in the Hartree-Fock theory, the second one
describes the contribution to the self-energy due to interactions
between the quasiparticle and its surrounding hole. As it is apparent
in Eqs.~(\ref{sex})-(\ref{coh}), the first term is
non-local, while the second one is local.

The idea behind the SEX+DMFT theory is to replace the static local
Coulomb hole self-energy by a dynamic one provided by the DMFT
solution of a \emph{downfolded} Hubbard model with retarded $U$, i.e. 
$\Sigma = \Sigma_\textrm{SEX} + \Sigma_\textrm{DMFT}$. The
clear advantage with respect to the COH self-energy is that the
DMFT one is dynamical and non-perturbative. Moreover, the validity of
the SEX+DMFT theory is supported by the observation, verified in the iron
pnictides and transition metal oxides such as SrVO$_3$, that at low-energy
scales the non-local contributions to the self-energy are essentially 
static, while the local ones are dynamic. The self-energy separation between
static non-local terms on the one side and dynamic local terms on the other
side is implemented in the SEX+DMFT by merging SEX and DMFT. 
This is done in the same spirit as in LDA+DMFT, except that the  
$H^\textrm{LDA}=H_0 + V^\text{xc}_\text{LDA}$ 
Hamiltonian is
replaced by  
$H_0 + \Sigma_\text{SEX}$. 
We note that SEX+DMFT is not double-counting error free, because the DMFT
Hamiltonian contains a local Hartree term already included in
$H_0$. This can be easily estimated as a mean-field approximation of
the Hubbard terms. 
Therefore, a double counting correction is needed as in regular
LDA+DMFT, or LDA+DMFT with dynamic $U$.

Despite this fact, SEX+DMFT improves upon the LDA+DMFT method
with dynamic $U$. The reason is that the SEX part yields wider bands
than LDA, which partially compensates the band narrowing produced 
by the frequency dependence of $U$. As found in the case of the compound
BaCo$_2$As$_2$ (Sec.~\ref{sec:sex+dmft_baco2as2}), which is isostructural to the more famous BaFe$_2$As$_2$ and
only moderately correlated, these opposite effects almost cancel each other in the
final result, and rather accurate quasiparticle energies are obtained.

In the practical implementation of Ref.~\cite{Roekeghem2014}, the
screened exchange contribution is calculated as a Fock exchange with the
screened potential in the limit of long wavelengths, i.e. $W \approx
W_\textrm{TF} = \frac{e^2}{q^2 + k_\textrm{TF}^2}$, where
$k_\textrm{TF}$ is the Thomas-Fermi wavevector or inverse screening length.  
In the Thomas-Fermi theory, its value depends on the density of states (DOS)
at the Fermi level. Thus, a self-consistent determination of
$W_\textrm{TF}$ can be devised, as a given $W_\textrm{TF}$ yields 
a new DOS, that implies a new $k_\textrm{TF}$, that closes the loop
by finally fixing a new $W_\textrm{TF}$, and so on. In the
actual calculation of Ref.~\cite{Roekeghem2014}, this 
self-consistency has been replaced by a simple manual inspection to
check that the guessed $k_\textrm{TF}$ is consistent with the final DOS.

The band widening produced by SEX has the same origin as the larger
bandwidth found in the non-local self-energy framework by a number of
authors \cite{Motoaki2013,Miyake2013,Sakuma2014,Schafer2015}. 
The simplified static non-local version bears the same physics as the
more involved full GW convolution. SEX+DMFT is the simplest
theory capable of including non-local correlation effects besides the
non-perturbative local ones provided by DMFT. Therefore, SEX+DMFT goes
in the direction of extending DMFT in a fully \emph{ab initio}
fashion, to include longer-range interactions beyond the Hubbard type.

\subsubsection{DMFT@nonlocal-GW}
\label{sec:dmft@nonlocal-GW}

The DMFT@nonlocal-GW approach \cite{Tomczak2014}, as the previous
SEX+DMFT method, is based on the observation (see Sec.~\ref{sec:widening}) that the local and non-local
self-energy contributions are dynamically separable, with the former one
frequency dependent and the latter one static. 
The 
DMFT self-consistency
condition for the one-body quantities requires the local Green's function
to satisfy 
\begin{equation}
G^\text{loc} (i \omega) = \sum_\vc{k} [i \omega + \mu - H_0(\vc{k})
- \Sigma_{GW}^\text{nonloc}(\vc{k}, i \omega) - \Sigma_\text{imp}(i \omega)]^{-1},
\label{Eq:sc-ksum}
\end{equation}
with $H_0=H_\text{LDA}-V^\text{xc}_\text{LDA}$, and
$\Sigma_{GW}^\text{nonloc}$ is the nonlocal part of the full GW t$_{2g}$
self-energy. 
As the nonlocal correlation self-energies are purely static in the
low-energy window,
i.e. $\Sigma_{GW}^\text{nonloc}(\vc{k},\omega)=\Sigma_{GW}^\text{nonloc}(\vc{k})$, one
can construct an effective quasi-particle Hamiltonian that also comprises
these correlation effects: 
\begin{equation}%
H^\text{qp}(\vc{k})=H_0(\vc{k})+\Re\Sigma_{GW}^\text{nonloc}(\vc{k}).%
\label{Hqp}%
\end{equation}
$H^\text{qp}$ is a simplified one-shot analogue of the QSGW
Hamiltonian $H^{\hbox{\tiny
    QSGW}}$ that was proposed in
the context of the QSGW+DMFT formalism \cite{jmt_pnict}.
Then the DMFT self-consistency is much simpler since quantities are
either frequency {\it or} momentum dependent, but not both, which
drastically reduces memory requirements.   
Of course, the simplified DMFT@nonlocal-GW scheme is not
expected to give quantitatively accurate results {\it outside} the
quasi-particle energy range. In particular the dispersion 
of collective excitations will not be captured. However, their
position in the local spectrum which is determined by the structure of
the dynamic interaction $\mathcal{U}(\omega)$ 
is still meaningful. The results of the DMFT@nonlocal-GW for 
SrVO$_3$ have been presented by Tomczak \emph{et al.} in
Ref.~\cite{Tomczak2014}.  
A previous attempt on the same compound was presented by
Taranto \emph{et al.} \cite{Taranto2013} by using a one-shot version of
QSGW. However, in the latter work the authors obtained quite different
results from Tomczak's, with a much stronger quasi-particle
renormalization ($Z$=0.36) when the $Z_B$ factor was used, probably due to a
different way of dealing with the local self-energy subtraction at the
QSGW level.

\subsubsection{QSGW+DMFT}
\label{sec:qsgw+dmft}

In the quasi-particle self-consistent GW (QSGW) + DMFT, one defines
an effective static Hamiltonian, which includes non-local and dynamic correlation
effects, through the fully self-consistent QSGW
construction \cite{vanSchilfgaarde2006,Kotani2007}.
In QSGW+DMFT, an additional self-consistency on the GW-level is
performed which circumvents the full GW+DMFT self-consistency that 
is computationally very demanding and has so far has only been 
achieved on the model level \cite{Ayral2012,Ayral2013,Huang2014},
and the simpler case of a two-dimensional system of adatoms on
surfaces \cite{Hansmann2013}. 
In the QSGW+DMFT framework,
once a self-consistent QSGW calculation is
performed and the quasiparticles energies $E_{\vc{k}n}$ are found, the local part
of the self-energy is subtracted to avoid double counting, and the
non-local self-energy is evaluated at the corresponding $E_{\vc{k}n}$
energies. In this way, a static correction to the initial Hamiltonian is
obtained, and incorporated into a modified Hamiltonian
$H^\text{nl}$. Then, a DMFT calculation with dynamical ${\mathcal U}$
  can be performed on top of $H^\text{nl}$. In contrast to 
  DMFT@nonlocal-GW, dynamical non-local corrections can be incorporated into the
  QSGW+DMFT through the iterative QSGW construction. 
Suggested in Ref.~\cite{jmt_pnict} and later in \cite{Tomczak2014}, a simplified 
variant of this approach has recently been
applied to the Mott insulator La$_2$CuO$_4$ \cite{Choi2015}.

\subsubsection{Concluding remarks on simplified GW+DMFT approaches}
\label{sec:gw+dmft_approximations_final}

All methods in Secs.~\ref{sec:sex+dmft}, \ref{sec:dmft@nonlocal-GW},
and \ref{sec:qsgw+dmft}, make the calculation of the non-local GW part
faster and more efficient. However, the self-consistency is performed at
the $G$ level only. Therefore, these schemes are not fully self-consistent GW+DMFT
approaches. However, in the SEX+DMFT approach, it is possible to
perform an additional self-consistency on the 
screening Fermi wave-vector, through the evaluation of the density of
states at the Fermi level. This replaces the
self-consistency at the polarization level in a purely static
screening approach. 

We remark that 
the Thomas-Fermi model used for the SEX part in Ref.~\cite{Roekeghem2014} has several
well-known limitations. 
Its exponential decay form is valid only for metals; in insulators
there is a longer-range decay instead \cite{PhysRevLett.62.2160}. 
Moreover,
the Thomas-Fermi model (as well as RPA) overestimates screening in metals. In fact, due to
the singularity at $2k_F$, there are Friedel oscillations with a $2k_F$
period, decaying as $1/r^3$, while in the Thomas-Fermi model the decay is
always exponential. 
On the other hand, SEX+DMFT solves one of the major problems of the
fully static COHSEX approximation, namely the lack of quasiparticle
renormalization coming from the $Z$ factor. Indeed,
in SEX+DMFT, the DMFT part provides a non-perturbative frequency-dependent local
self-energy, which usually yields a good estimate of $Z$. Therefore,
as future perspective, it is worth trying to implement better approximations
for the SEX part, to go beyond the Thomas-Fermi model, within the
SEX+DMFT framework.

\section{Analytical continuation}
\label{sec:maxent}

At present, the Monte Carlo technique discussed in Sec.~\ref{ctqmc} is
the method of choice for the solution of impurity problems with
dynamically screened interactions. For this reason, the extended DMFT
or GW+DMFT calculations are most conveniently implemented on the
Matsubara axis. For the interpretation of the results, it is however
often important to have access to spectral functions. The analytical
continuation from the Matsubara-frequency to the real-frequency axis
is a delicate problem and particularly challenging in the case where
high-energy features exist and need to be resolved. Both the Pad\'e
\cite{Vidberg1977} and maximum entropy methods \cite{Jarrell1996},
which usually work well for low-frequency features in systems at low
enough temperature, are not well-suited for capturing the
high-frequency part of the spectrum. A pre- and post-processing of the
data is needed in order to use these methods as part of a somewhat
elaborate analytical continuation scheme. In
Sec.~\ref{sec:fermi_spectral_function}, we explain the scheme for
calculating the fermionic spectral function corresponding to some
LDA+DMFT+$U(\omega)$ calculation. We will also briefly address the
calculation of the bosonic spectral functions from EDMFT or GW+DMFT
calculations, focusing in this case on models with static bare
interactions (Sec.~\ref{sec:bose_spectral_function}). 

\subsection{Fermionic spectral function}
\label{sec:fermi_spectral_function}

The idea proposed in Ref.~\cite{Casula2012maxent} is to split the Green function $G(\tau)$ into a product of a bosonic function $B(\tau)$ and an auxiliary fermionic Green function $G_\text{aux}(\tau)$:
\begin{equation}
G(\tau)=B(\tau)G_\text{aux}(\tau).
\label{bose_factorization}
\end{equation}
From the spectral functions $\rho_B$ and $\rho_\text{aux}$ of the two factors one can then obtain the spectral function $\rho(\omega)$ of the original Green function using the convolution
\begin{equation}
\rho(\omega)=\int_{-\infty}^\infty d\epsilon \rho_B(\epsilon)\frac{1+e^{-\beta\omega}}{(1+e^{\beta(\epsilon-\omega)})(1-e^{-\beta\omega})}\rho_\text{aux}(\omega-\epsilon).\label{conv}
\end{equation}
The bosonic function $B(\tau)$ can be chosen arbitrarily, as long as the factorization (\ref{bose_factorization}) does not lead to unphysical properties of $G_\text{aux}$. A natural choice, which often works in practice, is
\begin{equation}
B(\tau)=e^{-K(\tau)},
\end{equation}
where $K(\tau)$ is the twice-integrated screening function defined in Eq.~(\ref{K}). The rationale for this choice is that such a factorization holds in the atomic limit (see Sec.~\ref{holstein_hubbard}). 

Substituting $\tau=it$ 
in Eq.~(\ref{K}), and expressing the factor 
$\frac{\cosh(\omega(\beta/2-it))}{\sinh(\omega\beta/2)}$ as $e^{-it\omega}+\frac{2\cos(\omega t)}{e^{\beta\omega}-1}$, 
one finds
\begin{equation}
K(t) =  \int_0^\infty d\omega \frac{\text{Im}U(\omega)}{\pi\omega^2}\left(e^{-it\omega}+\frac{2\cos(\omega t)}{e^{\beta\omega}-1}-\frac{e^{\beta\omega}+1}{e^{\beta\omega-1}}\right)
\approx  \int_0^\infty d\omega \frac{\text{Im}U(\omega)}{\pi\omega^2} (e^{-it\omega}-1).
\end{equation}
In the last step we used the fact that at low temperatures, $e^{\beta\omega}\gg 1$, except near $\omega=0$, where $\text{Im}U(\omega)$ vanishes. At low enough temperature, we can therefore express the bosonic factor on the real-time axis as 
\begin{align}
B(t)=e^{-K(t)}&=
\underbrace{\exp\Big[\int_0^\infty d\omega\frac{\text{Im}U(\omega)}{\pi\omega^2}\Big]}_{Z_B}\exp\Big[-\int_0^\infty d\omega\frac{\text{Im}U(\omega)}{\pi\omega^2}e^{-i\omega t}\Big] 
\end{align}
The first term corresponds to the Bose factor defined in Eq.~(\ref{Z_B_general}), while the second term oscillates around $1$ as $t\rightarrow \infty$. We thus split off $Z_B$ and write 
\begin{equation}
B(t)=Z_B+B_\text{reg}(t), \label{Breg}
\end{equation}
with the regular term $B_\text{reg}(t)=Z_B (\exp[-\int_0^\infty d\omega\frac{\text{Im}U(\omega)}{\pi\omega^2}e^{-i\omega t}]-1)$. 
The corresponding spectral density $\rho^B_\text{reg}(\omega)$ can be obtained numerically from an appropriate Fourier transform. 
We have the spectral representation 
\begin{equation}
B_\text{reg}(t)=\int_{-\infty}^\infty d\omega'\rho^B_\text{reg}(\omega')\frac{e^{-i\omega't}}{1-e^{-\beta\omega'}},
\end{equation}
and therefore 
\begin{equation}
\int_{-\infty}^\infty dt e^{i\omega t} B_\text{reg}(t)=\int_{-\infty}^\infty d\omega' \underbrace{\int_{-\infty}^\infty dt e^{i(\omega-\omega') t}}_{=2\pi\delta(\omega-\omega')}\frac{\rho^B_\text{reg}(\omega')}{1-e^{-\beta\omega'}}=2\pi\frac{\rho^B_\text{reg}(\omega)}{1-e^{-\beta\omega}}.
\end{equation}
Using $B_\text{reg}(-t)=[B_\text{reg}(t)]^*$, we can write the left hand side as $2\text{Re}\int_0^\infty dt e^{i\omega t} B_\text{reg}(t)$, which finally yields the expression
\begin{equation}
\rho^B_\text{reg}(\omega)=\frac{1-e^{-\beta\omega}}{\pi}\text{Re}\int_0^\infty dt e^{i\omega t} B_\text{reg}(t).
\end{equation}
It immediately follows from this derivation that the constant $Z_B$ in Eq.~(\ref{Breg}) 
gives a non-regular contribution
\begin{equation}
\rho^B_\text{non-reg}(\omega)=Z_B(1-e^{-\beta\omega})\delta(\omega)\label{rhobnonreg}
\end{equation}
to the spectral density. Substitution of (\ref{rhobnonreg}) into Eq.~(\ref{conv}) gives 
\begin{equation}
\rho_\text{non-reg}(\omega)=Z_B\rho_\text{aux}(\omega), 
\end{equation}
which is the expected renormalization of the quasi-particle peak by the Bose factor (see Sec.~\ref{sec:static} and Eq.~(\ref{Z_B_general})).

In summary, the spectral function $\rho$ can be computed from the maximum entropy result for $\rho_\text{aux}(\omega)$ and either the cRPA result or some numerical estimate for $\text{Im}U(\omega)$ using the following formulas:
\begin{align}
\rho(\omega)&=Z_B\rho_\text{aux}(\omega)+\int_{-\infty}^\infty d\epsilon \rho^B_\text{reg}(\epsilon)\frac{1+e^{-\beta\omega}}{(1+e^{\beta(\epsilon-\omega)})(1-e^{-\beta\omega})}\rho_\text{aux}(\omega-\epsilon),\label{maxent_final}\\
\rho^B_\text{reg}(\omega)&=\frac{1-e^{-\beta\omega}}{\pi}\text{Re}\int_0^\infty dt e^{i\omega t} B_\text{reg}(t),\\
B_\text{reg}(t)&=Z_B \Big(\exp\Big[-\int_0^\infty d\omega\frac{\text{Im}U(\omega)}{\pi\omega^2}e^{-i\omega t}\Big]-1\Big).
\end{align}

\subsection{Bosonic spectral function}
\label{sec:bose_spectral_function}

Here we discuss a useful strategy for calculating the spectral function $\text{Im}U(\omega)$ corresponding to some retarded interaction $U_\text{ret}(\tau)$, as it is obtained for example in the self-consistency loop of an EDMFT or GW+DMFT simulation \cite{Huang2014}. In a maximum entropy approach, it is important to work with a properly normalized spectral function. In the bosonic case, we can use the relation
\begin{equation}
U_\text{scr}-U=2\int_0^\infty d\omega \frac{1}{\pi}\frac{\text{Im}U(\omega)}{\omega}
\end{equation}
to define such a normalized distribution function:  
\begin{align}
&B(\omega) = \frac{2}{\pi(U_\text{scr}-U)}\frac{\text{Im}U(\omega)}{\omega},\label{Bdef} \quad \int_0^\infty d\omega B(\omega)=1.
\end{align}
Equation (\ref{Uret}), which connects the retarded interaction to the spectral density can then be written in the form
\begin{align}
U_\text{ret}(\tau)&=\int_0^\infty d\omega K(\omega,\tau)B(\omega), 
\label{maxent_boson}
\end{align}
with the bosonic kernel
\begin{equation}
K(\omega,\tau)=\frac{\omega(U_\text{scr}-U)}{2}\frac{\cosh(\omega(\beta-\tau/2))}{\sinh(\omega\beta/2)}.
\end{equation}
(If the $\omega$ integration is taken from $-\infty$ to $\infty$, the kernel becomes $\tilde K(\omega,\tau)=\frac{e^{-\tau \omega}}{1-e^{-\beta\omega}}\frac{\omega(U_\text{scr}-U)}{2}$.)
The factor $(U_\text{scr}-U)$ is known from the solution on the Matsubara axis.

We can now use the maximum entropy method \cite{Jarrell1996} to solve Eq.~(\ref{maxent_boson}) for $B(\omega)$, and finally Eq.~(\ref{Bdef}) to find $\text{Im}U(\omega)$. The real part can as usual be obtained from the antisymmetry of $\text{Im}U(\omega)$ and the Kramers-Kronig relation
\begin{equation}
\text{Re}U_\text{ret}(\omega)=\frac{1}{\pi}P\int_{-\infty}^\infty d\omega'\frac{\text{Im}U(\omega')}{\omega'-\omega}=\frac{1}{\pi}P\int_{-\infty}^\infty d\omega'\Bigg(\frac{\text{Im}U(\omega')-\text{Im}U(\omega)}{\omega'-\omega}
\Bigg).
\end{equation} 
In the last step, we have re-expressed the integral in a form which is suitable for numerical treatment.

\section{Applications to model systems}
\label{sec:applications_models}

\subsection{Extended DMFT}
\label{sec:applications_models_edmft}

\subsubsection{$U$-$V$ Hubbard model}
\label{sec:applications_edmft_UVhubbard}

\begin{figure}[t]
\begin{center} 
\includegraphics[angle=0, width=0.49\columnwidth]{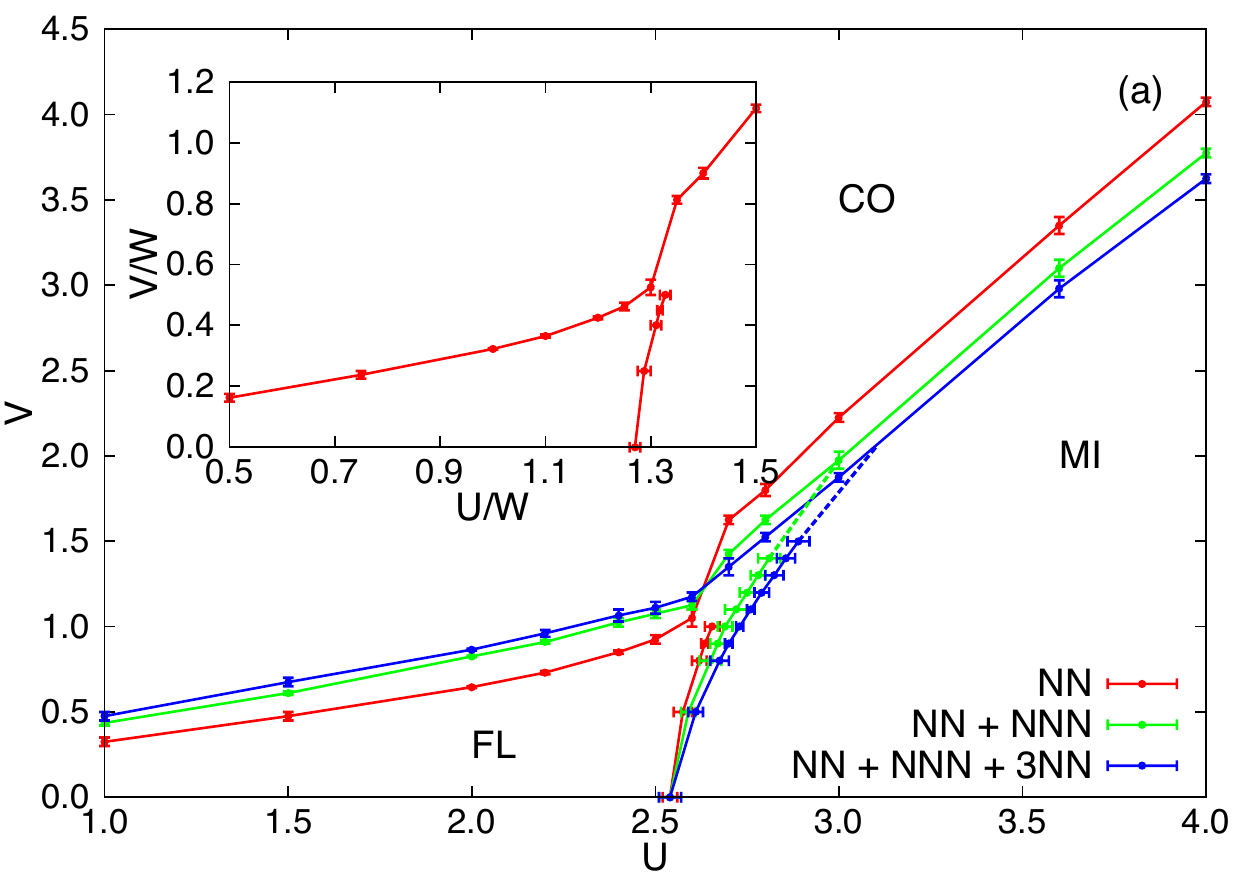}\hfill
\includegraphics[angle=0, width=0.49\columnwidth]{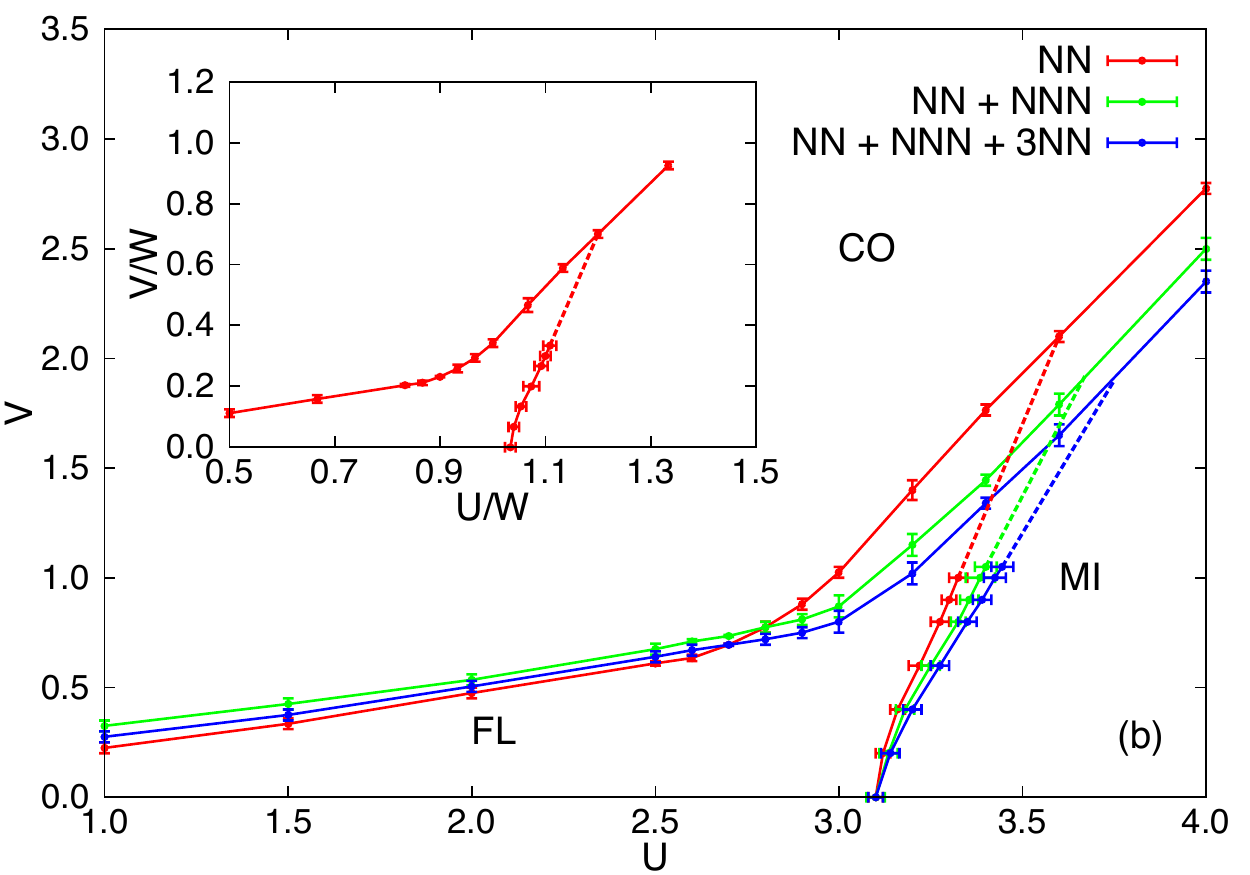}
\caption{ Paramagnetic phase diagram of the half-filled $U$-$V$ Hubbard model on the 2D square lattice (a) and 3D cubic lattice (b). Results for non-local interactions up to nearest neighbors (NN), next-nearest neighbors (NN+NNN) and third nearest neighbors (NN+NNN+3NN) are shown. FL denotes the metallic phase, CO the charge ordered insulating phase, and MI the Mott insulating phase. The insets show the phase diagrams of the models with NN interactions with axes rescaled by the bandwidth. (From Ref.~\cite{Huang2014}.)}   
\label{phasediagram_uv}
\end{center}
\end{figure}

The EDMFT solution for the half-filled $U$-$V$ Hubbard model (\ref{H_UV}) on the cubic lattice predicts a paramagnetic phase diagram with three phases: a metallic phase for small $U$ and small $V$, a Mott insulating phase for large $U$ and small $V$, and a charge ordered insulating phase for sufficiently large $V$ \cite{Sun2002}. Recently, the low-temperature phase diagrams on the 2D square and 3D cubic lattices have been mapped out using the efficient and unbiased impurity solver described in Sec.~\ref{ctqmc} \cite{Ayral2013,Huang2014}. In Fig.~\ref{phasediagram_uv} we show the results for generalizations of model (\ref{H_UV}) with nonlocal interactions up to the third-nearest neighbors. Here, the parameter $V$ encodes the strength of the nonlocal interactions, i.e. the non-local interactions are scaled as $V_{ij}=a/|r_i-r_j|$, with $a$ the lattice spacing and $r_{i,j}$ the positions of the nearest-neighbor, next-nearest neighbor or third nearest neighbor sites.  The unit of energy is the hopping. 

In the square-lattice case with only nearest-neighbor interactions, the $V_c(U)$ phase boundary jumps near the intersection with the Mott transition line. In the models with longer-ranged interactions, the metallic phase extends to larger values of $U$, forming a ``nose" which separates the Mott insulator and charge ordered phases at low temperature. The jump in $V_c(U)$ disappears, so that the phase diagram looks qualitatively similar to that of the Hubbard-Holstein model with large phonon frequency (Fig.~\ref{phasediagram_g}). However, in the Hubbard-Holstein case, the boundary to the charge ordered phase does not exhibit a slope change near the critical $U$ for the Mott transition, which indicates that the slope change in the EDMFT phase diagram originates from changes in the dominant screening modes near $U_c$ \cite{Huang2014}.

Overall, the phase diagrams for the 2D and 3D lattice are similar, with the main difference being the larger extent of the metallic nose in the 3D case, and a different dependence of the metal-charge order phase boundary on the interaction range. The latter can be explained by considering the lattice geometry and our definition of next-nearest and third-nearest neighbors \cite{Huang2014}: in the 2D case, the third-nearest neighbor interactions act between sites on the same sublattice, and hence frustrate the charge order, while in the 3D case, they act between sites on different sublattices. 

\begin{figure}[t]
\begin{center}
\includegraphics[width=0.325\textwidth]{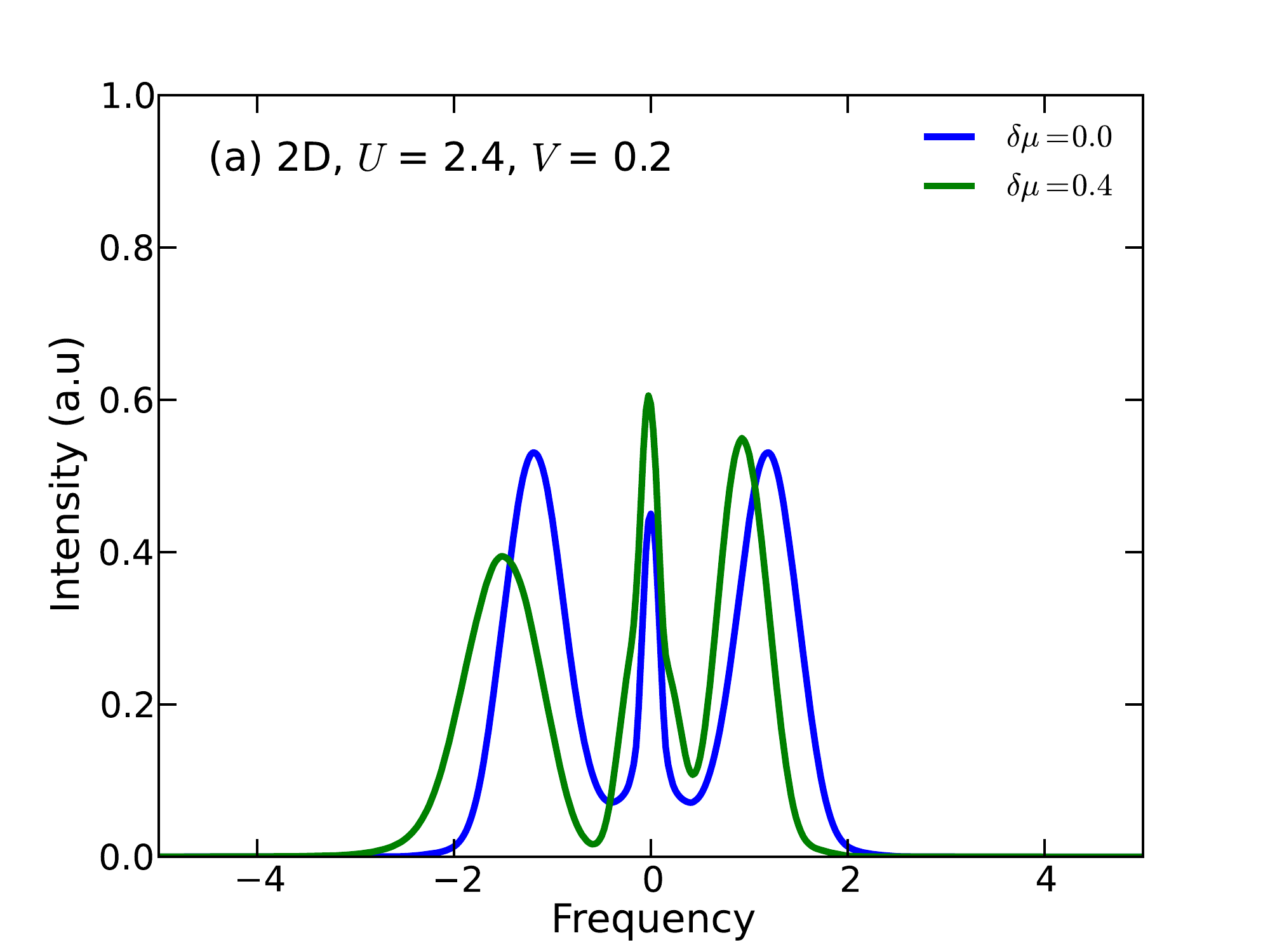}
\includegraphics[width=0.325\textwidth]{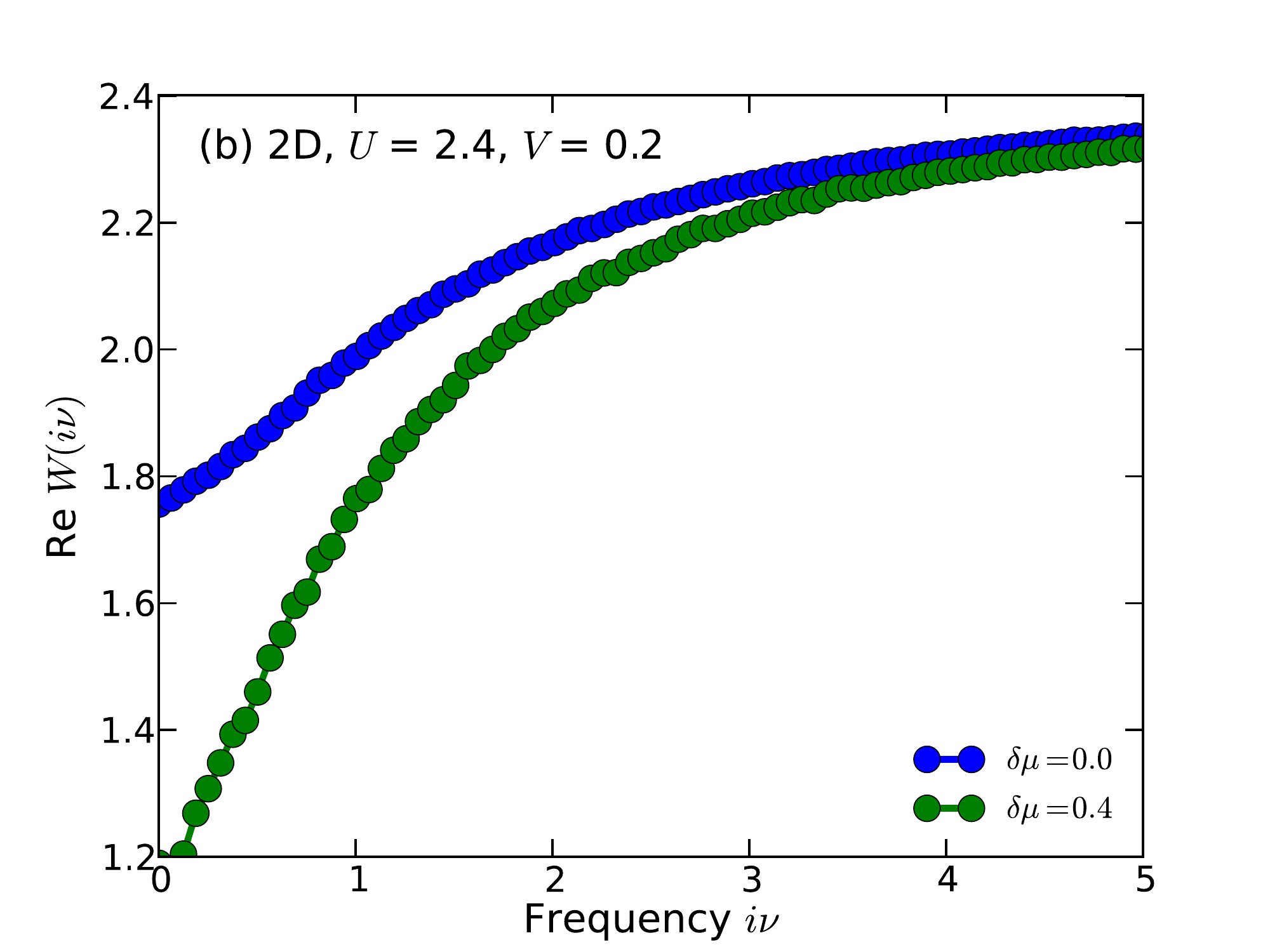}
\includegraphics[width=0.325\textwidth]{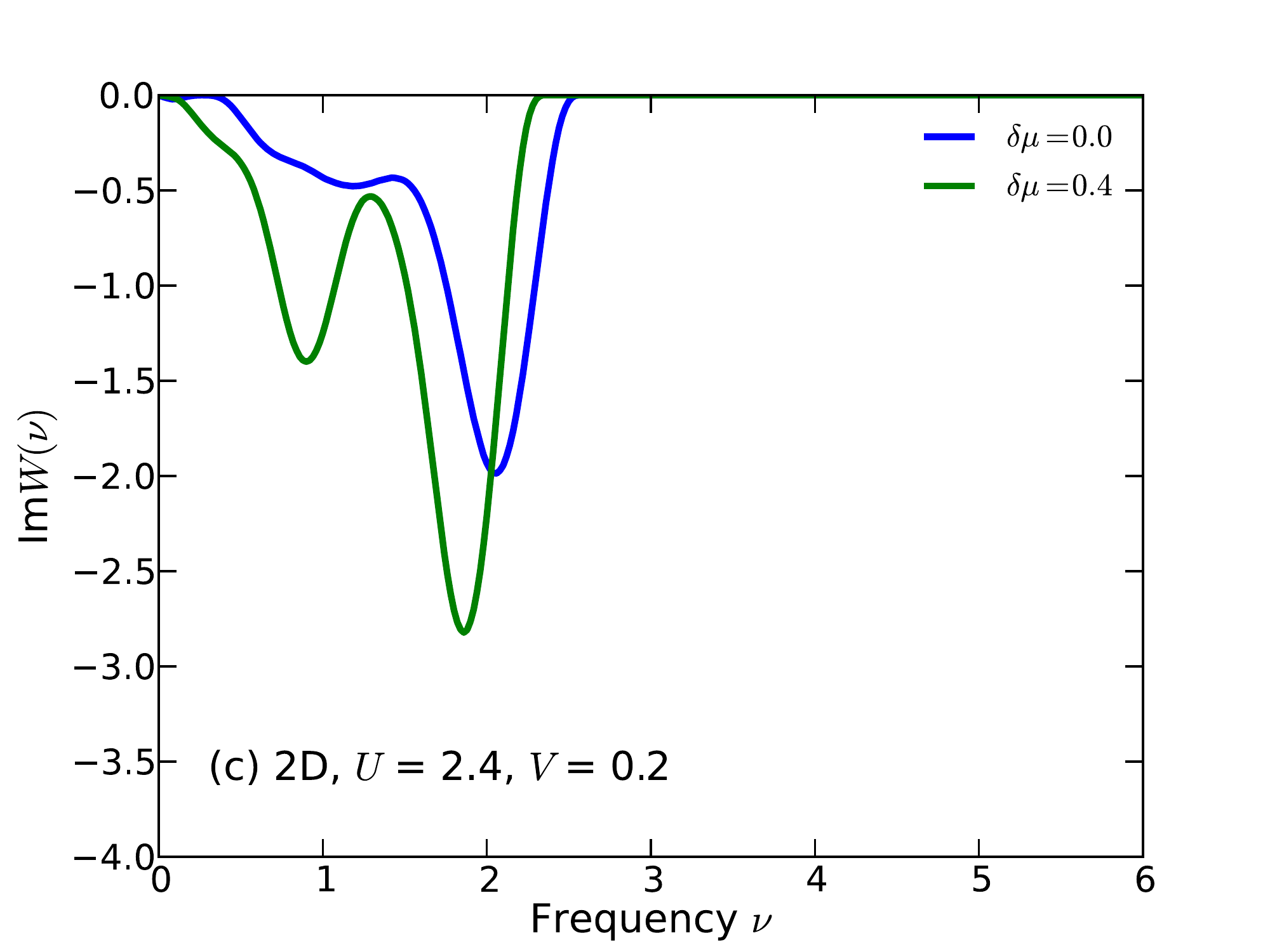}
\includegraphics[width=0.325\textwidth]{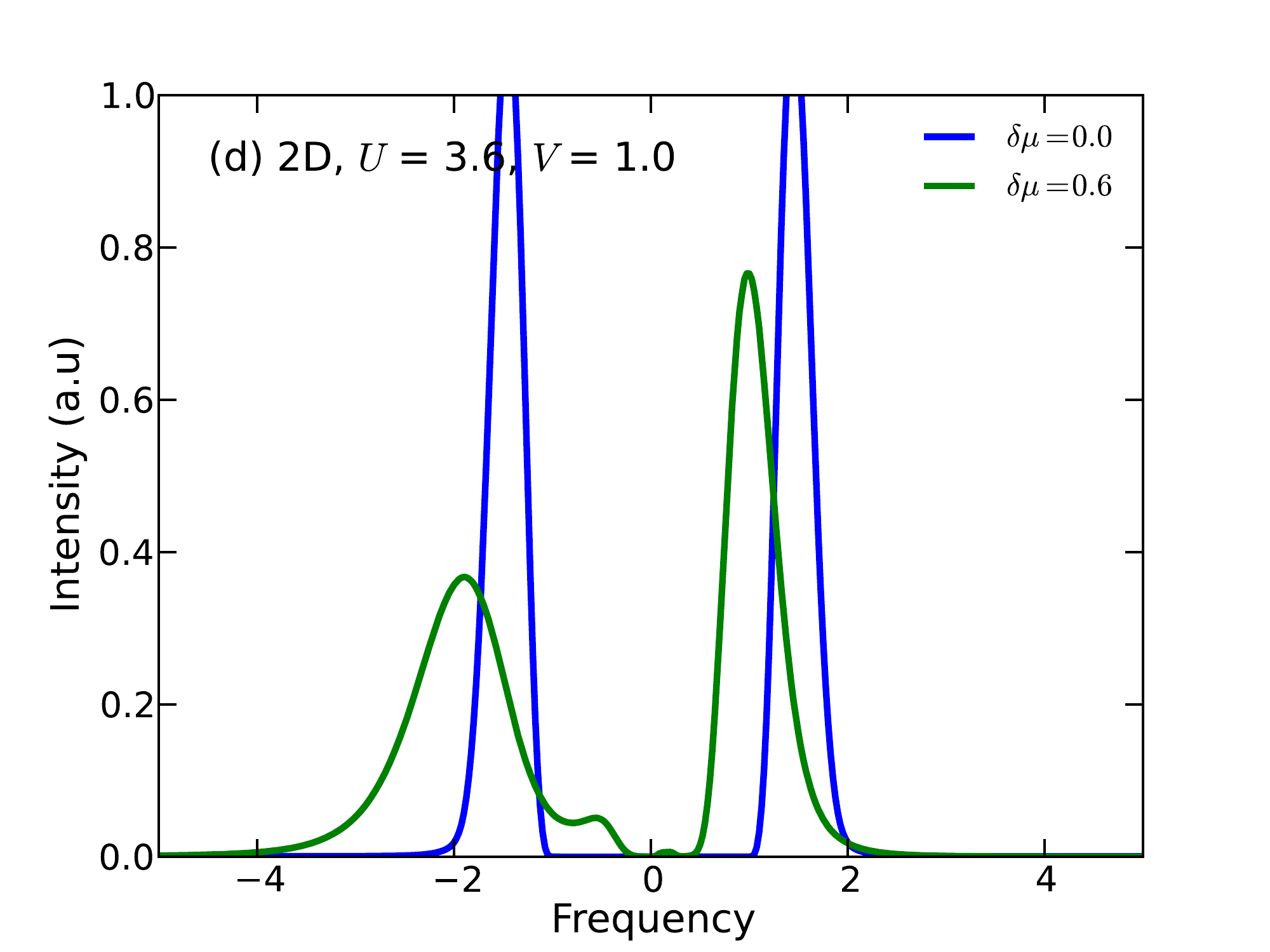}
\includegraphics[width=0.325\textwidth]{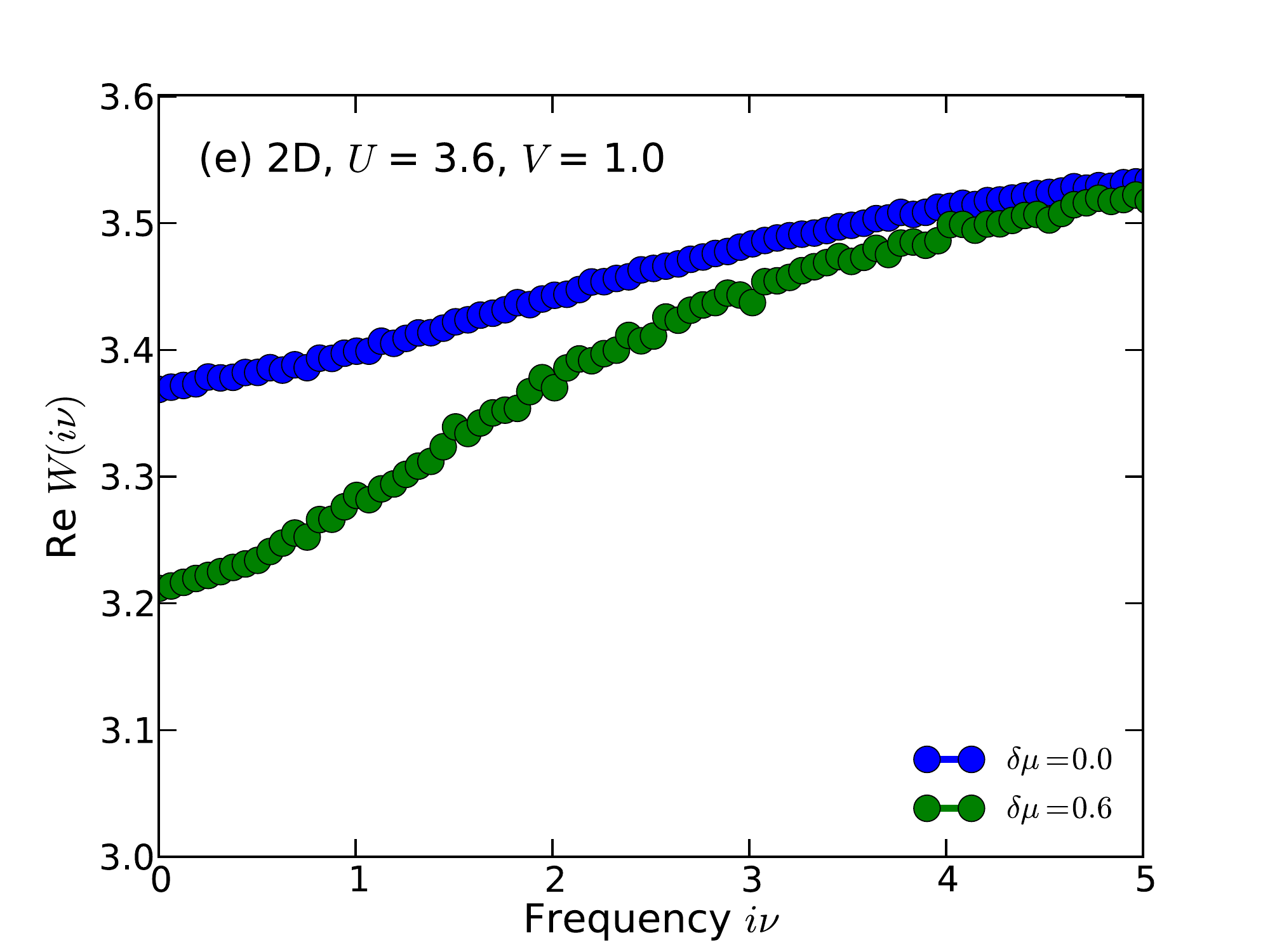}
\includegraphics[width=0.325\textwidth]{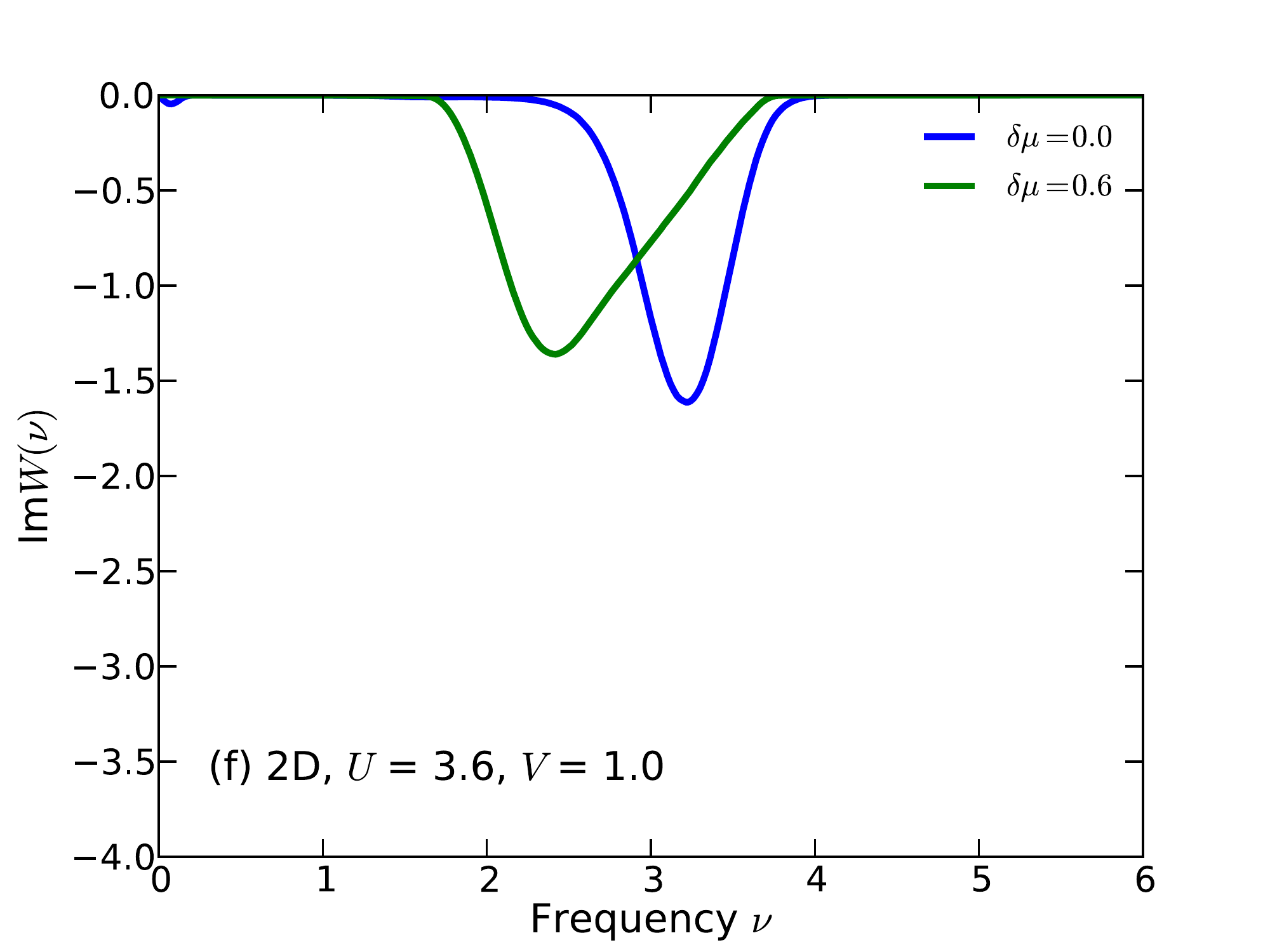}
\caption{ Fermionic and bosonic spectral functions for the
  paramagnetic  Hubbard model with NN interactions on the square
  lattice. The blue curves show the results for a half-filled system
  and the green curves for a particle-hole asymmetric system with a
  chemical potential shift of $\delta\mu=0.4$ and $0.6$ relative to the half-filled value $\mu=U/2$. Panels (a)-(c) show
  results for a metallic system with $U=2.4$ and $V=0.2$, while panels
  (d)-(f) show results for a Mott insulating system with $U=3.6$ and
  $V=1.0$. The left panels plot the fermionic spectral function
  $A(\omega)$, the middle panels the real part of the screened
  interaction $\text{Re}W(i\nu)$ on the Matsubara axis, and the right
  panels the bosonic spectral functions $\text{Im}W(\omega)$.  
(From Ref.~\cite{Huang2014}.)
}   

\label{screening_uv}
\end{center}
\end{figure}

In order to identify the dominant screening modes, and understand their origin, it is instructive to compute the bosonic and fermionic spectral functions, as illustrated in Fig.~\ref{screening_uv} for a metallic and Mott insulating system. Let us focus first on the half-filled case (blue lines and symbols). As seen in panels (c) and (f), $\text{Im}W(\omega)$, which is proportional to the square of the coupling strength of the screening modes with frequency $\omega$, exhibits two peaks near $U$ and $U/2$ in the metallic system, and a single peak near $U$ in the insulating case. The comparison to the fermionic spectral functions plotted in panels (a) and (d) suggests that the peak at $U$ is related to transitions between the Hubbard bands, while the peak at $U/2$ in the metallic system originates from transitions between the quasi-particle band and one of the Hubbard bands. 

If the chemical potential is shifted away from the particle-hole symmetric value (green curves), the spectral functions change. In the metallic system, the peak near $U/2$ grows and shifts to lower energies, consistent with the increased weight of the quasi-particle peak and the reduced separation between this peak and the upper Hubbard band in the fermionic spectral function. The high energy peak also grows, which suggests that the asymmetric shape and population of the Hubbard bands enhances the screening effect. 
Despite the uncertainties in the analytical continuation, it appears that one can even identify a third low-energy mode associated with transitions within the quasi-particle peak. In the insulating case, the reduction in the gap size and the broadening of the lower Hubbard band are reflected in a shift of the peak in $\text{Im}W(\omega)$ to lower energies, and a broader distribution of screening modes. 

In the middle panels, we plot the real part of the screened interactions on the Matsubara axis. In the high-frequency limit $\text{Re}W(i\nu_n)$ approaches the bare on-site interaction $U$, while below an energy scale determined by the dominant screening modes, the screened interaction is reduced. Here, we should recall the fact that this reduction is dominated by the low-energy modes (see Eq.~(\ref{Uscr})). Hence, $\text{Re}W(0)$ is not much smaller than $U$ in the insulator, while the low-energy screening modes in the metallic case lead to a substantially reduced static interaction.

\subsubsection{Dynamical screening in photo-doped Mott insulators}
\label{sec:applications_edmft_photo_doped_Mott}

\begin{figure}[t]
\begin{center}
\includegraphics[angle=0,width=0.49\textwidth]{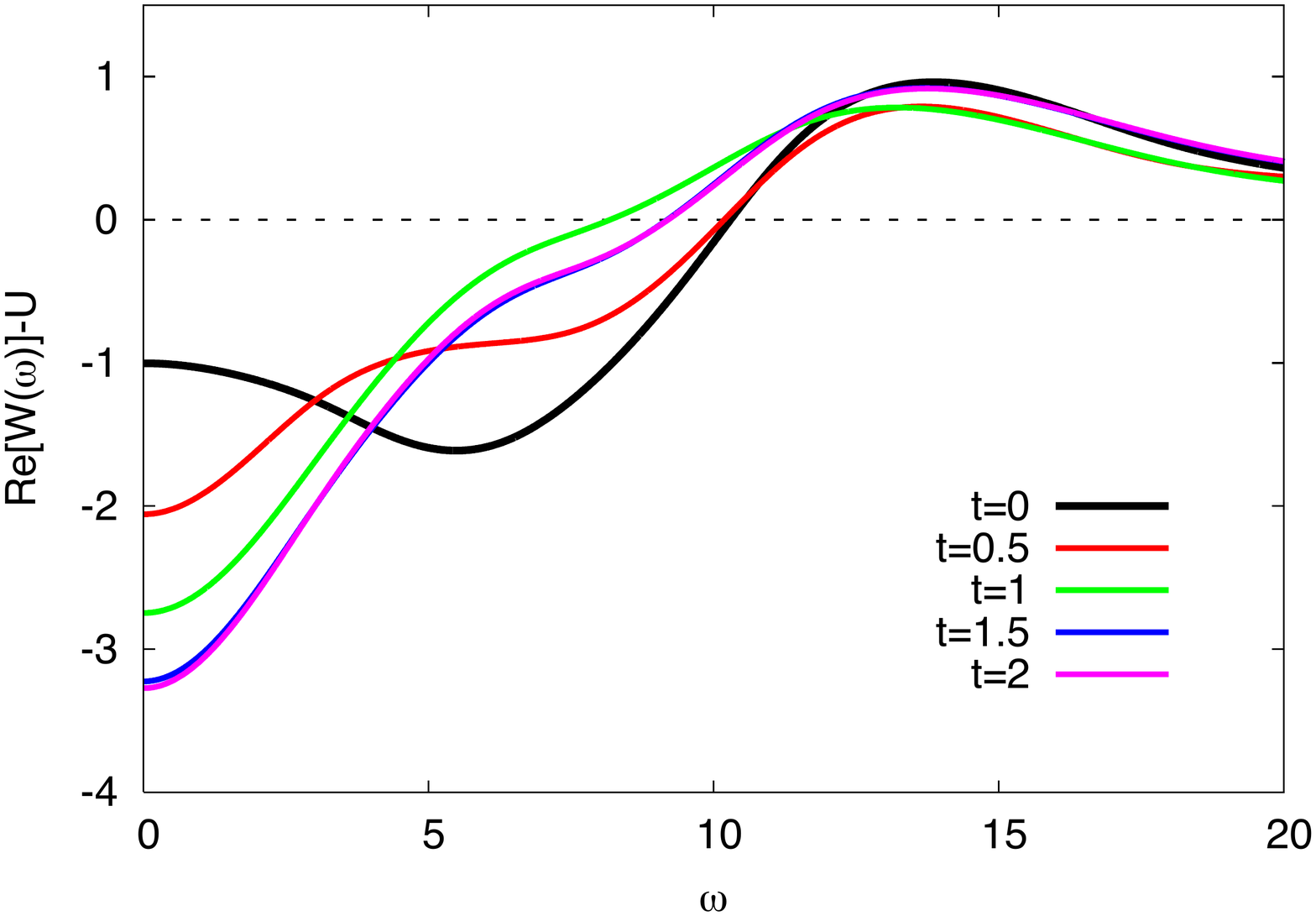}
\includegraphics[angle=0,width=0.49\textwidth]{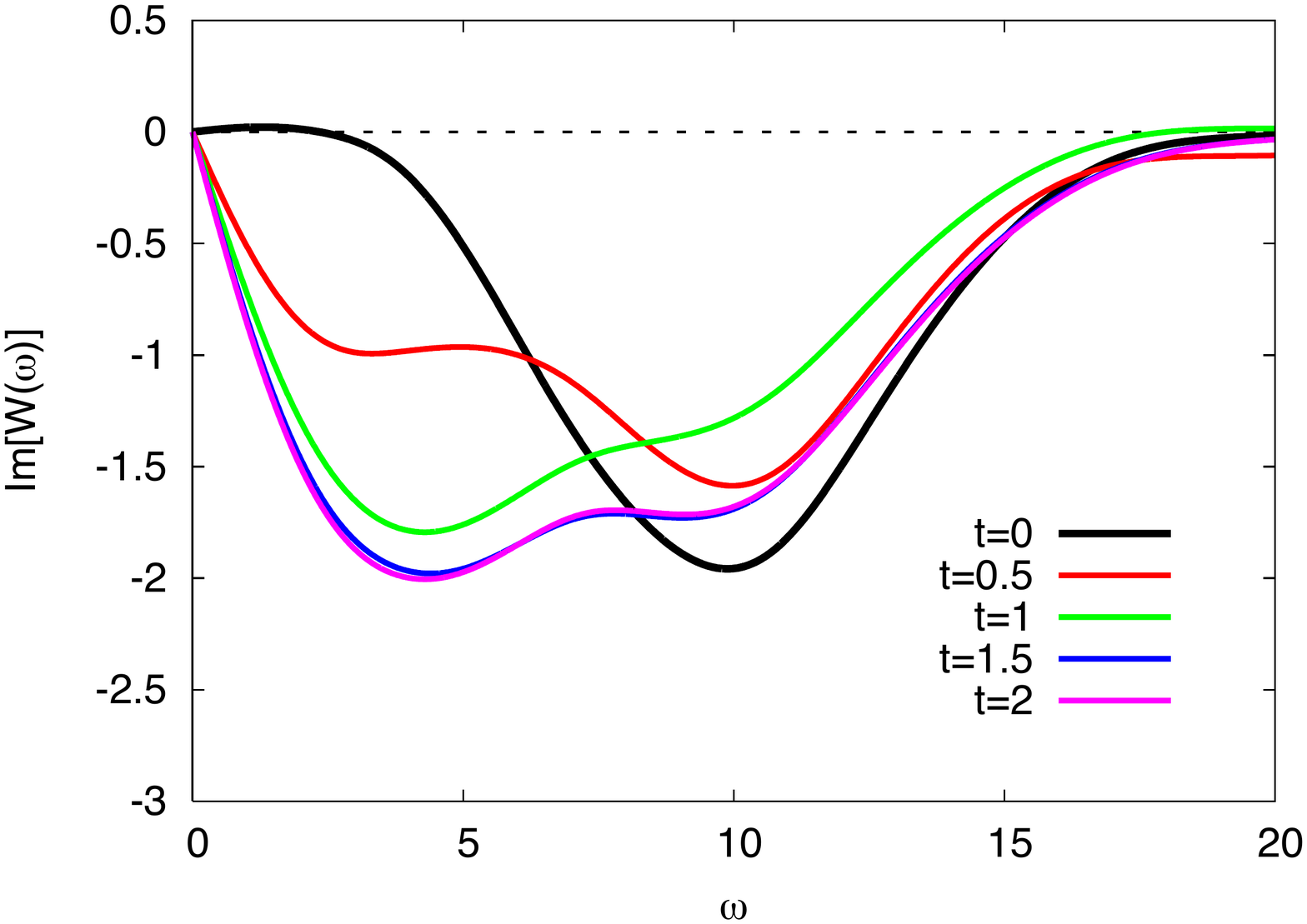}
\caption{Time-evolution of the screened interaction $W$ in a Mott insulator which is excited by a mono-cycle electric field pulse with $\Omega_\text{pulse}= U$. The electric field pulse has the form $E(t)=E_0\sin(\Omega_\text{pulse}(t-t_0))\exp(-4.6(t-t_0)^2/t_0^2)$ with $t_0=2\pi/\Omega_\text{pulse}$ and $E_0=9$. $U=10$, $V=2$, initial $\beta=5$.  
(From Ref.~\cite{Golez2015}.)}   
\label{fig_noneq_edmft}
\end{center}
\end{figure}

The real-time dynamics of screening has been recently investigated using a nonequilibrium generalization of EDMFT \cite{Golez2015}. As discussed in Ref.~\cite{Aoki2014}, the DMFT formalism can be applied to nonequilibrium problems by 
solving the DMFT equations on a 3-branch Kadanoff-Baym contour which runs from 0 to some maximum time $t_\text{max}$ along the real time axis, back to time 0 along the real time axis, and then to $-i\beta$ along the imaginary time axis. The solution on the imaginary-time branch corresponds to the usual DMFT solution for the initial equilibrium state (with $\beta$ the inverse temperature), while the solution on the real-time branches allows to describes the evolution of the system after some perturbation or in the presence of external fields. In a similar manner, the bosonic self-consistency loop of EDMFT can be solved on the Kadanoff-Baym contour and the resulting nonequilibrium EDMFT formalism then captures the changes in the screening properties resulting from the nonthermal state of the system. In Ref.~\cite{Golez2015}, the $U$-$V$ Hubbard model (\ref{H_UV}) on the square lattice was driven out of an initially Mott insulating state by a single-cycle electric-field pulse with frequency $\Omega_\text{pulse}\approx U$ directed along the lattice diagonal. As shown in the right hand panel of Fig.~\ref{fig_noneq_edmft}, $\text{Im}W(\omega)$, which represents the distribution of screening modes, initially exhibits a single broad peak centered at energy $U$, similar to Fig.~\ref{screening_uv}(f).  This is because in a Mott insulator, the screening processes involve particle excitations across the gap.    

However, as soon as the field pulse creates doublon-hole pairs, a second screening mode appears at lower energy, and the imaginary part of $W(\omega,t)=\int_t^{t+t_\text{cut}} dt' e^{i\omega (t'-t)}W^\text{ret}(t', t)$ (with Fourier cutoff $t_\text{cut}=10$) qualitatively resembles the equilibrium result for a metallic system (Fig.~\ref{screening_uv}(c)). 
An essential difference to the equilibrium metallic system however is the absence of a coherent quasi-particle band \cite{Eckstein2013}, which implies that the low-energy screening modes in the photo-doped system are not associated with transitions between quasi-particle and Hubbard bands, but rather with screening transitions within the photo-doped Hubbard bands. This also explains the broader energy distribution.  
As shown in Fig.~\ref{fig_noneq_edmft}, the low-energy mode grows while the pulse (which lasts up to $t\approx 1.5$) produces additional carriers, and then essentially saturates. The subsequent slower evolution of the bosonic spectral function (not shown) is governed by changes in the energy distribution of the photo-doped carriers, and on much longer timescales by doublon-hole recombination processes. 

Looking at the real part of the screened interaction, which is plotted in the left hand panel, we see that the high-energy mode produces only a small screening effect, as expected in a Mott insulator, while the low-energy screening modes linked to screening transitions within the photo-doped Hubbard bands lead to a significant reduction of $\text{Re}W(\omega=0,t)$. The evolution of the bosonic Weiss field $\mathcal U(\omega,t)$ looks qualitatively similar to that of $W(\omega,t)$, but with smaller screening effects. The reduction of $\mathcal{U}$ by the enhanced screening in the photo-doped Mott insulator results in a shrinking of the gap size and potentially even in a screening-induced transition to a transient metallic state.   
 
\subsection{GW+DMFT}
\label{sec:applications_models_gw+dmft}

\subsubsection{$U$-$V$ Hubbard model}
\label{sec:UVhubbard_gw+dmft}

The first self-consistent GW+DMFT calculations have been presented for the $U$-$V$ Hubbard model (\ref{H_UV}) in Refs.~\cite{Ayral2012} and \cite{Ayral2013}. The nonlocal component of the self-energy $\Sigma$ was found to be much smaller than the local contribution, even close to the Mott transition. This result is inconsistent with cluster DMFT data, which (for the half-filled Hubbard model) predict a strong momentum differentiation between the nodal and antinodal region, and even a momentum-selective metal-insulator transition \cite{Werner2009}. However, as mentioned in Sec.~\ref{sec:gw+dmft}, the main purpose of the GW+DMFT scheme is not to provide an accurate description of the momentum dependent self-energy, but to enable a self-consistent description of screening. Hence, it is more interesting to look at the polarization function $P(\vc{k},\omega)$. It was shown in Ref.~\cite{Ayral2013} that the nonlocal components of the polarization can be of the same order of magnitude as the local polarization if the system is close to the charge ordering instability. 

\begin{figure}[t]
\includegraphics[width=0.49\textwidth]{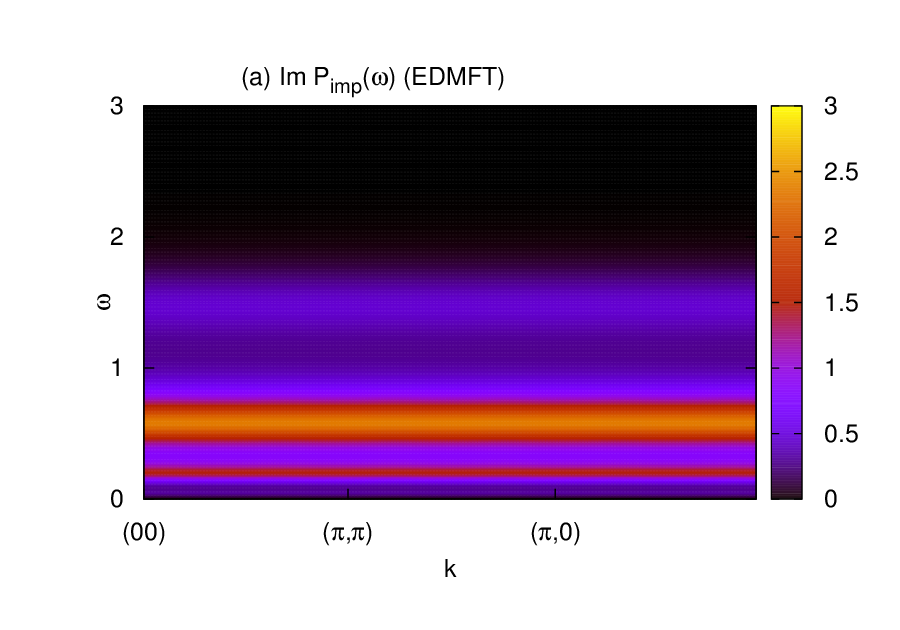}
\includegraphics[width=0.49\textwidth]{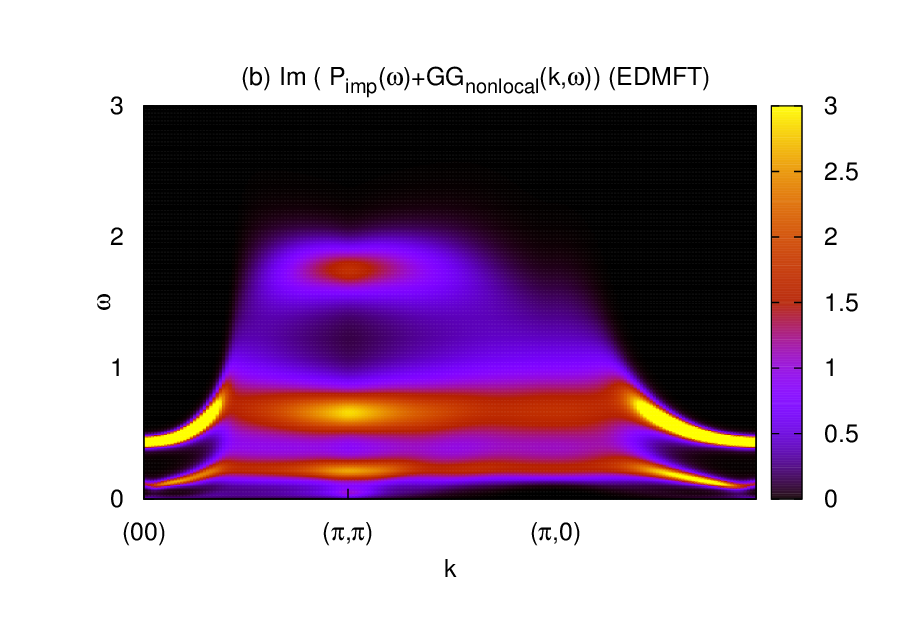}\\
\includegraphics[width=0.49\textwidth]{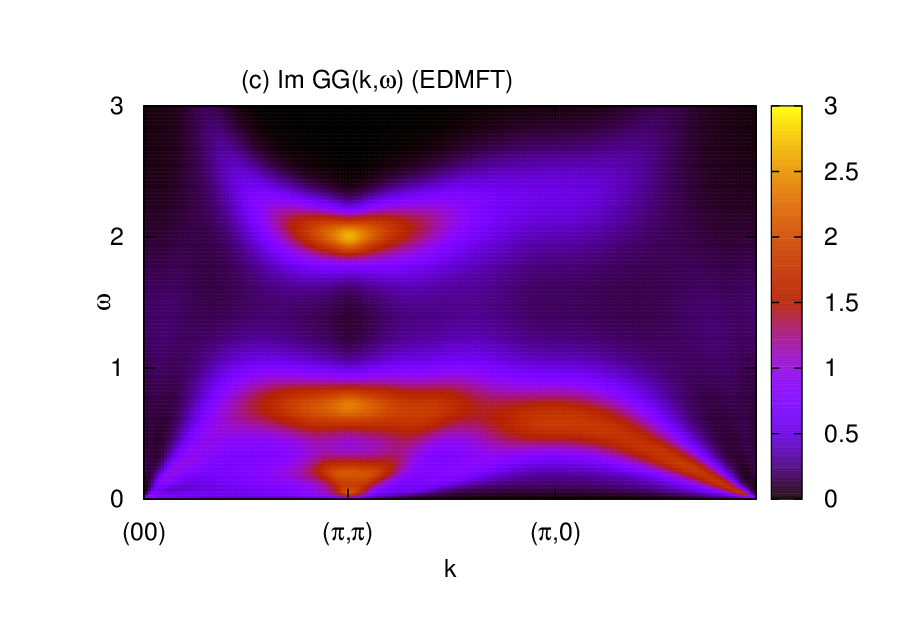}
\includegraphics[width=0.49\textwidth]{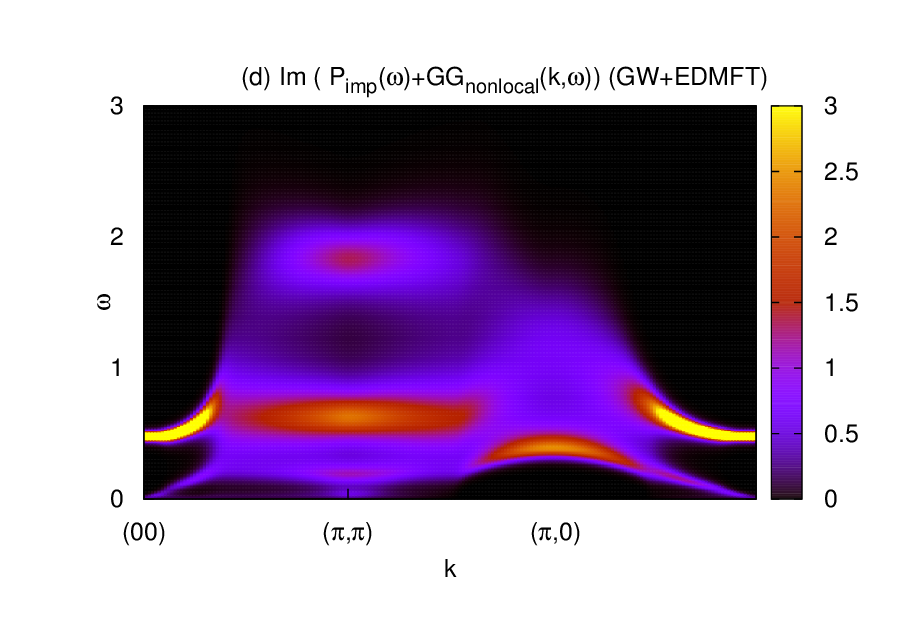}
\caption{Comparison of various approximations for the momentum dependent polarization function ($U = 1.5$, $V = 0.4$). (From Ref.~\cite{Ayral2013}.)}
\label{Fig_P}
\end{figure}

A relevant question is to what extent the self-consistent feedback of
the momentum-dependent polarization and self-energy affects the
converged result. To illustrate this, we compare in Fig.~\ref{Fig_P}
different approximations for the polarization \cite{Ayral2013}: the local polarization
obtained from EDMFT (panel (a)), the bubble $GG$ computed with EDMFT
lattice Green's functions (panel (b)), the sum of the local EDMFT
polarization and the nonlocal part of the bubble (panel (c)), and the
self-consistent GW+DMFT polarization (panel (d)). Let us first
consider panels (a) and (c). While the bubble diagram $GG$ yields a
$\vc{k}$-dependent polarization, it lacks vertex corrections beyond those
built into the EDMFT propagators. The local EDMFT polarization, which
is calculated from the density-density correlation function
(\ref{W_chi}), contains the local vertex. An advantage of the GW+DMFT
method is that it incorporates both this local vertex and the
momentum-dependence of the bubble in a self-consistent manner. The
effect of the self-consistent treatment becomes apparent by comparing
panels (b) and (d). While we can combine the local EDMFT polarization
and the nonlocal component of the EDMFT polarization bubble in the
spirit of GW+DMFT, such a calculation lacks a self-consistent feedback,
and as a result, the momentum-dependence of the polarization looks
quite similar to the bubble result, away from the $(0,0)$ point. In
the GW+DMFT polarization, some of the momentum-dependent structures
differ significantly from the ``one-shot'' result in panel (b),
especially near the $(\pi,0)$ point.      

The self-consistent feedback of the $\vc{k}$-dependent self-energy and polarization also has an effect on local observables, such as the local spectral function. In Refs.~\cite{Ayral2012}, \cite{Ayral2013} and \cite{Huang2014} it was found that in the half-filled $U$-$V$ Hubbard model, the self-consistent GW+DMFT calculation yields stronger correlation effects than DMFT, as exemplified by a larger mass enhancement or more pronounced Hubbard bands. ``One shot'' calculations, in which a nonlocal GW contribution is added to the converged EDMFT self-energy, produced the opposite effect. This result however appears not to be a generic one. For example, it was found in a recent study of the Hubbard model away from half-filling that the inclusion of the non-local GW diagrams reduces the correlation effects \cite{Gukelberger2015}. The latter study, which was restricted to the weak-coupling regime, concluded that GW+DMFT provides slightly more accurate results for the local self-energy than DMFT, while the non-local components are not improved with respect to pure GW. 

In the strong-coupling regime, where the failure of self-consistent GW to produce Hubbard bands is well known, the combination with DMFT enables physically meaningful self-consistent calculations \cite{Ayral2012}. Therefore, the main advantage of the GW+DMFT scheme is that it enables a self-consistent treatment of screening effects at arbitrary interaction strength. 

\begin{table}[t]
\caption{Bare [effective] on-site interactions $U^\text{bare}$ [$U$] and nearest-neighbor interactions $V^\text{bare}$ [$V$]. Also shown is the static value of the effective on-site interaction $\mathcal{U}(\omega=0)$ obtained from GW+DMFT. All values are in eV. (From Ref.~\cite{Hansmann2013}.)}
\begin{tabular*}{\columnwidth}{@{}l*{15}{@{\extracolsep{0pt plus 12pt}}l}}
\\
\hline\\[-0.8cm]
 &C&Si&Sn&Pb \\
\hline\\[-0.6cm] 
$U$&$1.4$&$1.1$&$1.0$&$0.9$\\ 
$V$&$0.5$&$0.5$&$0.5$&$0.5$ 
\\[0.2cm]
\hline\\[-0.6cm]
$U^\text{bare}$&$6.0$&$4.7$&$4.4$&$4.3$\\ 
$V^\text{bare}$&$2.8$&$2.8$&$2.7$&$2.8$
\\[0.3cm]
\hline\\[-0.6cm]
$\mathcal{U}(\omega=0)$&$1.3$&$0.94$&$0.84$&$0.67 \text{ (insulator)}$\\
 & & & &$0.54 \text{ (metal)}$
 \\[-0.25cm]
\end{tabular*}
\label{tab_surface}
\end{table}

\subsubsection{Adatom systems on semiconductor surfaces}
\label{sec:applications_gw+dmft_adatoms}

An interesting playground to explore correlation effects in two-dimensional lattice systems are periodic systems of adatoms on a semiconductor surface \cite{Tosatti1974}. A recent \emph{ab initio} study of Si(111):X (X=Sn, C, Si, or Pb) based on cRPA downfolding and a self-consistent GW+DMFT solution of the low-energy effective theory has revealed the importance of nonlocal Coulomb interactions and dynamical screening effects in these systems and consistently explained material trends for this series of adatoms \cite{Hansmann2013}. In this work, the one-particle part of the Hamiltonian was calculated in the LDA approximation, yielding a half-filled single band of predominantly $p_z$ character near the Fermi level, with a bandwidth of approximately 0.5 eV for all systems considered. The interaction parameters (partially screened Coulomb matrix elements) were calculated using cRPA and a low-energy window containing the surface band. The resulting {\it static} interaction parameters for the on-site ($U$) and nearest-neighbor ($V$) interactions are listed in Tab.~\ref{tab_surface}. Also shown for comparison are the bare interaction values, which ignore the screening effects from higher-energy bands. 

While the on-site interaction is large, about 2-3 times the bandwidth, the non-local interactions are also substantial, and the resulting nonlocal screening effects are essential for understanding the properties of the different adatom systems. It is furthermore evident that the nearest-neighbor interaction is almost independent of the adatom type. The reason is the relatively large distance of 6 \AA~between the adatoms, which implies that the intersite Coulomb energy is essentially that of two point charges. It was furthermore found that $V$ is very close to the value of $V^\text{bare}$ divided by the static dielectric constant of the silicon surface, which suggests that one can compute the longer-range interactions by rescaling $V$ with $a/r$, where $a$ is the nearest neighbor distance. In this sense, $V$ parametrizes the strength of all the non-local interactions. The $1/r$ tail can be treated by an Ewald summation. 

\begin{figure}[t]
\begin{center}
\includegraphics[height=0.32\textwidth]{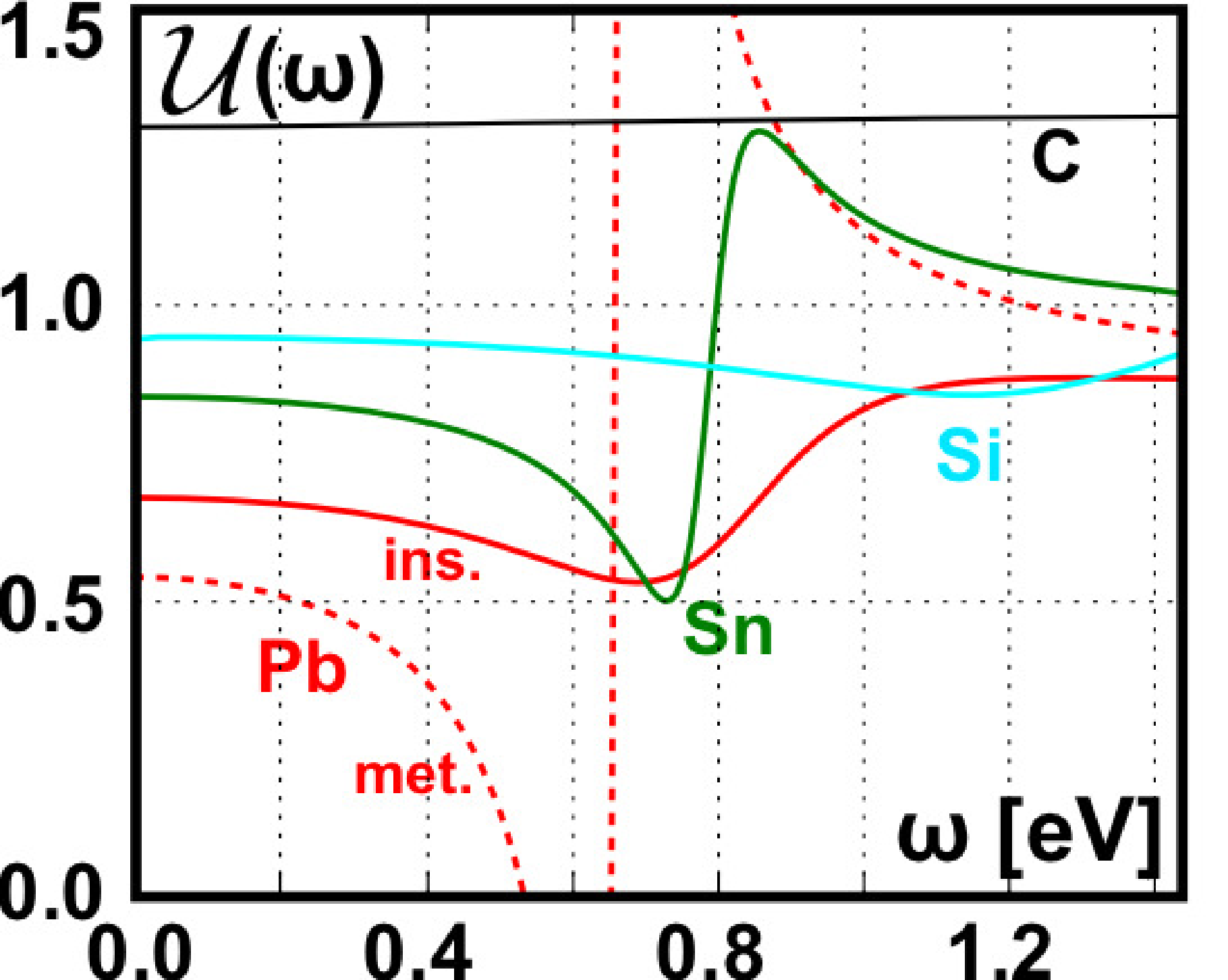}\vspace{5mm}\mbox{}
\includegraphics[height=0.32\textwidth]{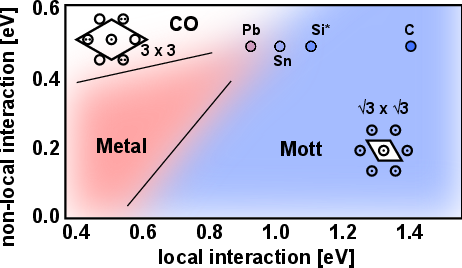}\hfill
\end{center}
\caption{Left panel: frequency-dependent on-site interactions for the
  adatom systems Si(111):X (X=Sn, C, Si, Pb) obtained within
  GW+DMFT. Right panel: Schematic phase diagram in the space of
  non-local and local interactions with Mott insulating, metallic and
  charge ordered (CO) regions. The surface unit cells of the insulating phases are sketched as in-sets.
  (From Ref.~\cite{Hansmann2013}.)}
\label{fig_phasediagram_adatom}
\end{figure}

The low-energy model with these static on-site and off-site interactions (a particular realization of the single-orbital $U$-$V$ Hubbard model) was solved using self-consistent GW+DMFT \cite{Hansmann2013}. The dynamical interaction $\mathcal U(\omega)$ obtained within this scheme is plotted in the left panel of Fig.~\ref{fig_phasediagram_adatom} and reflects the nonlocal screening effect on the local interaction. While at high frequencies, screening is not effective and $\mathcal U(\omega\rightarrow \infty)=U$, the static value of $\mathcal{U}$ can be substantially reduced, especially in a metallic system (Si(111):Pb). As a result, also the gap values are smaller than they would be in the absence of nonlocal screening. 

The simulation results, which reproduce the experimentally observed materials trends, are summarized in the schematic phase diagram of Fig.~\ref{fig_phasediagram_adatom} (right panel). While all the considered systems have the same strength of the nonlocal interactions, their on-site interactions differ. As a result of this, the screening effect in Si(111):C and Si(111):Si is small (see left panel), which places these systems in the Mott insulating region of the phase diagram, while Si(111):Sn and Si(111):Pb are close to a metallic solution (the latter one is in fact in a coexistence regime). It was also found that Si(111):Pb is close to a charge ordering instability.

\section{Applications to realistic materials}
\label{sec:realistic_applications}

\subsection{LDA+DMFT+$U(\omega)$}
\label{sec:lda+dmft+uomega}

The frequency-dependent interaction parameters derived from cRPA have
been employed in several recent \emph{ab initio} simulations based on
the LDA+DMFT framework
\cite{Werner2012,Huang2012,Sakuma2013_U,Werner2015}. These simulations
have produced high-energy satellites and enhanced correlations in
often good agreement with experiments. In this section, we illustrate
the important effects of the dynamical $U(\omega)$ by focusing on
three materials: SrVO$_3$ (Sec.~\ref{srvo3_lda+dmft}), hole-doped
BaFe$_2$As$_2$ (Sec.~\ref{bafe2as2}) and undoped La$_2$CuO$_4$
(Sec.~\ref{sec:mott_gap_cuprates}).   

\subsubsection{Mass enhancement and satellites in SrVO$_3$}
\label{srvo3_lda+dmft}

The correlated metal SrVO$_3$, with an undistorted perovskite
structure, has been studied extensively within LDA+DMFT,
LDA+DMFT+$U(\omega)$ and variants of GW+DMFT. It is a suitable test
material for new computational schemes, because of its relatively
simple bandstructure, with a well-defined low-energy window containing
the three $t_{2g}$ bands. The LDA bandwidth is 2.6 eV. Experiments
indicate a substantial narrowing of these bands, by about a factor of
two, and the appearance of satellites below and above the renormalized
quasi-particle band \cite{Yoshida2010}. Conventionally, these
satellites have been interpreted as Hubbard bands, but recent
theoretical results force us to reconsider this interpretation. 

LDA+DMFT calculations with a static $U=5$ eV and a Hund coupling
parameter $J=0.68$ eV were shown in Ref.~\cite{Pavarini2004} to
produce the correct band renormalization. The resulting $\vc{k}$-integrated spectrum 
features a pronounced lower and upper Hubbard band at $-1.8$ and $3$
eV, respectively. A more recent study \cite{Nekrasov2006} employed
$U=5.5$ eV, which results in an even larger splitting between the
Hubbard bands. LDA+DMFT calculations with a dynamically screened
$U(\omega)$ have been performed within the DALA approximation
(Sec.~\ref{dala})  in Ref.~\cite{Casula2012maxent} and using the full
cRPA interaction and the CTQMC scheme of Sec.~\ref{ctqmc} in
Refs.~\cite{Huang2012,Sakuma2013}. The frequency dependent $U$ used in
Ref.~\cite{Sakuma2013} is plotted in the left panel of  
Fig.~\ref{Uw_srvo3_crpa}. 
It has a
static value of $3.4$ eV and a dominant pole structure near
$\omega=14$ eV. The width of the renormalized quasiparticle band
predicted by the LDA+DMFT+$U(\omega)$ calculation is 0.9 eV, which is too narrow compared to
experiment. As we will argue in Sec.~\ref{sec:widening}, the missing ingredient
in this calculation is the widening of the band due to the momentum
dependence of the self-energy. In this section, we would like to comment on the proper
static-$U$ description (Sec.~\ref{sec:static}), and the implications for the position of the Hubbard bands. 

\begin{figure}[t]
\begin{center}
\includegraphics[angle=0,width=0.49\textwidth]{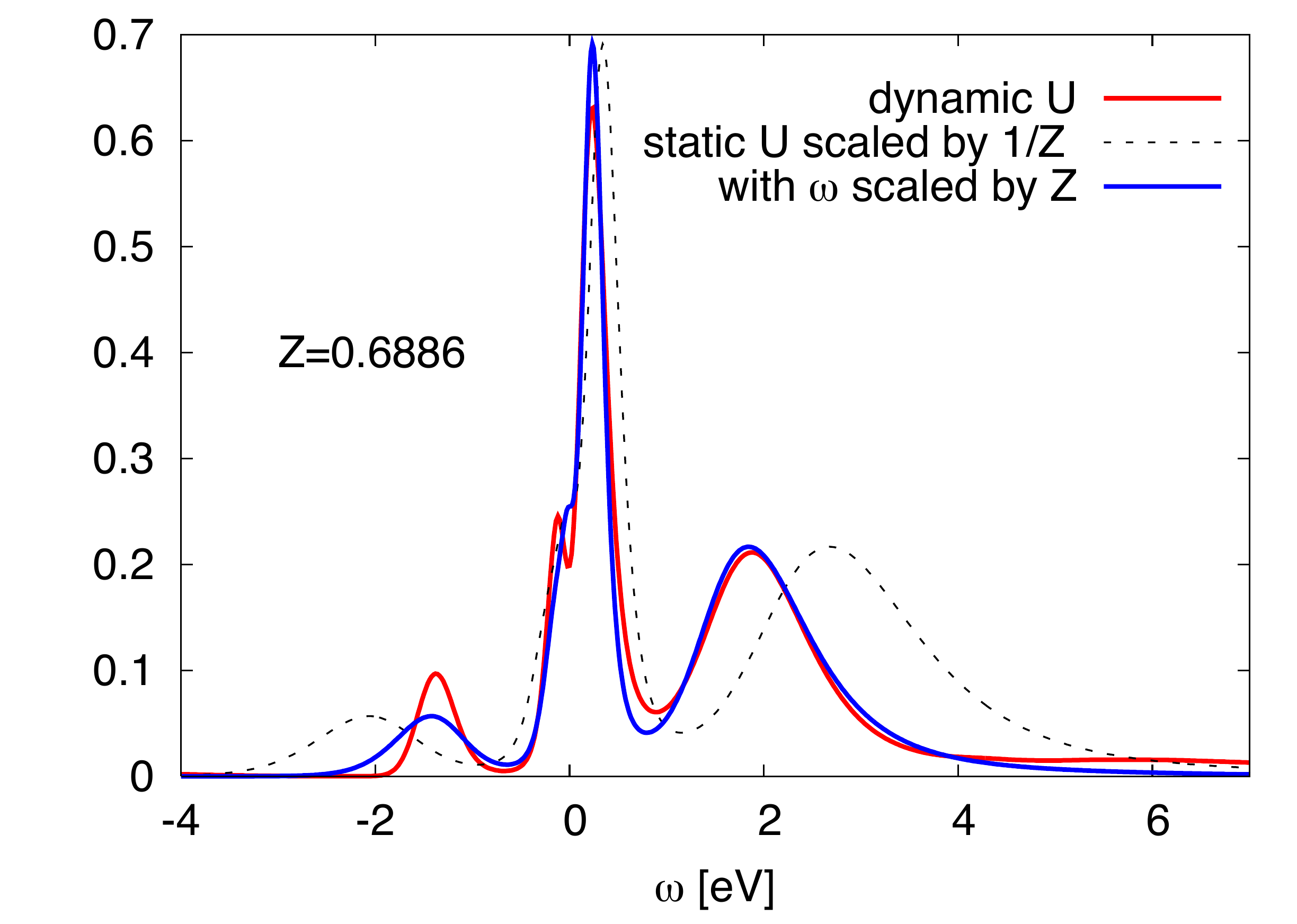} \hfill
\includegraphics[angle=0,width=0.49\textwidth]{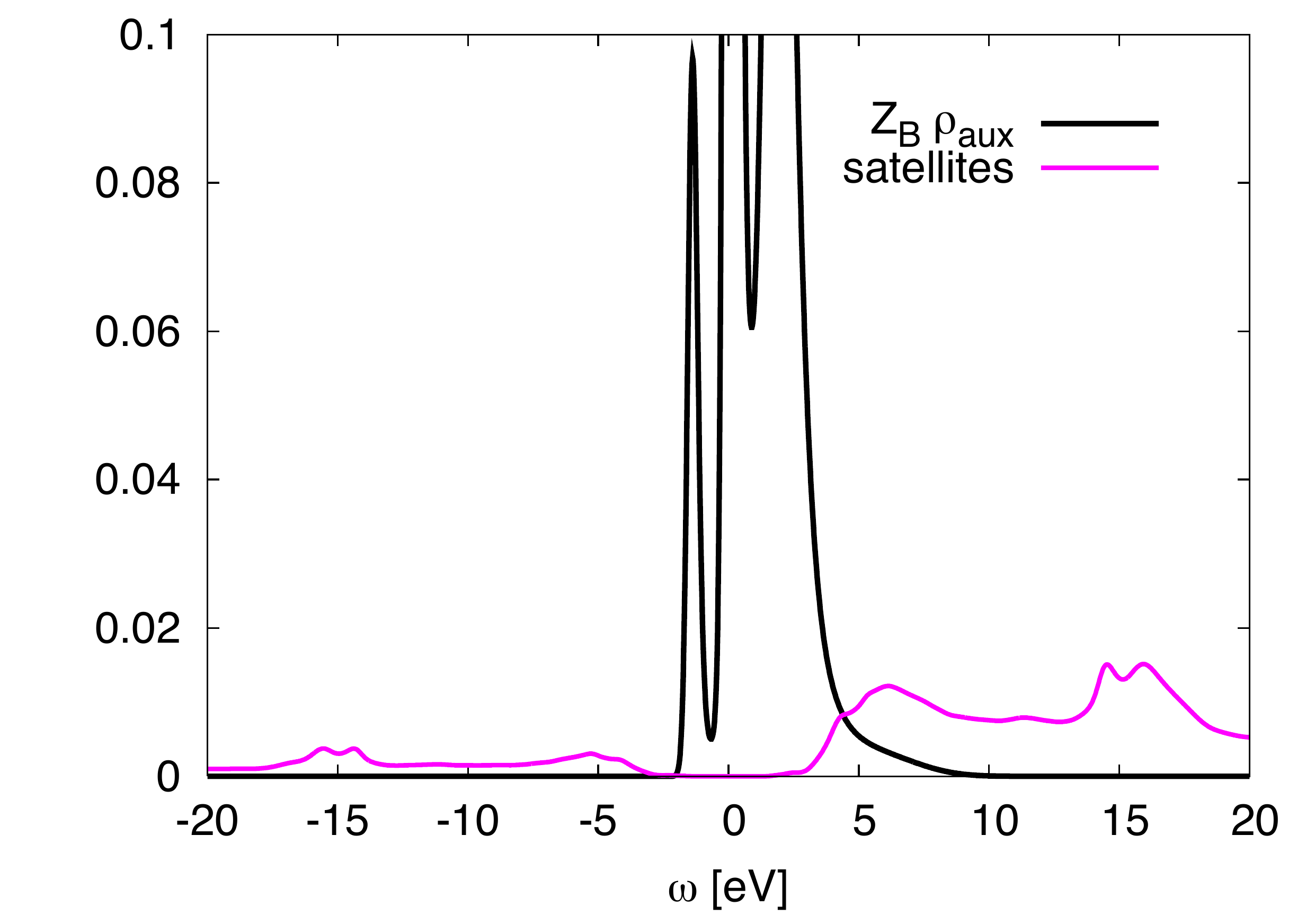} 
\end{center}
\caption{
Left panel: local spectral function of SrVO$_3$ from the dynamic-$U$ simulation (red) and the effective static model of Sec.~\ref{sec:static} (blue). Right panel: Low-energy and satellite contributions to the local spectral function (Eq.~(\ref{maxent_final})). }
\label{Uw_srvo3_lda}
\end{figure}

If the band-widening effect of the $\vc{k}$-dependent self-energy is
neglected, the static description should involve the cRPA
$U(\omega=0)$ and a bandwidth reduced by the Bose factor
(\ref{Z_B_general}), which for SrVO$_3$ is $Z_B=0.689$. Since the material 
has three equivalent $t_{2g}$ bands, shrinking the bandwidth by $Z_B$ is equivalent to
increasing the static interaction to $3.4/Z_B=4.93$ eV and rescaling
the frequency axis by $Z_B$. 

The left panel of Fig.~\ref{Uw_srvo3_lda} compares the local spectral function from the dynamic-$U$ simulation (red line) to a static-$U$ simulation with enhanced on-site interaction before (dashed black line) and after (blue line) the rescaling of the frequency axis. The dashed line is essentially the result of the previous LDA+DMFT simulations \cite{Pavarini2004}. After the rescaling of the frequency axis by the factor $Z_B$, one recovers the mass renormalization of the quasi-particle band and the Hubbard band positions of the $U(\omega)$ calculation. The
upper Hubbard band is placed at $2$ eV, instead of $3$ eV, 
and the lower Hubbard band is shifted from about $-2$ eV to $-1.6$ eV. The latter position is consistent with photo-emission experiments \cite{Yoshida2010}. The spectral weight which is lost by the
rescaling of the frequency axis is shifted to high-energy satellites near $\pm 15$ eV, and also at $\pm 5$ eV (see right panel), which cannot be extracted from a static-$U$ simulation. Looking at the cRPA result for $U(\omega)$ (Fig.~\ref{Uw_srvo3_crpa}), we see that the energies of these satellites are determined by the dominant screening modes (peaks in $\text{Im}U(\omega)$).  
In fact, the analytical continuation procedure discussed in Sec.~\ref{sec:maxent} by construction leads to satellite features at the corresponding energy off-sets. 

This example illustrates the general fact that LDA+DMFT+$U(\omega)$ simulations can produce the same mass renormalizations as static-$U$ LDA+DMFT simulations, but with a lower $U(\omega=0)$ (here, 3.4 eV instead of 5 eV). As a consequence, the splitting between the Hubbard bands is reduced, or the Hubbard bands may not even be well defined anymore, as in the example discussed in the following subsection.

\subsubsection{Spin-freezing crossover in hole-doped BaFe$_2$As$_2$}
\label{bafe2as2}

BaFe$_2$As$_2$ is a prototypical compound of the so-called 122 family
of iron based superconductors. It becomes superconducting under
pressure, or by hole- and electron-doping. The hole-doped compound
exhibits nontrivial correlation effects, even in the normal phase
(above the maximum $T_c$ of 
38 K \cite{Rotter2008}), as exemplified by the widely varying
experimental estimates of the mass enhancement
\cite{Yi2009,Brouet2013}. 
The remarkable sensitivity of the electronic structure to changes in temperature, pressure or doping was shown \cite{Werner2012} to be related to the proximity of the optimally hole-doped compound to a spin-freezing crossover. Spin-freezing \cite{Werner2008} has recently been recognized as a generic and important phenomenon affecting the properties of multi-orbital systems with Hund coupling in a certain regime of filling, interaction strength and temperature \cite{Georges2013}. Inside the spin-frozen regime, long-lived magnetic moments appear, which leads to strong scattering and bad metallic behavior. The boundary of the spin freezing regime is characterized by fluctuating local moments and non-Fermi liquid properties, in particular a self-energy which varies as a square-root of frequency (rather than linearly with frequency) on the Matsubara axis \cite{Werner2008}. Such a square-root self-energy leads to strong band renormalizations at low energy, even in system which do not exhibit Hubbard bands. Due to the dramatic changes in the electronic structure in the spin-freezing crossover regime, an \emph{ab initio} simulation of hole-doped BaFe$_2$As$_2$ requires an accurate estimation of the interaction parameters. 

\begin{figure}[t]
\begin{center}
\includegraphics[angle=0, width=1\columnwidth]{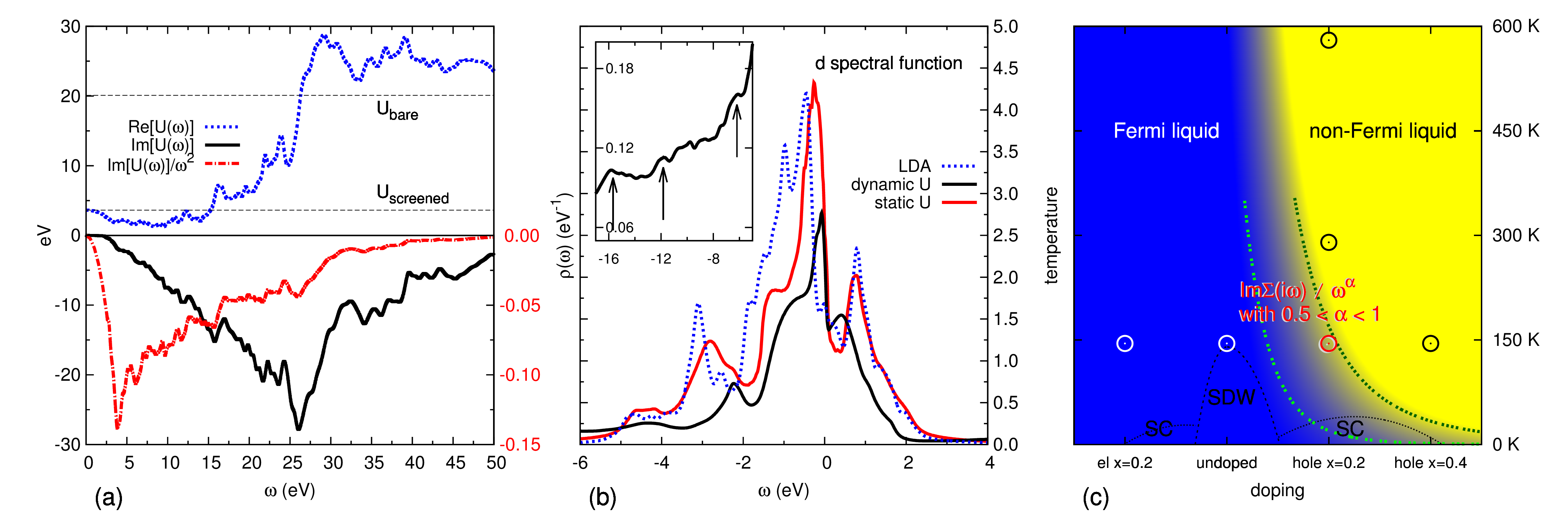}
\caption{Panel (a): cRPA result for the partially screened $U$ of the 5-band model of BaFe$_2$As$_2$. Panel (b): $d$-electron spectral function from LDA (blue), LDA+DMFT with static $U=U_\text{scr}$ (red), and LDA+DMFT+$U(\omega)$ calculations (black). The inset shows the high-energy features in the dynamic-$U$ spectral function. Panel (c): Sketch of the phase diagram of BaFe$_2$As$_2$ in the space of temperature and doping.  The spin-frozen region is indicated by yellow, whereas blue corresponds to Fermi liquid behavior. The border of the spin-frozen regime is characterized by a self-energy which varies like $\text{Im}\Sigma(i\omega_n)\sim \omega_n^\alpha$, with $0.5<\alpha<1$. (Adapted from Ref.~\cite{Werner2012}.)}   
\label{fig:bafe2as2}
\end{center}
\end{figure}

The left panel of Fig.~\ref{fig:bafe2as2} shows the real and imaginary parts of the partially screened interaction $U(\omega)$ for the Fe-$d$ states estimated from cRPA \cite{Werner2012}. The real part varies from the static value $U(\omega=0)=3.6$ eV to a bare value $U_\text{bare}\approx 20$ eV. The plasmon excitation near $\omega\approx 26$ eV overlaps with single-particle excitations, which results in a broad peak in Im$U(\omega)$. In order to properly judge the importance of the different features we also plot Im$U(\omega)/\omega^2$. As discussed in Sec.~\ref{sec:generalU} it is the twice-integrated retarded interaction, or Im$U(\omega)/\omega^2$, which enters into the calculation of the diagram weights in a hybridization expansion solver (see Eq.~(\ref{K})). Apart from a dominant peak at 3.8 eV, which results from the lack of ``high-energy" screening processes below this frequency, there are additional peaks at 6.1, 12 and 16 eV. 
As in the case of SrVO$_3$ discussed in the previous subsection, 
such sharp structures can be expected to lead to side-bands in the $d$-electron spectral function at the corresponding energies. Indeed, as shown in the middle panel (inset), we can identify these peaks in the spectral function computed with the procedure described in Sec.~\ref{sec:maxent}. It is interesting to note that a satellite at approximately $-6.5$ eV has been seen in photoemission experiments \cite{Ding2011,DeJong2009}.

The middle panel also shows a comparison of the LDA+DMFT+$U(\omega)$ spectral function for optimally doped BaFe$_2$As$_2$ to the result obtained with LDA+DMFT using the static interaction parameter $U_\text{scr}=U(\omega=0)$ and to the LDA density of states. (A static Hund coupling parameter $J=0.675$ was used in the DMFT calculations.) The much stronger renormalization of the quasi-particle peak in the dynamic-$U$ calculation is due to the non-Fermi liquid self-energy in the spin-freezing crossover regime: Near optimal doping and for the simulation temperature of 145 K, the effective increase of the Coulomb interaction due to the barely screened fast charge fluctuations pushes the system closer to the spin-freezing region, resulting in large mass enhancements. In the underdoped regime, the behavior is more Fermi liquid like and the effect of the dynamic $U$ on the electronic structure is less pronounced, while in the overdoped region, the spin-freezing leads to very short quasi-particle life-times. The location of the spin-freezing region in the space of doping and temperature is sketched in the right-hand panel of Fig.~\ref{fig:bafe2as2} (yellow region). 

\begin{figure}[t]
\begin{center}
\includegraphics[angle=0, width=1\columnwidth]{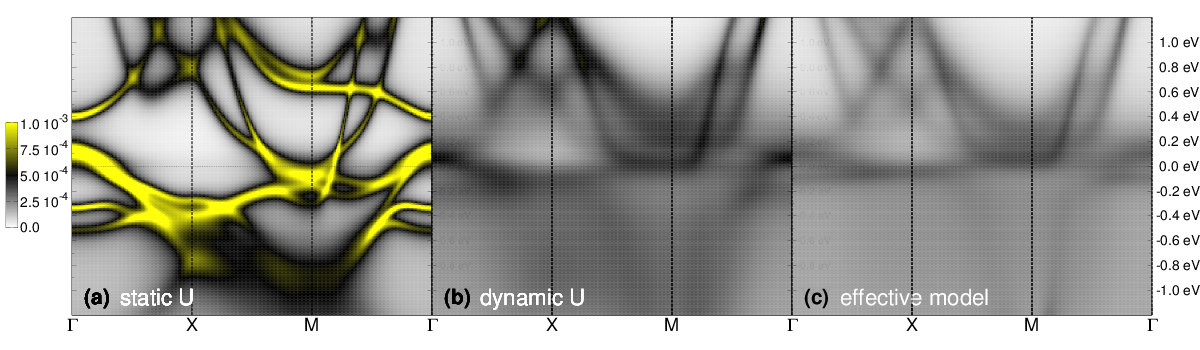}
\caption{Momentum-resolved spectral function for K$_x$Ba$_{(1-x)}$Fe$_2$As$_2$ at optimal doping $x = 0.4$ and temperature $T=600$ K. The left panel shows the LDA+DMFT result for the static interaction $U=U_\text{scr}$, while the middle panel has been obtained using the frequency-dependent $U(\omega)$. The right panel shows the spectral function obtained in a simulation with $U=U_\text{scr}$ and a bandwidth which is reduced by $Z_B=0.59$. 
(From Ref.~\cite{Casula2012static}.) }   
\label{bafe2as2_bands}
\end{center}
\end{figure}

The calculated renormalized band structure of optimally doped BaFe$_2$As$_2$ is shown in Fig.~\ref{bafe2as2_bands} with the left panel plotting the result from the static-$U$ approximation and the middle panel the much more strongly renormalized and smeared out bands obtained in the dynamic-$U$ simulation. As discussed in Sec.~\ref{sec:static} the proper static model involves an effectively reduced bandwidth. In the case of BaFe$_2$As$_2$ the renormalization factor $Z_B=0.59$ is rather low \cite{Casula2012static}. The right hand panel of  Fig.~\ref{bafe2as2_bands} shows the renormalized low-energy bands from a static-$U$ calculation with such a renormalized bandwidth. Due to the enhanced correlation effects, we now get the proper mass enhancement and also the broadening of the bands due to the scattering with local moments. While the low-energy physics is correctly reproduced by the model with reduced bandwidth, this static description will of course not produce any high-energy satellites.

\subsubsection{Mott gap and $-13$ eV satellite in La$_2$CuO$_4$}
\label{sec:mott_gap_cuprates}

Low energy models of cuprates usually involve the Cu $d_{x^2-y^2}$ and O $p_x$ and $p_y$ orbitals. The one-band description considers the anti-bonding combination of these orbitals, while the three-band model also takes into account the bonding combination of  Cu $d_{x^2-y^2}$ and O $p_{x,y}$, as well as the non-bonding $p$ orbital. LDA+DMFT+$U(\omega)$ calculations based on the cRPA estimate of $U$ have recently been analyzed for both models in Ref.~\cite{Werner2015}. The conclusion of this study was that in both models, a static approximation $U=U_\text{scr}$ fails to open a Mott gap, while in the three-band model the dynamic-$U$ calculations yields a gap of approximately 1.9 eV, in good agreement with experiment \cite{Ginder1988}, {\it provided} that $p$-$d$ interactions are accounted for within the DMFT self-consistency loop, at least at the Hartree level. The important role of interatomic Hartree potentials in the correct positioning of the $p$-bands and the opening of the gap is consistent with previous results \cite{Hansmann2014} for a related three-band model.   

\begin{figure}[t]
\begin{center}
\includegraphics[angle=0, width=0.49\columnwidth]{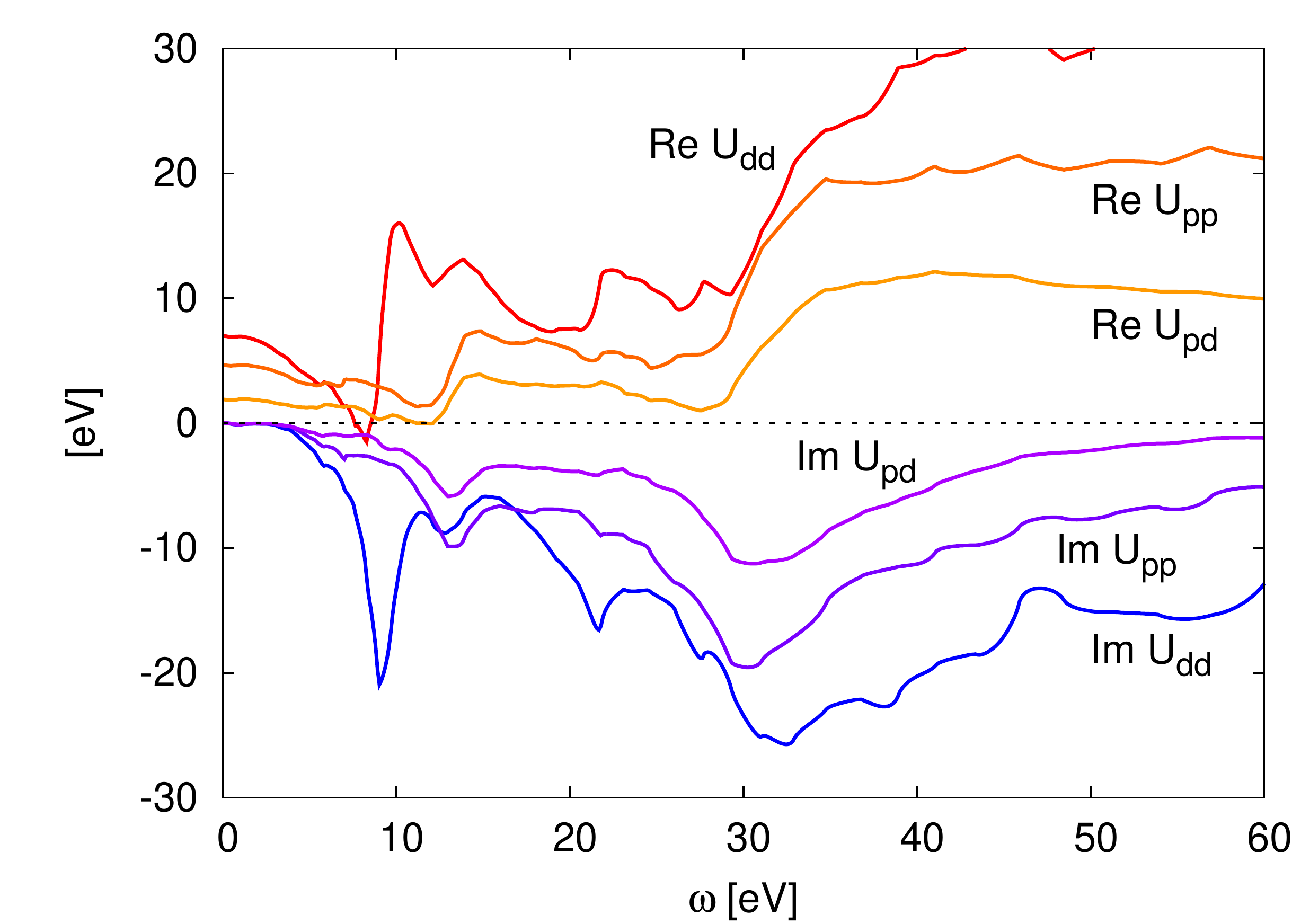}\hfill
\includegraphics[angle=0, width=0.49\columnwidth]{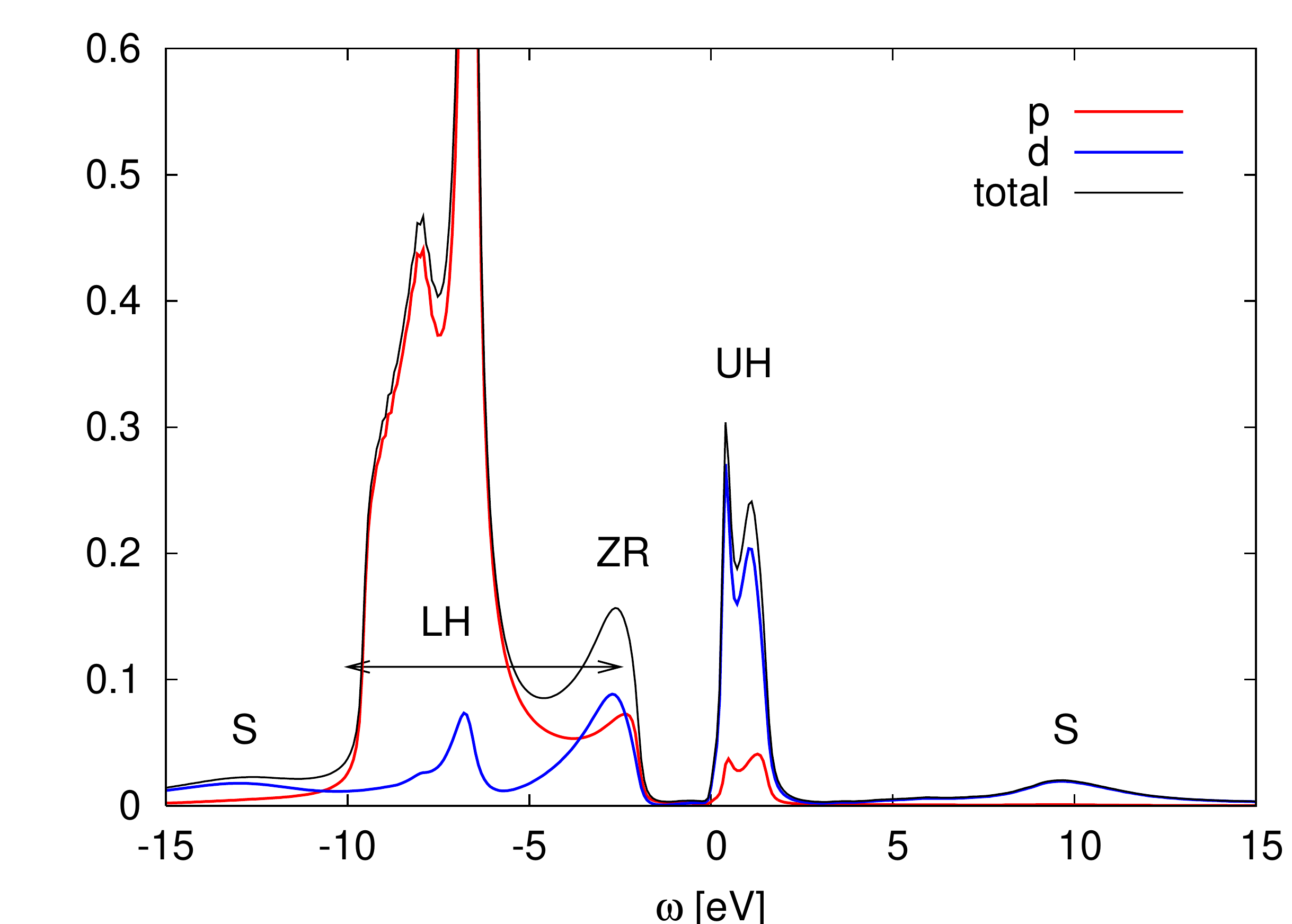}
\caption{ Partially screened interactions from cRPA for the 3-band model of La$_2$CuO$_4$ (left panel) and spectral function at $T=1200$ K from a LDA+DMFT+$U(\omega)$ simulation, which takes into account the interatomic Hartree potential (right panel). In the spectral function, we indicate the following features: satellites (S), lower Hubbard band (LH), upper Hubbard band (UH), and the Zhang-Rice singlet band (ZR). (Adapted from Ref.~\cite{Werner2015}.)}   
\label{la2cuo4}
\end{center}
\end{figure}

The partially screened $U(\omega)$ for the three-band model of La$_2$CuO$_4$ is plotted in the left panel of Fig.~\ref{la2cuo4}. The broad peak in $\text{Im} U(\omega)$ centered at $\omega=30$ eV is a plasmon excitation coupled to single-particle excitations. At lower energies, $\text{Im}U_{dd}$ exhibits a sharp peak at $\omega=9$, which is absent in $\text{Im}U_{pp}$ and $\text{Im}U_{pd}$. This indicates that the collective excitation associated with the 9 eV peak is localized on the Cu site. The prominent 9 eV peak is primarily responsible for the low band renormalization factor $Z_B^{dd}=0.52$, which suggests important screening effects  in La$_2$CuO$_4$.

The $\vc{k}$-integrated spectral function and its $p$-electron and $d$-electron contributions are plotted in the right hand panel of Fig.~\ref{la2cuo4}. At the simulated temperature of $1200$ K, antiferromagnetic correlations do not play an important role, so that the gap in the spectral function is a Mott gap. The upper Hubbard band has a width of about 2 eV, while the lower Hubbard band obtained in this simulation covers the same energy range as the $p$-bands and is hence not easily identified. The states near the lower gap edge, which have mixed $p$-$d$ character, may be interpreted as Zhang-Rice singlet states \cite{Medici2009}. The most interesting feature in this spectral function, as far as dynamical screening is concerned, are the two satellites at $-13$ eV and $+10$ eV. They originate from the sharp 9 eV peak in $\text{Im}U(\omega)$, and should not be confused with the Hubbard bands. A satellite feature at $-13$ eV has indeed been observed in photoemission measurements \cite{Shen1987}.

\subsection{SEX+DMFT results for BaCo$_2$As$_2$}
\label{sec:sex+dmft_baco2as2}

In the quest for new iron-based high-T$_c$
superconductors, a significant effort has been devoted to alloying
iron with other ferromagnetic $3d$ elements, such as Co and
Ni. Ba(Fe,Co)$_2$As$_2$ and Ba(Fe,Ni)$_2$As$_2$ are superconducting,
but with a lower T$_c$ than the ``parent'' BaFe$_2$As$_2$
compound. While pristine BaNi$_2$As$_2$ is still a superconductor, BaCo$_2$As$_2$
is a paramagnetic metal close to a ferromagnetic instability, as
suggested by a high Wilson ratio. 
The latter material is particularly interesting from the theoretical
viewpoint, because it challenges
\emph{ab initio} methods to predict the correct magnetism and ARPES
data. Density functional theory in the standard local spin density approximation
(LSDA) gives a ferromagnetic ground state, which
is not consistent with experiment. Moreover, the LDA bandwidth in the
paramagnetic solution is a factor 1.5 too large compared to ARPES.
The LDA density of states is peaked near the Fermi level, with a quite
large value of 2.12 states/Co/spin/eV, which fulfills the Stoner
criterion for ferromagnetism \cite{Sefat2009}.

\begin{figure}[t]
\begin{center}
\includegraphics[angle=0, width=\textwidth]{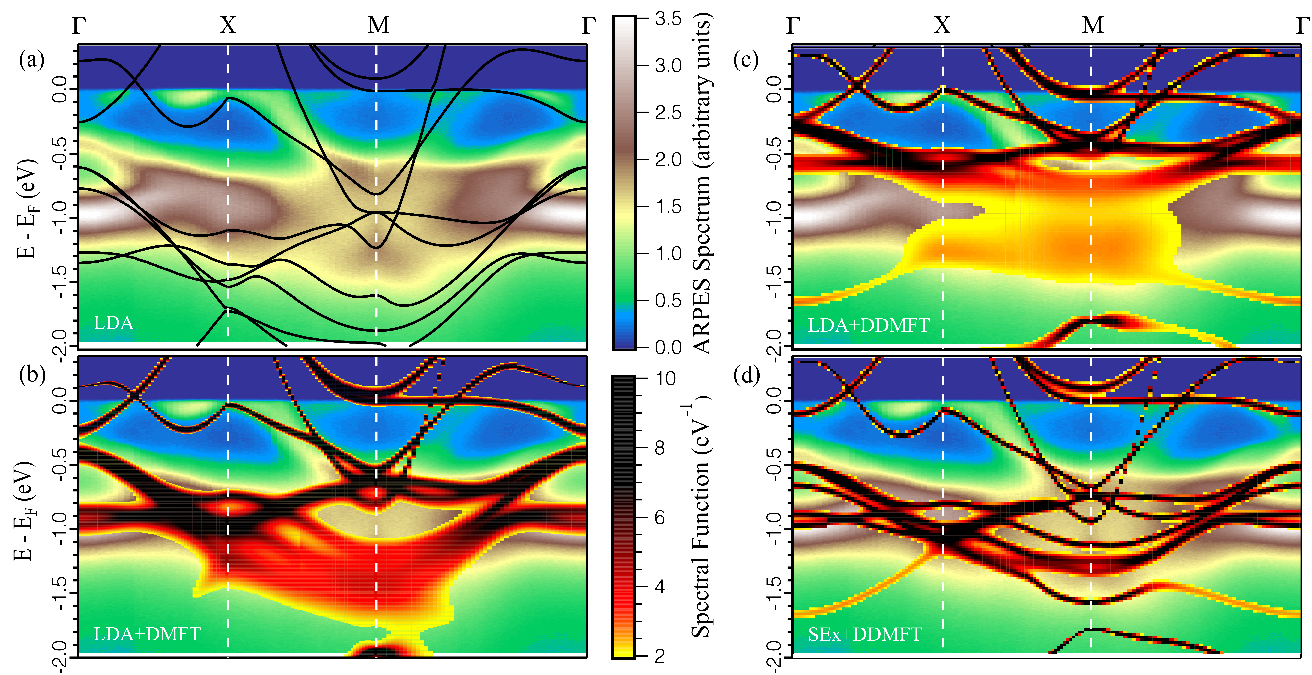}
\caption{LDA band structure of BaCo$_2$As$_2$ (panel (a)), and $\vc{k}$-resolved spectral functions from 
  LDA+DMFT (panel (b)), LDA+DMFT+$U(\omega)$ (dubbed LDA+DDMFT in
  panel (c)), and SEX+DMFT+$U(\omega)$ (dubbed SEx+DDMFT in panel
  (d)), overlaid on the experimental photoemission spectra. (Data
  taken from Ref.~\cite{Roekeghem2014}.)
}
\label{baco2as2_fig}
\end{center}
\end{figure}

In Ref.~\cite{Roekeghem2014}, the SEX+DMFT scheme, presented in
Sec.~\ref{sec:sex+dmft}, has been applied to BaCo$_2$As$_2$, 
in order to see whether a more refined treatment of electron
correlations results in a better
agreement with the experimental situation. On theoretical
grounds, we expect BaCo$_2$As$_2$ to be less correlated than the
superconducting BaFe$_2$As$_2$, 
because the
nominal occupation of the $d$-orbital lattice site goes from the
ideal spin-freezing value of 6 electrons
in BaFe$_2$As$_2$ \cite{Werner2012} to 7 in BaCo$_2$As$_2$. By moving farther away from
half-filling, the onset of strong electron correlations is pushed up to larger
values of $U$. Therefore, the electronic structure of BaCo$_2$As$_2$ is supposed to be characterized by coherent 
quasiparticles, in contrast to BaFe$_2$As$_2$,
where coherence is lost very quickly as the temperature increases,
particularly in the hole-doped compounds 
 (see Sec.~\ref{bafe2as2}).

ARPES data confirm this scenario \cite{Xu2013}, by reporting quite sharp low-energy
spectra for BaCo$_2$As$_2$, typical of a weakly-to-moderately correlated
material, as shown in Fig.~\ref{baco2as2_fig}. A
peculiar feature of its band structure is represented by the $d_{x^2-y^2}$
band, which is very flat along the $\Gamma$-M direction just below the
empty part of the spectrum. This band contributes
to a quite tall peak in the LDA density of states, exactly located at the Fermi level, 
and to the ferromagnetic nature of the LSDA solution.

\begin{figure}[t]
\begin{center}
\includegraphics[angle=0, width=0.5\textwidth]{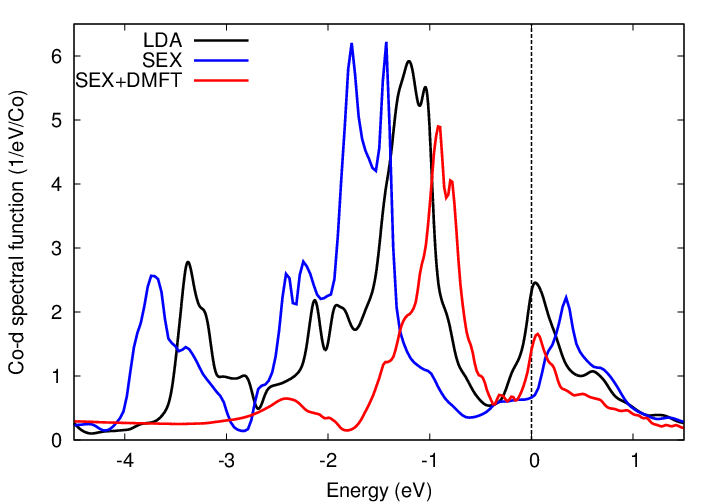}
\caption{
Cobalt-$d$ spectral function of of BaCo$_2$As$_2$ from SEX+DMFT+$U(\omega)$ (red), compared to the
Co-$d$ density of states within LDA (black), and SEX (solid blue
line). (Data taken from Ref.~\cite{Roekeghem2014}.)
}
\label{baco2as2_dos}
\end{center}
\end{figure}

The situation encountered here is common in iron-based superconductors, where
the tendency to magnetism is largely overestimated by density functional theory.
In this sense, it is useful to perform a detailed study of
BaCo$_2$As$_2$ as a benchmark system. 
The authors of 
Ref.~\cite{Roekeghem2014} tested LDA+DMFT calculations of different
flavors, namely LDA+DMFT with the cRPA static $U$, LDA+DMFT+$U(\omega)$, and 
SEX+DMFT+$U(\omega)$. The resulting spectral functions are shown in
Fig.~\ref{baco2as2_fig}. It was found that the regular DFT+DMFT
performs quite well, whereas the dynamic $U$ worsens the DFT+DMFT results
particularly around the M $\mathbf{k}$-point and below $-0.5$ eV, due to a too large band 
renormalization. On the other hand, these features are corrected by 
the SEX+DMFT+$U(\omega)$, as the $Z_B$ band narrowing is
compensated by non-local correlation effects included in the SEX part,
which yields instead a band widening, as it is apparent in Fig.~\ref{baco2as2_dos}.

Therefore, both ingredients, the dynamically
screened $U$ and the non-local screened exchange, are essential for a
consistent description of this moderately correlated material. Electron
correlations included in the non-perturbative solution of the impurity
problem with retarded $U$ are key to the broadening of the spectral function peak at the Fermi
level (Fig.~\ref{baco2as2_dos}), which results in a density 
of states below the Stoner threshold. One thus recovers the correct
paramagnetic phase in agreement with experimental conditions.
SEX+DMFT is a promising scheme that should be applied to other strongly
correlated compounds for further benchmarks and predictions.

\subsection{GW+DMFT results for SrVO$_3$}
\label{sub:gw+dmft_srvo3}

\subsubsection{Band widening from $k$-dependent $\Sigma$}
\label{sec:widening}

\begin{figure}[t]
\begin{center}
\includegraphics[height=0.45\textwidth]{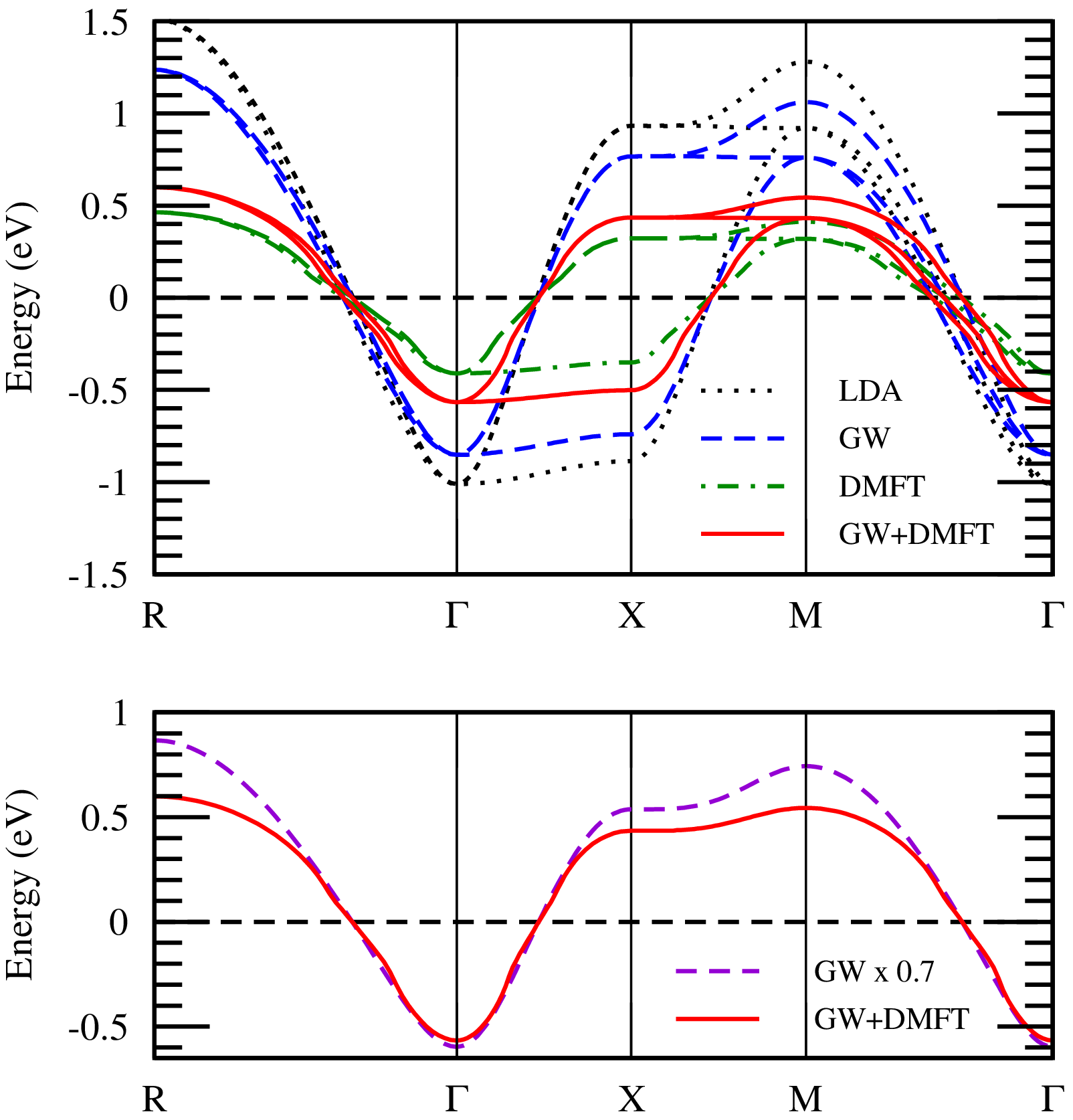}
\end{center}
\caption{Bare and renormalized quasi-particle bandstructures of SrVO$_3$ obtained from different approximate methods. DMFT here means the LDA+DMFT+$U(\omega)$ scheme, GW+DMFT the one-shot result. (From Ref.~\cite{Sakuma2013}.)}
\label{Uw_srvo3}
\end{figure}

The simplest way of combining GW and DMFT is to add the local self-energy of a converged LDA+DMFT+$U(\omega)$ calculation to the nonlocal self-energy from a separate GW calculation \cite{Sakuma2013}. To avoid a double-counting of interaction effects, the local component of the GW self-energy must be subtracted. 
Figure~\ref{Uw_srvo3} compares the resulting quasi-particle bandstructure of SrVO$_3$, obtained as
the solution of the quasi-particle equation $E_{\alpha,\vc{k}} - \epsilon_{\alpha,\vc{k}} - \text{Re}\Sigma_{\alpha,\alpha}(k,E_{\alpha,\vc{k}})=0$,
to LDA, GW and LDA+DMFT+$U(\omega)$ data.
The bandwidth of 1.2 eV predicted by the one-shot GW+DMFT scheme is
in good agreement with photoemission data \cite{Yoshida2010}. In particular, 
the one-shot GW+DMFT corrects the overestimation of the
correlation effect in LDA+DMFT+$U(\omega)$ (Sec.~\ref{srvo3_lda+dmft}), while producing a
substantial renormalization of the GW bandstructure, especially in the
unoccupied part of the spectrum. 
This is illustrated in the lower panel, where a rescaled GW quasiparticle bandstructure is
compared to the one-shot GW+DMFT result. Note that the stronger
renormalization of the bands in the unoccupied part is the result of
the {\it local} self-energy, while 
the overall GW+DMFT band structure is wider than the DMFT one, 
because of the nonlocal self-energy components coming from the GW part.

\begin{figure*}%
\includegraphics[width=0.475\columnwidth,angle=0]{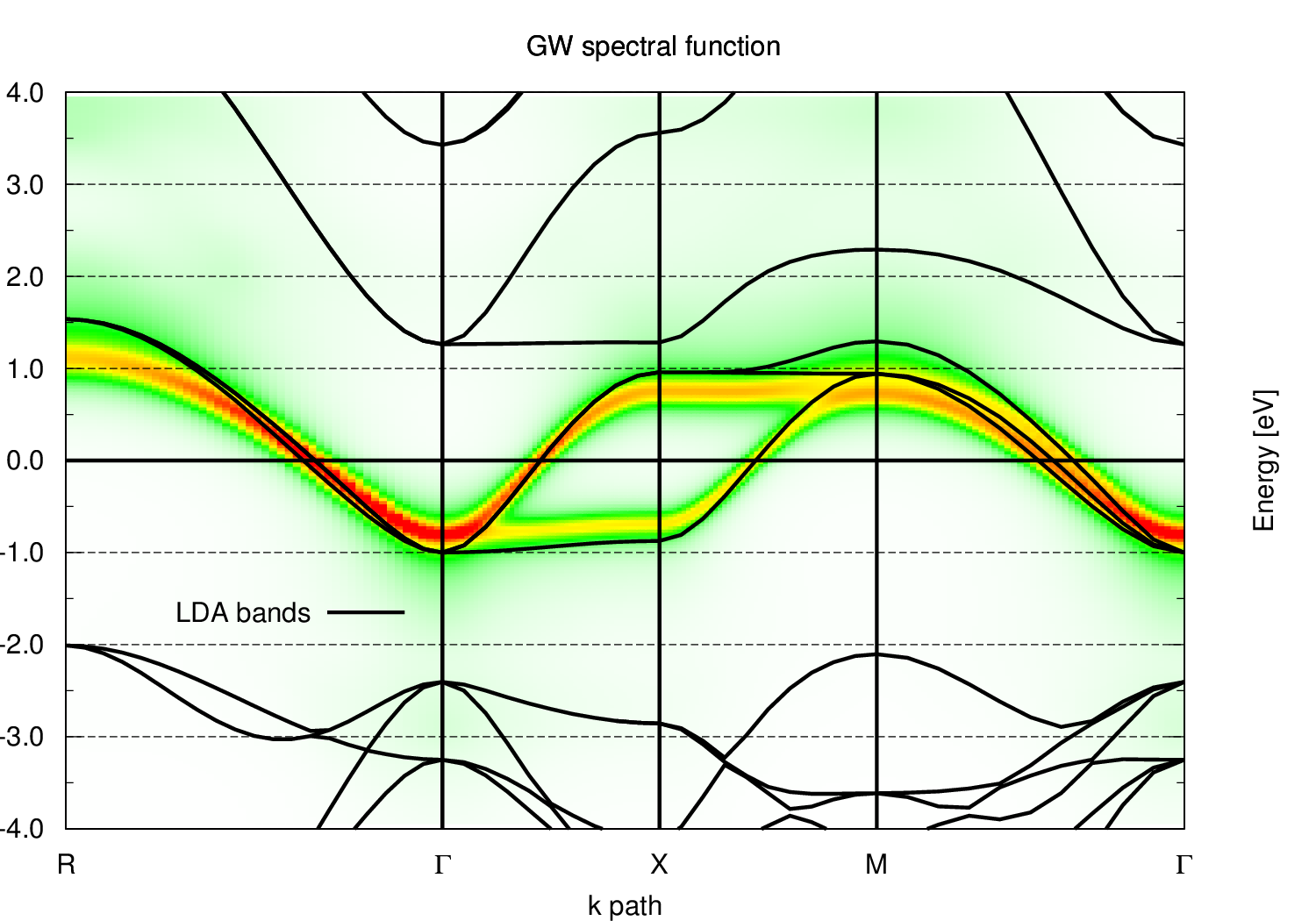}\hfill
\includegraphics[width=0.475\columnwidth,angle=0]{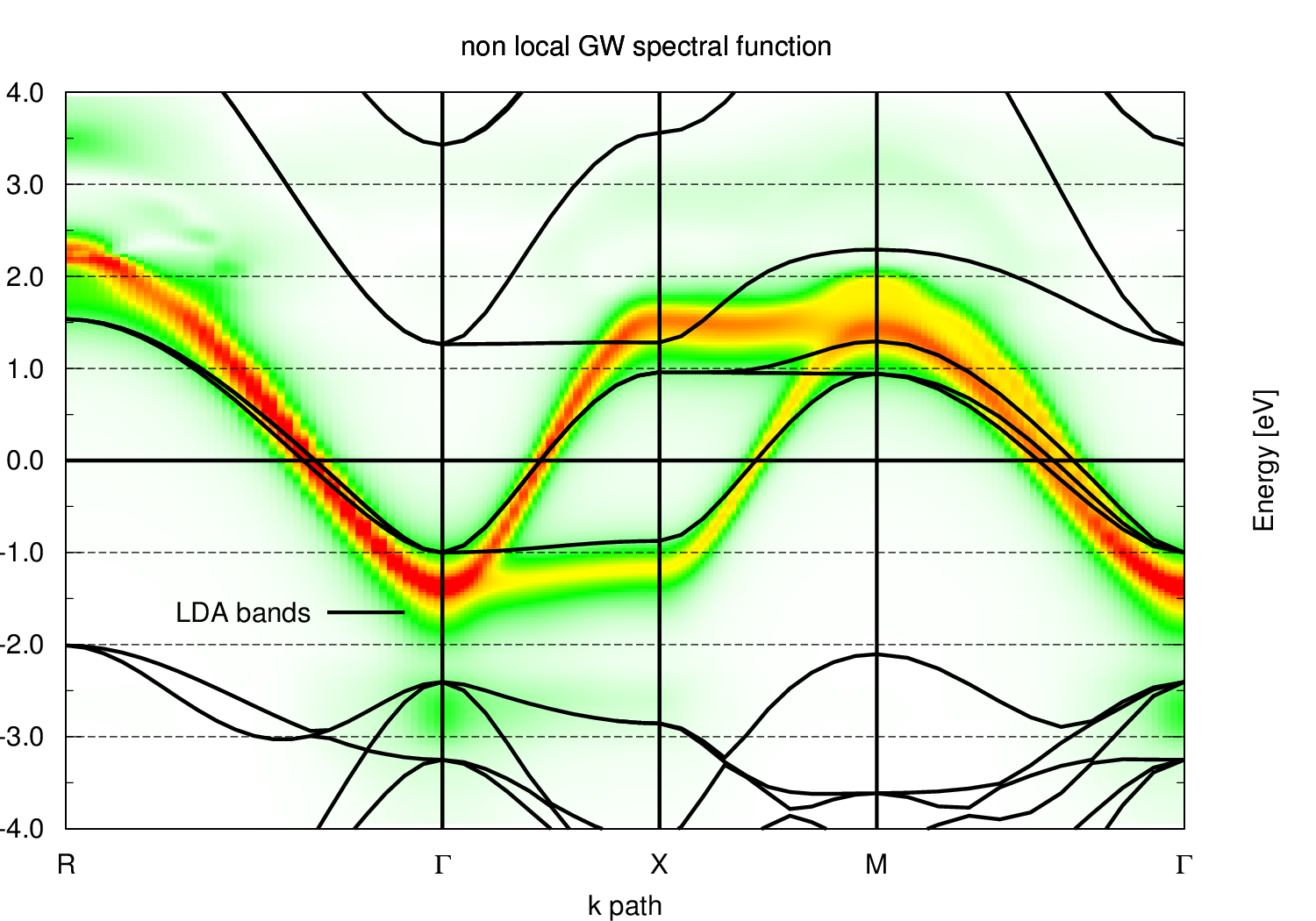}
\caption{Momentum resolved spectral function of SrVO$_3$ obtained  
from the GW approximation (left panel) and by 
taking into account only the {\it nonlocal} part of the GW self-energy
(right panel).
Superimposed is the LDA band structure. (From Ref. ~\cite{Tomczak2014}.)
} %
\label{fig:akwgw}%
\end{figure*}

It has been argued in Ref.~\cite{Miyake2013} that the partial
cancellation between the band-narrowing effect of a dynamically
screened interaction (Sec.~\ref{sec:static}) and the band-widening
effect of the nonlocal self-energy contributions is a generic
phenomenon, which highlights the advantage of formalisms
such as GW+DMFT, which incorporate both effects. 
In fact, it is useful and instructive to analyze the
band widening effect due to the nonlocal exchange at the GW level.  
A one-shot GW calculation of SrVO$_3$
starting from the LDA band structure and wavefunctions produces the
$t_\textrm{2g}$ spectral function shown in the left panel of
Fig.~\ref{fig:akwgw}. The GW bandwidth is renormalized with respect to LDA,
such that the effective mass is increased by a factor
$m_\textrm{GW}/m_\textrm{LDA}=1.3$. However, the quasiparticle weight  
$Z_{\mathbf{k}_F}=1/\left(1-\partial_\omega\Re\Sigma_\textrm{GW}
(\mathbf{k}_F,\omega)\right)_{\omega=0}$ is $\sim 0.53$, 
where the
self-energy is defined with respect to the LDA exchange-correlation 
potential: $\Sigma_{\hbox{\tiny GW}}=\Sigma^\text{xc}_{\hbox{\tiny GW}}-V^\text{xc}_{\hbox{\tiny LDA}}$.
This would give a
mass enhancement of $\sim 2$ in the absence of non-local self-energy
effects. 
However, the expression for the group velocity within GW, from where
the total mass enhancement is extracted, reads:
\begin{equation}
\frac{dE_{\vc{k}i}}{dk_\alpha}=\left.\frac{
\jmbra{\Psi_{\vc{k}i}}   \partial_{k_\alpha}\left( H_{\hbox{\tiny
      {\it LDA}}}(\vc{k})+\Re\Sigma_{\hbox{\tiny {\it GW}}}
(\vc{k},\omega)\right)
\jmket{\Psi_{\vc{k}i}} } 
{\left( 1- \jmbra{\Psi_{\vc{k}i}} \partial_\omega\Re\Sigma_{\hbox{\tiny {\it GW}}}(\vc{k},\omega) 
\jmket{\Psi_{\vc{k}i}}\right)}\right|_{k=k_F,\omega=0},
\label{landau}
\end{equation}
where the additional renormalization via the {\it nonlocality} of the
self-energy, $\partial_{k_\alpha}\Re\Sigma(\vc{k},\omega)$, must be
taken into account.  

In order to illustrate and quantify the effect of the nonlocal components,  one can remove the local
part from  $\Sigma^\text{xc}_{\hbox{\tiny GW}}$ by defining
$\Sigma^\textrm{nonloc}_\textrm{GW}(\mathbf{k},\omega)=\Sigma^\textrm{xc}_\textrm{GW}(\mathbf{k},\omega)-\sum_\mathbf{k} 
\Sigma^\textrm{xc}_\textrm{GW}(\mathbf{k},\omega)$. The spectral function of the corresponding Green's
function 
$\tilde{G}(\mathbf{k},\omega)=1/[\omega+\tilde{\mu}-H_\textrm{LDA}(\mathbf{k})+V^\text{xc}_{\hbox{\tiny
    LDA}}-\Sigma^\textrm{nonloc}_\textrm{GW}(\mathbf{k},\omega)]$
is plotted in the right panel of Fig.~\ref{fig:akwgw}. It is apparent
that non-local contributions, which mainly come from the exchange part, 
lead to a significant band
widening, yielding a bandwidth $\approx 1.5$ times larger than LDA. 
We also note  a small asymmetry in the band-widening, that is, 
the effect is more prominent in the empty part of the
spectrum, 
as pointed out in Refs.~\cite{Tomczak2014} and \cite{Roekeghem2014}.

\begin{figure}[b]
\begin{center}
\includegraphics[width=0.45\columnwidth,angle=0]{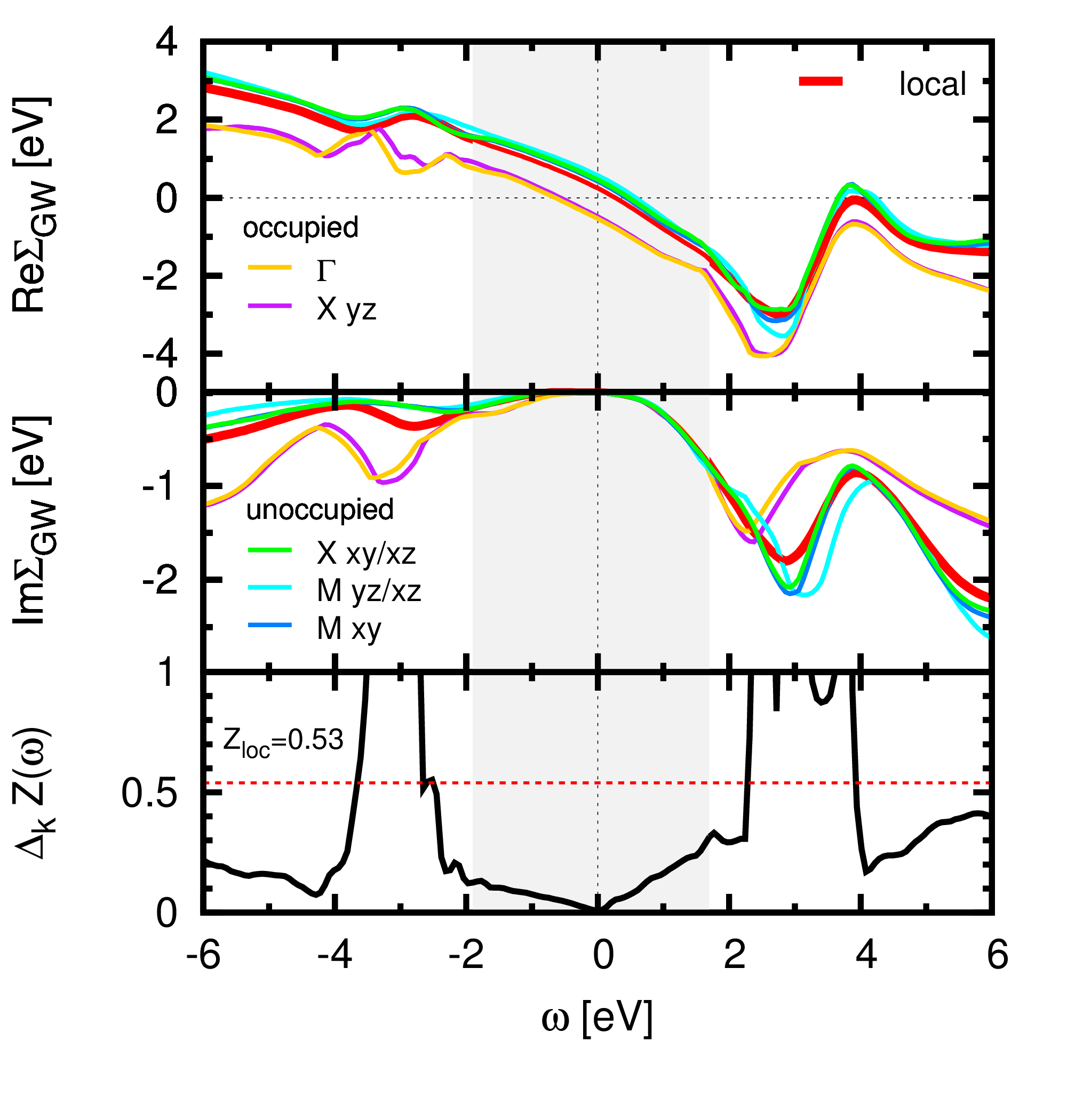}%
\caption{
The GW self-energy of SrVO$_3$ at several high symmetry points resolved into the three t$_{2g}$ (Wannier) orbital contributions as a function of frequency. 
Also shown is the local projection (real parts: top panel, imaginary parts: middle panel).
The lower panel displays the momentum dependence of the frequency dependent generalization of the quasi-particle weight.
The origin of energy corresponds to the Fermi level and the shaded
area roughly indicates the Fermi liquid regime within GW.
(From Ref. ~\cite{Tomczak2014}.)
} %
\label{fig:Zw}%
\end{center}
\end{figure}

Another noteworthy feature of the GW self-energy
of SrVO$_3$ is the separation of the local and non-local self-energy
effects into dynamic and static ones, respectively. A way to quantify the
non-locality of dynamical renormalizations is to compute the generalized
$\vc{k}$-dependent quasiparticle weight
\begin{equation}
Z_\vc{k} (\omega) = \left[ 1-\frac{\partial \Re\Sigma(\vc{k}, \omega)}{\partial \omega} \right]^{-1},
\label{Zk}
\end{equation}
and its $\vc{k}$-fluctuations, defined by 
\begin{equation}
\Delta_k Z=\sqrt{\sum_\vc{k} \text{Tr}|Z_\vc{k}{(\omega)}-Z^\text{loc}(\omega)|^2},
\label{DZk}
\end{equation}
where the local quantity is, as usual, $Z^\text{loc}(\omega)=\sum_\vc{k}
Z_\vc{k}{(\omega)}$. $\Delta_k Z$ and $Z^\text{loc}(0)$ are plotted in the
bottom panel of Fig.~\ref{fig:Zw}. It turns out that the $\vc{k}$-dispersion of $Z_\vc{k}$ around $Z^\text{loc}$ is very weak in the frequency window from -2 to 2 eV, signaling that dynamical effects are
\emph{local} at low-energy. Conversely, \emph{non-local} effects are
static in the same energy window. This is confirmed by the inspection
of the real and imaginary parts of the GW self-energy, plotted for some
selected high symmetry $\vc{k}$-points in the
top and middle panels of Fig.~\ref{fig:Zw}, respectively. One
easily sees that at low energy the frequency dependence is $\vc{k}$-insensitive in
both the real and imaginary parts, while the $\vc{k}$-dependence is
$\omega$-independent in the real part, leading to a $\vc{k}$-dependent
rigid shift of the self-energy curves. This effect has not only been
observed in SrVO$_3$ \cite{Tomczak2014,vanRoekeghem2014}, but also in the iron
pnictides and chalcogenides \cite{jmt_pnict}. 

We come back to the asymmetry
between the occupied and empty parts of the spectrum, which results from the
non-local part of the self-energy. 
This asymmetry is not only present
in the band widening, as seen in Fig.~\ref{fig:akwgw}, but most prominently in
the quasiparticle lifetimes. Indeed, 
the imaginary part of the GW self-energy is largest on the
unoccupied orbitals for $\omega > 0$, as shown in the middle panel of
Fig.~\ref{fig:Zw}, which implies a stronger effect of
electron-electron scattering in the empty part of the spectrum.
However, this large imaginary part lies outside the energy window where the dynamic and static
effects are separable in frequency. Hence, the asymmetry of the
quasiparticle lifetime, a dynamic quantity in nature, \emph{is}
$\vc{k}$-dependent.

\subsubsection{Self-consistent calculation of $\Sigma$}
\label{srvo3:self-consistent}

The partially self-consistent implementation of
Refs.~\cite{Tomczak2012,Tomczak2014} goes a step beyond the 
one-shot calculation discussed in the previous subsection by performing a self-consistency loop
for $\Sigma$ and $G$. However, compared to the full GW+DMFT scheme discussed in Sec.~\ref{sec:gw+dmft},
a number of approximations have been made. The bosonic Weiss field
$\mathcal U$ was kept frozen to the $U_{mn}(\vc{0},\omega)$ cRPA
value, as described in Sec.~\ref{sec:URPA}. In other words, the self-consistency loop for 
$P$ and $W$ was not performed. However, 
the RPA polarization function computed from the LDA band
structure of SrVO$_3$, and its corresponding
dielectric function, compare favorably to electron energy loss spectroscopy measurements of SrTiO$_3$,
an isostructural compound with a $d^0$ occupation,
where data of this kind are available up to 40 eV. 

\begin{figure*}[t]%
\includegraphics[width=0.32\columnwidth,angle=0]{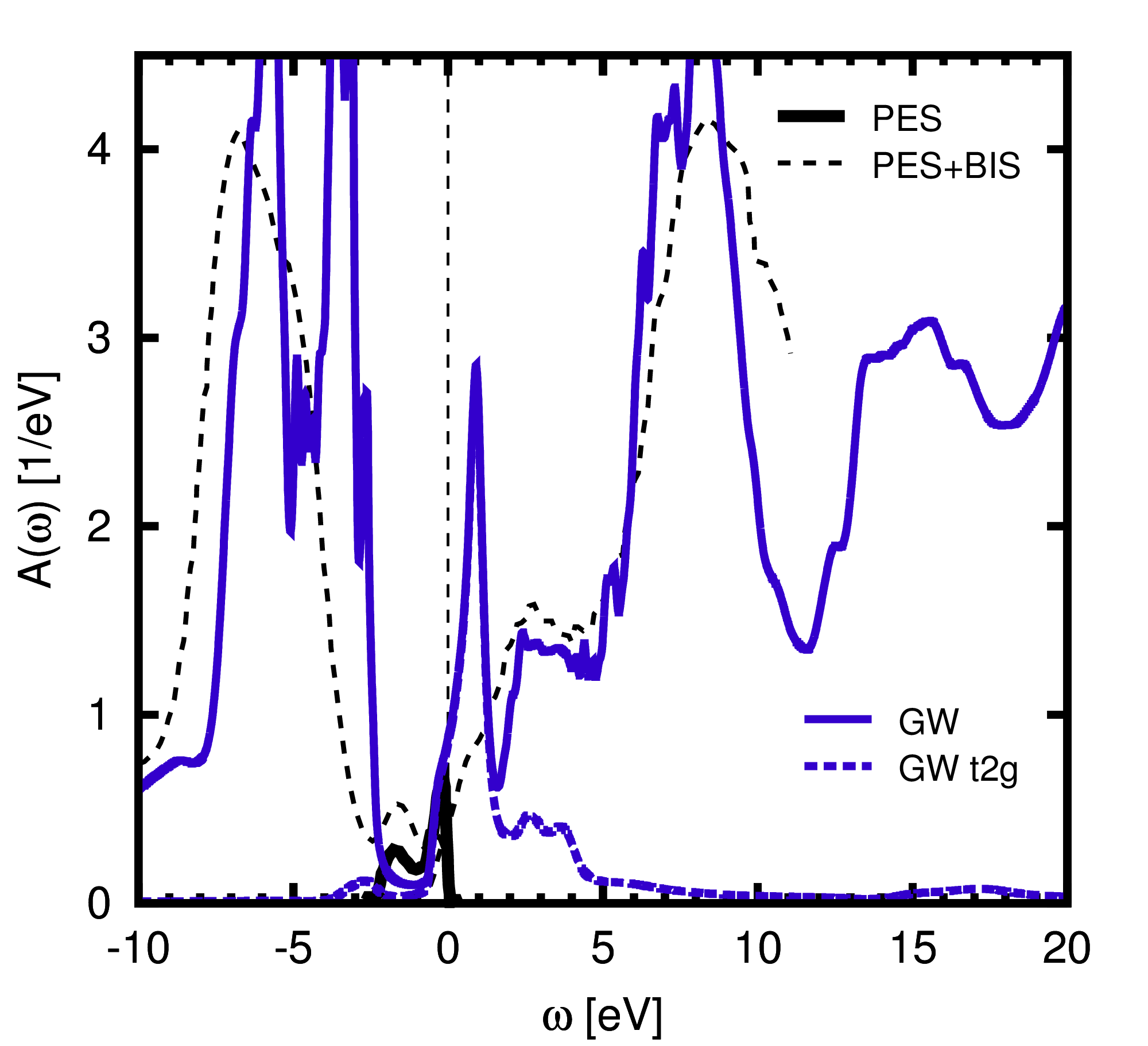}\hfill
\includegraphics[width=0.32\columnwidth,angle=0]{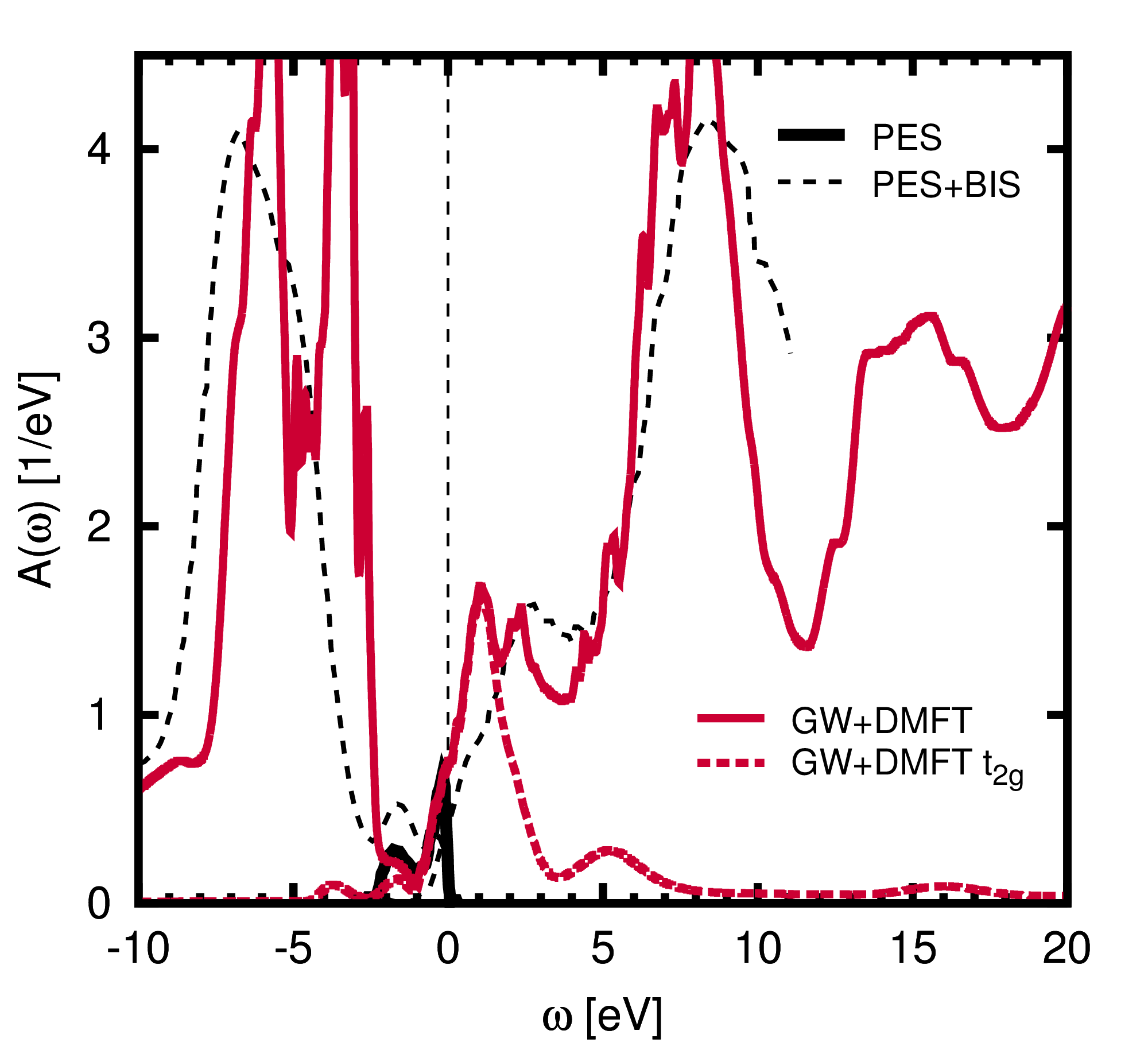}\hfill
\includegraphics[width=0.32\columnwidth,angle=0]{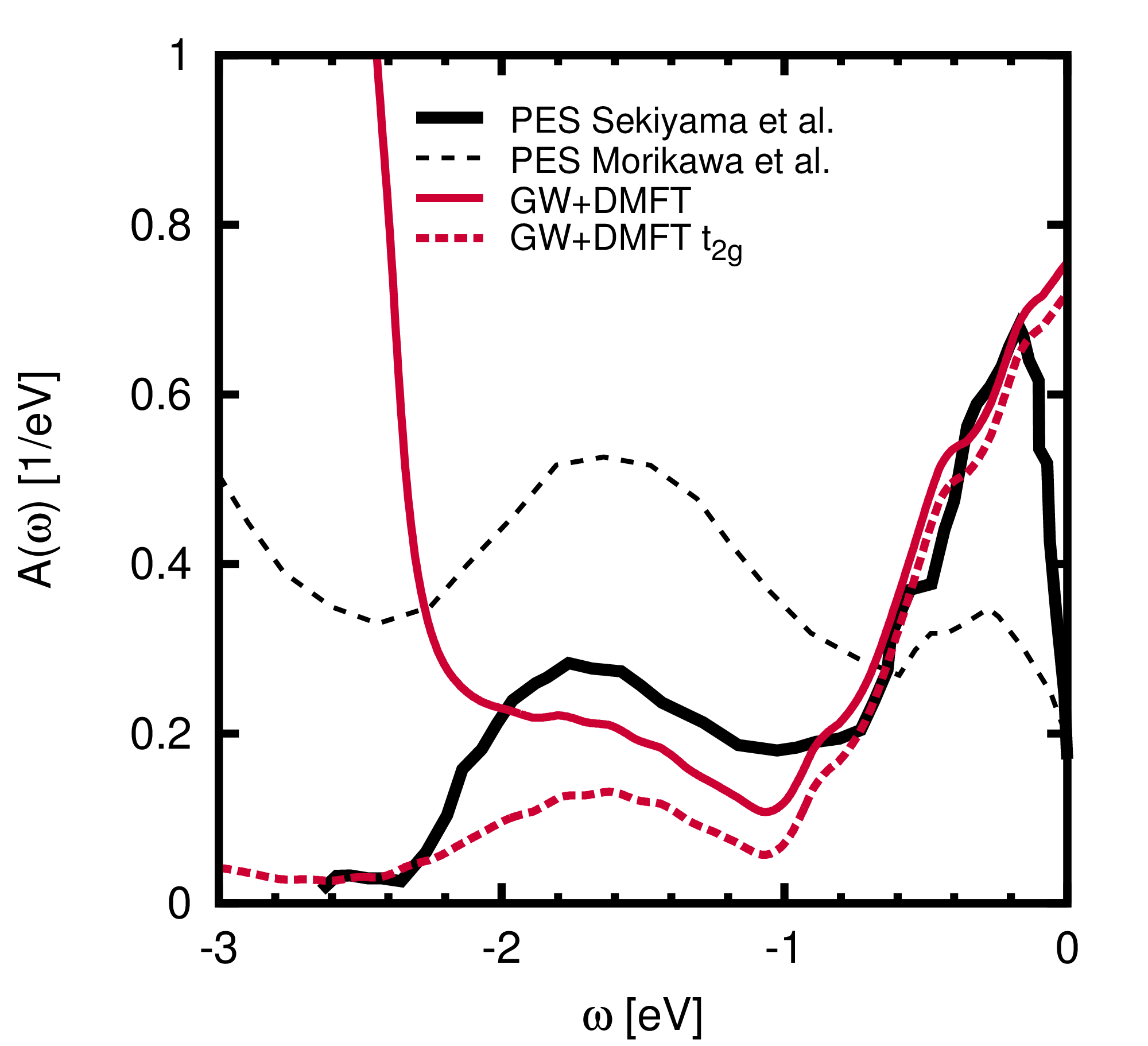}
\caption{
GW (left panel) and GW+DMFT (middle panel) spectral functions of SrVO$_3$
in comparison to photoemission and inverse photoemission spectra.
Right panel: Zoom of the middle panel around the lower Hubbard band region.
(From Ref.~\cite{Tomczak2014}.)
} %
\label{fig:GW+DMFT}%
\end{figure*}

The results for the GW and the GW+DMFT spectra are shown in
Fig.~\ref{fig:GW+DMFT}. 
The dynamically screened 
impurity problem of GW+DMFT 
has been solved by DALA. 
Compared to
LDA, 
the GW spectral function, plotted in the
leftmost panel of Fig.~\ref{fig:GW+DMFT}, gives a
better position of the O-$2p$ and Sr-$4d$ states, which are closer to the
experimental photoemission spectroscopy (PES) and Bremsstrahl-Isochromat spectroscopy (BIS)  
curves taken from Refs.~\cite{PhysRevLett.93.156402} and
\cite{PhysRevB.52.13711}. However, GW yields a too strong
quasiparticle peak at the Fermi level, of full $t_\textrm{2g}$
character, and a too weak mass enhancement. GW+DMFT, reported in the
middle panel of Fig.~\ref{fig:GW+DMFT}, corrects for these deficiencies, by 
strongly renormalizing the height of this peak, and producing a mass
enhancement of $\sim 2$. The renormalization is accompanied by a
spectral weight transfer from the quasiparticle peak to the lower
Hubbard band, correctly located at $-1.6$ eV (see rightmost panel in
Fig.~\ref{fig:GW+DMFT}), and to plasmon satellites, which can be identified in the figure at 
$-4$ eV, 5 eV, and 15 eV. Therefore, dynamic screening and correlation effects 
play a major role in determining the renormalized low-energy properties
of the material.

Surprisingly, the upper Hubbard band is not visible in the GW+DMFT
spectral function computed in Refs.~\cite{Tomczak2012,Tomczak2014}. It may be hidden 
by non-local self-energy effects, particularly large in the empty
part of the spectrum. However, other calculations, such as the one-shot GW+DMFT
scheme of  Ref.~\cite{Sakuma2013}, or the simplified SEX+DMFT \cite{vanRoekeghem2014},
both using a CTQMC solver, show a quite sharp upper Hubbard band well
separated from the quasiparticle peak. 
The missing Hubbard band could thus also be a consequence of an underestimation of the bandwidth reduction by DALA
with respect to QMC (reported in Ref.~\cite{Huang2012}), or by non-local
dynamical self-energy effects just above the Fermi liquid regime,
which are neglected in Ref.~\cite{vanRoekeghem2014} and treated 
non-selfconsistently 
in Ref.~\cite{Sakuma2013}. 

In any case, 
within the partially self-consistent GW+DMFT scheme of Refs.~\cite{Tomczak2012,Tomczak2014} 
the upper Hubbard band should be located around 2 eV, as its
separation from the lower one must be the \emph{static} value of $U$
(here 3.3 eV). 
This implies that the BIS signal detected at 2.7 eV is \emph{not} the
upper Hubbard band of the $t_\textrm{2g}$ manifold, as previously
identified. Overlapping the GW+DMFT spectrum with the BIS data shows
that the peak just above the Fermi level has an $e_\textrm{g}$ origin,
instead, 
as seen in Fig.~\ref{fig:GW+DMFT}.

\section{Conclusions and Outlook}
\label{conclusions}

We have discussed the effect of dynamical screening in strongly correlated lattice systems and materials, and detailed some recently developed or implemented techniques based on extensions of the DMFT framework. At this point, the most advanced scheme which allows a self-consistent treatment of screening and correlation effects in materials is the combination of the GW \emph{ab initio} method and DMFT \cite{Biermann2003}. A fully self-consistent GW+DMFT calculation yields retarded interaction parameters for the DMFT impurity problem, which incorporate the effect of local and nonlocal screening processes in the correlated system. This method also captures the momentum dependence of the self-energy (at the GW level), and hence the competition between the band-narrowing effect of the frequency-dependent interaction and the band-widening effect of the non-local self-energy components. Self-consistent GW+DMFT calculations of multi-band systems and realistic materials have only been implemented very recently. While these results demonstrate prominent non-local screening effects, more work on a broad range of materials is needed to judge the reliability and predictive power of the scheme. Also, it should be emphasized that in current implementations, the self-consistent calculation is performed within a low-energy window containing just a few bands, after a cRPA downfolding or one-shot GW calculation. In the near future, this scheme should be extended to a multi-scale approach with three energy windows, a large energy window for the initial cRPA calculation, an intermediate-energy window for self-consistent GW calculations, and a low-energy window  for the GW+DMFT calculations. In such a scheme, the GW+DMFT estimate of the self-energy of the low-energy space has to be merged with the GW self-energy for the remaining orbitals in the intermediate-energy window, and the calculations within the intermediate and low-energy space should be iterated until a self-consistent solution for $G$, $W$, and the auxiliary impurity problem is obtained. 

Apart from the implementation of these hopefully accurate, but numerically demanding schemes, the further exploration of simplified versions of the GW+DMFT framework will produce useful insights into the effects of screening and nonlocal correlations. We have discussed the partially self-consistent formulation of Tomczak {\it et al.}, which performs a self-consistency loop on $\Sigma$ and $G$, while approximating the effective interaction by the cRPA estimate \cite{Tomczak2014}. If the self-energy is furthermore approximated as local, one ends up with the LDA+DMFT+$U(\omega)$ method, which has been used in recent years by several groups to study plasmon satellites in strongly correlated materials \cite{Werner2012,Huang2012,Sakuma2013,Werner2015}. Another recently proposed variant, which has been applied to correlated materials with promising results, is SEX+DMFT+$U(\omega)$, which takes into account the band-widening effect of the nonlocal screened exchange \cite{Roekeghem2014}. Further applications and comparisons of these different simplified schemes will provide valuable insights into the roles played by the different ingredients of the GW+DMFT formalism, and the importance of self-consistency.   

It is also essential to clarify the accuracy and limitations of the cRPA method, or related RPA schemes, which underpin the material simulations discussed in this review. The RPA polarization lacks vertex corrections, and in principle, a downfolding onto a low-energy subspace should generate higher-order interaction terms which are not contained in an effective Hubbard model description. Some of these issues have recently been addressed in simple model set-ups containing a small number of screening bands \cite{Shinaoka2015,Kinza2015}, but more work is needed to properly judge the realistic situation with a large number of high-energy screening bands.   

It should be kept in mind that GW+DMFT and related methods, which combine the local self-energy from an effective impurity model with the nonlocal self-energy of some weak-coupling perturbation theory, cannot be expected to capture the strong nonlocal correlation effects in low dimensional systems. This has been explicitly demonstrated for the two-dimensional Hubbard model in the weak-coupling regime \cite{Gukelberger2015}, while at intermediate coupling, there are obvious inconsistencies between the weakly momentum-dependent GW+DMFT results, and the strong momentum-variation predicted by cluster DMFT simulations \cite{Parcollet2004,Werner2009}. Hence, we should view GW+DMFT primarily as a method which is suitable for capturing dynamical screening effects in three dimensional compounds. For two-dimensional systems, the combination of (extended) cluster DMFT and many-body perturbation theory may be a promising strategy. However, efficient impurity solvers capable of handling dynamically screened interactions in cluster impurity problems have yet to be developed.   

In fact, progress in this field is intricately linked to futher improvements of the available impurity solvers and analytical continuation techniques. The methods reviewed in this paper enable an efficient simulation of single-site, multi-orbital impurity problems with dynamically screened monopole interactions. Retarded spin-flip terms cannot be handled efficiently with methods based on a Lang-Firsov decoupling of the electron-boson interaction and must be dealt with in a double-expansion approach \cite{Otsuki2013,Steiner2015}, which may suffer from a sign problem. The analytical continuation procedure explained in Sec.~\ref{sec:maxent} is also based on a Lang-Firsov picture, and works reliably only in systems with a clear energy separation between low-energy and satellite features. More flexible and powerful methods will be essential in particular for the eventual application of more advanced schemes, such as TRILEX \cite{Ayral2015}, dual bosons \cite{Rubtsov2012}, or extended cluster methods to realistic materials. 

Another new and interesting research direction is the extension of the methods described here to nonequilibrium systems \cite{Aoki2014,Golez2015}. A proper description of dynamical screening effects should be very important in materials perturbed by intense laser fields, especially if the laser pulse produces inter-band transitions. GW+DMFT is a promising starting point for the development of a formalism, which  enables \emph{ab initio} predictions of nonequilibrium phenomena in strongly correlated materials.

\section*{Acknowledgements}

We would like to thank F. Aryasetiawan, T. Ayral, S. Biermann,  L. Boehnke, V. Brouet, M. Eckstein, D. Golez, J.~Gukelberger, L. Huang, A. J. Millis, F. Nilsson, T. Miyake, Y. Nomura, R. Sakuma, A. van Roekeghem, H. Shinaoka, J. Tomczak, and L. Vaugier for helpful discussions and close collaborations on the topics covered in this review. PW acknowledges support from ERC starting grant 278023. Some of the results presented in this review were obtained using HPC resources from GENCI 096493 allocation.

\bibliography{screening}

\begin{thebibliography}{152}
\expandafter\ifx\csname natexlab\endcsname\relax\def\natexlab#1{#1}\fi
\expandafter\ifx\csname bibnamefont\endcsname\relax
  \def\bibnamefont#1{#1}\fi
\expandafter\ifx\csname bibfnamefont\endcsname\relax
  \def\bibfnamefont#1{#1}\fi
\expandafter\ifx\csname citenamefont\endcsname\relax
  \def\citenamefont#1{#1}\fi
\expandafter\ifx\csname url\endcsname\relax
  \def\url#1{\texttt{#1}}\fi
\expandafter\ifx\csname urlprefix\endcsname\relax\def\urlprefix{URL }\fi
\providecommand{\bibinfo}[2]{#2}
\providecommand{\eprint}[2][]{\url{#2}}

\bibitem[{\citenamefont{Hedin}(1965)}]{Hedin1965}
\bibinfo{author}{\bibfnamefont{L.}~\bibnamefont{Hedin}},
  \bibinfo{journal}{Phys. Rev.} \textbf{\bibinfo{volume}{139}},
  \bibinfo{pages}{A796} (\bibinfo{year}{1965}).

\bibitem[{\citenamefont{Aryasetiawan and Gunnarsson}(1998)}]{ferdi_gw}
\bibinfo{author}{\bibfnamefont{F.}~\bibnamefont{Aryasetiawan}}
  \bibnamefont{and}
  \bibinfo{author}{\bibfnamefont{O.}~\bibnamefont{Gunnarsson}},
  \bibinfo{journal}{Rep. Prog. Phys.} \textbf{\bibinfo{volume}{61}},
  \bibinfo{pages}{237} (\bibinfo{year}{1998}).

\bibitem[{\citenamefont{Hedin}(1999)}]{Hedin1999}
\bibinfo{author}{\bibfnamefont{L.}~\bibnamefont{Hedin}}, \bibinfo{journal}{J.
  Phys.: Condens. Matter} \textbf{\bibinfo{volume}{11}}, \bibinfo{pages}{R489}
  (\bibinfo{year}{1999}).

\bibitem[{\citenamefont{Onida et~al.}(2002)\citenamefont{Onida, Reining, and
  Rubio}}]{Onida2002}
\bibinfo{author}{\bibfnamefont{G.}~\bibnamefont{Onida}},
  \bibinfo{author}{\bibfnamefont{L.}~\bibnamefont{Reining}}, \bibnamefont{and}
  \bibinfo{author}{\bibfnamefont{A.}~\bibnamefont{Rubio}},
  \bibinfo{journal}{Rev. Mod. Phys.} \textbf{\bibinfo{volume}{74}},
  \bibinfo{pages}{601} (\bibinfo{year}{2002}).

\bibitem[{\citenamefont{Jeng et~al.}(1991)\citenamefont{Jeng, Lad, and
  Henrich}}]{PhysRevB.43.11971}
\bibinfo{author}{\bibfnamefont{S.-P.} \bibnamefont{Jeng}},
  \bibinfo{author}{\bibfnamefont{R.~J.} \bibnamefont{Lad}}, \bibnamefont{and}
  \bibinfo{author}{\bibfnamefont{V.~E.} \bibnamefont{Henrich}},
  \bibinfo{journal}{Phys. Rev. B} \textbf{\bibinfo{volume}{43}},
  \bibinfo{pages}{11971} (\bibinfo{year}{1991}).

\bibitem[{\citenamefont{Marton et~al.}(1962)\citenamefont{Marton, Simpson,
  Fowler, and Swanson}}]{marton1962}
\bibinfo{author}{\bibfnamefont{L.}~\bibnamefont{Marton}},
  \bibinfo{author}{\bibfnamefont{J.~A.} \bibnamefont{Simpson}},
  \bibinfo{author}{\bibfnamefont{H.~A.} \bibnamefont{Fowler}},
  \bibnamefont{and} \bibinfo{author}{\bibfnamefont{N.}~\bibnamefont{Swanson}},
  \bibinfo{journal}{Phys. Rev.} \textbf{\bibinfo{volume}{126}},
  \bibinfo{pages}{182} (\bibinfo{year}{1962}).

\bibitem[{\citenamefont{Aryasetiawan et~al.}(1996)\citenamefont{Aryasetiawan,
  Hedin, and Karlsson}}]{PhysRevLett.77.2268}
\bibinfo{author}{\bibfnamefont{F.}~\bibnamefont{Aryasetiawan}},
  \bibinfo{author}{\bibfnamefont{L.}~\bibnamefont{Hedin}}, \bibnamefont{and}
  \bibinfo{author}{\bibfnamefont{K.}~\bibnamefont{Karlsson}},
  \bibinfo{journal}{Phys. Rev. Lett.} \textbf{\bibinfo{volume}{77}},
  \bibinfo{pages}{2268} (\bibinfo{year}{1996}).

\bibitem[{\citenamefont{Guzzo et~al.}(2011)\citenamefont{Guzzo, Lani, Sottile,
  Romaniello, Gatti, Kas, Rehr, Silly, Sirotti, and Reining}}]{Guzzo2011}
\bibinfo{author}{\bibfnamefont{M.}~\bibnamefont{Guzzo}},
  \bibinfo{author}{\bibfnamefont{G.}~\bibnamefont{Lani}},
  \bibinfo{author}{\bibfnamefont{F.}~\bibnamefont{Sottile}},
  \bibinfo{author}{\bibfnamefont{P.}~\bibnamefont{Romaniello}},
  \bibinfo{author}{\bibfnamefont{M.}~\bibnamefont{Gatti}},
  \bibinfo{author}{\bibfnamefont{J.~J.} \bibnamefont{Kas}},
  \bibinfo{author}{\bibfnamefont{J.~J.} \bibnamefont{Rehr}},
  \bibinfo{author}{\bibfnamefont{M.~G.} \bibnamefont{Silly}},
  \bibinfo{author}{\bibfnamefont{F.}~\bibnamefont{Sirotti}}, \bibnamefont{and}
  \bibinfo{author}{\bibfnamefont{L.}~\bibnamefont{Reining}},
  \bibinfo{journal}{Phys. Rev. Lett.} \textbf{\bibinfo{volume}{107}},
  \bibinfo{pages}{166401} (\bibinfo{year}{2011}).

\bibitem[{\citenamefont{Gull et~al.}(2011)\citenamefont{Gull, Millis,
  Lichtenstein, Rubtsov, Troyer, and Werner}}]{Gull2011}
\bibinfo{author}{\bibfnamefont{E.}~\bibnamefont{Gull}},
  \bibinfo{author}{\bibfnamefont{A.~J.} \bibnamefont{Millis}},
  \bibinfo{author}{\bibfnamefont{A.~I.} \bibnamefont{Lichtenstein}},
  \bibinfo{author}{\bibfnamefont{A.~N.} \bibnamefont{Rubtsov}},
  \bibinfo{author}{\bibfnamefont{M.}~\bibnamefont{Troyer}}, \bibnamefont{and}
  \bibinfo{author}{\bibfnamefont{P.}~\bibnamefont{Werner}},
  \bibinfo{journal}{Rev. Mod. Phys.} \textbf{\bibinfo{volume}{83}},
  \bibinfo{pages}{349} (\bibinfo{year}{2011}).

\bibitem[{\citenamefont{Biermann et~al.}(2003)\citenamefont{Biermann,
  Aryasetiawan, and Georges}}]{Biermann2003}
\bibinfo{author}{\bibfnamefont{S.}~\bibnamefont{Biermann}},
  \bibinfo{author}{\bibfnamefont{F.}~\bibnamefont{Aryasetiawan}},
  \bibnamefont{and} \bibinfo{author}{\bibfnamefont{A.}~\bibnamefont{Georges}},
  \bibinfo{journal}{Phys. Rev. Lett.} \textbf{\bibinfo{volume}{90}},
  \bibinfo{pages}{086402} (\bibinfo{year}{2003}).

\bibitem[{\citenamefont{Zgid and Chan}(2011)}]{Zgid2011}
\bibinfo{author}{\bibfnamefont{D.}~\bibnamefont{Zgid}} \bibnamefont{and}
  \bibinfo{author}{\bibfnamefont{G.~K.-L.} \bibnamefont{Chan}},
  \bibinfo{journal}{The Journal of Chemical Physics}
  \textbf{\bibinfo{volume}{134}}, \bibinfo{eid}{094115} (\bibinfo{year}{2011}).

\bibitem[{\citenamefont{Kananenka et~al.}(2015)\citenamefont{Kananenka, Gull,
  and Zgid}}]{Kananenka2015}
\bibinfo{author}{\bibfnamefont{A.~A.} \bibnamefont{Kananenka}},
  \bibinfo{author}{\bibfnamefont{E.}~\bibnamefont{Gull}}, \bibnamefont{and}
  \bibinfo{author}{\bibfnamefont{D.}~\bibnamefont{Zgid}},
  \bibinfo{journal}{Phys. Rev. B} \textbf{\bibinfo{volume}{91}},
  \bibinfo{pages}{121111} (\bibinfo{year}{2015}).

\bibitem[{\citenamefont{Lan et~al.}(2015)\citenamefont{Lan, Kananenka, and
  Zgid}}]{Lan2015}
\bibinfo{author}{\bibfnamefont{T.~N.} \bibnamefont{Lan}},
  \bibinfo{author}{\bibfnamefont{A.~A.} \bibnamefont{Kananenka}},
  \bibnamefont{and} \bibinfo{author}{\bibfnamefont{D.}~\bibnamefont{Zgid}},
  \bibinfo{journal}{The Journal of Chemical Physics}
  \textbf{\bibinfo{volume}{143}}, \bibinfo{eid}{241102} (\bibinfo{year}{2015}).

\bibitem[{\citenamefont{Zgid et~al.}(2012)\citenamefont{Zgid, Gull, and
  Chan}}]{Zgid2012}
\bibinfo{author}{\bibfnamefont{D.}~\bibnamefont{Zgid}},
  \bibinfo{author}{\bibfnamefont{E.}~\bibnamefont{Gull}}, \bibnamefont{and}
  \bibinfo{author}{\bibfnamefont{G.~K.-L.} \bibnamefont{Chan}},
  \bibinfo{journal}{Phys. Rev. B} \textbf{\bibinfo{volume}{86}},
  \bibinfo{pages}{165128} (\bibinfo{year}{2012}).

\bibitem[{\citenamefont{Go and Millis}(2015)}]{Go2015}
\bibinfo{author}{\bibfnamefont{A.}~\bibnamefont{Go}} \bibnamefont{and}
  \bibinfo{author}{\bibfnamefont{A.~J.} \bibnamefont{Millis}},
  \bibinfo{journal}{Phys. Rev. Lett.} \textbf{\bibinfo{volume}{114}},
  \bibinfo{pages}{016402} (\bibinfo{year}{2015}).

\bibitem[{\citenamefont{Maier et~al.}(2005)\citenamefont{Maier, Jarrell,
  Pruschke, and Hettler}}]{Maier2005}
\bibinfo{author}{\bibfnamefont{T.}~\bibnamefont{Maier}},
  \bibinfo{author}{\bibfnamefont{M.}~\bibnamefont{Jarrell}},
  \bibinfo{author}{\bibfnamefont{T.}~\bibnamefont{Pruschke}}, \bibnamefont{and}
  \bibinfo{author}{\bibfnamefont{M.~H.} \bibnamefont{Hettler}},
  \bibinfo{journal}{Rev. Mod. Phys.} \textbf{\bibinfo{volume}{77}},
  \bibinfo{pages}{1027} (\bibinfo{year}{2005}).

\bibitem[{\citenamefont{Tocchio et~al.}(2008)\citenamefont{Tocchio, Becca,
  Parola, and Sorella}}]{Sorella2008}
\bibinfo{author}{\bibfnamefont{L.~F.} \bibnamefont{Tocchio}},
  \bibinfo{author}{\bibfnamefont{F.}~\bibnamefont{Becca}},
  \bibinfo{author}{\bibfnamefont{A.}~\bibnamefont{Parola}}, \bibnamefont{and}
  \bibinfo{author}{\bibfnamefont{S.}~\bibnamefont{Sorella}},
  \bibinfo{journal}{Phys. Rev. B} \textbf{\bibinfo{volume}{78}},
  \bibinfo{pages}{041101} (\bibinfo{year}{2008}).

\bibitem[{\citenamefont{Werner et~al.}(2009)\citenamefont{Werner, Gull,
  Parcollet, and Millis}}]{Werner2009}
\bibinfo{author}{\bibfnamefont{P.}~\bibnamefont{Werner}},
  \bibinfo{author}{\bibfnamefont{E.}~\bibnamefont{Gull}},
  \bibinfo{author}{\bibfnamefont{O.}~\bibnamefont{Parcollet}},
  \bibnamefont{and} \bibinfo{author}{\bibfnamefont{A.~J.}
  \bibnamefont{Millis}}, \bibinfo{journal}{Phys. Rev. B}
  \textbf{\bibinfo{volume}{80}}, \bibinfo{pages}{045120}
  (\bibinfo{year}{2009}).

\bibitem[{\citenamefont{Gull et~al.}(2009)\citenamefont{Gull, Parcollet,
  Werner, and Millis}}]{Gull2009}
\bibinfo{author}{\bibfnamefont{E.}~\bibnamefont{Gull}},
  \bibinfo{author}{\bibfnamefont{O.}~\bibnamefont{Parcollet}},
  \bibinfo{author}{\bibfnamefont{P.}~\bibnamefont{Werner}}, \bibnamefont{and}
  \bibinfo{author}{\bibfnamefont{A.~J.} \bibnamefont{Millis}},
  \bibinfo{journal}{Phys. Rev. B} \textbf{\bibinfo{volume}{80}},
  \bibinfo{pages}{245102} (\bibinfo{year}{2009}).

\bibitem[{\citenamefont{Chang and Zhang}(2010)}]{Chang2010}
\bibinfo{author}{\bibfnamefont{C.-C.} \bibnamefont{Chang}} \bibnamefont{and}
  \bibinfo{author}{\bibfnamefont{S.}~\bibnamefont{Zhang}},
  \bibinfo{journal}{Phys. Rev. Lett.} \textbf{\bibinfo{volume}{104}},
  \bibinfo{pages}{116402} (\bibinfo{year}{2010}).

\bibitem[{\citenamefont{Shi and Zhang}(2013)}]{Zhang2013}
\bibinfo{author}{\bibfnamefont{H.}~\bibnamefont{Shi}} \bibnamefont{and}
  \bibinfo{author}{\bibfnamefont{S.}~\bibnamefont{Zhang}},
  \bibinfo{journal}{Phys. Rev. B} \textbf{\bibinfo{volume}{88}},
  \bibinfo{pages}{125132} (\bibinfo{year}{2013}).

\bibitem[{\citenamefont{Gukelberger et~al.}(2015)\citenamefont{Gukelberger,
  Huang, and Werner}}]{Gukelberger2015}
\bibinfo{author}{\bibfnamefont{J.}~\bibnamefont{Gukelberger}},
  \bibinfo{author}{\bibfnamefont{L.}~\bibnamefont{Huang}}, \bibnamefont{and}
  \bibinfo{author}{\bibfnamefont{P.}~\bibnamefont{Werner}},
  \bibinfo{journal}{Phys. Rev. B} \textbf{\bibinfo{volume}{91}},
  \bibinfo{pages}{235114} (\bibinfo{year}{2015}).

\bibitem[{\citenamefont{Zheng and Chan}(2015)}]{Chan2015}
\bibinfo{author}{\bibfnamefont{B.-X.} \bibnamefont{Zheng}} \bibnamefont{and}
  \bibinfo{author}{\bibfnamefont{G.~K.-L.} \bibnamefont{Chan}},
  \bibinfo{journal}{arXiv:1504.01784}  (\bibinfo{year}{2015}).

\bibitem[{\citenamefont{LeBlanc and
  et~al.}(2015)}]{simons_foundation_2D_Hubbard}
\bibinfo{author}{\bibfnamefont{J.~P.~F.} \bibnamefont{LeBlanc}}
  \bibnamefont{and} \bibinfo{author}{\bibnamefont{et~al.}},
  \bibinfo{journal}{arXiv:1505.02290}  (\bibinfo{year}{2015}).

\bibitem[{\citenamefont{Imada and Miyake}(2010)}]{Imada_Miyake_2010}
\bibinfo{author}{\bibfnamefont{M.}~\bibnamefont{Imada}} \bibnamefont{and}
  \bibinfo{author}{\bibfnamefont{T.}~\bibnamefont{Miyake}},
  \bibinfo{journal}{Journal of the Physical Society of Japan}
  \textbf{\bibinfo{volume}{79}}, \bibinfo{pages}{112001}
  (\bibinfo{year}{2010}).

\bibitem[{\citenamefont{Hirayama et~al.}(2013)\citenamefont{Hirayama, Miyake,
  and Imada}}]{Motoaki2013}
\bibinfo{author}{\bibfnamefont{M.}~\bibnamefont{Hirayama}},
  \bibinfo{author}{\bibfnamefont{T.}~\bibnamefont{Miyake}}, \bibnamefont{and}
  \bibinfo{author}{\bibfnamefont{M.}~\bibnamefont{Imada}},
  \bibinfo{journal}{Phys. Rev. B} \textbf{\bibinfo{volume}{87}},
  \bibinfo{pages}{195144} (\bibinfo{year}{2013}).

\bibitem[{\citenamefont{Hohenberg and Kohn}(1964)}]{Hohenberg64}
\bibinfo{author}{\bibfnamefont{P.}~\bibnamefont{Hohenberg}} \bibnamefont{and}
  \bibinfo{author}{\bibfnamefont{W.}~\bibnamefont{Kohn}},
  \bibinfo{journal}{Phys. Rev.} \textbf{\bibinfo{volume}{136}},
  \bibinfo{pages}{B864} (\bibinfo{year}{1964}).

\bibitem[{\citenamefont{Kohn and Sham}(1965)}]{Kohn65}
\bibinfo{author}{\bibfnamefont{W.}~\bibnamefont{Kohn}} \bibnamefont{and}
  \bibinfo{author}{\bibfnamefont{L.~J.} \bibnamefont{Sham}},
  \bibinfo{journal}{Phys. Rev.} \textbf{\bibinfo{volume}{140}},
  \bibinfo{pages}{A1133} (\bibinfo{year}{1965}).

\bibitem[{\citenamefont{Jones}(2015)}]{Jones15}
\bibinfo{author}{\bibfnamefont{R.~O.} \bibnamefont{Jones}},
  \bibinfo{journal}{Rev. Mod. Phys.} \textbf{\bibinfo{volume}{87}},
  \bibinfo{pages}{897} (\bibinfo{year}{2015}).

\bibitem[{\citenamefont{Marzari and Vanderbilt}(1997)}]{PhysRevB.56.12847}
\bibinfo{author}{\bibfnamefont{N.}~\bibnamefont{Marzari}} \bibnamefont{and}
  \bibinfo{author}{\bibfnamefont{D.}~\bibnamefont{Vanderbilt}},
  \bibinfo{journal}{Phys. Rev. B} \textbf{\bibinfo{volume}{56}},
  \bibinfo{pages}{12847} (\bibinfo{year}{1997}).

\bibitem[{\citenamefont{Miyake and Aryasetiawan}(2008)}]{PhysRevB.77.085122}
\bibinfo{author}{\bibfnamefont{T.}~\bibnamefont{Miyake}} \bibnamefont{and}
  \bibinfo{author}{\bibfnamefont{F.}~\bibnamefont{Aryasetiawan}},
  \bibinfo{journal}{Phys. Rev. B} \textbf{\bibinfo{volume}{77}},
  \bibinfo{pages}{085122} (\bibinfo{year}{2008}).

\bibitem[{\citenamefont{Aryasetiawan et~al.}(2004)\citenamefont{Aryasetiawan,
  Imada, Georges, Kotliar, Biermann, and Lichtenstein}}]{PhysRevB.70.195104}
\bibinfo{author}{\bibfnamefont{F.}~\bibnamefont{Aryasetiawan}},
  \bibinfo{author}{\bibfnamefont{M.}~\bibnamefont{Imada}},
  \bibinfo{author}{\bibfnamefont{A.}~\bibnamefont{Georges}},
  \bibinfo{author}{\bibfnamefont{G.}~\bibnamefont{Kotliar}},
  \bibinfo{author}{\bibfnamefont{S.}~\bibnamefont{Biermann}}, \bibnamefont{and}
  \bibinfo{author}{\bibfnamefont{A.~I.} \bibnamefont{Lichtenstein}},
  \bibinfo{journal}{Phys. Rev. B} \textbf{\bibinfo{volume}{70}},
  \bibinfo{pages}{195104} (\bibinfo{year}{2004}).

\bibitem[{\citenamefont{Aryasetiawan et~al.}(2006)\citenamefont{Aryasetiawan,
  Karlsson, Jepsen, and Sch\"onberger}}]{PhysRevB.74.125106}
\bibinfo{author}{\bibfnamefont{F.}~\bibnamefont{Aryasetiawan}},
  \bibinfo{author}{\bibfnamefont{K.}~\bibnamefont{Karlsson}},
  \bibinfo{author}{\bibfnamefont{O.}~\bibnamefont{Jepsen}}, \bibnamefont{and}
  \bibinfo{author}{\bibfnamefont{U.}~\bibnamefont{Sch\"onberger}},
  \bibinfo{journal}{Phys. Rev. B} \textbf{\bibinfo{volume}{74}},
  \bibinfo{pages}{125106} (\bibinfo{year}{2006}).

\bibitem[{\citenamefont{Springer and Aryasetiawan}(1998)}]{Springer1998}
\bibinfo{author}{\bibfnamefont{M.}~\bibnamefont{Springer}} \bibnamefont{and}
  \bibinfo{author}{\bibfnamefont{F.}~\bibnamefont{Aryasetiawan}},
  \bibinfo{journal}{Phys. Rev. B} \textbf{\bibinfo{volume}{57}},
  \bibinfo{pages}{4364} (\bibinfo{year}{1998}).

\bibitem[{\citenamefont{\ifmmode \mbox{\c{S}}\else \c{S}\fi{}a\ifmmode
  \mbox{\c{s}}\else \c{s}\fi{}\ifmmode \imath \else \i
  \fi{}o\ifmmode~\breve{g}\else \u{g}\fi{}lu
  et~al.}(2011)\citenamefont{\ifmmode \mbox{\c{S}}\else \c{S}\fi{}a\ifmmode
  \mbox{\c{s}}\else \c{s}\fi{}\ifmmode \imath \else \i
  \fi{}o\ifmmode~\breve{g}\else \u{g}\fi{}lu, Friedrich, and
  Bl\"ugel}}]{Sasioglu2011}
\bibinfo{author}{\bibfnamefont{E.}~\bibnamefont{\ifmmode \mbox{\c{S}}\else
  \c{S}\fi{}a\ifmmode \mbox{\c{s}}\else \c{s}\fi{}\ifmmode \imath \else \i
  \fi{}o\ifmmode~\breve{g}\else \u{g}\fi{}lu}},
  \bibinfo{author}{\bibfnamefont{C.}~\bibnamefont{Friedrich}},
  \bibnamefont{and} \bibinfo{author}{\bibfnamefont{S.}~\bibnamefont{Bl\"ugel}},
  \bibinfo{journal}{Phys. Rev. B} \textbf{\bibinfo{volume}{83}},
  \bibinfo{pages}{121101} (\bibinfo{year}{2011}).

\bibitem[{\citenamefont{Sakuma and
  Aryasetiawan}(2013{\natexlab{a}})}]{Sakuma_private}
\bibinfo{author}{\bibfnamefont{R.}~\bibnamefont{Sakuma}} \bibnamefont{and}
  \bibinfo{author}{\bibfnamefont{F.}~\bibnamefont{Aryasetiawan}},
  \bibinfo{journal}{private communication}
  (\bibinfo{year}{2013}{\natexlab{a}}).

\bibitem[{\citenamefont{Miyake et~al.}(2008)\citenamefont{Miyake, Pourovskii,
  Vildosola, Biermann, and Georges}}]{Miyake2008}
\bibinfo{author}{\bibfnamefont{T.}~\bibnamefont{Miyake}},
  \bibinfo{author}{\bibfnamefont{L.}~\bibnamefont{Pourovskii}},
  \bibinfo{author}{\bibfnamefont{V.}~\bibnamefont{Vildosola}},
  \bibinfo{author}{\bibfnamefont{S.}~\bibnamefont{Biermann}}, \bibnamefont{and}
  \bibinfo{author}{\bibfnamefont{A.}~\bibnamefont{Georges}},
  \bibinfo{journal}{Journal of the Physical Society of Japan}
  \textbf{\bibinfo{volume}{77}}, \bibinfo{pages}{99} (\bibinfo{year}{2008}).

\bibitem[{\citenamefont{Nakamura et~al.}(2008)\citenamefont{Nakamura, Arita,
  and Imada}}]{nakamura-2008}
\bibinfo{author}{\bibfnamefont{K.}~\bibnamefont{Nakamura}},
  \bibinfo{author}{\bibfnamefont{R.}~\bibnamefont{Arita}}, \bibnamefont{and}
  \bibinfo{author}{\bibfnamefont{M.}~\bibnamefont{Imada}},
  \bibinfo{journal}{Journal of the Physical Society of Japan}
  \textbf{\bibinfo{volume}{77}}, \bibinfo{pages}{093711}
  (\bibinfo{year}{2008}).

\bibitem[{\citenamefont{Tomczak et~al.}(2009)\citenamefont{Tomczak, Miyake,
  Sakuma, and Aryasetiawan}}]{Tomczak2009}
\bibinfo{author}{\bibfnamefont{J.~M.} \bibnamefont{Tomczak}},
  \bibinfo{author}{\bibfnamefont{T.}~\bibnamefont{Miyake}},
  \bibinfo{author}{\bibfnamefont{R.}~\bibnamefont{Sakuma}}, \bibnamefont{and}
  \bibinfo{author}{\bibfnamefont{F.}~\bibnamefont{Aryasetiawan}},
  \bibinfo{journal}{Phys. Rev. B} \textbf{\bibinfo{volume}{79}},
  \bibinfo{eid}{235133} (pages~\bibinfo{numpages}{8}) (\bibinfo{year}{2009}).

\bibitem[{\citenamefont{Miyake et~al.}(2010)\citenamefont{Miyake, Nakamura,
  Arita, and Imada}}]{Miyake2010}
\bibinfo{author}{\bibfnamefont{T.}~\bibnamefont{Miyake}},
  \bibinfo{author}{\bibfnamefont{K.}~\bibnamefont{Nakamura}},
  \bibinfo{author}{\bibfnamefont{R.}~\bibnamefont{Arita}}, \bibnamefont{and}
  \bibinfo{author}{\bibfnamefont{M.}~\bibnamefont{Imada}},
  \bibinfo{journal}{Journal of the Physical Society of Japan}
  \textbf{\bibinfo{volume}{79}}, \bibinfo{pages}{044705}
  (\bibinfo{year}{2010}).

\bibitem[{\citenamefont{Tomczak et~al.}(2010)\citenamefont{Tomczak, Miyake, and
  Aryasetiawan}}]{Tomczak2010}
\bibinfo{author}{\bibfnamefont{J.~M.} \bibnamefont{Tomczak}},
  \bibinfo{author}{\bibfnamefont{T.}~\bibnamefont{Miyake}}, \bibnamefont{and}
  \bibinfo{author}{\bibfnamefont{F.}~\bibnamefont{Aryasetiawan}},
  \bibinfo{journal}{Phys. Rev. B} \textbf{\bibinfo{volume}{81}},
  \bibinfo{pages}{115116} (\bibinfo{year}{2010}).

\bibitem[{\citenamefont{Nomura et~al.}(2012{\natexlab{a}})\citenamefont{Nomura,
  Nakamura, and Arita}}]{Nomura2012}
\bibinfo{author}{\bibfnamefont{Y.}~\bibnamefont{Nomura}},
  \bibinfo{author}{\bibfnamefont{K.}~\bibnamefont{Nakamura}}, \bibnamefont{and}
  \bibinfo{author}{\bibfnamefont{R.}~\bibnamefont{Arita}},
  \bibinfo{journal}{Phys. Rev. B} \textbf{\bibinfo{volume}{85}},
  \bibinfo{pages}{155452} (\bibinfo{year}{2012}{\natexlab{a}}).

\bibitem[{\citenamefont{Vaugier et~al.}(2012)\citenamefont{Vaugier, Jiang, and
  Biermann}}]{Vaugier2012}
\bibinfo{author}{\bibfnamefont{L.}~\bibnamefont{Vaugier}},
  \bibinfo{author}{\bibfnamefont{H.}~\bibnamefont{Jiang}}, \bibnamefont{and}
  \bibinfo{author}{\bibfnamefont{S.}~\bibnamefont{Biermann}},
  \bibinfo{journal}{Phys. Rev. B} \textbf{\bibinfo{volume}{86}},
  \bibinfo{pages}{165105} (\bibinfo{year}{2012}).

\bibitem[{\citenamefont{Sakuma and
  Aryasetiawan}(2013{\natexlab{b}})}]{Sakuma2013_U}
\bibinfo{author}{\bibfnamefont{R.}~\bibnamefont{Sakuma}} \bibnamefont{and}
  \bibinfo{author}{\bibfnamefont{F.}~\bibnamefont{Aryasetiawan}},
  \bibinfo{journal}{Phys. Rev. B} \textbf{\bibinfo{volume}{87}},
  \bibinfo{pages}{165118} (\bibinfo{year}{2013}{\natexlab{b}}).

\bibitem[{\citenamefont{Souza et~al.}(2001)\citenamefont{Souza, Marzari, and
  Vanderbilt}}]{PhysRevB.65.035109}
\bibinfo{author}{\bibfnamefont{I.}~\bibnamefont{Souza}},
  \bibinfo{author}{\bibfnamefont{N.}~\bibnamefont{Marzari}}, \bibnamefont{and}
  \bibinfo{author}{\bibfnamefont{D.}~\bibnamefont{Vanderbilt}},
  \bibinfo{journal}{Phys. Rev. B} \textbf{\bibinfo{volume}{65}},
  \bibinfo{pages}{035109} (\bibinfo{year}{2001}).

\bibitem[{\citenamefont{Miyake et~al.}(2009)\citenamefont{Miyake, Aryasetiawan,
  and Imada}}]{miyake:155134}
\bibinfo{author}{\bibfnamefont{T.}~\bibnamefont{Miyake}},
  \bibinfo{author}{\bibfnamefont{F.}~\bibnamefont{Aryasetiawan}},
  \bibnamefont{and} \bibinfo{author}{\bibfnamefont{M.}~\bibnamefont{Imada}},
  \bibinfo{journal}{Phys. Rev. B} \textbf{\bibinfo{volume}{80}},
  \bibinfo{eid}{155134} (\bibinfo{year}{2009}).

\bibitem[{\citenamefont{Kinza and Honerkamp}(2015)}]{Kinza2015}
\bibinfo{author}{\bibfnamefont{M.}~\bibnamefont{Kinza}} \bibnamefont{and}
  \bibinfo{author}{\bibfnamefont{C.}~\bibnamefont{Honerkamp}},
  \bibinfo{journal}{Phys. Rev. B} \textbf{\bibinfo{volume}{92}},
  \bibinfo{pages}{045113} (\bibinfo{year}{2015}).

\bibitem[{\citenamefont{Shinaoka et~al.}(2015)\citenamefont{Shinaoka, Troyer,
  and Werner}}]{Shinaoka2015}
\bibinfo{author}{\bibfnamefont{H.}~\bibnamefont{Shinaoka}},
  \bibinfo{author}{\bibfnamefont{M.}~\bibnamefont{Troyer}}, \bibnamefont{and}
  \bibinfo{author}{\bibfnamefont{P.}~\bibnamefont{Werner}},
  \bibinfo{journal}{Phys. Rev. B} \textbf{\bibinfo{volume}{91}},
  \bibinfo{pages}{245156} (\bibinfo{year}{2015}).

\bibitem[{\citenamefont{Cococcioni and de~Gironcoli}(2005)}]{Cococcioni2005}
\bibinfo{author}{\bibfnamefont{M.}~\bibnamefont{Cococcioni}} \bibnamefont{and}
  \bibinfo{author}{\bibfnamefont{S.}~\bibnamefont{de~Gironcoli}},
  \bibinfo{journal}{Phys. Rev. B} \textbf{\bibinfo{volume}{71}},
  \bibinfo{pages}{035105} (\bibinfo{year}{2005}).

\bibitem[{\citenamefont{Hsu et~al.}(2009)\citenamefont{Hsu, Umemoto,
  Cococcioni, and Wentzcovitch}}]{Hsu2009}
\bibinfo{author}{\bibfnamefont{H.}~\bibnamefont{Hsu}},
  \bibinfo{author}{\bibfnamefont{K.}~\bibnamefont{Umemoto}},
  \bibinfo{author}{\bibfnamefont{M.}~\bibnamefont{Cococcioni}},
  \bibnamefont{and}
  \bibinfo{author}{\bibfnamefont{R.}~\bibnamefont{Wentzcovitch}},
  \bibinfo{journal}{Phys. Rev. B} \textbf{\bibinfo{volume}{79}},
  \bibinfo{eid}{125124} (\bibinfo{year}{2009}).

\bibitem[{\citenamefont{Anisimov and Gunnarsson}(1991)}]{constrainedLDA}
\bibinfo{author}{\bibfnamefont{V.~I.} \bibnamefont{Anisimov}} \bibnamefont{and}
  \bibinfo{author}{\bibfnamefont{O.}~\bibnamefont{Gunnarsson}},
  \bibinfo{journal}{Phys. Rev. B} \textbf{\bibinfo{volume}{43}},
  \bibinfo{pages}{7570} (\bibinfo{year}{1991}).

\bibitem[{\citenamefont{Anisimov et~al.}(2009)\citenamefont{Anisimov, Korotin,
  Korotin, Kozhevnikov, Kunes, Shorikov, Skornyakov, and
  Streltsov}}]{Anisimov2009}
\bibinfo{author}{\bibfnamefont{V.~I.} \bibnamefont{Anisimov}},
  \bibinfo{author}{\bibfnamefont{D.~M.} \bibnamefont{Korotin}},
  \bibinfo{author}{\bibfnamefont{M.~A.} \bibnamefont{Korotin}},
  \bibinfo{author}{\bibfnamefont{A.~V.} \bibnamefont{Kozhevnikov}},
  \bibinfo{author}{\bibfnamefont{J.}~\bibnamefont{Kunes}},
  \bibinfo{author}{\bibfnamefont{A.~O.} \bibnamefont{Shorikov}},
  \bibinfo{author}{\bibfnamefont{S.~L.} \bibnamefont{Skornyakov}},
  \bibnamefont{and} \bibinfo{author}{\bibfnamefont{S.~V.}
  \bibnamefont{Streltsov}}, \bibinfo{journal}{Journal of Physics: Condensed
  Matter} \textbf{\bibinfo{volume}{21}}, \bibinfo{pages}{075602}
  (\bibinfo{year}{2009}).

\bibitem[{\citenamefont{Kutepov et~al.}(2010)\citenamefont{Kutepov, Haule,
  Savrasov, and Kotliar}}]{Kutepov2010}
\bibinfo{author}{\bibfnamefont{A.}~\bibnamefont{Kutepov}},
  \bibinfo{author}{\bibfnamefont{K.}~\bibnamefont{Haule}},
  \bibinfo{author}{\bibfnamefont{S.~Y.} \bibnamefont{Savrasov}},
  \bibnamefont{and} \bibinfo{author}{\bibfnamefont{G.}~\bibnamefont{Kotliar}},
  \bibinfo{journal}{Phys. Rev. B} \textbf{\bibinfo{volume}{82}},
  \bibinfo{pages}{045105} (\bibinfo{year}{2010}).

\bibitem[{\citenamefont{Nomura et~al.}(2012{\natexlab{b}})\citenamefont{Nomura,
  Kaltak, Nakamura, Taranto, Sakai, Toschi, Arita, Held, Kresse, and
  Imada}}]{Nomura2012_U}
\bibinfo{author}{\bibfnamefont{Y.}~\bibnamefont{Nomura}},
  \bibinfo{author}{\bibfnamefont{M.}~\bibnamefont{Kaltak}},
  \bibinfo{author}{\bibfnamefont{K.}~\bibnamefont{Nakamura}},
  \bibinfo{author}{\bibfnamefont{C.}~\bibnamefont{Taranto}},
  \bibinfo{author}{\bibfnamefont{S.}~\bibnamefont{Sakai}},
  \bibinfo{author}{\bibfnamefont{A.}~\bibnamefont{Toschi}},
  \bibinfo{author}{\bibfnamefont{R.}~\bibnamefont{Arita}},
  \bibinfo{author}{\bibfnamefont{K.}~\bibnamefont{Held}},
  \bibinfo{author}{\bibfnamefont{G.}~\bibnamefont{Kresse}}, \bibnamefont{and}
  \bibinfo{author}{\bibfnamefont{M.}~\bibnamefont{Imada}},
  \bibinfo{journal}{Phys. Rev. B} \textbf{\bibinfo{volume}{86}},
  \bibinfo{pages}{085117} (\bibinfo{year}{2012}{\natexlab{b}}).

\bibitem[{\citenamefont{Anisimov et~al.}(1991)\citenamefont{Anisimov, Zaanen,
  and Andersen}}]{Anisimov1991}
\bibinfo{author}{\bibfnamefont{V.~I.} \bibnamefont{Anisimov}},
  \bibinfo{author}{\bibfnamefont{J.}~\bibnamefont{Zaanen}}, \bibnamefont{and}
  \bibinfo{author}{\bibfnamefont{O.~K.} \bibnamefont{Andersen}},
  \bibinfo{journal}{Phys. Rev. B} \textbf{\bibinfo{volume}{44}},
  \bibinfo{pages}{943} (\bibinfo{year}{1991}).

\bibitem[{\citenamefont{Anisimov et~al.}(1997)\citenamefont{Anisimov,
  Aryasetiawan, and Lichtenstein}}]{Anisimov1997}
\bibinfo{author}{\bibfnamefont{V.~I.} \bibnamefont{Anisimov}},
  \bibinfo{author}{\bibfnamefont{F.}~\bibnamefont{Aryasetiawan}},
  \bibnamefont{and} \bibinfo{author}{\bibfnamefont{A.~I.}
  \bibnamefont{Lichtenstein}}, \bibinfo{journal}{Journal of Physics: Condensed
  Matter} \textbf{\bibinfo{volume}{9}}, \bibinfo{pages}{767}
  (\bibinfo{year}{1997}).

\bibitem[{\citenamefont{Kotliar et~al.}(2006)\citenamefont{Kotliar, Savrasov,
  Haule, Oudovenko, Parcollet, and Marianetti}}]{Kotliar2006}
\bibinfo{author}{\bibfnamefont{G.}~\bibnamefont{Kotliar}},
  \bibinfo{author}{\bibfnamefont{S.~Y.} \bibnamefont{Savrasov}},
  \bibinfo{author}{\bibfnamefont{K.}~\bibnamefont{Haule}},
  \bibinfo{author}{\bibfnamefont{V.~S.} \bibnamefont{Oudovenko}},
  \bibinfo{author}{\bibfnamefont{O.}~\bibnamefont{Parcollet}},
  \bibnamefont{and} \bibinfo{author}{\bibfnamefont{C.~A.}
  \bibnamefont{Marianetti}}, \bibinfo{journal}{Rev. Mod. Phys.}
  \textbf{\bibinfo{volume}{78}}, \bibinfo{pages}{865} (\bibinfo{year}{2006}).

\bibitem[{\citenamefont{Aichhorn et~al.}(2011)\citenamefont{Aichhorn,
  Pourovskii, and Georges}}]{Aichhorn2011}
\bibinfo{author}{\bibfnamefont{M.}~\bibnamefont{Aichhorn}},
  \bibinfo{author}{\bibfnamefont{L.}~\bibnamefont{Pourovskii}},
  \bibnamefont{and} \bibinfo{author}{\bibfnamefont{A.}~\bibnamefont{Georges}},
  \bibinfo{journal}{Phys. Rev. B} \textbf{\bibinfo{volume}{84}},
  \bibinfo{pages}{054529} (\bibinfo{year}{2011}).

\bibitem[{\citenamefont{Haule}(2015)}]{Haule2015}
\bibinfo{author}{\bibfnamefont{K.}~\bibnamefont{Haule}},
  \bibinfo{journal}{Phys. Rev. Lett.} \textbf{\bibinfo{volume}{115}},
  \bibinfo{pages}{196403} (\bibinfo{year}{2015}).

\bibitem[{\citenamefont{Haule et~al.}(2010)\citenamefont{Haule, Yee, and
  Kim}}]{Haule2010}
\bibinfo{author}{\bibfnamefont{K.}~\bibnamefont{Haule}},
  \bibinfo{author}{\bibfnamefont{C.-H.} \bibnamefont{Yee}}, \bibnamefont{and}
  \bibinfo{author}{\bibfnamefont{K.}~\bibnamefont{Kim}},
  \bibinfo{journal}{Phys. Rev. B} \textbf{\bibinfo{volume}{81}},
  \bibinfo{pages}{195107} (\bibinfo{year}{2010}).

\bibitem[{\citenamefont{Hansmann et~al.}(2013)\citenamefont{Hansmann, Ayral,
  Vaugier, Werner, and Biermann}}]{Hansmann2013}
\bibinfo{author}{\bibfnamefont{P.}~\bibnamefont{Hansmann}},
  \bibinfo{author}{\bibfnamefont{T.}~\bibnamefont{Ayral}},
  \bibinfo{author}{\bibfnamefont{L.}~\bibnamefont{Vaugier}},
  \bibinfo{author}{\bibfnamefont{P.}~\bibnamefont{Werner}}, \bibnamefont{and}
  \bibinfo{author}{\bibfnamefont{S.}~\bibnamefont{Biermann}},
  \bibinfo{journal}{Phys. Rev. Lett.} \textbf{\bibinfo{volume}{110}},
  \bibinfo{pages}{166401} (\bibinfo{year}{2013}).

\bibitem[{\citenamefont{Imada and Hatsugai}(1989)}]{Imada1989}
\bibinfo{author}{\bibfnamefont{M.}~\bibnamefont{Imada}} \bibnamefont{and}
  \bibinfo{author}{\bibfnamefont{Y.}~\bibnamefont{Hatsugai}},
  \bibinfo{journal}{Journal of the Physical Society of Japan}
  \textbf{\bibinfo{volume}{58}}, \bibinfo{pages}{3752} (\bibinfo{year}{1989}).

\bibitem[{\citenamefont{Zhang and Krakauer}(2003)}]{Zhang2003}
\bibinfo{author}{\bibfnamefont{S.}~\bibnamefont{Zhang}} \bibnamefont{and}
  \bibinfo{author}{\bibfnamefont{H.}~\bibnamefont{Krakauer}},
  \bibinfo{journal}{Phys. Rev. Lett.} \textbf{\bibinfo{volume}{90}},
  \bibinfo{pages}{136401} (\bibinfo{year}{2003}).

\bibitem[{\citenamefont{Sorella}(2005)}]{Sorella2005}
\bibinfo{author}{\bibfnamefont{S.}~\bibnamefont{Sorella}},
  \bibinfo{journal}{Phys. Rev. B} \textbf{\bibinfo{volume}{71}},
  \bibinfo{pages}{241103} (\bibinfo{year}{2005}).

\bibitem[{\citenamefont{Tahara and Imada}(2008)}]{Tahara2008}
\bibinfo{author}{\bibfnamefont{D.}~\bibnamefont{Tahara}} \bibnamefont{and}
  \bibinfo{author}{\bibfnamefont{M.}~\bibnamefont{Imada}},
  \bibinfo{journal}{Journal of the Physical Society of Japan}
  \textbf{\bibinfo{volume}{77}}, \bibinfo{pages}{114701}
  (\bibinfo{year}{2008}).

\bibitem[{\citenamefont{Sorella et~al.}(2012)\citenamefont{Sorella, Otsuka, and
  Yunoki}}]{Sorella2012}
\bibinfo{author}{\bibfnamefont{S.}~\bibnamefont{Sorella}},
  \bibinfo{author}{\bibfnamefont{Y.}~\bibnamefont{Otsuka}}, \bibnamefont{and}
  \bibinfo{author}{\bibfnamefont{S.}~\bibnamefont{Yunoki}},
  \bibinfo{journal}{Scientific Reports} \textbf{\bibinfo{volume}{2}},
  \bibinfo{pages}{992} (\bibinfo{year}{2012}).

\bibitem[{\citenamefont{Jordan et~al.}(2008)\citenamefont{Jordan, Or\'us,
  Vidal, Verstraete, and Cirac}}]{Jordan2008}
\bibinfo{author}{\bibfnamefont{J.}~\bibnamefont{Jordan}},
  \bibinfo{author}{\bibfnamefont{R.}~\bibnamefont{Or\'us}},
  \bibinfo{author}{\bibfnamefont{G.}~\bibnamefont{Vidal}},
  \bibinfo{author}{\bibfnamefont{F.}~\bibnamefont{Verstraete}},
  \bibnamefont{and} \bibinfo{author}{\bibfnamefont{J.~I.} \bibnamefont{Cirac}},
  \bibinfo{journal}{Phys. Rev. Lett.} \textbf{\bibinfo{volume}{101}},
  \bibinfo{pages}{250602} (\bibinfo{year}{2008}).

\bibitem[{\citenamefont{Corboz et~al.}(2014)\citenamefont{Corboz, Rice, and
  Troyer}}]{Corboz2014}
\bibinfo{author}{\bibfnamefont{P.}~\bibnamefont{Corboz}},
  \bibinfo{author}{\bibfnamefont{T.~M.} \bibnamefont{Rice}}, \bibnamefont{and}
  \bibinfo{author}{\bibfnamefont{M.}~\bibnamefont{Troyer}},
  \bibinfo{journal}{Phys. Rev. Lett.} \textbf{\bibinfo{volume}{113}},
  \bibinfo{pages}{046402} (\bibinfo{year}{2014}).

\bibitem[{\citenamefont{Corboz}(2015)}]{Corboz2015}
\bibinfo{author}{\bibfnamefont{P.}~\bibnamefont{Corboz}},
  \bibinfo{journal}{arxiv:1508.04003}  (\bibinfo{year}{2015}).

\bibitem[{\citenamefont{Potthoff et~al.}(2003)\citenamefont{Potthoff, Aichhorn,
  and Dahnken}}]{Potthoff2003}
\bibinfo{author}{\bibfnamefont{M.}~\bibnamefont{Potthoff}},
  \bibinfo{author}{\bibfnamefont{M.}~\bibnamefont{Aichhorn}}, \bibnamefont{and}
  \bibinfo{author}{\bibfnamefont{C.}~\bibnamefont{Dahnken}},
  \bibinfo{journal}{Phys. Rev. Lett.} \textbf{\bibinfo{volume}{91}},
  \bibinfo{pages}{206402} (\bibinfo{year}{2003}).

\bibitem[{\citenamefont{Kotliar et~al.}(2001)\citenamefont{Kotliar, Savrasov,
  P\'alsson, and Biroli}}]{Kotliar2001}
\bibinfo{author}{\bibfnamefont{G.}~\bibnamefont{Kotliar}},
  \bibinfo{author}{\bibfnamefont{S.~Y.} \bibnamefont{Savrasov}},
  \bibinfo{author}{\bibfnamefont{G.}~\bibnamefont{P\'alsson}},
  \bibnamefont{and} \bibinfo{author}{\bibfnamefont{G.}~\bibnamefont{Biroli}},
  \bibinfo{journal}{Phys. Rev. Lett.} \textbf{\bibinfo{volume}{87}},
  \bibinfo{pages}{186401} (\bibinfo{year}{2001}).

\bibitem[{\citenamefont{Hettler et~al.}(1998)\citenamefont{Hettler,
  Tahvildar-Zadeh, Jarrell, Pruschke, and Krishnamurthy}}]{Hettler1998}
\bibinfo{author}{\bibfnamefont{M.~H.} \bibnamefont{Hettler}},
  \bibinfo{author}{\bibfnamefont{A.~N.} \bibnamefont{Tahvildar-Zadeh}},
  \bibinfo{author}{\bibfnamefont{M.}~\bibnamefont{Jarrell}},
  \bibinfo{author}{\bibfnamefont{T.}~\bibnamefont{Pruschke}}, \bibnamefont{and}
  \bibinfo{author}{\bibfnamefont{H.~R.} \bibnamefont{Krishnamurthy}},
  \bibinfo{journal}{Phys. Rev. B} \textbf{\bibinfo{volume}{58}},
  \bibinfo{pages}{R7475} (\bibinfo{year}{1998}).

\bibitem[{\citenamefont{Georges et~al.}(1996)\citenamefont{Georges, Kotliar,
  Krauth, and Rozenberg}}]{Georges1996}
\bibinfo{author}{\bibfnamefont{A.}~\bibnamefont{Georges}},
  \bibinfo{author}{\bibfnamefont{G.}~\bibnamefont{Kotliar}},
  \bibinfo{author}{\bibfnamefont{W.}~\bibnamefont{Krauth}}, \bibnamefont{and}
  \bibinfo{author}{\bibfnamefont{M.~J.} \bibnamefont{Rozenberg}},
  \bibinfo{journal}{Rev. Mod. Phys.} \textbf{\bibinfo{volume}{68}},
  \bibinfo{pages}{13} (\bibinfo{year}{1996}).

\bibitem[{\citenamefont{Georges and Kotliar}(1992)}]{Georges1992}
\bibinfo{author}{\bibfnamefont{A.}~\bibnamefont{Georges}} \bibnamefont{and}
  \bibinfo{author}{\bibfnamefont{G.}~\bibnamefont{Kotliar}},
  \bibinfo{journal}{Phys. Rev. B} \textbf{\bibinfo{volume}{45}},
  \bibinfo{pages}{6479} (\bibinfo{year}{1992}).

\bibitem[{\citenamefont{Metzner and Vollhardt}(1989)}]{Metzner1989}
\bibinfo{author}{\bibfnamefont{W.}~\bibnamefont{Metzner}} \bibnamefont{and}
  \bibinfo{author}{\bibfnamefont{D.}~\bibnamefont{Vollhardt}},
  \bibinfo{journal}{Phys. Rev. Lett.} \textbf{\bibinfo{volume}{62}},
  \bibinfo{pages}{324} (\bibinfo{year}{1989}).

\bibitem[{\citenamefont{M\"uller-Hartmann}(1989)}]{MuellerHartmann1989}
\bibinfo{author}{\bibfnamefont{E.}~\bibnamefont{M\"uller-Hartmann}},
  \bibinfo{journal}{Zeitschrift f\"ur Physik B Condensed Matter}
  \textbf{\bibinfo{volume}{74}}, \bibinfo{pages}{507} (\bibinfo{year}{1989}).

\bibitem[{\citenamefont{Koller et~al.}(2004{\natexlab{a}})\citenamefont{Koller,
  Meyer, Ōno, and Hewson}}]{Koller2004}
\bibinfo{author}{\bibfnamefont{W.}~\bibnamefont{Koller}},
  \bibinfo{author}{\bibfnamefont{D.}~\bibnamefont{Meyer}},
  \bibinfo{author}{\bibfnamefont{Y.}~\bibnamefont{Ōno}}, \bibnamefont{and}
  \bibinfo{author}{\bibfnamefont{A.~C.} \bibnamefont{Hewson}},
  \bibinfo{journal}{EPL} \textbf{\bibinfo{volume}{66}}, \bibinfo{pages}{559}
  (\bibinfo{year}{2004}{\natexlab{a}}).

\bibitem[{\citenamefont{Koller et~al.}(2004{\natexlab{b}})\citenamefont{Koller,
  Meyer, and Hewson}}]{Koller2004_prb}
\bibinfo{author}{\bibfnamefont{W.}~\bibnamefont{Koller}},
  \bibinfo{author}{\bibfnamefont{D.}~\bibnamefont{Meyer}}, \bibnamefont{and}
  \bibinfo{author}{\bibfnamefont{A.~C.} \bibnamefont{Hewson}},
  \bibinfo{journal}{Phys. Rev. B} \textbf{\bibinfo{volume}{70}},
  \bibinfo{pages}{155103} (\bibinfo{year}{2004}{\natexlab{b}}).

\bibitem[{\citenamefont{Jeon et~al.}(2004)\citenamefont{Jeon, Park, Han, Lee,
  and Choi}}]{GunSang2004}
\bibinfo{author}{\bibfnamefont{G.~S.} \bibnamefont{Jeon}},
  \bibinfo{author}{\bibfnamefont{T.-H.} \bibnamefont{Park}},
  \bibinfo{author}{\bibfnamefont{J.~H.} \bibnamefont{Han}},
  \bibinfo{author}{\bibfnamefont{H.~C.} \bibnamefont{Lee}}, \bibnamefont{and}
  \bibinfo{author}{\bibfnamefont{H.-Y.} \bibnamefont{Choi}},
  \bibinfo{journal}{Phys. Rev. B} \textbf{\bibinfo{volume}{70}},
  \bibinfo{pages}{125114} (\bibinfo{year}{2004}).

\bibitem[{\citenamefont{Koller et~al.}(2005)\citenamefont{Koller, Meyer,
  Hewson, and Ōno}}]{Koller2005}
\bibinfo{author}{\bibfnamefont{W.}~\bibnamefont{Koller}},
  \bibinfo{author}{\bibfnamefont{D.}~\bibnamefont{Meyer}},
  \bibinfo{author}{\bibfnamefont{A.}~\bibnamefont{Hewson}}, \bibnamefont{and}
  \bibinfo{author}{\bibfnamefont{Y.}~\bibnamefont{Ōno}},
  \bibinfo{journal}{Physica B: Condensed Matter}
  \textbf{\bibinfo{volume}{359–361}}, \bibinfo{pages}{795 }
  (\bibinfo{year}{2005}), ISSN \bibinfo{issn}{0921-4526},
  \bibinfo{note}{proceedings of the International Conference on Strongly
  Correlated Electron Systems}.

\bibitem[{\citenamefont{Barone et~al.}(2006)\citenamefont{Barone, Raimondi,
  Capone, and Castellani}}]{Barone2006}
\bibinfo{author}{\bibfnamefont{P.}~\bibnamefont{Barone}},
  \bibinfo{author}{\bibfnamefont{R.}~\bibnamefont{Raimondi}},
  \bibinfo{author}{\bibfnamefont{M.}~\bibnamefont{Capone}}, \bibnamefont{and}
  \bibinfo{author}{\bibfnamefont{C.}~\bibnamefont{Castellani}},
  \bibinfo{journal}{Phys. Rev. B} \textbf{\bibinfo{volume}{73}},
  \bibinfo{pages}{085120} (\bibinfo{year}{2006}).

\bibitem[{\citenamefont{Freericks and Jarrell}(1994)}]{Freericks1993}
\bibinfo{author}{\bibfnamefont{J.~K.} \bibnamefont{Freericks}}
  \bibnamefont{and} \bibinfo{author}{\bibfnamefont{M.}~\bibnamefont{Jarrell}},
  \bibinfo{journal}{Phys. Rev. B} \textbf{\bibinfo{volume}{50}},
  \bibinfo{pages}{6939} (\bibinfo{year}{1994}).

\bibitem[{\citenamefont{Murakami et~al.}(2013)\citenamefont{Murakami, Werner,
  Tsuji, and Aoki}}]{Murakami2013}
\bibinfo{author}{\bibfnamefont{Y.}~\bibnamefont{Murakami}},
  \bibinfo{author}{\bibfnamefont{P.}~\bibnamefont{Werner}},
  \bibinfo{author}{\bibfnamefont{N.}~\bibnamefont{Tsuji}}, \bibnamefont{and}
  \bibinfo{author}{\bibfnamefont{H.}~\bibnamefont{Aoki}},
  \bibinfo{journal}{Phys. Rev. B} \textbf{\bibinfo{volume}{88}},
  \bibinfo{pages}{125126} (\bibinfo{year}{2013}).

\bibitem[{\citenamefont{Murakami et~al.}(2014)\citenamefont{Murakami, Werner,
  Tsuji, and Aoki}}]{Murakami2014}
\bibinfo{author}{\bibfnamefont{Y.}~\bibnamefont{Murakami}},
  \bibinfo{author}{\bibfnamefont{P.}~\bibnamefont{Werner}},
  \bibinfo{author}{\bibfnamefont{N.}~\bibnamefont{Tsuji}}, \bibnamefont{and}
  \bibinfo{author}{\bibfnamefont{H.}~\bibnamefont{Aoki}},
  \bibinfo{journal}{Phys. Rev. Lett.} \textbf{\bibinfo{volume}{113}},
  \bibinfo{pages}{266404} (\bibinfo{year}{2014}).

\bibitem[{\citenamefont{Casula et~al.}(2012{\natexlab{a}})\citenamefont{Casula,
  Rubtsov, and Biermann}}]{Casula2012maxent}
\bibinfo{author}{\bibfnamefont{M.}~\bibnamefont{Casula}},
  \bibinfo{author}{\bibfnamefont{A.}~\bibnamefont{Rubtsov}}, \bibnamefont{and}
  \bibinfo{author}{\bibfnamefont{S.}~\bibnamefont{Biermann}},
  \bibinfo{journal}{Phys. Rev. B} \textbf{\bibinfo{volume}{85}},
  \bibinfo{pages}{035115} (\bibinfo{year}{2012}{\natexlab{a}}).

\bibitem[{\citenamefont{Lang and Firsov}(1962)}]{Lang1962}
\bibinfo{author}{\bibfnamefont{I.~G.} \bibnamefont{Lang}} \bibnamefont{and}
  \bibinfo{author}{\bibfnamefont{Y.~A.} \bibnamefont{Firsov}},
  \bibinfo{journal}{Sov. Phys. JETP} \textbf{\bibinfo{volume}{16}},
  \bibinfo{pages}{1301} (\bibinfo{year}{1962}).

\bibitem[{\citenamefont{Werner et~al.}(2006)\citenamefont{Werner, Comanac, de'
  Medici, Troyer, and Millis}}]{Werner2006}
\bibinfo{author}{\bibfnamefont{P.}~\bibnamefont{Werner}},
  \bibinfo{author}{\bibfnamefont{A.}~\bibnamefont{Comanac}},
  \bibinfo{author}{\bibfnamefont{L.}~\bibnamefont{de' Medici}},
  \bibinfo{author}{\bibfnamefont{M.}~\bibnamefont{Troyer}}, \bibnamefont{and}
  \bibinfo{author}{\bibfnamefont{A.~J.} \bibnamefont{Millis}},
  \bibinfo{journal}{Phys. Rev. Lett.} \textbf{\bibinfo{volume}{97}},
  \bibinfo{pages}{076405} (\bibinfo{year}{2006}).

\bibitem[{\citenamefont{Werner and Millis}(2006)}]{Werner2006matrix}
\bibinfo{author}{\bibfnamefont{P.}~\bibnamefont{Werner}} \bibnamefont{and}
  \bibinfo{author}{\bibfnamefont{A.~J.} \bibnamefont{Millis}},
  \bibinfo{journal}{Phys. Rev. B} \textbf{\bibinfo{volume}{74}},
  \bibinfo{pages}{155107} (\bibinfo{year}{2006}).

\bibitem[{\citenamefont{Werner and Millis}(2007)}]{Werner2007}
\bibinfo{author}{\bibfnamefont{P.}~\bibnamefont{Werner}} \bibnamefont{and}
  \bibinfo{author}{\bibfnamefont{A.~J.} \bibnamefont{Millis}},
  \bibinfo{journal}{Phys. Rev. Lett.} \textbf{\bibinfo{volume}{99}},
  \bibinfo{pages}{146404} (\bibinfo{year}{2007}).

\bibitem[{\citenamefont{Mahan}(1990)}]{mahan}
\bibinfo{author}{\bibfnamefont{G.~D.} \bibnamefont{Mahan}},
  \emph{\bibinfo{title}{Many-particle Physics}} (\bibinfo{publisher}{Plenum
  Press}, \bibinfo{year}{1990}).

\bibitem[{\citenamefont{Hafermann et~al.}(2012)\citenamefont{Hafermann, Patton,
  and Werner}}]{Hafermann2012}
\bibinfo{author}{\bibfnamefont{H.}~\bibnamefont{Hafermann}},
  \bibinfo{author}{\bibfnamefont{K.~R.} \bibnamefont{Patton}},
  \bibnamefont{and} \bibinfo{author}{\bibfnamefont{P.}~\bibnamefont{Werner}},
  \bibinfo{journal}{Phys. Rev. B} \textbf{\bibinfo{volume}{85}},
  \bibinfo{pages}{205106} (\bibinfo{year}{2012}).

\bibitem[{\citenamefont{Hafermann}(2014)}]{Hafermann2014}
\bibinfo{author}{\bibfnamefont{H.}~\bibnamefont{Hafermann}},
  \bibinfo{journal}{Phys. Rev. B} \textbf{\bibinfo{volume}{89}},
  \bibinfo{pages}{235128} (\bibinfo{year}{2014}).

\bibitem[{\citenamefont{Werner and Millis}(2010)}]{Werner2010}
\bibinfo{author}{\bibfnamefont{P.}~\bibnamefont{Werner}} \bibnamefont{and}
  \bibinfo{author}{\bibfnamefont{A.~J.} \bibnamefont{Millis}},
  \bibinfo{journal}{Phys. Rev. Lett.} \textbf{\bibinfo{volume}{104}},
  \bibinfo{pages}{146401} (\bibinfo{year}{2010}).

\bibitem[{\citenamefont{Ayral et~al.}(2013)\citenamefont{Ayral, Biermann, and
  Werner}}]{Ayral2013}
\bibinfo{author}{\bibfnamefont{T.}~\bibnamefont{Ayral}},
  \bibinfo{author}{\bibfnamefont{S.}~\bibnamefont{Biermann}}, \bibnamefont{and}
  \bibinfo{author}{\bibfnamefont{P.}~\bibnamefont{Werner}},
  \bibinfo{journal}{Phys. Rev. B} \textbf{\bibinfo{volume}{87}},
  \bibinfo{pages}{125149} (\bibinfo{year}{2013}).

\bibitem[{\citenamefont{Casula et~al.}(2012{\natexlab{b}})\citenamefont{Casula,
  Werner, Vaugier, Aryasetiawan, Miyake, Millis, and
  Biermann}}]{Casula2012static}
\bibinfo{author}{\bibfnamefont{M.}~\bibnamefont{Casula}},
  \bibinfo{author}{\bibfnamefont{P.}~\bibnamefont{Werner}},
  \bibinfo{author}{\bibfnamefont{L.}~\bibnamefont{Vaugier}},
  \bibinfo{author}{\bibfnamefont{F.}~\bibnamefont{Aryasetiawan}},
  \bibinfo{author}{\bibfnamefont{T.}~\bibnamefont{Miyake}},
  \bibinfo{author}{\bibfnamefont{A.~J.} \bibnamefont{Millis}},
  \bibnamefont{and} \bibinfo{author}{\bibfnamefont{S.}~\bibnamefont{Biermann}},
  \bibinfo{journal}{Phys. Rev. Lett.} \textbf{\bibinfo{volume}{109}},
  \bibinfo{pages}{126408} (\bibinfo{year}{2012}{\natexlab{b}}).

\bibitem[{\citenamefont{Capone et~al.}(2009)\citenamefont{Capone, Fabrizio,
  Castellani, and Tosatti}}]{Capone2009}
\bibinfo{author}{\bibfnamefont{M.}~\bibnamefont{Capone}},
  \bibinfo{author}{\bibfnamefont{M.}~\bibnamefont{Fabrizio}},
  \bibinfo{author}{\bibfnamefont{C.}~\bibnamefont{Castellani}},
  \bibnamefont{and} \bibinfo{author}{\bibfnamefont{E.}~\bibnamefont{Tosatti}},
  \bibinfo{journal}{Rev. Mod. Phys.} \textbf{\bibinfo{volume}{81}},
  \bibinfo{pages}{943} (\bibinfo{year}{2009}).

\bibitem[{\citenamefont{Steiner et~al.}(2015)\citenamefont{Steiner, Nomura, and
  Werner}}]{Steiner2015}
\bibinfo{author}{\bibfnamefont{K.}~\bibnamefont{Steiner}},
  \bibinfo{author}{\bibfnamefont{Y.}~\bibnamefont{Nomura}}, \bibnamefont{and}
  \bibinfo{author}{\bibfnamefont{P.}~\bibnamefont{Werner}},
  \bibinfo{journal}{Phys. Rev. B} \textbf{\bibinfo{volume}{92}},
  \bibinfo{pages}{115123} (\bibinfo{year}{2015}).

\bibitem[{\citenamefont{Sengupta and Georges}(1995)}]{Sengupta1995}
\bibinfo{author}{\bibfnamefont{A.~M.} \bibnamefont{Sengupta}} \bibnamefont{and}
  \bibinfo{author}{\bibfnamefont{A.}~\bibnamefont{Georges}},
  \bibinfo{journal}{Phys. Rev. B} \textbf{\bibinfo{volume}{52}},
  \bibinfo{pages}{10295} (\bibinfo{year}{1995}).

\bibitem[{\citenamefont{Si and Smith}(1996)}]{Si1996}
\bibinfo{author}{\bibfnamefont{Q.}~\bibnamefont{Si}} \bibnamefont{and}
  \bibinfo{author}{\bibfnamefont{J.~L.} \bibnamefont{Smith}},
  \bibinfo{journal}{Phys. Rev. Lett.} \textbf{\bibinfo{volume}{77}},
  \bibinfo{pages}{3391} (\bibinfo{year}{1996}).

\bibitem[{\citenamefont{Sun and Kotliar}(2002)}]{Sun2002}
\bibinfo{author}{\bibfnamefont{P.}~\bibnamefont{Sun}} \bibnamefont{and}
  \bibinfo{author}{\bibfnamefont{G.}~\bibnamefont{Kotliar}},
  \bibinfo{journal}{Phys. Rev. B} \textbf{\bibinfo{volume}{66}},
  \bibinfo{pages}{085120} (\bibinfo{year}{2002}).

\bibitem[{\citenamefont{Negele and Orland}(1988)}]{negele}
\bibinfo{author}{\bibfnamefont{J.~W.} \bibnamefont{Negele}} \bibnamefont{and}
  \bibinfo{author}{\bibfnamefont{H.}~\bibnamefont{Orland}},
  \emph{\bibinfo{title}{Quantum Many-Particle Systems}}
  (\bibinfo{publisher}{Westview Press}, \bibinfo{year}{1988}).

\bibitem[{\citenamefont{Ayral et~al.}(2012)\citenamefont{Ayral, Werner, and
  Biermann}}]{Ayral2012}
\bibinfo{author}{\bibfnamefont{T.}~\bibnamefont{Ayral}},
  \bibinfo{author}{\bibfnamefont{P.}~\bibnamefont{Werner}}, \bibnamefont{and}
  \bibinfo{author}{\bibfnamefont{S.}~\bibnamefont{Biermann}},
  \bibinfo{journal}{Phys. Rev. Lett.} \textbf{\bibinfo{volume}{109}},
  \bibinfo{pages}{226401} (\bibinfo{year}{2012}).

\bibitem[{\citenamefont{Huang et~al.}(2014)\citenamefont{Huang, Ayral,
  Biermann, and Werner}}]{Huang2014}
\bibinfo{author}{\bibfnamefont{L.}~\bibnamefont{Huang}},
  \bibinfo{author}{\bibfnamefont{T.}~\bibnamefont{Ayral}},
  \bibinfo{author}{\bibfnamefont{S.}~\bibnamefont{Biermann}}, \bibnamefont{and}
  \bibinfo{author}{\bibfnamefont{P.}~\bibnamefont{Werner}},
  \bibinfo{journal}{Phys. Rev. B} \textbf{\bibinfo{volume}{90}},
  \bibinfo{pages}{195114} (\bibinfo{year}{2014}).

\bibitem[{\citenamefont{Prokof'ev and Svistunov}(2007)}]{Prokofev2007}
\bibinfo{author}{\bibfnamefont{N.}~\bibnamefont{Prokof'ev}} \bibnamefont{and}
  \bibinfo{author}{\bibfnamefont{B.}~\bibnamefont{Svistunov}},
  \bibinfo{journal}{Phys. Rev. Lett.} \textbf{\bibinfo{volume}{99}},
  \bibinfo{pages}{250201} (\bibinfo{year}{2007}).

\bibitem[{\citenamefont{Rubtsov et~al.}(2012)\citenamefont{Rubtsov, Katsnelson,
  and Lichtenstein}}]{Rubtsov2012}
\bibinfo{author}{\bibfnamefont{A.}~\bibnamefont{Rubtsov}},
  \bibinfo{author}{\bibfnamefont{M.}~\bibnamefont{Katsnelson}},
  \bibnamefont{and}
  \bibinfo{author}{\bibfnamefont{A.}~\bibnamefont{Lichtenstein}},
  \bibinfo{journal}{Annals of Physics} \textbf{\bibinfo{volume}{327}},
  \bibinfo{pages}{1320} (\bibinfo{year}{2012}).

\bibitem[{\citenamefont{Stepanov et~al.}(2015)\citenamefont{Stepanov, van Loon,
  Katanin, Lichtenstein, Katsnelson, and Rubtsov}}]{Stepanov2015}
\bibinfo{author}{\bibfnamefont{E.}~\bibnamefont{Stepanov}},
  \bibinfo{author}{\bibfnamefont{E.}~\bibnamefont{van Loon}},
  \bibinfo{author}{\bibfnamefont{A.}~\bibnamefont{Katanin}},
  \bibinfo{author}{\bibfnamefont{A.}~\bibnamefont{Lichtenstein}},
  \bibinfo{author}{\bibfnamefont{M.}~\bibnamefont{Katsnelson}},
  \bibnamefont{and} \bibinfo{author}{\bibfnamefont{A.}~\bibnamefont{Rubtsov}},
  \bibinfo{journal}{arxiv:1508.07237}  (\bibinfo{year}{2015}).

\bibitem[{\citenamefont{Rubtsov et~al.}(2008)\citenamefont{Rubtsov, Katsnelson,
  and Lichtenstein}}]{Rubtsov2008}
\bibinfo{author}{\bibfnamefont{A.~N.} \bibnamefont{Rubtsov}},
  \bibinfo{author}{\bibfnamefont{M.~I.} \bibnamefont{Katsnelson}},
  \bibnamefont{and} \bibinfo{author}{\bibfnamefont{A.~I.}
  \bibnamefont{Lichtenstein}}, \bibinfo{journal}{Phys. Rev. B}
  \textbf{\bibinfo{volume}{77}}, \bibinfo{pages}{033101}
  (\bibinfo{year}{2008}).

\bibitem[{\citenamefont{Tomczak et~al.}(2014)\citenamefont{Tomczak, Casula,
  Miyake, and Biermann}}]{Tomczak2014}
\bibinfo{author}{\bibfnamefont{J.~M.} \bibnamefont{Tomczak}},
  \bibinfo{author}{\bibfnamefont{M.}~\bibnamefont{Casula}},
  \bibinfo{author}{\bibfnamefont{T.}~\bibnamefont{Miyake}}, \bibnamefont{and}
  \bibinfo{author}{\bibfnamefont{S.}~\bibnamefont{Biermann}},
  \bibinfo{journal}{Phys. Rev. B} \textbf{\bibinfo{volume}{90}},
  \bibinfo{pages}{165138} (\bibinfo{year}{2014}).

\bibitem[{\citenamefont{Tomczak
  et~al.}(2012{\natexlab{a}})\citenamefont{Tomczak, Casula, Miyake,
  Aryasetiawan, and Biermann}}]{Tomczak2012}
\bibinfo{author}{\bibfnamefont{J.~M.} \bibnamefont{Tomczak}},
  \bibinfo{author}{\bibfnamefont{M.}~\bibnamefont{Casula}},
  \bibinfo{author}{\bibfnamefont{T.}~\bibnamefont{Miyake}},
  \bibinfo{author}{\bibfnamefont{F.}~\bibnamefont{Aryasetiawan}},
  \bibnamefont{and} \bibinfo{author}{\bibfnamefont{S.}~\bibnamefont{Biermann}},
  \bibinfo{journal}{EPL} \textbf{\bibinfo{volume}{100}}, \bibinfo{pages}{67001}
  (\bibinfo{year}{2012}{\natexlab{a}}).

\bibitem[{\citenamefont{Werner et~al.}(2012)\citenamefont{Werner, Casula,
  Miyake, Aryasetiawan, Millis, and Biermann}}]{Werner2012}
\bibinfo{author}{\bibfnamefont{P.}~\bibnamefont{Werner}},
  \bibinfo{author}{\bibfnamefont{M.}~\bibnamefont{Casula}},
  \bibinfo{author}{\bibfnamefont{T.}~\bibnamefont{Miyake}},
  \bibinfo{author}{\bibfnamefont{F.}~\bibnamefont{Aryasetiawan}},
  \bibinfo{author}{\bibfnamefont{A.~J.} \bibnamefont{Millis}},
  \bibnamefont{and} \bibinfo{author}{\bibfnamefont{S.}~\bibnamefont{Biermann}},
  \bibinfo{journal}{Nat. Phys.} pp. \bibinfo{pages}{1745--2481}
  (\bibinfo{year}{2012}).

\bibitem[{\citenamefont{Biermann}(2014)}]{Biermann2014}
\bibinfo{author}{\bibfnamefont{S.}~\bibnamefont{Biermann}},
  \bibinfo{journal}{Journal of Physics: Condensed Matter}
  \textbf{\bibinfo{volume}{26}}, \bibinfo{pages}{173202}
  (\bibinfo{year}{2014}).

\bibitem[{\citenamefont{van Roekeghem et~al.}(2014)\citenamefont{van Roekeghem,
  Ayral, Tomczak, Casula, Xu, Ding, Ferrero, Parcollet, Jiang, and
  Biermann}}]{Roekeghem2014}
\bibinfo{author}{\bibfnamefont{A.}~\bibnamefont{van Roekeghem}},
  \bibinfo{author}{\bibfnamefont{T.}~\bibnamefont{Ayral}},
  \bibinfo{author}{\bibfnamefont{J.~M.} \bibnamefont{Tomczak}},
  \bibinfo{author}{\bibfnamefont{M.}~\bibnamefont{Casula}},
  \bibinfo{author}{\bibfnamefont{N.}~\bibnamefont{Xu}},
  \bibinfo{author}{\bibfnamefont{H.}~\bibnamefont{Ding}},
  \bibinfo{author}{\bibfnamefont{M.}~\bibnamefont{Ferrero}},
  \bibinfo{author}{\bibfnamefont{O.}~\bibnamefont{Parcollet}},
  \bibinfo{author}{\bibfnamefont{H.}~\bibnamefont{Jiang}}, \bibnamefont{and}
  \bibinfo{author}{\bibfnamefont{S.}~\bibnamefont{Biermann}},
  \bibinfo{journal}{Phys. Rev. Lett.} \textbf{\bibinfo{volume}{113}},
  \bibinfo{pages}{266403} (\bibinfo{year}{2014}).

\bibitem[{\citenamefont{Miyake et~al.}(2013)\citenamefont{Miyake, Martins,
  Sakuma, and Aryasetiawan}}]{Miyake2013}
\bibinfo{author}{\bibfnamefont{T.}~\bibnamefont{Miyake}},
  \bibinfo{author}{\bibfnamefont{C.}~\bibnamefont{Martins}},
  \bibinfo{author}{\bibfnamefont{R.}~\bibnamefont{Sakuma}}, \bibnamefont{and}
  \bibinfo{author}{\bibfnamefont{F.}~\bibnamefont{Aryasetiawan}},
  \bibinfo{journal}{Phys. Rev. B} \textbf{\bibinfo{volume}{87}},
  \bibinfo{pages}{115110} (\bibinfo{year}{2013}).

\bibitem[{\citenamefont{Sakuma et~al.}(2014)\citenamefont{Sakuma, Martins,
  Miyake, and Aryasetiawan}}]{Sakuma2014}
\bibinfo{author}{\bibfnamefont{R.}~\bibnamefont{Sakuma}},
  \bibinfo{author}{\bibfnamefont{C.}~\bibnamefont{Martins}},
  \bibinfo{author}{\bibfnamefont{T.}~\bibnamefont{Miyake}}, \bibnamefont{and}
  \bibinfo{author}{\bibfnamefont{F.}~\bibnamefont{Aryasetiawan}},
  \bibinfo{journal}{Phys. Rev. B} \textbf{\bibinfo{volume}{89}},
  \bibinfo{pages}{235119} (\bibinfo{year}{2014}).

\bibitem[{\citenamefont{Sch\"afer et~al.}(2015)\citenamefont{Sch\"afer, Toschi,
  and Tomczak}}]{Schafer2015}
\bibinfo{author}{\bibfnamefont{T.}~\bibnamefont{Sch\"afer}},
  \bibinfo{author}{\bibfnamefont{A.}~\bibnamefont{Toschi}}, \bibnamefont{and}
  \bibinfo{author}{\bibfnamefont{J.~M.} \bibnamefont{Tomczak}},
  \bibinfo{journal}{Phys. Rev. B} \textbf{\bibinfo{volume}{91}},
  \bibinfo{pages}{121107} (\bibinfo{year}{2015}).

\bibitem[{\citenamefont{Tomczak
  et~al.}(2012{\natexlab{b}})\citenamefont{Tomczak, van Schilfgaarde, and
  Kotliar}}]{jmt_pnict}
\bibinfo{author}{\bibfnamefont{J.~M.} \bibnamefont{Tomczak}},
  \bibinfo{author}{\bibfnamefont{M.}~\bibnamefont{van Schilfgaarde}},
  \bibnamefont{and} \bibinfo{author}{\bibfnamefont{G.}~\bibnamefont{Kotliar}},
  \bibinfo{journal}{Phys. Rev. Lett.} \textbf{\bibinfo{volume}{109}},
  \bibinfo{pages}{237010} (\bibinfo{year}{2012}{\natexlab{b}}).

\bibitem[{\citenamefont{Taranto et~al.}(2013)\citenamefont{Taranto, Kaltak,
  Parragh, Sangiovanni, Kresse, Toschi, and Held}}]{Taranto2013}
\bibinfo{author}{\bibfnamefont{C.}~\bibnamefont{Taranto}},
  \bibinfo{author}{\bibfnamefont{M.}~\bibnamefont{Kaltak}},
  \bibinfo{author}{\bibfnamefont{N.}~\bibnamefont{Parragh}},
  \bibinfo{author}{\bibfnamefont{G.}~\bibnamefont{Sangiovanni}},
  \bibinfo{author}{\bibfnamefont{G.}~\bibnamefont{Kresse}},
  \bibinfo{author}{\bibfnamefont{A.}~\bibnamefont{Toschi}}, \bibnamefont{and}
  \bibinfo{author}{\bibfnamefont{K.}~\bibnamefont{Held}},
  \bibinfo{journal}{Phys. Rev. B} \textbf{\bibinfo{volume}{88}},
  \bibinfo{pages}{165119} (\bibinfo{year}{2013}).

\bibitem[{\citenamefont{van Schilfgaarde et~al.}(2006)\citenamefont{van
  Schilfgaarde, Kotani, and Faleev}}]{vanSchilfgaarde2006}
\bibinfo{author}{\bibfnamefont{M.}~\bibnamefont{van Schilfgaarde}},
  \bibinfo{author}{\bibfnamefont{T.}~\bibnamefont{Kotani}}, \bibnamefont{and}
  \bibinfo{author}{\bibfnamefont{S.}~\bibnamefont{Faleev}},
  \bibinfo{journal}{Phys. Rev. Lett.} \textbf{\bibinfo{volume}{96}},
  \bibinfo{eid}{226402} (\bibinfo{year}{2006}).

\bibitem[{\citenamefont{Kotani et~al.}(2007)\citenamefont{Kotani, van
  Schilfgaarde, and Faleev}}]{Kotani2007}
\bibinfo{author}{\bibfnamefont{T.}~\bibnamefont{Kotani}},
  \bibinfo{author}{\bibfnamefont{M.}~\bibnamefont{van Schilfgaarde}},
  \bibnamefont{and} \bibinfo{author}{\bibfnamefont{S.~V.}
  \bibnamefont{Faleev}}, \bibinfo{journal}{Phys. Rev. B}
  \textbf{\bibinfo{volume}{76}}, \bibinfo{eid}{165106} (\bibinfo{year}{2007}).

\bibitem[{\citenamefont{Choi et~al.}(2015)\citenamefont{Choi, Kutepov, Haule,
  van Schilfgaarde, and Kotliar}}]{Choi2015}
\bibinfo{author}{\bibfnamefont{S.}~\bibnamefont{Choi}},
  \bibinfo{author}{\bibfnamefont{A.}~\bibnamefont{Kutepov}},
  \bibinfo{author}{\bibfnamefont{K.}~\bibnamefont{Haule}},
  \bibinfo{author}{\bibfnamefont{M.}~\bibnamefont{van Schilfgaarde}},
  \bibnamefont{and} \bibinfo{author}{\bibfnamefont{G.}~\bibnamefont{Kotliar}},
  \bibinfo{journal}{arXiv:1504.07569v1}  (\bibinfo{year}{2015}).

\bibitem[{\citenamefont{Gygi and Baldereschi}(1989)}]{PhysRevLett.62.2160}
\bibinfo{author}{\bibfnamefont{F.}~\bibnamefont{Gygi}} \bibnamefont{and}
  \bibinfo{author}{\bibfnamefont{A.}~\bibnamefont{Baldereschi}},
  \bibinfo{journal}{Phys. Rev. Lett.} \textbf{\bibinfo{volume}{62}},
  \bibinfo{pages}{2160} (\bibinfo{year}{1989}).

\bibitem[{\citenamefont{Vidberg and Serene}(1977)}]{Vidberg1977}
\bibinfo{author}{\bibfnamefont{H.}~\bibnamefont{Vidberg}} \bibnamefont{and}
  \bibinfo{author}{\bibfnamefont{J.}~\bibnamefont{Serene}},
  \bibinfo{journal}{J. Low Temp. Phys.} \textbf{\bibinfo{volume}{29}},
  \bibinfo{pages}{179} (\bibinfo{year}{1977}).

\bibitem[{\citenamefont{Jarrell and Gubernatis}(1996)}]{Jarrell1996}
\bibinfo{author}{\bibfnamefont{M.}~\bibnamefont{Jarrell}} \bibnamefont{and}
  \bibinfo{author}{\bibfnamefont{J.~E.} \bibnamefont{Gubernatis}},
  \bibinfo{journal}{Physics Reports} \textbf{\bibinfo{volume}{269}},
  \bibinfo{pages}{133} (\bibinfo{year}{1996}).

\bibitem[{\citenamefont{Gole\ifmmode~\check{z}\else \v{z}\fi{}
  et~al.}(2015)\citenamefont{Gole\ifmmode~\check{z}\else \v{z}\fi{}, Eckstein,
  and Werner}}]{Golez2015}
\bibinfo{author}{\bibfnamefont{D.}~\bibnamefont{Gole\ifmmode~\check{z}\else
  \v{z}\fi{}}}, \bibinfo{author}{\bibfnamefont{M.}~\bibnamefont{Eckstein}},
  \bibnamefont{and} \bibinfo{author}{\bibfnamefont{P.}~\bibnamefont{Werner}},
  \bibinfo{journal}{Phys. Rev. B} \textbf{\bibinfo{volume}{92}},
  \bibinfo{pages}{195123} (\bibinfo{year}{2015}).

\bibitem[{\citenamefont{Aoki et~al.}(2014)\citenamefont{Aoki, Tsuji, Eckstein,
  Kollar, Oka, and Werner}}]{Aoki2014}
\bibinfo{author}{\bibfnamefont{H.}~\bibnamefont{Aoki}},
  \bibinfo{author}{\bibfnamefont{N.}~\bibnamefont{Tsuji}},
  \bibinfo{author}{\bibfnamefont{M.}~\bibnamefont{Eckstein}},
  \bibinfo{author}{\bibfnamefont{M.}~\bibnamefont{Kollar}},
  \bibinfo{author}{\bibfnamefont{T.}~\bibnamefont{Oka}}, \bibnamefont{and}
  \bibinfo{author}{\bibfnamefont{P.}~\bibnamefont{Werner}},
  \bibinfo{journal}{Rev. Mod. Phys.} \textbf{\bibinfo{volume}{86}},
  \bibinfo{pages}{779} (\bibinfo{year}{2014}).

\bibitem[{\citenamefont{Eckstein and Werner}(2013)}]{Eckstein2013}
\bibinfo{author}{\bibfnamefont{M.}~\bibnamefont{Eckstein}} \bibnamefont{and}
  \bibinfo{author}{\bibfnamefont{P.}~\bibnamefont{Werner}},
  \bibinfo{journal}{Phys. Rev. Lett.} \textbf{\bibinfo{volume}{110}},
  \bibinfo{pages}{126401} (\bibinfo{year}{2013}).

\bibitem[{\citenamefont{Tosatti and Anderson}(1974)}]{Tosatti1974}
\bibinfo{author}{\bibfnamefont{E.}~\bibnamefont{Tosatti}} \bibnamefont{and}
  \bibinfo{author}{\bibfnamefont{P.~W.} \bibnamefont{Anderson}},
  \bibinfo{journal}{Jpn. J. Appl. Phys.} \textbf{\bibinfo{volume}{13}},
  \bibinfo{pages}{381} (\bibinfo{year}{1974}).

\bibitem[{\citenamefont{Huang and Wang}(2012)}]{Huang2012}
\bibinfo{author}{\bibfnamefont{L.}~\bibnamefont{Huang}} \bibnamefont{and}
  \bibinfo{author}{\bibfnamefont{Y.}~\bibnamefont{Wang}},
  \bibinfo{journal}{EPL} \textbf{\bibinfo{volume}{99}}, \bibinfo{pages}{67003}
  (\bibinfo{year}{2012}).

\bibitem[{\citenamefont{Werner et~al.}(2015)\citenamefont{Werner, Sakuma,
  Nilsson, and Aryasetiawan}}]{Werner2015}
\bibinfo{author}{\bibfnamefont{P.}~\bibnamefont{Werner}},
  \bibinfo{author}{\bibfnamefont{R.}~\bibnamefont{Sakuma}},
  \bibinfo{author}{\bibfnamefont{F.}~\bibnamefont{Nilsson}}, \bibnamefont{and}
  \bibinfo{author}{\bibfnamefont{F.}~\bibnamefont{Aryasetiawan}},
  \bibinfo{journal}{Phys. Rev. B} \textbf{\bibinfo{volume}{91}},
  \bibinfo{pages}{125142} (\bibinfo{year}{2015}).

\bibitem[{\citenamefont{Yoshida et~al.}(2010)\citenamefont{Yoshida, Hashimoto,
  Takizawa, Fujimori, Kubota, Ono, and Eisaki}}]{Yoshida2010}
\bibinfo{author}{\bibfnamefont{T.}~\bibnamefont{Yoshida}},
  \bibinfo{author}{\bibfnamefont{M.}~\bibnamefont{Hashimoto}},
  \bibinfo{author}{\bibfnamefont{T.}~\bibnamefont{Takizawa}},
  \bibinfo{author}{\bibfnamefont{A.}~\bibnamefont{Fujimori}},
  \bibinfo{author}{\bibfnamefont{M.}~\bibnamefont{Kubota}},
  \bibinfo{author}{\bibfnamefont{K.}~\bibnamefont{Ono}}, \bibnamefont{and}
  \bibinfo{author}{\bibfnamefont{H.}~\bibnamefont{Eisaki}},
  \bibinfo{journal}{Phys. Rev. B} \textbf{\bibinfo{volume}{82}},
  \bibinfo{pages}{085119} (\bibinfo{year}{2010}).

\bibitem[{\citenamefont{Pavarini et~al.}(2004)\citenamefont{Pavarini, Biermann,
  Poteryaev, Lichtenstein, Georges, and Andersen}}]{Pavarini2004}
\bibinfo{author}{\bibfnamefont{E.}~\bibnamefont{Pavarini}},
  \bibinfo{author}{\bibfnamefont{S.}~\bibnamefont{Biermann}},
  \bibinfo{author}{\bibfnamefont{A.}~\bibnamefont{Poteryaev}},
  \bibinfo{author}{\bibfnamefont{A.~I.} \bibnamefont{Lichtenstein}},
  \bibinfo{author}{\bibfnamefont{A.}~\bibnamefont{Georges}}, \bibnamefont{and}
  \bibinfo{author}{\bibfnamefont{O.~K.} \bibnamefont{Andersen}},
  \bibinfo{journal}{Phys. Rev. Lett.} \textbf{\bibinfo{volume}{92}},
  \bibinfo{pages}{176403} (\bibinfo{year}{2004}).

\bibitem[{\citenamefont{Nekrasov et~al.}(2006)\citenamefont{Nekrasov, Held,
  Keller, Kondakov, Pruschke, Kollar, Andersen, Anisimov, and
  Vollhardt}}]{Nekrasov2006}
\bibinfo{author}{\bibfnamefont{I.~A.} \bibnamefont{Nekrasov}},
  \bibinfo{author}{\bibfnamefont{K.}~\bibnamefont{Held}},
  \bibinfo{author}{\bibfnamefont{G.}~\bibnamefont{Keller}},
  \bibinfo{author}{\bibfnamefont{D.~E.} \bibnamefont{Kondakov}},
  \bibinfo{author}{\bibfnamefont{T.}~\bibnamefont{Pruschke}},
  \bibinfo{author}{\bibfnamefont{M.}~\bibnamefont{Kollar}},
  \bibinfo{author}{\bibfnamefont{O.~K.} \bibnamefont{Andersen}},
  \bibinfo{author}{\bibfnamefont{V.~I.} \bibnamefont{Anisimov}},
  \bibnamefont{and}
  \bibinfo{author}{\bibfnamefont{D.}~\bibnamefont{Vollhardt}},
  \bibinfo{journal}{Phys. Rev. B} \textbf{\bibinfo{volume}{73}},
  \bibinfo{pages}{155112} (\bibinfo{year}{2006}).

\bibitem[{\citenamefont{Sakuma et~al.}(2013)\citenamefont{Sakuma, Werner, and
  Aryasetiawan}}]{Sakuma2013}
\bibinfo{author}{\bibfnamefont{R.}~\bibnamefont{Sakuma}},
  \bibinfo{author}{\bibfnamefont{P.}~\bibnamefont{Werner}}, \bibnamefont{and}
  \bibinfo{author}{\bibfnamefont{F.}~\bibnamefont{Aryasetiawan}},
  \bibinfo{journal}{Phys. Rev. B} \textbf{\bibinfo{volume}{88}},
  \bibinfo{pages}{235110} (\bibinfo{year}{2013}).

\bibitem[{\citenamefont{Rotter et~al.}(2008)\citenamefont{Rotter, Tegel, and
  Johrendt}}]{Rotter2008}
\bibinfo{author}{\bibfnamefont{M.}~\bibnamefont{Rotter}},
  \bibinfo{author}{\bibfnamefont{M.}~\bibnamefont{Tegel}}, \bibnamefont{and}
  \bibinfo{author}{\bibfnamefont{D.}~\bibnamefont{Johrendt}},
  \bibinfo{journal}{Phys. Rev. Lett.} \textbf{\bibinfo{volume}{101}},
  \bibinfo{pages}{107006} (\bibinfo{year}{2008}).

\bibitem[{\citenamefont{Yi et~al.}(2009)\citenamefont{Yi, Lu, Analytis, Chu,
  Mo, He, Moore, Zhou, Chen, Luo et~al.}}]{Yi2009}
\bibinfo{author}{\bibfnamefont{M.}~\bibnamefont{Yi}},
  \bibinfo{author}{\bibfnamefont{D.~H.} \bibnamefont{Lu}},
  \bibinfo{author}{\bibfnamefont{J.~G.} \bibnamefont{Analytis}},
  \bibinfo{author}{\bibfnamefont{J.-H.} \bibnamefont{Chu}},
  \bibinfo{author}{\bibfnamefont{S.-K.} \bibnamefont{Mo}},
  \bibinfo{author}{\bibfnamefont{R.-H.} \bibnamefont{He}},
  \bibinfo{author}{\bibfnamefont{R.~G.} \bibnamefont{Moore}},
  \bibinfo{author}{\bibfnamefont{X.~J.} \bibnamefont{Zhou}},
  \bibinfo{author}{\bibfnamefont{G.~F.} \bibnamefont{Chen}},
  \bibinfo{author}{\bibfnamefont{J.~L.} \bibnamefont{Luo}},
  \bibnamefont{et~al.}, \bibinfo{journal}{Phys. Rev. B}
  \textbf{\bibinfo{volume}{80}}, \bibinfo{pages}{024515}
  (\bibinfo{year}{2009}).

\bibitem[{\citenamefont{Brouet et~al.}(2013)\citenamefont{Brouet, Lin, Texier,
  Bobroff, Taleb-Ibrahimi, Le~F\`evre, Bertran, Casula, Werner, Biermann
  et~al.}}]{Brouet2013}
\bibinfo{author}{\bibfnamefont{V.}~\bibnamefont{Brouet}},
  \bibinfo{author}{\bibfnamefont{P.-H.} \bibnamefont{Lin}},
  \bibinfo{author}{\bibfnamefont{Y.}~\bibnamefont{Texier}},
  \bibinfo{author}{\bibfnamefont{J.}~\bibnamefont{Bobroff}},
  \bibinfo{author}{\bibfnamefont{A.}~\bibnamefont{Taleb-Ibrahimi}},
  \bibinfo{author}{\bibfnamefont{P.}~\bibnamefont{Le~F\`evre}},
  \bibinfo{author}{\bibfnamefont{F.}~\bibnamefont{Bertran}},
  \bibinfo{author}{\bibfnamefont{M.}~\bibnamefont{Casula}},
  \bibinfo{author}{\bibfnamefont{P.}~\bibnamefont{Werner}},
  \bibinfo{author}{\bibfnamefont{S.}~\bibnamefont{Biermann}},
  \bibnamefont{et~al.}, \bibinfo{journal}{Phys. Rev. Lett.}
  \textbf{\bibinfo{volume}{110}}, \bibinfo{pages}{167002}
  (\bibinfo{year}{2013}).

\bibitem[{\citenamefont{Werner et~al.}(2008)\citenamefont{Werner, Gull, Troyer,
  and Millis}}]{Werner2008}
\bibinfo{author}{\bibfnamefont{P.}~\bibnamefont{Werner}},
  \bibinfo{author}{\bibfnamefont{E.}~\bibnamefont{Gull}},
  \bibinfo{author}{\bibfnamefont{M.}~\bibnamefont{Troyer}}, \bibnamefont{and}
  \bibinfo{author}{\bibfnamefont{A.~J.} \bibnamefont{Millis}},
  \bibinfo{journal}{Phys. Rev. Lett.} \textbf{\bibinfo{volume}{101}},
  \bibinfo{pages}{166405} (\bibinfo{year}{2008}).

\bibitem[{\citenamefont{Georges et~al.}(2013)\citenamefont{Georges, de' Medici,
  and Mravlje}}]{Georges2013}
\bibinfo{author}{\bibfnamefont{A.}~\bibnamefont{Georges}},
  \bibinfo{author}{\bibfnamefont{L.}~\bibnamefont{de' Medici}},
  \bibnamefont{and} \bibinfo{author}{\bibfnamefont{J.}~\bibnamefont{Mravlje}},
  \bibinfo{journal}{Annual Review of Condensed Matter Physics}
  \textbf{\bibinfo{volume}{4}}, \bibinfo{pages}{137} (\bibinfo{year}{2013}).

\bibitem[{\citenamefont{Ding et~al.}(2011)\citenamefont{Ding, Nakayama,
  Richard, Souma, Sato, Takahashi, Neupane, Xu, Pan, Fedorov
  et~al.}}]{Ding2011}
\bibinfo{author}{\bibfnamefont{H.}~\bibnamefont{Ding}},
  \bibinfo{author}{\bibfnamefont{K.}~\bibnamefont{Nakayama}},
  \bibinfo{author}{\bibfnamefont{P.}~\bibnamefont{Richard}},
  \bibinfo{author}{\bibfnamefont{S.}~\bibnamefont{Souma}},
  \bibinfo{author}{\bibfnamefont{T.}~\bibnamefont{Sato}},
  \bibinfo{author}{\bibfnamefont{T.}~\bibnamefont{Takahashi}},
  \bibinfo{author}{\bibfnamefont{M.}~\bibnamefont{Neupane}},
  \bibinfo{author}{\bibfnamefont{Y.-M.} \bibnamefont{Xu}},
  \bibinfo{author}{\bibfnamefont{Z.-H.} \bibnamefont{Pan}},
  \bibinfo{author}{\bibfnamefont{A.~V.} \bibnamefont{Fedorov}},
  \bibnamefont{et~al.}, \bibinfo{journal}{Journal of Physics: Condensed Matter}
  \textbf{\bibinfo{volume}{23}}, \bibinfo{pages}{135701}
  (\bibinfo{year}{2011}).

\bibitem[{\citenamefont{de~Jong et~al.}(2009)\citenamefont{de~Jong, Huang,
  Huisman, Massee, Thirupathaiah, Gorgoi, Schaefers, Follath, Goedkoop, and
  Golden}}]{DeJong2009}
\bibinfo{author}{\bibfnamefont{S.}~\bibnamefont{de~Jong}},
  \bibinfo{author}{\bibfnamefont{Y.}~\bibnamefont{Huang}},
  \bibinfo{author}{\bibfnamefont{R.}~\bibnamefont{Huisman}},
  \bibinfo{author}{\bibfnamefont{F.}~\bibnamefont{Massee}},
  \bibinfo{author}{\bibfnamefont{S.}~\bibnamefont{Thirupathaiah}},
  \bibinfo{author}{\bibfnamefont{M.}~\bibnamefont{Gorgoi}},
  \bibinfo{author}{\bibfnamefont{F.}~\bibnamefont{Schaefers}},
  \bibinfo{author}{\bibfnamefont{R.}~\bibnamefont{Follath}},
  \bibinfo{author}{\bibfnamefont{J.~B.} \bibnamefont{Goedkoop}},
  \bibnamefont{and} \bibinfo{author}{\bibfnamefont{M.~S.}
  \bibnamefont{Golden}}, \bibinfo{journal}{Phys. Rev. B}
  \textbf{\bibinfo{volume}{79}}, \bibinfo{pages}{115125}
  (\bibinfo{year}{2009}).

\bibitem[{\citenamefont{Ginder et~al.}(1988)\citenamefont{Ginder, Roe, Song,
  McCall, Gaines, Ehrenfreund, and Epstein}}]{Ginder1988}
\bibinfo{author}{\bibfnamefont{J.~M.} \bibnamefont{Ginder}},
  \bibinfo{author}{\bibfnamefont{M.~G.} \bibnamefont{Roe}},
  \bibinfo{author}{\bibfnamefont{Y.}~\bibnamefont{Song}},
  \bibinfo{author}{\bibfnamefont{R.~P.} \bibnamefont{McCall}},
  \bibinfo{author}{\bibfnamefont{J.~R.} \bibnamefont{Gaines}},
  \bibinfo{author}{\bibfnamefont{E.}~\bibnamefont{Ehrenfreund}},
  \bibnamefont{and} \bibinfo{author}{\bibfnamefont{A.~J.}
  \bibnamefont{Epstein}}, \bibinfo{journal}{Phys. Rev. B}
  \textbf{\bibinfo{volume}{37}}, \bibinfo{pages}{7506} (\bibinfo{year}{1988}).

\bibitem[{\citenamefont{Hansmann et~al.}(2014)\citenamefont{Hansmann, Parragh,
  Toschi, Sangiovanni, and Held}}]{Hansmann2014}
\bibinfo{author}{\bibfnamefont{P.}~\bibnamefont{Hansmann}},
  \bibinfo{author}{\bibfnamefont{N.}~\bibnamefont{Parragh}},
  \bibinfo{author}{\bibfnamefont{A.}~\bibnamefont{Toschi}},
  \bibinfo{author}{\bibfnamefont{G.}~\bibnamefont{Sangiovanni}},
  \bibnamefont{and} \bibinfo{author}{\bibfnamefont{K.}~\bibnamefont{Held}},
  \bibinfo{journal}{New Journal of Physics} \textbf{\bibinfo{volume}{16}},
  \bibinfo{pages}{033009} (\bibinfo{year}{2014}).

\bibitem[{\citenamefont{de' Medici et~al.}(2009)\citenamefont{de' Medici, Wang,
  Capone, and Millis}}]{Medici2009}
\bibinfo{author}{\bibfnamefont{L.}~\bibnamefont{de' Medici}},
  \bibinfo{author}{\bibfnamefont{X.}~\bibnamefont{Wang}},
  \bibinfo{author}{\bibfnamefont{M.}~\bibnamefont{Capone}}, \bibnamefont{and}
  \bibinfo{author}{\bibfnamefont{A.~J.} \bibnamefont{Millis}},
  \bibinfo{journal}{Phys. Rev. B} \textbf{\bibinfo{volume}{80}},
  \bibinfo{pages}{054501} (\bibinfo{year}{2009}).

\bibitem[{\citenamefont{Shen et~al.}(1987)\citenamefont{Shen, Allen, Yeh, Kang,
  Ellis, Spicer, Lindau, Maple, Dalichaouch, Torikachvili et~al.}}]{Shen1987}
\bibinfo{author}{\bibfnamefont{Z.-X.} \bibnamefont{Shen}},
  \bibinfo{author}{\bibfnamefont{J.~W.} \bibnamefont{Allen}},
  \bibinfo{author}{\bibfnamefont{J.~J.} \bibnamefont{Yeh}},
  \bibinfo{author}{\bibfnamefont{J.~S.} \bibnamefont{Kang}},
  \bibinfo{author}{\bibfnamefont{W.}~\bibnamefont{Ellis}},
  \bibinfo{author}{\bibfnamefont{W.}~\bibnamefont{Spicer}},
  \bibinfo{author}{\bibfnamefont{I.}~\bibnamefont{Lindau}},
  \bibinfo{author}{\bibfnamefont{M.~B.} \bibnamefont{Maple}},
  \bibinfo{author}{\bibfnamefont{Y.~D.} \bibnamefont{Dalichaouch}},
  \bibinfo{author}{\bibfnamefont{M.~S.} \bibnamefont{Torikachvili}},
  \bibnamefont{et~al.}, \bibinfo{journal}{Phys. Rev. B}
  \textbf{\bibinfo{volume}{36}}, \bibinfo{pages}{8414} (\bibinfo{year}{1987}).

\bibitem[{\citenamefont{Sefat et~al.}(2009)\citenamefont{Sefat, Singh, Jin,
  McGuire, Sales, and Mandrus}}]{Sefat2009}
\bibinfo{author}{\bibfnamefont{A.~S.} \bibnamefont{Sefat}},
  \bibinfo{author}{\bibfnamefont{D.~J.} \bibnamefont{Singh}},
  \bibinfo{author}{\bibfnamefont{R.}~\bibnamefont{Jin}},
  \bibinfo{author}{\bibfnamefont{M.~A.} \bibnamefont{McGuire}},
  \bibinfo{author}{\bibfnamefont{B.~C.} \bibnamefont{Sales}}, \bibnamefont{and}
  \bibinfo{author}{\bibfnamefont{D.}~\bibnamefont{Mandrus}},
  \bibinfo{journal}{Phys. Rev. B} \textbf{\bibinfo{volume}{79}},
  \bibinfo{pages}{024512} (\bibinfo{year}{2009}).

\bibitem[{\citenamefont{Xu et~al.}(2013)\citenamefont{Xu, Richard, van
  Roekeghem, Zhang, Miao, Zhang, Qian, Ferrero, Sefat, Biermann
  et~al.}}]{Xu2013}
\bibinfo{author}{\bibfnamefont{N.}~\bibnamefont{Xu}},
  \bibinfo{author}{\bibfnamefont{P.}~\bibnamefont{Richard}},
  \bibinfo{author}{\bibfnamefont{A.}~\bibnamefont{van Roekeghem}},
  \bibinfo{author}{\bibfnamefont{P.}~\bibnamefont{Zhang}},
  \bibinfo{author}{\bibfnamefont{H.}~\bibnamefont{Miao}},
  \bibinfo{author}{\bibfnamefont{W.-L.} \bibnamefont{Zhang}},
  \bibinfo{author}{\bibfnamefont{T.}~\bibnamefont{Qian}},
  \bibinfo{author}{\bibfnamefont{M.}~\bibnamefont{Ferrero}},
  \bibinfo{author}{\bibfnamefont{A.~S.} \bibnamefont{Sefat}},
  \bibinfo{author}{\bibfnamefont{S.}~\bibnamefont{Biermann}},
  \bibnamefont{et~al.}, \bibinfo{journal}{Phys. Rev. X}
  \textbf{\bibinfo{volume}{3}}, \bibinfo{pages}{011006} (\bibinfo{year}{2013}).

\bibitem[{\citenamefont{{van Roekeghem, Ambroise} and {Biermann,
  Silke}}(2014)}]{vanRoekeghem2014}
\bibinfo{author}{\bibnamefont{{van Roekeghem, Ambroise}}} \bibnamefont{and}
  \bibinfo{author}{\bibnamefont{{Biermann, Silke}}}, \bibinfo{journal}{EPL}
  \textbf{\bibinfo{volume}{108}}, \bibinfo{pages}{57003}
  (\bibinfo{year}{2014}).

\bibitem[{\citenamefont{Sekiyama et~al.}(2004)\citenamefont{Sekiyama, Fujiwara,
  Imada, Suga, Eisaki, Uchida, Takegahara, Harima, Saitoh, Nekrasov
  et~al.}}]{PhysRevLett.93.156402}
\bibinfo{author}{\bibfnamefont{A.}~\bibnamefont{Sekiyama}},
  \bibinfo{author}{\bibfnamefont{H.}~\bibnamefont{Fujiwara}},
  \bibinfo{author}{\bibfnamefont{S.}~\bibnamefont{Imada}},
  \bibinfo{author}{\bibfnamefont{S.}~\bibnamefont{Suga}},
  \bibinfo{author}{\bibfnamefont{H.}~\bibnamefont{Eisaki}},
  \bibinfo{author}{\bibfnamefont{S.~I.} \bibnamefont{Uchida}},
  \bibinfo{author}{\bibfnamefont{K.}~\bibnamefont{Takegahara}},
  \bibinfo{author}{\bibfnamefont{H.}~\bibnamefont{Harima}},
  \bibinfo{author}{\bibfnamefont{Y.}~\bibnamefont{Saitoh}},
  \bibinfo{author}{\bibfnamefont{I.~A.} \bibnamefont{Nekrasov}},
  \bibnamefont{et~al.}, \bibinfo{journal}{Phys. Rev. Lett.}
  \textbf{\bibinfo{volume}{93}}, \bibinfo{pages}{156402}
  (\bibinfo{year}{2004}).

\bibitem[{\citenamefont{Morikawa et~al.}(1995)\citenamefont{Morikawa, Mizokawa,
  Kobayashi, Fujimori, Eisaki, Uchida, Iga, and Nishihara}}]{PhysRevB.52.13711}
\bibinfo{author}{\bibfnamefont{K.}~\bibnamefont{Morikawa}},
  \bibinfo{author}{\bibfnamefont{T.}~\bibnamefont{Mizokawa}},
  \bibinfo{author}{\bibfnamefont{K.}~\bibnamefont{Kobayashi}},
  \bibinfo{author}{\bibfnamefont{A.}~\bibnamefont{Fujimori}},
  \bibinfo{author}{\bibfnamefont{H.}~\bibnamefont{Eisaki}},
  \bibinfo{author}{\bibfnamefont{S.}~\bibnamefont{Uchida}},
  \bibinfo{author}{\bibfnamefont{F.}~\bibnamefont{Iga}}, \bibnamefont{and}
  \bibinfo{author}{\bibfnamefont{Y.}~\bibnamefont{Nishihara}},
  \bibinfo{journal}{Phys. Rev. B} \textbf{\bibinfo{volume}{52}},
  \bibinfo{pages}{13711} (\bibinfo{year}{1995}).

\bibitem[{\citenamefont{Parcollet et~al.}(2004)\citenamefont{Parcollet, Biroli,
  and Kotliar}}]{Parcollet2004}
\bibinfo{author}{\bibfnamefont{O.}~\bibnamefont{Parcollet}},
  \bibinfo{author}{\bibfnamefont{G.}~\bibnamefont{Biroli}}, \bibnamefont{and}
  \bibinfo{author}{\bibfnamefont{G.}~\bibnamefont{Kotliar}},
  \bibinfo{journal}{Phys. Rev. Lett.} \textbf{\bibinfo{volume}{92}},
  \bibinfo{pages}{226402} (\bibinfo{year}{2004}).

\bibitem[{\citenamefont{Otsuki}(2013)}]{Otsuki2013}
\bibinfo{author}{\bibfnamefont{J.}~\bibnamefont{Otsuki}},
  \bibinfo{journal}{Phys. Rev. B} \textbf{\bibinfo{volume}{87}},
  \bibinfo{pages}{125102} (\bibinfo{year}{2013}).

\bibitem[{\citenamefont{Ayral and Parcollet}(2015)}]{Ayral2015}
\bibinfo{author}{\bibfnamefont{T.}~\bibnamefont{Ayral}} \bibnamefont{and}
  \bibinfo{author}{\bibfnamefont{O.}~\bibnamefont{Parcollet}},
  \bibinfo{journal}{Phys. Rev. B} \textbf{\bibinfo{volume}{92}},
  \bibinfo{pages}{115109} (\bibinfo{year}{2015}).

\end{thebibliography}

\end{document}